\documentclass[traditabstract]{aa}  
\usepackage{graphicx}
\usepackage{txfonts, color}
\usepackage{float}

\newcommand{\less}{\raisebox{-1.1mm}{$\stackrel{<}{\sim}$}} 
 
\newcommand{\msol}{\mbox{$M_{\odot}$}}

\newcommand{\lsol}{\mbox{$L_{\odot}$}}

\begin{document}

\title{Luminosities and infrared excess in Type II and anomalous Cepheids in the Large and Small Magellanic Clouds\thanks{
Table~A1 is only available in electronic form at the CDS via anonymous ftp to 
cdsarc.u-strasbg.fr (130.79.128.5) or via http://cdsweb.u-strasbg.fr/cgi-bin/qcat?J/A+A/
}}


   \author{M.~A.~T. Groenewegen\inst{1}
          \and
          M.~I. Jurkovic\inst{2,3}
          }

   \institute{
              Koninklijke Sterrenwacht van Belgi\"e (KSB), 
              Ringlaan 3, B-1180 Brussels, Belgium \\
              \email{martin.groenewegen@oma.be}
         \and
              Astronomical Observatory of Belgrade (AOB),
              Volgina 7, 11 060 Belgrade, Serbia \\
              \email{mojur@aob.rs}
         \and
                         Konkoly Observatory, Research Centre for Astronomy and Earth Sciences, 
                         Hungarian Academy of Sciences,
                         H-1121 Budapest, Konkoly Thege Mikl\'{o}s \'{u}t 15-17., Hungary\\            
             }

   \date{Received ...; accepted ...}

\abstract
{Type II and anomalous Cepheids (ACs) are useful distance indicators when there are too few classical Cepheids or when RR Lyrae stars are too faint. 
Type II and ACs follow a period-luminosity relation as well, but they are less well-studied classes of objects.
In this paper we study the sample of 335 Type II and ACs in the Small and Large Magellanic Clouds detected in OGLE-III data.
The spectral energy distributions (SEDs) are constructed from photometric data available in the literature and fitted with a 
dust radiative transfer model, thereby leading to a determination of luminosity and effective temperature.
In addition, a subsample of targets is investigated for possible binarity by looking for the light-time travel effect (LITE).
Hertzsprung-Russell diagrams (HRD) are constructed and compared to evolutionary tracks and theoretical instability strips (ISs).
In agreement with previous suggestions, the BL Her subclass can be explained by the evolution of $\sim$0.5-0.6~\msol\ stars evolving 
off the zero-age horizontal branch and the ACs can be explained by the evolution of $\sim$1.1-2.3~\msol\ stars.
The evolution of the W Vir subclass is not clear. These objects are at higher luminosities than ACs and evolutionary 
tracks of $\sim$2.5-4~\msol\ stars cross this region in the HRD, but the periods of the W Vir are longer than those of
the short period classical Cepheids at these luminosities, which indicates the former have lower masses. 
A low-mass star experiencing a thermal pulse when the envelope mass is small can make a blue loop into 
the IS region of the W Vir stars. But the timescale is extremely short, so this is also no explanation for the W Vir as a class.
A relation to binarity might be at the origin of the W Vir stars, which has already been explicitly suggested for the peculiar W Vir stars.
For $\sim60\%$ of the RV Tau and $\sim 10\%$ of the W Vir objects an infrared excess is detected from the SED fitting.
A recent result is confirmed that stars exist with luminosities below that predicted from 
single-star evolution, which show a clear infrared excess, and the shape of the excess suggests a connection to binary evolution. 
The investigation of the LITE effect revealed 20 systems that appear to show periodic variations and may be new binaries, 
although this study requires follow-up. About 40 stars show significant period changes.
}


        \authorrunning{Groenewegen \& Jurkovic}
        \titlerunning{Luminosities and infrared excess in T2Cs and ACs in the LMC and SMC}
    \maketitle
%

\section{Introduction}
\label{Sect:Intro}

Type II Cepheids (T2Cs) are pulsating low-mass stars stars that are usually associated with the old Population II (hence the name).
They are typically separated into three subgroups according to their pulsation periods, 
but the exact definition of the dividing periods varies; see \citet{Welch_2012}. 
In the OGLE-{\sc III} samples that we use in this paper, the BL Herculis (BLH) 
stars have pulsation periods of $1-4$~days, the W Virginis (WVir) $4-20$~days, 
and the RV Tauris (RVT) $20-70$~days. 
This classification is based on the sample in the Large and Small Magellanic Clouds
(LMC and SMC), as described in \citet{Soszynski2008} and \citet{Soszynski2010}. 
So far, only fundamental mode (FU) pulsators have been discovered.

Evolutionary modelling of T2Cs has been pioneered by \citet{Gingold_1976} 
and \citet{Gingold_1985}, who established that T2Cs are low-mass stars 
"evolving from the blue horizontal branch (HB) through the instability strip (IS) 
to the asymptotic giant branch (AGB) for the short-period stars, blue loops off the 
AGB for the stars of intermediate period, and post-AGB (PAGB) evolution for
the longest period" as in \citet{Wallerstein_2002}. 
\citet{Bono_1997_T2CEP} came to the conclusion that these should be low-mass 
($0.52$ to $0.8~{\msol}$), low-metallicity ([Fe/H] = $-1.3$ to $-2.3$) objects. 
\citet{Wallerstein_2002} remarks that the evolution of $1.0~{\msol}$ star
at [Fe/H] = $-1.3$ by \citet{Vassiliadis_Wood_1993} closely describes 
the T2C parameters, in particular the transition from the AGB to PAGB phase, 
and the occurrence of RVT stars. 

Anomalous Cepheids (ACs) are also pulsating stars that overlap in period range 
with RR Lyrae (RRL) and BLH stars. Although ACs share the IS with RRL and BLH stars, these pulsating stars seem to be stars 
with higher masses that have evolved to the IS. 
Anomalous Cepheids pulsate in the fundamental mode (FU) and first overtone (FO) as well. 

Looking at the $PL$ relations of different pulsating stars in the LMC and SMC, they definitely form a separate relation 
distinguishing themselves from RRL, classical Cepheids, and T2Cs. \citet{Fiorentino_Monelli_2012} argue 
that if the ACs were single stars evolving in the Hertzsprung-Russell diagram (HRD), i.e. not resulting from binary interaction, 
they should be metal-poor stars that do not cross into the classical Cepheid IS, while \citet{Caputo_2004} 
investigated the possibility that they continue 
to the $PL$ relation of classical Cepheids, which was not substantiated significantly. 
 
Previously, \citet{Bono_1997_ANCEP} and \citet{Marconi_2004} have modelled these stars. 
\citet{Bono_1997_ANCEP} treat the evolutionary status of these stars as if they were evolving 
from the horizontal branch (HB), but they state that it is hard to
distinguish if they have evolved as a single star or as a result
of mass transfer in a binary system \citep{Renzini_1977}. 
According to \citet{Fiorentino_Monelli_2012}, their mean mass is around $1.2 \pm 0.2~{\msol}$. 
\cite{MartinezVazques_2016} derive an average mass of $1.5~{\msol}$ for four ACs in Sculptor.

Recently, \cite{Kamath2016} presented a newly discovered class of low-luminosity, dusty, 
evolved objects in the Magellanic Clouds (MCs). These objects have dust excesses, 
stellar parameters, and spectral energy distributions (SEDs) similar to those of dusty PAGB stars. 
However, they have lower luminosities and hence lower masses. 
\cite{Kamath2016} suggests that these objects have evolved off the red giant branch (RGB) instead 
of the AGB as a result of binary interaction.  Interestingly, many of their 
post-RGB/AGB candidates are known T2Cs. Their initial search was based on optically visible PAGB stars (or candidates) 
in the MCs \citep{Kamath2014, Kamath2015}, so we considered that it would be interesting to look 
specifically for infrared excess starting from a sample of T2Cs. 
The OGLE-{\sc III} catalogues (\citet{Soszynski2008}, \citet{Soszynski2010}, \citet{Soszynski2010_ANC}) 
provide us with a unique sample of 335 T2Cs and ACs in the SMC 
and LMC\footnote{After the analysis in this paper was completed we 
became aware of \citet{Soszynski_2015}, which increased the number of known ACs in the MCs 
based on OGLE-{\sc IV} data. 
We did not include these additional stars; see Sect.~\ref{Sect:sample} for details.}.
In the present paper we determine the luminosity and effective temperature of the T2Cs and ACs by fitting the 
SEDs and we identify stars with an infrared excess.
In addition we look for the light-time travel (LITE) effect in (O-C) diagrams to identify candidate binary systems.
In an accompanying paper we will investigate the period-luminosity and period-radius relation of T2Cs and ACs and we will estimate the masses of these objects.
In section~\ref{Sect:sample} we introduce the sample of T2Cs and ACs.
Section~\ref{Sect:SEDconstruct} presents the photometric data used to construct the SEDs and the model and procedure to fit it.
In section~\ref{Sect:Bin} we investigate the LITE effect in selected systems.
Section~\ref{Sect:Results} discusses the results, and our conclusions are presented in 
Section~\ref{Sect:Conclusion}.
\section{The sample}
\label{Sect:sample}

We used the OGLE-{\sc III} sample of T2Cs and ACs introduced 
by \citet{Soszynski2008} and \citet{Soszynski2010}.
\citet{Soszynski2008} classify 197 T2Cs, which was updated in 2009 to 203 objects 
in the online catalogue\footnote{Available via ftp://ftp.astrouw.edu.pl/ogle/ogle3/OIII-CVS/} 
(64 BLH, 97 WVir, and 42 RVT stars), and 83 ACs (62 FU and 21 FO mode) in the LMC.
\citet{Soszynski2010} classify 43 T2Cs (17 BLH, 17 WVir, and 9 RVT stars), 
while \citet{Soszynski2010_ANC} identify 6 (candidate) ACs (3 FU and 3 FO mode) in the 
SMC\footnote{\citet{Soszynski2010_ANC} in their Table~1 originally listed these stars with a classical 
cepheid identification number. In the OGLE-III Variable Stars Database (http://ogledb.astrouw.edu.pl/$\sim$ogle/CVS/ they 
were subsequently listed under the names that we use in the present paper, OGLE-SMC-ACEP 01...06.}. 
Some WVir objects have been labelled as "peculiar W Virginis" (pWVir) objects, which in most cases is an indication 
that the pulsating star is in a binary system.
This is the sample of 335 stars considered in the paper.
The periods are taken from the OGLE catalogues. Specifically for the RVT, which show light curves (LCs)
with alternating deep and shallow minima, this implies that the periods refer to the time between 
consecutive minima (thought to be the true pulsation period), 
not the time between deep (or shallow) minima\footnote{see http://www.sai.msu.su/gcvs/gcvs/iii/vartype.txt}.
After this work was completed \citet{Soszynski_2015} provided a new
catalogue of 250 ACs, 141 in the LMC and 109 in the SMC, based on OGLE-{\sc IV} data. 
This larger sample will be considered in a future paper where the
analysis presented in this paper will be expanded to classical
Cepheids and RR Lyrae stars. The increase of the number of ACs in the SMC is
considerable, but all 6 SMC ACs considered here remain in the new
catalogue (with the same FU/FO classification).
There are some implications for the present paper however.
Two of the LMC ACs have been reclassified in the new catalogue: OGLE-LMC-ACEP-022 and -083 
are now considered RRL, while OGLE-LMC-T2CEP-114 (formerly a BLH) is considered an AC now.
For completeness, the stellar parameters that have been determined and the fits to the SEDs 
are presented in the Appendices, but the two reclassified RRL are not discussed in the 
main text and figures, and are excluded in fitting any relations.
The numbering scheme remains the same between the OGLE-{\sc III} and OGLE-{\sc IV} catalogues for stars 1-83.
OGLE-LMC-T2CEP-114 is kept under its OGLE-{\sc III} name in the present paper but in the 
figures is plotted as a FU mode AC (OGLE-{\sc IV} name: OGLE-LMC-ACEP-114).
For the SMC the numbering scheme did change with respect to that introduced in the 
online OGLE-III Variable Stars Database, although all 6 ACs are in fact 
confirmed with the original FU/FO classification.
We keep the OGLE-{\sc III} numbering in the tables and figures in the present paper.
OGLE-SMC-ACEP 01...06 are numbered 32, 41, 57, 62, 68, and 81 in \citet{Soszynski_2015}.
\section{Constructing and fitting the spectral energy distributions}
\label{Sect:SEDconstruct}

\begin{figure}
\centering
\includegraphics[width=0.95\hsize]{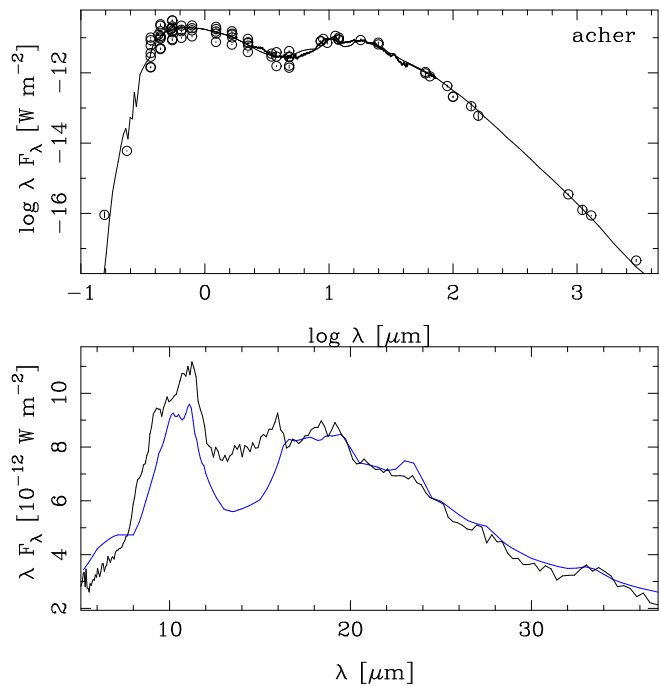}
\caption{
Fit to the SED and ISO SWS spectrum \citep{Sloan2003} of AC Her, with a two-component wind, 
illustrating the dust mix that has been used in fitting the stars, which show a clear MIR excess.
In the bottom panel the model is scaled to the observed flux in the 20.5-22.5 $\mu$m region to facilitate the comparison of the dust features.
}
\label{figACHER}
\end{figure}

The SEDs were constructed using photometry retrieved mostly, but not exclusively, 
via the VizieR web interface\footnote{http://vizier.u-strasbg.fr/viz-bin/VizieR}.
In the optical, the mean-magnitudes from OGLE, EROS, and MACHO (when available) are given the highest weight 
(by assigning a photometric error of 0.01 mag). Additional optical photometry comes from, for example 
the Magellanic Cloud Photometric Survey (MCPS) (\cite{Zaritsky_2002, Zaritsky_2004}), 
\cite{Massey_2002}, and \cite{Sebo_2002}.
In the near-infrared (NIR) and longer wavelengths generally no light-curve averaged mean magnitudes exist, but the 
photometric amplitudes decrease with increasing wavelength and so the effect of the variability on the derived luminosity is lower. 
There are a fair number of independent measurements available, especially in the NIR,  for example 
DENIS, 2MASS, 2MASS 6X, IRSF \citep{Kato_IRSF}, and the LMC Synoptic Survey \citep{Macri_2015}; 
we also considered NIR photometry from \cite{Ciechanowska_2010} and \cite{Ripepi_2015} 
and other publicly available data from the VMC survey \citep{Cioni_VMC}.
At longer wavelengths WISE data were available for the majority of objects \citep{Cutri_Allwise}.
Akari data were available from \cite{Ita_AkariSMC} and \cite{Kato_AkariLMC}.
We retrieved IRAC and sometimes MIPS (multi-epoch) data from the 
{\it NASA/IPAC} Infrared Science Archive\footnote{irsa.ipac.caltech.edu/}.
The smallest number of photometric points is 4 (OGLE-LMC-T2CEP-142) and there are seven other stars 
with 9 or fewer data points. On the other hand there are 10 stars with 40 or more data points. 
Ninety percent of the 335 Cepheids have between 11 and 36 data points.
The SEDs are fitted with More of DUSTY (MoD; \cite{Gr_MOD}), an extension of the DUSTY 
radiative transfer code DUSTY \citep{Ivezic_D}. For a given set of photometry, and/or spectra, visibility data, and intensity profiles, 
as input data the programme determines the best-fitting luminosity ($L$), dust optical depth ($\tau$, at 0.55 $\mu$m), 
dust temperature at the inner radius ($T_{\rm c}$), 
and slope of the density profile ($\rho \sim r^{-p}$). Any of these parameters can also be fixed.
The SEDs are fitted under the assumption of being representative of a single star.
The influence of any unresolved binary on the photometry depends on the luminosity ratio and difference in 
spectral type and hence the resulting effective temperature and luminosity.
Canonical distances of 50.0 and 61.0 kpc are adopted for the LMC and SMC, respectively, 
corresponding to distance moduli of 18.49 and 18.93, well within the error bars of the 
current best estimates \citep{deGrijs2014,deGrijs2015}.
We decided against fitting the $E(B-V)$ as an additional free parameter 
and adopted $E(B-V) = 0.15$ for all stars. Although this simplification ignores the spatial variation 
seen in the MCs (e.g. \cite{HaschkeLMC_2012,HaschkeSMC_2012}, \cite{Inno_2016}), the effect on the 
derived luminosity and effective 
temperature should be small in these stars with SEDs that peak in the NIR.  
The MARCS model atmospheres are used as input \citep{Gustafsson_MARCS} for most stars with 
overall metallicities of [Fe/H]= $-0.50$ and $-0.75$ dex 
for LMC and SMC stars, respectively. The model grid is available at 250~K intervals for the effective temperature 
range of interest and we used adjacent model atmospheres to interpolate models at 125~K intervals, 
which reflects better the accuracy in $T_{\rm eff}$ that can be achieved.
A few model atmospheres with temperatures that are hotter than the upper limit of 8000~K available in the MARCS model grid had to 
be considered and for those we used PHOENIX model atmospheres \citep{Hauschildt1999}. In the end, only one star has a 
best-fitting temperature above 8000~K.
Most stars have no dust and are best represented by a naked star. In those cases the dust optical depth is 
fixed to a very small number (and $T_{\rm c}$ and $p$ are also fixed to standard values of 1000~K and 2, respectively).
For every model atmosphere ($T_{\rm eff}$) a best-fitting luminosity (with its [internal] error) is derived 
with the corresponding $\chi^2$ of the fit. 
The model with the lowest $\chi^2$ then gives the best-fitting effective temperature. 
Considering models within a certain range above the minimum $\chi^2$ then gives the error in the 
effective temperature and luminosity. For the luminosity this error is added in quadrature to the internal error in $L$.
For some stars a better fit is achieved by adding a dust component. 
The Bayesian information criterion (BIC; see \cite{Schwarz1978}) is used to verify if the lower $\chi^2$ that 
is obviously obtained when adding additional parameters is, in fact, statistically significant.
Most of the objects that have an excess are RVT stars, some of which even have a {\it Spitzer} IRS spectrum 
available (see below). It is not our aim to investigate the dust content of RVT stars in the MCs in this paper.
As most RVT stars that have an IR excess show SEDs that point to a disc structure rather than an expanding 
outflow the use of a 1-D code is also limited. For the purpose of the present paper we included 
a realistic dust component to get a more realistic estimate of the luminosity.
The dust component that we used is based on a fit to the SED and dust spectrum of the well-known 
Galactic RVT star AC Her, which was recently discussed in \cite{Hillen_2015}. 
The fit obtained with MoD (using two shells) is shown in Figure~\ref{figACHER} and is remarkably good.
The dust is a combination of amorphous silicates, corundum, crystalline silicates, and metallic iron, 
similar to that in \cite{Hillen_2015}.
We used this species in all of the SED fits to the MC stars.
The derived luminosities, effective temperatures, and dust optical depth are listed in Table~\ref{table:params} 
and the fitted SEDs are shown in Figure~\ref{Fig_SEDs}.
The stars for which an IRS spectrum is available are explicitly discussed in Section~\ref{Sect:RVT}.

\section{A search for binarity}
\label{Sect:Bin}

Ten T2C or ACs in the present sample have been identified by the OGLE team to be in eclipsing binary systems. 
We searched for additional candidate binary systems by investigating the presence of the 
light-travel time effect or light-time effect (LITE) \citep{Irwin52} in 
so-called {\it observed minus calculated} (O-C) diagrams. Traditionally, the (O-C) diagram is constructed from 
timing measurements of, typically, maximum light and plotted versus time. Here, we considered and 
implemented the method outlined in \citet{Hajdu2015}, which uses a template LC constructed from the data to determine the (O-C) values.
This method is extremely well suited in the case of the long time series available in the OGLE database, 
where it is impractical to determine individual times of maximum, but where exquisite phased LCs can be constructed. 
Details are given in Appendix~\ref{AppLITE} but in essence the method is as follows:

\begin{itemize}

\item A Fourier series is fitted to part of the data to define the template LC. 
This also defines the reference epoch and pulsation period relative to which the (O-C) diagram is constructed.

\item Sections of, typically, 50, consecutive data points are taken and fitted to the 
template LC and from that the (O-C) value is determined. 
As an extension to the method proposed by \cite{Hajdu2015} we consider the fact that the observed LC might 
show amplitude modulations and therfore allow for a possible scaling of the Fourier amplitudes to determine the best fit.
\end{itemize}

The derived (O-C) diagram might diagnose several physical effects (e.g. see \cite{Sterken2005}) but the most relevant here 
are a change of the pulsation period with time (resulting in a parabolic shape) and/or the presence of a binary.
In the numerical code we fit a function of the form

\begin{equation}
(O - C) (t)= c_0 + c_1 t + c_2 t^2 + c_3 t^3 + ...\label{parabola}
\end{equation}
\begin{equation}
\hspace{30mm}               ... + (a \sin i) \, \frac{1-e^2}{1+e\cos(\nu)} \, \sin(\nu+\omega), \label{lteffect}
\end{equation}
\noindent where $a$ is the semi-major axis, $i$ is the inclination, $e$ is the eccentricity, and $\omega$ is the
argument of the periastron. The true anomaly $\nu$ is a function of the time $t$,
the orbital period $P_{\rm orb}$, the time of periastron passage $T_{\rm peri}$, and $e$.
All parameters of the model ($c_0, c_1, c_2, c_3$, $P_{\rm orb}$, $T_{\rm peri}$, $a \sin i$, $e$, $\omega$) can be fitted or fixed.
Since the procedure is time consuming\ as it requires manual supervision not all objects were investigated.
In total, we considered 133 systems. These were the systems classified by OGLE to show eclipsing or 
ellipsoidal variations, the pWVir stars that have been suggested to be in binary systems, and stars that showed 
unusual scatter in the rising branch of the published phased LC that hints at changes in period or binarity. 
In Section~\ref{Sect:CandBinarity} we discuss candidate binary systems in more detail. In Appendix~\ref{AppLITE} we 
additionally present systems that
appear to show significant changes in period, and list, for reference, the remainder of the sources for which 
the data are inconclusive or that do not show any significant variations in the (O-C) diagram.

In all cases it will be essential to reinvestigate the results when the OGLE-IV data become available, which will 
extend the time series, allowing us to improve on these findings.

\section{Results and discussion}
\label{Sect:Results}

\begin{figure}
\centering
\includegraphics[width=0.99\hsize]{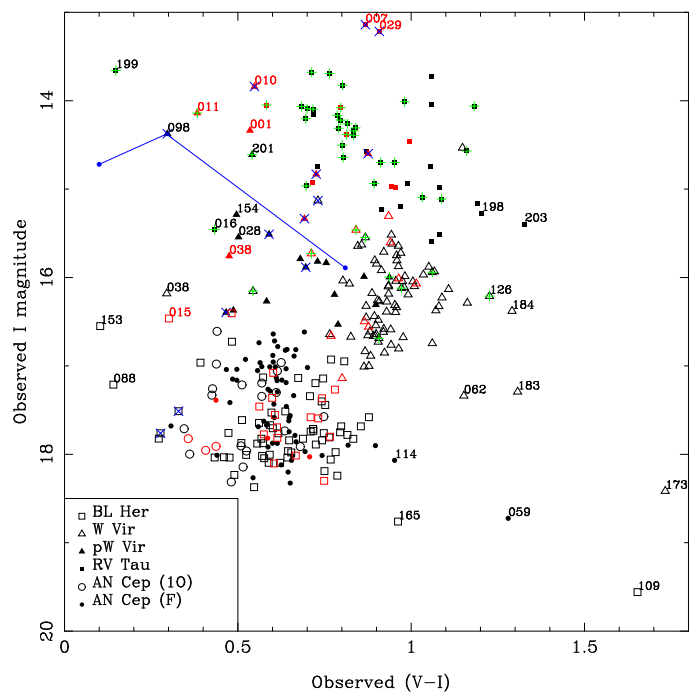}
\caption{
Plot of observed $I, (V-I)$ CMD.
Stars in the SMC are plotted in red and are shifted by $-0.432$ mag in $I$ to account 
for the difference in adopted distance (50 vs. 61 kpc).
For the pWVir object OGLE-LMC-T2CEP-098 (MACHO 6.6454.5), a known eclipsing system \citep{MACHO02}, 
the location of the two components is indicated by the dark blue lines. 
The pulsating star is the fainter, redder object (see text).
Stars with an IR excess are denoted by a green plus sign.
Stars that show eclipsing or ellipsoidal variations according to OGLE are denoted by a blue cross.
Some stars are labelled by their identifier.
}
\label{figHRDVI}
\end{figure}


\subsection{Hertzsprung-Russell diagram}
\label{Sect:HRD}

\begin{figure}
\centering
\includegraphics[width=0.99\hsize]{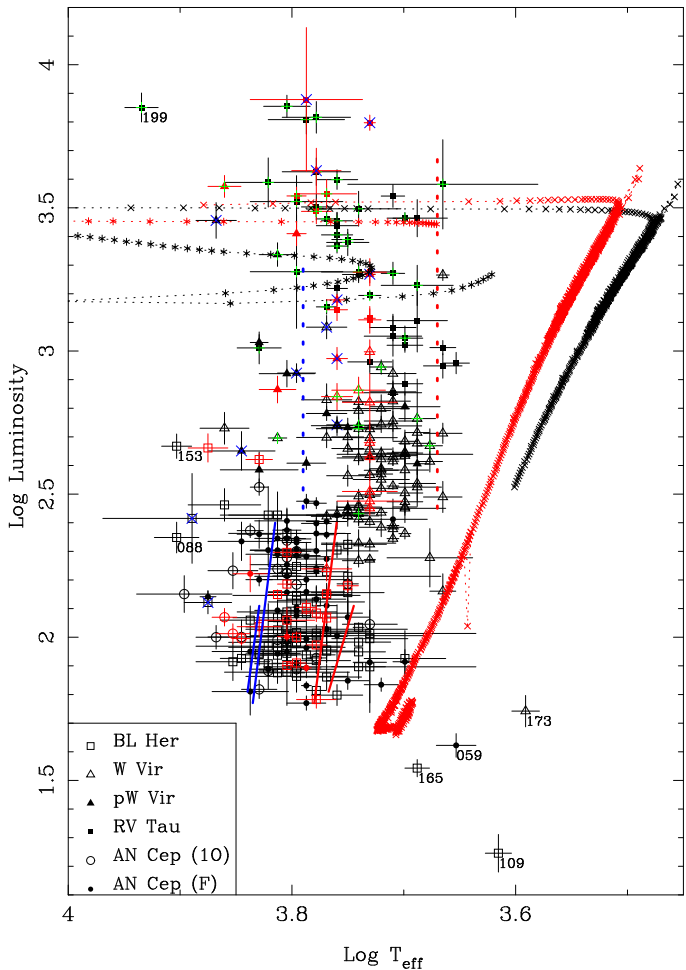}
\caption{
Plot of physical Hertzsprung-Russell diagram.
Stars in the SMC are plotted in red.
Stars with an IR excess are indicated by a green plus sign.
Stars that show eclipsing or ellipsoidal variations according to OGLE are  indicated by a blue cross.
The tracks plotted with $\times$ are the lowest initial mass tracks of \citet{Vassiliadis_Wood_1993}
for the LMC ($0.945~{\msol}$), and SMC ($0.89~{\msol}$, in red). 
The final masses are $0.555$ and $0.558~{\msol}$, respectively.
The tracks plotted with a $\ast$ are the lowest initial mass tracks of \citet{MB16} for 
a metallicity of $Z= 0.01$ (1.0~\msol, current mass 0.534~\msol), and $Z= 0.001$ (0.9~\msol, current mass 0.536~\msol, in red).
In both tracks every tick marked represents 500 years of evolution.
The blue and red edge of the fundamental mode IS of 
BLH (between $\log L \sim 1.81-2.1$, for a mass of 0.65~\msol) and 
FU ACs (between $\log L \sim 1.77-2.4$) are indicated by the solid line (see text). 
The vertical dotted lines indicate the location of most of the variables at higher luminosity (see text). 
Some stars are labelled by their identifier.
}
\label{figHRD}
\end{figure}

\begin{figure}
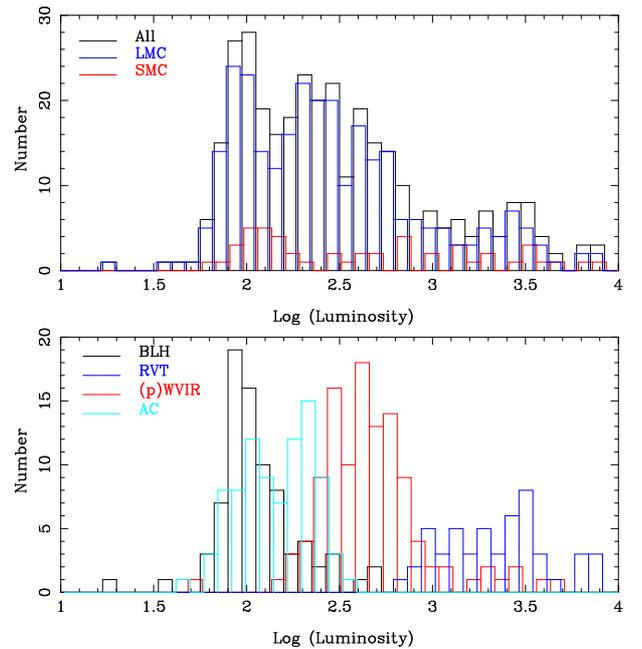

\centering
\includegraphics[width=0.89\hsize]{LumDistr_ALL.ps}

\includegraphics[width=0.89\hsize]{LumDistrType.ps}

\caption{
Histogram of the distribution over luminosity.
In the top panel the black, red, and blue histograms represent all the SMC, and LMC objects, respectively.
The latter two are slightly offset by $\pm$0.01 in $\log L$ for clarity.
The bottom panel shows the distributions for SMC and LMC objects according to type: 
BLH (black), ACs (light blue), (p)WVir (red), and RVT (dark blue).
The histograms are offset by $\pm$0.01 in $\log L$ for clarity.
}
\label{fig-LumDistAll}
\end{figure}


\begin{figure}
\centering
\includegraphics[width=0.95\hsize]{HRD_Lum_Teff_BLHer.ps}
\caption{
As Figure~\ref{figHRD} but only for the BLH objects.
The blue and red edge of the fundamental mode IS of BLH (between $\log L \sim 1.81-2.1$, for a mass of 0.65~\msol) 
are indicated by the solid line \citep{DiCriscienzo_2007}. The dashed lines indicate the location of the first overtone blue edge
and fundamental mode red edge of RR Lyrae (see text).
PARSEC horizontal branch models \citep{Bressan2012} are plotted in light blue ($Z= 0.001$), magenta  ($Z= 0.004$), and 
yellow ($Z= 0.008$) for, respectively, 0.515, 0.56, 0.62~\msol\ (decreasing in luminosity),  0.505, 0.53, 0.57~\msol, 
and 0.50, 0.52, 0.57~\msol\ (see text).
A point is plotted for every 1 Myr of evolution.
}
\label{figHRD-BLH}
\end{figure}


\begin{figure}
\centering
\includegraphics[width=0.95\hsize]{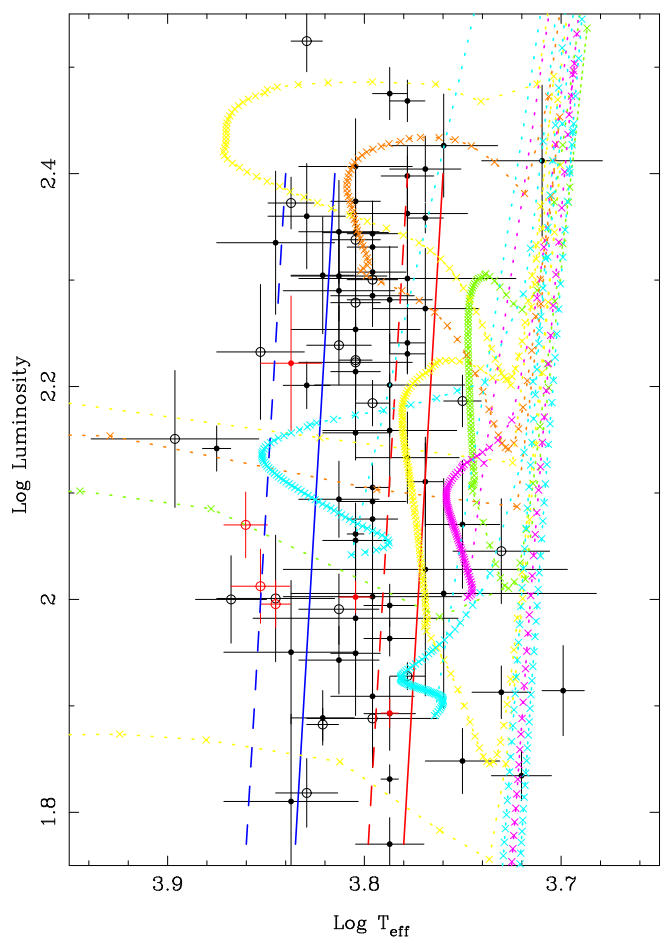}
\caption{
As Figure~\ref{figHRD} but only for the AC objects.
The fundamental mode (solid line) and first overtone (dashed line) blue and red edge 
of the IS of AC are indicated \citep{Fiorentino2006}.
Tracks from the BaSTI database \citep{Pietrinferni2004} are plotted in light blue ($Z= 0.0001$), magenta  ($Z= 0.0003$), yellow  ($Z= 0.0006$), 
brown ($Z= 0.001$), green ($Z= 0.002$) for stars of 1.1~\msol\ (1 model, bluest extension near $\log L\sim 1.9$),
1.5~\msol\ (2 models, bluest extension near $\log L\sim 2.1$), 1.9~\msol\ (1 model, bluest extension near $\log L\sim 2.2$), and 
2.3~\msol\ (3 models, bluest extension between $\log L= 2.25-2.45$).
A point is plotted for every 1 Myr of evolution.
}
\label{figHRD-AC}
\end{figure}


\begin{figure}
\centering
\includegraphics[width=0.95\hsize]{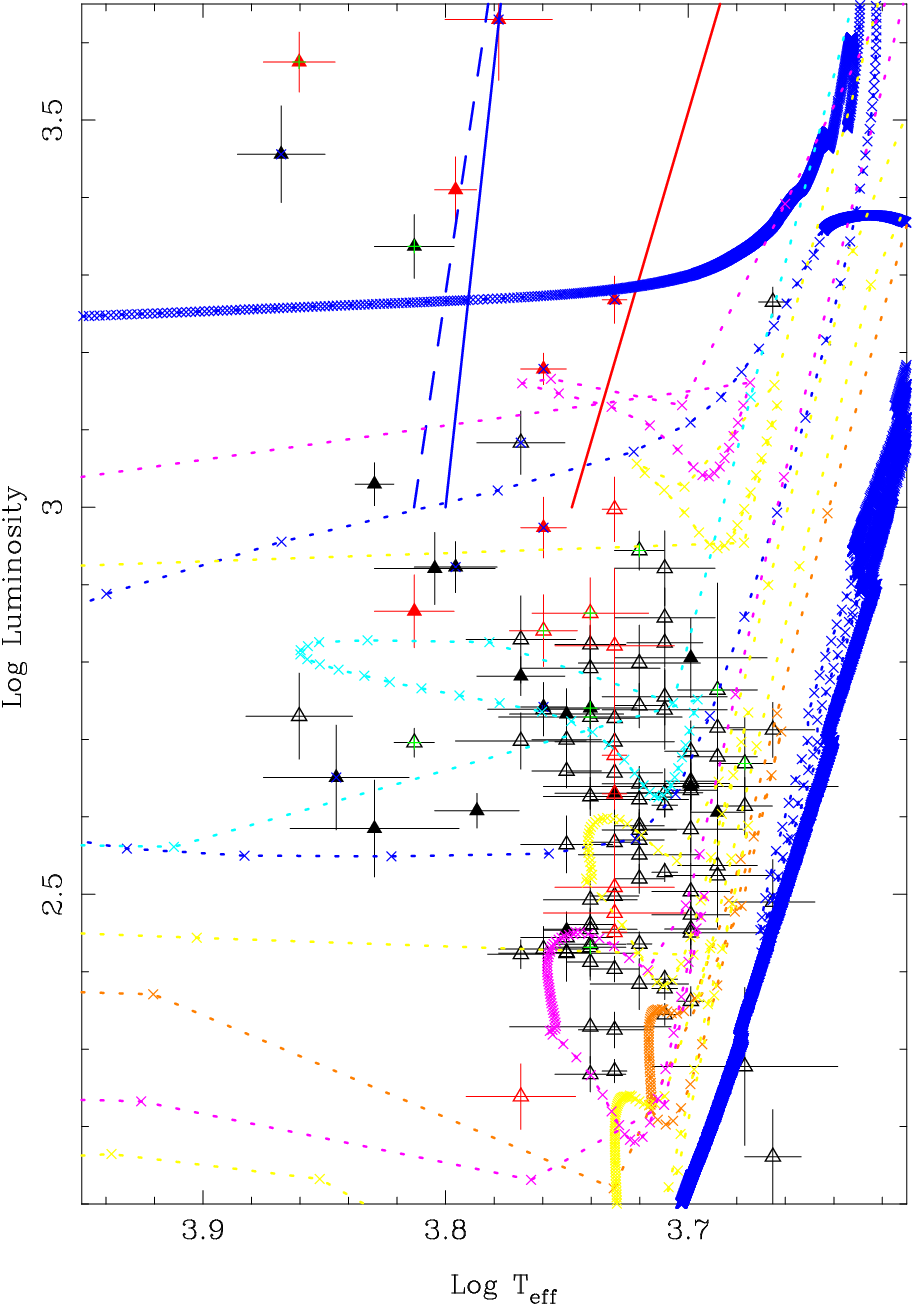}
\caption{
As Figure~\ref{figHRD} but only for the (p)WVir.
For reference, the fundamental mode (solid line) and first overtone (dashed line) blue (for $Z= 0.004$) and red edge (for $Z= 0.008$) 
of the IS of classical Cepheids are indicated \citet{Bono_2000}.
Tracks from the BaSTI database \citep{Pietrinferni2004} are plotted in light blue ($Z= 0.001$), magnenta  ($Z= 0.002$), yellow  ($Z= 0.004$), and 
brown ($Z= 0.008$) for stars with initial masses of 2.5~\msol\ (2 models, bluest extension between $\log L= 2.2-2.4$),
3.0~\msol\ (3 models, bluest extension between $\log L= 2.35-2.8$), and 4.0~\msol\ (2 models, bluest extension 
between $\log L= 3.0-3.2$).
A point is plotted for every 1 Myr of evolution.
The dark blue track represents the evolutionary track (Miller Bertolami, private communication) of a star 
with 0.60~\msol\ at the ZAHB that evolved through the AGB and experiences a TP, which results in the star crossing the IS. 
The evolution in that region is very fast. Each points represents 10 years of evolution.
}
\label{figHRD-WVIR}
\end{figure}


Figure~\ref{figHRDVI} shows the observed $I, (V-I)$ colour-magnitude diagram (CMD) essentially combining 
Figure~2 in \citet{Soszynski2008} and Figure~2 in \citet{Soszynski2010} into one. 
They did not comment on the apparent outliers at the time.

LMC-T2CEP-098 is an interesting case. It is classified as a pWVir and is an eclipsing binary (EB). 
It has been analysed by \cite{MACHO02} who derived the magnitudes of the components. 
These are also plotted in Figure~\ref{figHRDVI} and illustrate how binarity can influence the appearance on the CMD. 
The pulsating star is located close to the bulk of the WVir stars, while its bright non-pulsating 
component makes the system appear rather blue.

LMC-T2CEP-088 and -153 (near $I \sim 17, (V-I)= 0.1$) are indicated as "blended" by OGLE, which probably explains their blue colour.
The blue colour of LMC-T2CEP-199 (near $I \sim 14$) is probably real. It has the hottest effective temperature (8600~K) 
based on the fits to the SED of all stars and it has been classified as a B2III object (see below).
The four stars that are fainter than the other stars are not remarkable in any way (in terms of the SED or their LC), 
which would explain their position in the CMD. An anonymous referee for this paper suggested that three objects (T2CEP-165, -173, ACEP-059) are in
the direction of the Tarantula Nebula (30 Dor) and that high reddening could play a role. This is certainly possible as 
the distance from -165 to 30 Dor is about 20\arcmin\ and that of the other two objects is about 45\arcmin. 
However there are a few T2Cs and ACs located even closer than 20\arcmin\ and many objects within 45\arcmin\ that are not outliers.

Figure~\ref{figHRD} shows the physical Hertzsprung-Russell diagram (HRD). 
The same morphology as in the $I, (V-I)$ CMD is seen.
We treated stars as single stars in the SED fitting so that binaries would appear more luminous and hotter 
(assuming the companion is an unevolved object) than the parameters the pulsating component would have.
The evolutionary tracks from \cite{Vassiliadis_Wood_1993} of a 0.945~\msol\ (for LMC abundance) 
and 0.89~\msol\ (for SMC abundance, in red) initial mass star are plotted for reference as the crosses.
Recently, \cite{MB16} published new PAGB tracks. The lowest available initial mass tracks for a metallicity 
of $Z= 0.01$ (1.0~\msol, current mass 0.534~\msol) and $Z= 0.001$ (0.9~\msol, current mass 0.536~\msol, plotted in red) 
are plotted with a $\ast$.
For both sets every tick mark represents 500 years of stellar evolution.
The important thing to note is that stars with IR excess (denoted by the green plus sign) occur at luminosities below
those allowed for by single-star evolutionary tracks, as found by \citet{Kamath2016}.

The IS of the BLH and ACs are indicated in Fig.~\ref{figHRD} and are further explored in Figs.~\ref{figHRD-BLH} 
and \ref{figHRD-AC}, which show the HRD for these classes of objects in detail.
At higher luminosities, covering the WVir (including the pWVir objects), and the lower luminosity RVT stars, an IS is drawn by eye 
in Fig.~\ref{figHRD} that covers most variables. 
The red edge is drawn at $\log T_{\rm eff} = 3.67$ ($T_{\rm eff} = 4680$~K), 
the blue edge at $\log T_{\rm eff} = 3.79$ ($T_{\rm eff} = 6160$~K). The red edge seems to cover the variables for all luminosities 
in the present sample, for the blue edge a simple constant effective temperature seems to be a good approximation 
only up to $\log L \sim 3.3$. 

Figure~\ref{fig-LumDistAll} shows the projection of the HRD onto the luminosity axis. 
The luminosity distribution is qualitatively similar to that of T2C in globular clusters shown in \citet{Gingold_1985} with 
a minimum in the distribution around $\log L \sim 2.2$ (but at slightly larger luminosity in the SMC, although there are fewer stars).

Figure~\ref{figHRD-BLH} shows the blue and red edge of the FU IS of BLH \citep{DiCriscienzo_2007} 
(between $\log L \sim 1.81-2.1$) for a mass of 0.65~\msol.
The IS of the BLH depends little on mass; changing it to 0.55~\msol\ would make the IS bluer 
by a very small amount of $\Delta \log T_{\rm eff} = 0.009$. 
The BLH models were calculated for metallicities between $Z= 0.0001$ and $0.004$ and the IS does not appear 
to depend on metallicity within this range \citep{DiCriscienzo_2007}. The observations are in good agreement 
with the theoretical blue edge, but there are some stars that are cooler by up to 500~K than the theoretical red edge.
For comparison, the range in effective temperature covered by similar mass RR Lyrae stars is illustrated by the FU blue edge and 
the FU red edge based on \cite{Marconi_2015} for a metallicity of [Fe/H]= $-1.5$. 
The figure also shows selected horizontal branch models from the PARSEC tracks\footnote{http://people.sissa.it/~sbressan/parsec.html}
 \citep{Bressan2012} that encompass the observations.
The most luminous stars could be explained by the lowest masses (0.50-0.515~\msol, depending on metallicity), 
while the bulk of the stars would have masses in the range 0.52-0.56~\msol.

Figure~\ref{figHRD-AC} shows the FU (solid line) and FO (dashed line) blue and red edge of ACs \citep{Fiorentino2006}. 
The models, calculated for $Z= 0.0001$, show a remarkable agreement with the observations.
The overtone pulsators are preferentially located at hotter temperatures consistent with the IS.
With two exceptions the first overtone pulsators have temperatures between 5625 and 7375~K with a median of 6500~K and 
with two exceptions. The FU pulsators have temperatures between 5000 and 7000~K with a median of 6250~K.
Figure~\ref{figHRD-AC} also includes evolutionary tracks from the 
BaSTI library\footnote{http://albione.oa-teramo.inaf.it/} \citep{Pietrinferni2004} for masses and metallicities that 
cover the region in the HRD where the ACs are found. The lower luminosites would be explained by 
low-mass ($\sim 1.1~\msol$) low-metallicity ($Z= 0.0001$) stars, the upper luminosity range by stars of 
$\sim$ 2.3~\msol\ at 10 times that metallicity. The few ACs observed at the lowest luminosities ($\log L \sim 1.8~\lsol$) 
could be explained by the fast evolution of such stars crossing the IS on their way to the base of the RGB.

Figure~\ref{figHRD-WVIR} shows the WVir (including pWVir) objects, which are located at relatively high luminosity.
The pWVir objects are plotted as filled symbols and most of these are hotter than the bulk of the WVir objects.
For reference, the IS of classical Cepheids is shown \citep{Bono_2000}.
It shows that the (p)WVir are not simply the extension of classical Cepheids to lower masses, even though
evolutionary tracks can be found that cover the observed location in the HRD.

The figure shows BaSTI tracks \citep{Pietrinferni2004} for selected models with metallicities between $Z= 0.001$ and
$Z= 0.008$ (see caption for details) for stars of 2.5, 3.0, and 4.0~\msol.

However the WVir objects as a class cannot be counterparts of such stars.
The FU pulsation period of classical Cepheids of 5~\msol\ (the lowest mass considered in \cite{Bono_2000}) is of order 5 days, 
while the typical period of the bulk of the WVir with $\log L = 2.8 - 3$ and $\log T_{\rm eff} < 3.78$ is 15 days.
Even a rough application of the period-luminosity-mass relation of classical Cepheids by \cite{Bono_2000} 
indicates that the bulk of the (p)WVir objects should have masses of order 1~\msol\ (near $\log L = 3$) 
or less (at lower luminosities).
However, single stars of 1~\msol\ or lower do not cross the IS at these luminosities.

Another argument are the evolutionary timescales. Along the tracks a point is plotted for every million years. 
The 4~\msol\ star that loops in to the classical Cepheid IS spends a few million years there.
In the OGLE-IV catalogues of SMC and LMC \citep{Soszynski_2015Cep} about 4100 of the 5100 FU mode classical Cepheids 
have periods below five days, while there are only about 110 WVir in our sample, which, according to the 
tracks plotted in Fig.~\ref{figHRD-WVIR} spend many millions of years in the IS. 
This calculation is very rough, ignoring the details of the star formation rate, the initial mass function, 
and incompleteness, but the mismatch is evident. 

One scenario to explain the WVir stars is that these stars experience a thermal pulse on the AGB when the 
envelope mass is low enough for the star to make an excursion to lower luminosities and higher effective 
temperatures \citep{Gingold_1976, Gingold_1985}.
A modern view on this scenario is also presented in Fig.~\ref{figHRD-WVIR}. 
The evolutionary track is shown of a star of 0.60~\msol\ (and [Fe/H]$ = -1$) evolving off the  zero-age horizontal branch (ZAHB) 
(based on the \citet{MB16} models, private communication).
This example was picked to show that these excursions are possible. The TP occurs when the envelope mass is 0.0073~\msol.
The star ends as a 0.522~\msol\ white dwarf which, through the initial final mass relation \citep{Gesicki_ea_2014}, 
suggests an initial mass of \less 1.25~\msol.
The evolution is extremely fast however, the time spent in the IS region is only of order 100 years.
This shows that this scenario may apply to very rare individual cases of WVir stars, which would then show 
very large period changes, but cannot explain the WVir as a class.
 
In conclusion, the evolutionary status of these stars remains unclear. As single-star evolution appears not to be able
to explain the WVir, the binary hypothesis must be considered in more detail, as has been suggested specifically 
for the pWVir class.

%

\subsection{Comparison to the literature}
\label{Sect:CMP}

\begin{table*}[ht]
\setlength{\tabcolsep}{0.8mm}
\caption{Stellar parameters from the literature for the sample.}
\centering
\begin{tabular}{lccrllcclcc}
\hline
Name    & $T_{\rm eff}$ & $\log g$ & [Fe/H] &  $L$          & Classification & $M$        & Ref. & OGLE Name  & $T_{\rm eff}$ & $L$ \\ 
        &    (K)      &          &        &  ($L_\odot$)   &                 & ($M_\odot$) &      &            &    (K)      & ($L_\odot$)\\
\hline

 \multicolumn{11}{c}{SMC} \\

J005107.19-734133.3\tablefootmark{a} & 5767 & 0.72 & -1.56 & 3465  & P-RGB Disc &  0.43 & 1 & T2CEP-018 (RVT) & 5875 $\pm$ 375 & 3539 $\pm$  166  \\ 
\\
 \multicolumn{11}{c}{LMC} \\

MACHO 47.2496.8     & 4900 & 0.0 & -1.5 & 5000 &   -     &  -         & 4 & T2CEP-015 (RVT) & 5000 $\pm$  125 & 2910 $\pm$  \phantom{0}53 \\

J050304.95-684024.7 & 5586 & 0.5 & -2.3 & 3251           & P-RGB Disc &   0.44 & 2 & T2CEP-029 (RVT)  & 5750 $\pm$ 188 & 2851 $\pm$ \phantom{0}79  \\
J050738.94-682005.9 & 5420 & 1.5 & -1.0 & \phantom{0}859 & P-RGB Disc &   0.37 & 2 & T2CEP-046 (WVIR) & 5250 $\pm$ 125 & \phantom{0}879 $\pm$  \phantom{0}21 \\

J051418.09-691234.9 & 6112 & 0.5 &  -1.6  & 6667           & P-AGB Disc & 0.57 & 2 & T2CEP-067 (RVT) & 6125 $\pm$ 500 & 6429 $\pm$ 305 \\
  (idem)            & 5750 & 0.5 &  -2.0  & 5000 $\pm$ 500 & -          & -    & 3 & T2CEP-067 (RVT) & ...            & ....     \\

J051845.47-690321.8 & 5860 & 1.5 & -0.8                  & 4001  & YSO        &   -  & 2 & T2CEP-091 (RVT) & 6625 $\pm$ 625 & 3880 $\pm$ 319   \\
J052519.48-705410.0 & 8117 & 1.0 & -0.5\tablefootmark{b} & 3219  & P-AGB Disc & 0.58 & 2 & T2CEP-119 (RVT) & 6250 $\pm$ 625 & 3325 $\pm$ 290 \\

J053150.9-691146    & 6000 & 0.5 & -2.5 & 4000 $\pm$ 500 &   -    & -                & 3 & T2CEP-147 (RVT) & 6375 $\pm$ 312 & 7160 $\pm$ 259 \\
  (idem)            & 6000 & 0.5 & -2.5 & -              &   -    & -                & 5 & T2CEP-147 (RVT) & ...            & ....  \\ 

J053254.5-693513    & 6250 & 1.0 & -1.5 & 4200 $\pm$ 500 &   -    & -                & 3 & T2CEP-149 (RVT) & 5750 $\pm$ 250 & 2741 $\pm$ 117 \\ 
J054000.5-694214    & 5103 & 1.5 & -1.9 & 4200 $\pm$ 500 &   -    & -                & 3 & T2CEP-174 (RVT) & 6000 $\pm$ 438 & 6549 $\pm$ 342 \\ 

J054312.86-683357.1 & 5103 & 1.5 & -1.9 & 3085           & YSO        &   -   & 2 & T2CEP-180 (RVT) & 5500 $\pm$  500 & 3139 $\pm$ 182 \\  
J055122.52-695351.4 & 6237 & 1.5 & -2.5 & 3780           & P-AGB Disc & 0.56  & 2 & T2CEP-191 (RVT) & 5750 $\pm$  250 & 3969 $\pm$ 127 \\

\hline
\end{tabular}
\tablebib{
(1)~\citet{Kamath2014}; (2)~\citet{Kamath2015}; (3)~\citet{Gielen2009}; 
(4)~\citet{Reyniers2007}, also see \citet{Pollard2000}; (5)~\citet{RW2007}.
}
\tablefoot{
\tablefoottext{a}{This star was studied by \citet{Boyer_2011} who classified it (erroneously) as an extreme-AGB star.}
\tablefoottext{b}{[Fe/H] assumed.}
}
\label{tab:pagb_param}
\end{table*}

Some of the stars in our sample have been analysed previously using high-resolution spectroscopy and Table~\ref{tab:pagb_param} contains the stellar 
parameters derived in the literature.
Columns 1-5 list the name, effective temperature, gravity, and total luminosity, with an error bar when explicitly given, from the literature.
For the stars in common with \citet{Kamath2014,Kamath2015} the next two columns show their classification: 
two stars are classified as young stellar objects (YSO), three stars as post-RGB, and three as PAGB stars, and 
all six show disc-like SED (contrary to shell-like), and current mass. The last three columns give the 
OGLE name, and the effective temperature and luminosity derived in the present paper (from Table~\ref{table:params}).

In most cases the agreement is very good (within our 2$\sigma$ error bars), keeping in mind as well that 
the spectra were taken during one phase in the pulsation cycle while the stellar parameters derived 
here are based on all available photometry and so should represent the star at mean light.

The most discrepant values are the luminosity of LMC-T2CEP-015 and the effective temperature of LMC-T2CEP-119.
There are no obvious reasons for the discrepancies. The fits to both SEDs are good. Interestingly, both stars 
have an IR excess (see Sect.~\ref{Sect:RVT}), which may have influenced the analysis in Kamath et al.  in some way.
\citet{Reyniers2007} fitted the SED of LMC-T2CEP-015  but only had photometry up to the $K$-band available, 
not noticing the excess emission.  
From fitting the SED, \cite{vanAarle11} find $L= 3080 \pm 150$~\lsol, and $T_{\rm eff}= 4750 \pm 250$~K in good agreement with our 
determination and quote a spectral type G2-8(R)Ibe, which also points to a temperature around 5000~K. 
For LMC-T2CEP-119 \cite{vanAarle11} give a spectral type of G0Ib suggesting a temperature slightly below 6000~K, 
which is closer to our determination than the high value by Kamath et al.

%

\subsection{Far-infrared excess, and the RV Tau phenomenon}
\label{Sect:RVT}

For 10 stars {\it Spitzer Space telescope} IRS spectra \citep{Houck2004} are available, which are all in the LMC and are 
all classified as RV Tau stars. The spectra have been discussed in detail in \citet{Gielen2011} in the context of 
silicate features in Galactic and extragalactic PAGB stars.
Figure~\ref{Fig_SED_RVT} shows the fitted SEDs of the RVT stars with their IRS spectra.
The spectrum was not used in the SED fitting; the spectra are overplotted on top of the best model 
fit to the overall SED. In the bottom panel the model is scaled to the observed flux in the 20.5-22.5 $\mu$m region to facilitate the comparison of the dust features.

It is not the aim of this paper to discuss in detail the spectral features of the RVT stars 
in the MCs \citep[see][]{Gielen2011}. In most cases the spectral shapes are well fitted and very similar to the 
template of the Galactic RV Tau star AC Her. If there are larger differences then these are always in the sense 
that the silicate feature appears weaker w.r.t. the dust continuum (stars 067, 104, 174, 180).

\begin{figure*}
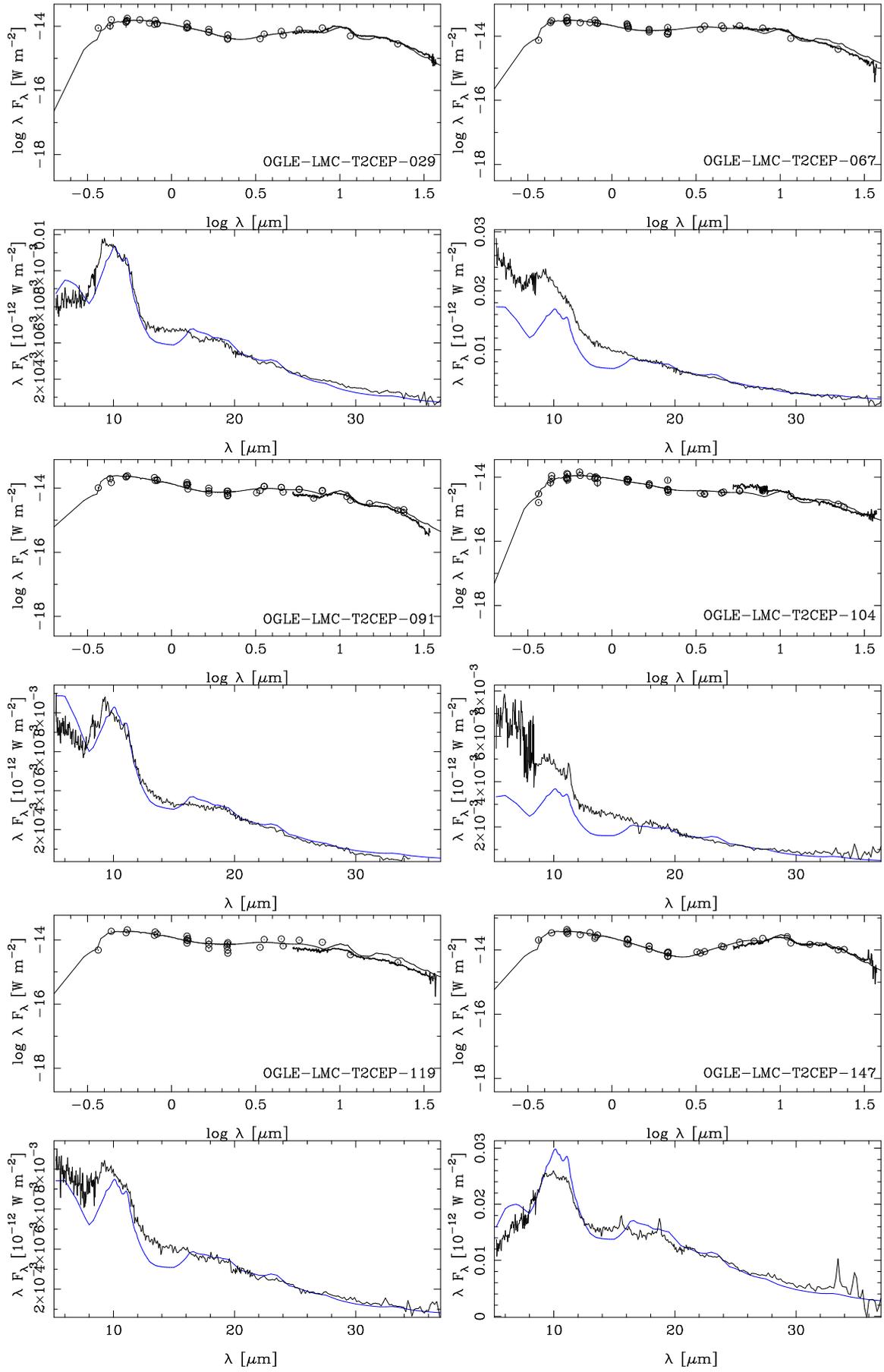

\centering

\begin{minipage}{0.41\textwidth}
\resizebox{\hsize}{!}{\includegraphics[angle=-0]{OGLE-LMC-T2CEP-029_sed_IRS.ps}} 
\end{minipage}
\begin{minipage}{0.41\textwidth}
\resizebox{\hsize}{!}{\includegraphics[angle=-0]{OGLE-LMC-T2CEP-067_sed_IRS.ps}} 
\end{minipage}

\begin{minipage}{0.41\textwidth}
\resizebox{\hsize}{!}{\includegraphics[angle=-0]{OGLE-LMC-T2CEP-091_sed_IRS.ps}} 
\end{minipage}
\begin{minipage}{0.41\textwidth}
\resizebox{\hsize}{!}{\includegraphics[angle=-0]{OGLE-LMC-T2CEP-104_sed_IRS.ps}} 
\end{minipage}

\begin{minipage}{0.41\textwidth}
\resizebox{\hsize}{!}{\includegraphics[angle=-0]{OGLE-LMC-T2CEP-119_sed_IRS.ps}} 
\end{minipage}
\begin{minipage}{0.41\textwidth}
\resizebox{\hsize}{!}{\includegraphics[angle=-0]{OGLE-LMC-T2CEP-147_sed_IRS.ps}} 
\end{minipage}

\caption{Fits to the SEDs of the RV Tau stars that have an IRS spectrum. 
In the bottom panel the model is scaled to the observed flux in the 20.5-22.5 $\mu$m region  
to facilitate the comparison of the dust features.
}
\label{Fig_SED_RVT}
\end{figure*}

\setcounter{figure}{7}
\begin{figure*}
\centering

\begin{minipage}{0.41\textwidth}
\resizebox{\hsize}{!}{\includegraphics[angle=-0]{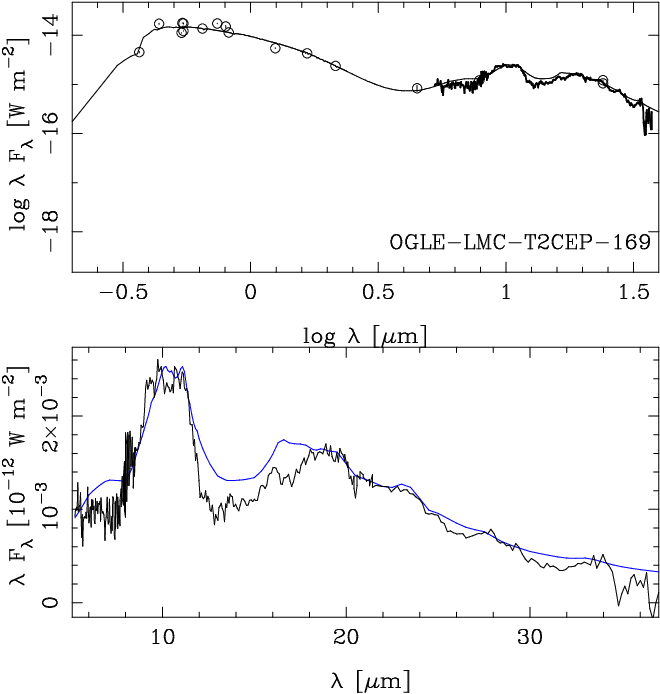}} 
\end{minipage}
\begin{minipage}{0.41\textwidth}
\resizebox{\hsize}{!}{\includegraphics[angle=-0]{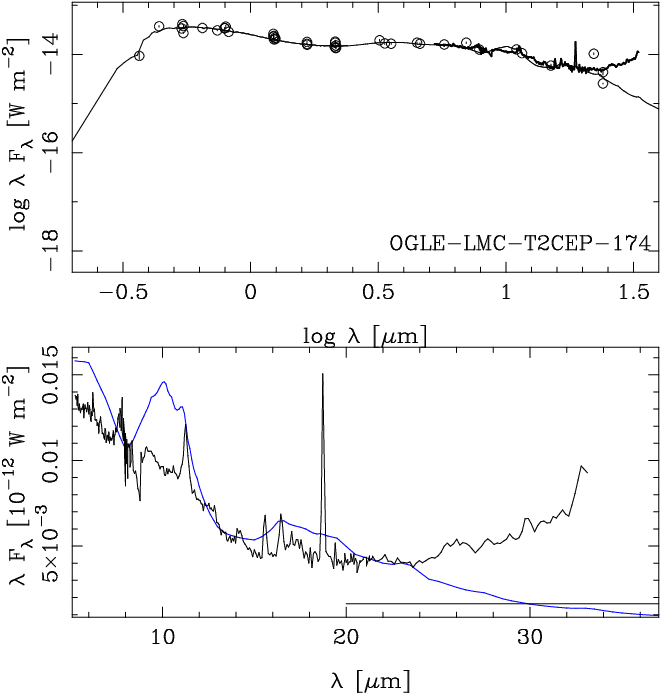}} 
\end{minipage}

\begin{minipage}{0.41\textwidth}
\resizebox{\hsize}{!}{\includegraphics[angle=-0]{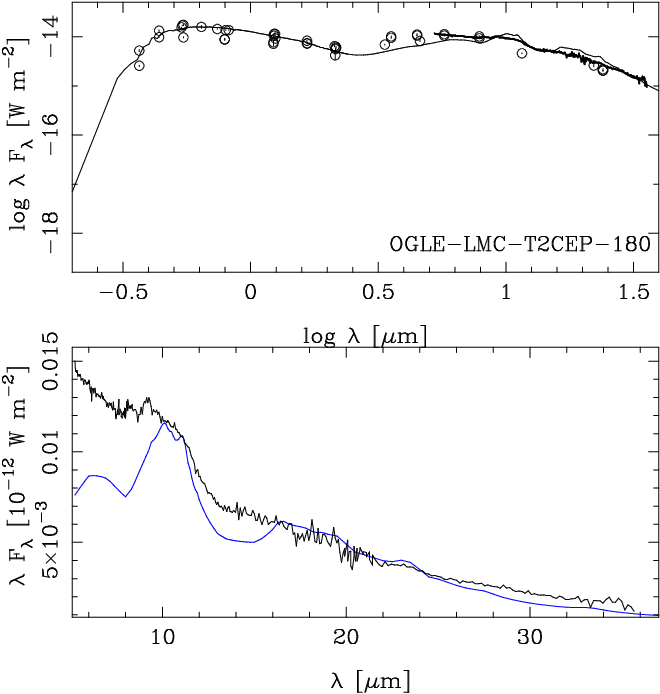}} 
\end{minipage}
\begin{minipage}{0.41\textwidth}
\resizebox{\hsize}{!}{\includegraphics[angle=-0]{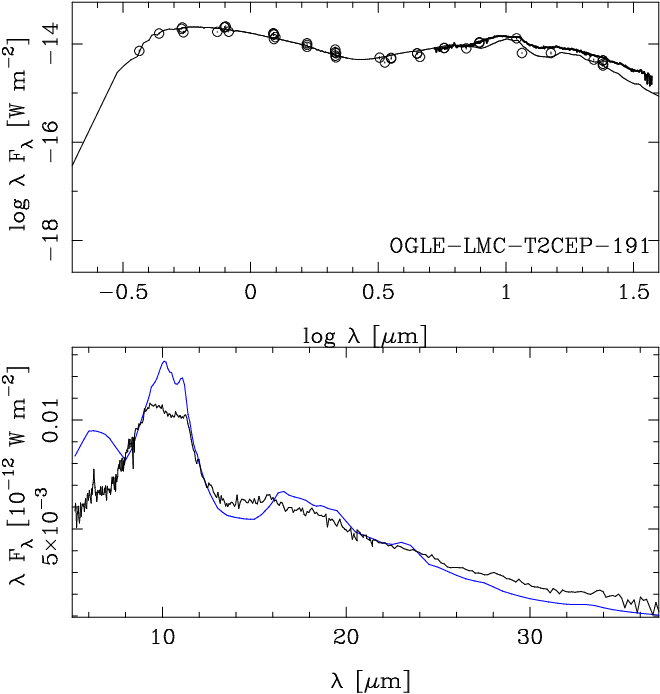}} 
\end{minipage}

\caption{Continued}
\end{figure*}

\begin{figure}
\centering
\includegraphics[width=0.95\hsize]{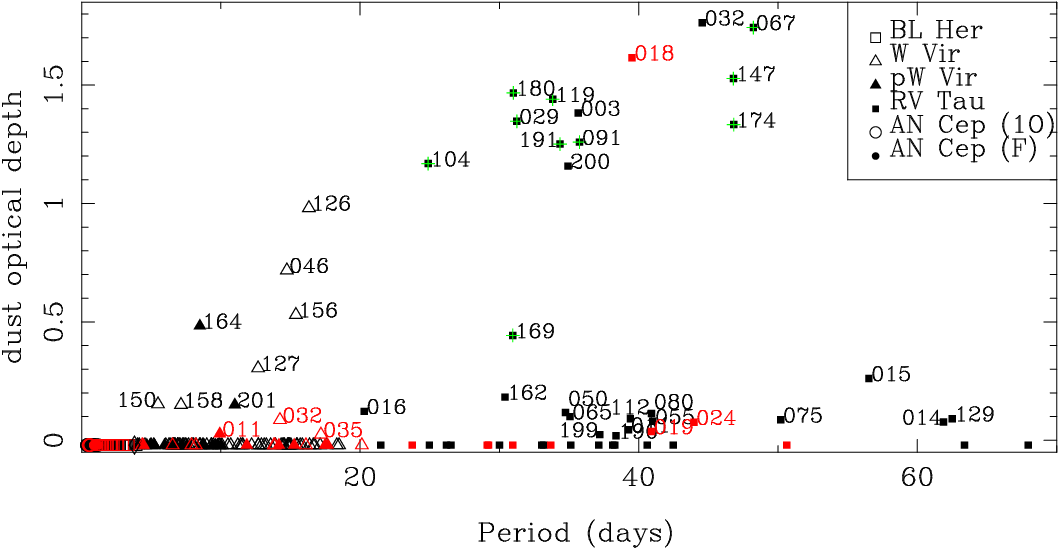}

\includegraphics[width=0.95\hsize]{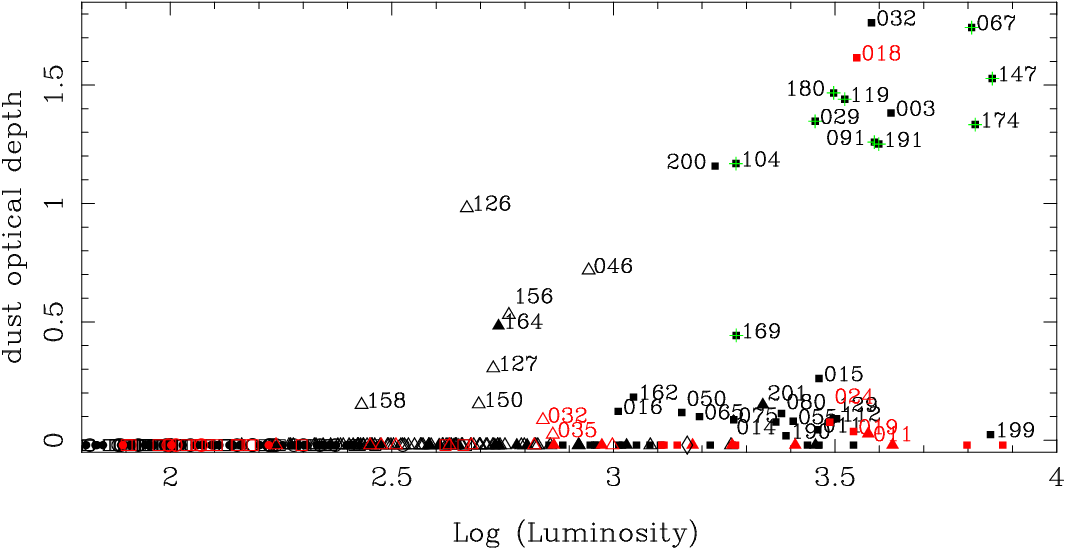}

\caption{Dependence of the derived dust optical depth on pulsation period and $\log L$. 
Stars in the SMC are plotted in red.
Stars with an IRS spectrum are indicated by a green plus sign.
Stars with a detectable IR excess are labelled by their identifier.
}
\label{fig:RVT_P_massloss}
\end{figure}

There are more stars that show (weak) excess IR emission than the 10 for which  IRS spectra are available.
The results are summarised in Fig.~\ref{fig:RVT_P_massloss}, where the derived dust optical depth is plotted 
against period and luminosity with more detailed information available in the Appendices.
In some cases it is difficult to exclude excess emission definitively. 
Table~\ref{table:params} includes 4 objects for which the WISE W4 filter is unreliable, 
but where the W3 filter lies above the model atmosphere.  These stars are marked ("W3 excess"). 
If the excess were real it would imply very low optical depths and these sources could be examples of where 
the dust shell has expanded away from the star, contrary to most SEDs of stars with excess emission that 
point to hot dust and a disc  structure.
Another four objects are marked ("IR excess?"), where there could possibly be weak IR excess; 
one of these is the RVT object OGLE-LMC-T2CEP-025, which has the longest period in the sample at 68 days.

Infrared excess is absent (or undetectable) in AC (0:76) and BL Her (0:81) objects 
and present in $\sim 10\%$ of the W Vir objects (3:24 pWVir, and 8:90 WVir).
In RVT the phenomenon is common with 30:52 objects showing detectable emission (plus 3 possible).
Of the 41 stars with detectable IR emission only 6 are located in the SMC. 

The SEDs of PAGB and RVT objects are typically divided between sources showing hot dust, 
which are interpreted as dust in a circumbinary system, and those showing cold dust, which are interpreted as dust in an expanding
shell; see for example the recent work by \citet{Gezer2015} (their Figure~2) and references therein.
This distinction is also seen in Fig.~\ref{fig:RVT_P_massloss}. All objects with a derived optical depth above unity 
show SEDs characteristic of disc sources. For optical depths \less~0.2 the SEDs are all consistent with an 
expanding shell model. For the half dozen stars with intermediate optical depths one could argue either way.

As stated above, the disc sources are thought to be binary objects. The OGLE catalogues identify additional 
eclipsing variability in the LCs of 10 of the stars in the sample (marked "EB" in Table~\ref{table:params}, 
one of which is an RVT star), but excess emission is detected in none of them.

\subsection{Candidate binary T2C}
\label{Sect:CandBinarity}

Table~\ref{Tab-LITE-BIN} lists the systems where our investigation suggested a possible binary system.
The (phased) LC and the (O-C) diagrams and model fits are shown in Figure~\ref{Fig-LITE-BIN}.
The Table lists the binary period, time of periastron, $a \sin i$, and the change in period, $\dot{P}$. 
Pulsation type and period are repeated from Table~\ref{table:params}. The last column adds some remarks.
The eccentricity was fixed to zero, except in one case, where the data were good enough and required a non-zero eccentricity.
The LITE could only be established in one known EB, and only marginally in two others. For LMC-T2CEP-021 the derived binary 
period of 172 days is in good agreement with the period of 174.8 days that was photometrically derived by the OGLE team.

Most of the candidate binary systems hosts are (p)WVir stars. As mentioned in Sect.~\ref{Sect:Bin} the main criterion for investigating
a possible LITE was unusual scatter in the raising branch of the phased LC. This was not the case for the BLH and ACs, and only 13 out of
170 such objects were investigated for the LITE with only 2 candidate binaries (and 2 known BLH where the LITE was not found).
For the (p)WVir and RVT about half of the stars in the sample were investigated for the LITE with only a few candidates among the RVT, 
and about 1/3 among (p)WVir. On the other hand, one RVT and 5 (p)WVir in known EB systems did not show the LITE using the present data.

There is no correlation between candidate binary systems and infrared excess. Only 3 of the 23 candidates also shows IR excess in 
their SEDs.

\begin{table*}
\setlength{\tabcolsep}{1.25mm}
\caption{T2C candidate binaries from the LITE.}

\begin{tabular}{lrrrrrrlrr}
\hline
Name          &    $P_{\rm bin}$    &   $T_{\rm peri}$  &     $a \sin i$    &     $\dot{P}$        &  Type  & Period & Remarks \\  %
              &      (d)          &   (JD-2450000)   &       (AU)       &      (s/yr)          &        &   (d)  &         \\
\hline
LMC-ACEP-050  & 2468 $\pm$   68   & 4770 $\pm$  30 &   5.33 $\pm$   0.37 & $  -1.06 \pm   0.06$ & ANCep &  1.0 & \\%
LMC-T2CEP-011 & 1657 $\pm$  255   & 4252 $\pm$ 133 & 500.   $\pm$  88.   & 0 (fixed)            & RVTau & 39.3 & IR excess \\%
LMC-T2CEP-021 &  172 $\pm$    4.2 & 3447 $\pm$  12 &  57.   $\pm$  27.   & $ 505    \pm 108   $ & pWVir &  9.8 & known EB \\ %
LMC-T2CEP-033 & 2643 $\pm$   92   & 5775 $\pm$  46 &  90.   $\pm$  10.   & 0 (fixed)            & pWVir &  9.4 &  \\ %
LMC-T2CEP-040 & 3552 $\pm$ 2226   & 7207 $\pm$  51 & 166.   $\pm$ 234.   & 0 (fixed)            & pWVir &  9.6 & marginal LITE \\ %
LMC-T2CEP-044 & 2784 $\pm$  246   & 5047 $\pm$  82 & 163.   $\pm$  15.   & $ 233    \pm  62   $ &  WVir & 13.3 &  \\ 
LMC-T2CEP-062 & 2622 $\pm$  140   & 2446 $\pm$  47 &  56.0  $\pm$   5.1  & $ -31.2  \pm   6.5 $ &  WVir &  6.0 & \\  
LMC-T2CEP-079 & 1568 $\pm$  109   & 5688 $\pm$  64 &  64.4  $\pm$   6.4  & $ 631    \pm  58   $ &  WVir & 14.9 & \\  
LMC-T2CEP-087 & 2951 $\pm$   85   & 6669 $\pm$  79 &  41.   $\pm$  12.   & 0 (fixed)            &  WVir &  5.2 & $e= 0.74^{+0.23}_{-0.17}$, $\omega= 18 \pm 12\deg$ & \\
LMC-T2CEP-097 & 2221 $\pm$   81   & 4741 $\pm$  48 &  90.   $\pm$  12.   & $-107    \pm  21   $ &  WVir & 10.5 & \\
LMC-T2CEP-098 &  397 (fixed)      & 2970 $\pm$  22 &   8.7  $\pm$   2.6  & 0 (fixed)           & pWVir &  5.0 & known EB, marginal LITE \\
LMC-T2CEP-100 & 1905 $\pm$  171   & 4114 $\pm$  82 &  31.7  $\pm$   4.5  & $ -31    \pm  15   $ &  WVir &  7.4 & Fit for JD $>$1600, \\ 
              &                   &                 &                   &                        &      &      &  Very different period before \\
LMC-T2CEP-106 & 1617 $\pm$   38   & 3819 $\pm$  30 &  39.3  $\pm$   6.1  & 0 (fixed)            &  WVir &  6.7 &  \\  
LMC-T2CEP-127 & 3352 $\pm$  319   & 7068 $\pm$  82 & 408    $\pm$  71    & $-884    \pm 230   $ &  WVir & 12.7 & IR excess  \\  
LMC-T2CEP-132 & 2632 $\pm$  102   & 6576 $\pm$  83 &  94    $\pm$  11    & 0 (fixed)            & pWVir & 10.0 & amplitude variations    \\
LMC-T2CEP-137 & 2417 $\pm$   84   & 7406 $\pm$ 116 &  60.6  $\pm$   6.6  & $  24.7  \pm   7.1 $ &  WVir &  6.4 & \\
LMC-T2CEP-168 & 1668 $\pm$  171   & 6016 $\pm$ 120 &  42.5  $\pm$   5.5  & $ 145    \pm  69   $ &  WVir & 15.7 & \\
LMC-T2CEP-172 & 1983 $\pm$  101   & 5379 $\pm$  69 &  69.8  $\pm$  10.8  & $  78    \pm  29   $ &  WVir & 11.2 &  \\
LMC-T2CEP-177 & 2060 $\pm$   68   & 6229 $\pm$  37 &  83.8    $\pm$  2.8    &  0 (fixed)        &  WVir & 15.0 & \\

SMC-T2CEP-001 & 2019 $\pm$   84   & 5029 $\pm$  41 &  85.1  $\pm$  11.5  & 0 (fixed)           & pWVir & 11.9 & \\
SMC-T2CEP-018 & 1404 $\pm$   61   & 5554 $\pm$  48 & 368    $\pm$  80    & 0 (fixed)           & RVTau & 39.5 & IR excess \\ %
SMC-T2CEP-029 &  609 (fixed)      & 3569 $\pm$  46 &  25    $\pm$  18    & $-960    \pm  75  $ & RVTau & 33.7 & known EB, marginal LITE \\
SMC-T2CEP-030 & 2354 $\pm$  226   & 5929 $\pm$ 103 &   5.02 $\pm$   0.93 & $   1.6  \pm   0.9$ & BLHer &  3.9 & \\

\hline
\end{tabular}

\label{Tab-LITE-BIN}
\end{table*}

Another way to investigate binarity is to look for unusually low pulsation amplitudes. 
If a pulsating star is in a physical binary with a (assumed to be non-pulsating) secondary, the amplitude of the system is smaller.
Also the magnitude and colour are affected, which was already illustrated in Fig.~\ref{figHRDVI} for the system 
pWVir object LMC-T2CEP-098.

\begin{figure}
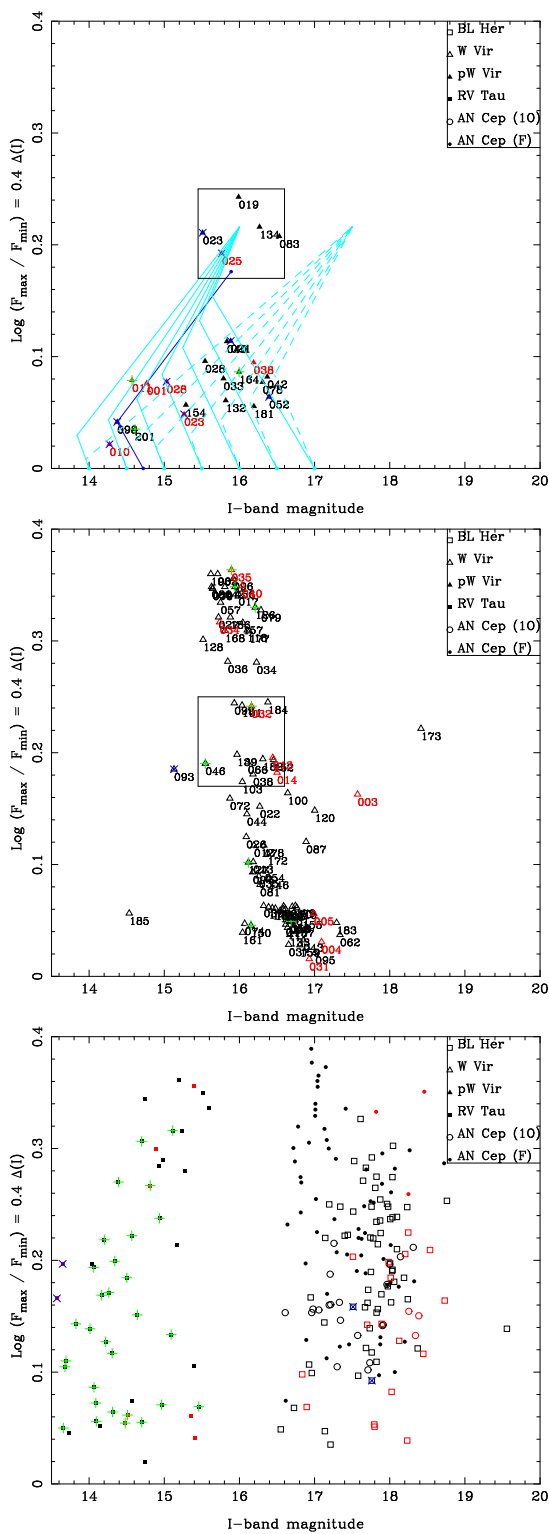

\centering

\begin{minipage}{0.385\textwidth}
\resizebox{\hsize}{!}{\includegraphics[angle=-0]{FluxRatios_pWVir.ps}} 
\end{minipage}
\begin{minipage}{0.385\textwidth}
\resizebox{\hsize}{!}{\includegraphics[angle=-0]{FluxRatios_WVir.ps}} 
\end{minipage}
\begin{minipage}{0.385\textwidth}
\resizebox{\hsize}{!}{\includegraphics[angle=-0]{FluxRatios_Other.ps}} 
\end{minipage}

\caption{Logarithm of the ratio between the flux at maximum and minimum light, i.e. $0.4\times$ the peak-to-peak amplitude, 
vs. mean $I$-band magnitudes %
for the pWVir (top), WVir (middle), and all other (bottom) objects.
Stars with an IR excess are indicated by a green plus sign.
Stars that show eclipsing or ellipsoidal variations according to OGLE are indicated by a blue cross.
For the pWVir object -098 (MACHO 6.6454.5), a known eclipsing system, the location of the two components is also indicated 
by dark blue lines.
The light blue lines indicate the location of two hypothetical binary systems, consisting of a pulsating component 
with a fixed amplitude and an increasingly fainter non-pulsating component.
The pWVir stars that appear like normal WVir are enclosed by a box that is repeated in the middle panel.
}
\label{figFLUXR}
\end{figure}

In Fig.~\ref{figFLUXR} the logarithm of the ratio between the flux at maximum and minimum light (or $0.4\times$ the 
peak-to-peak amplitude) is plotted against $I$ magnitudes for the pWVir (top), WVir (middle), and all other 
objects (bottom panel). For LMC-T2CEP-098 the position of the two components and the position of the system as it is 
observed are shown. The pulsating component is located in a region occupied by WVir stars. 
Interestingly, some other objects classified as pWVir (those enclosed in the box) show amplitudes and magnitudes that are 
comparable to WVir stars.
The unusual location of LMC-T2CEP-185 can be explained by the fact that it is listed as a blended object.
The light blue lines indicate the locations of hypothetical systems consisting of a 16.0 and 17.5 mag pulsating variable 
with a peak-to-peak amplitude of 0.54 mag and a non-pulsating companion of various magnitudes. The group of 
small amplitudes pWVir can indeed be explained by such binary systems.

\subsection{Period changes}
\label{Sect:periodchange}

The analysis of the LCs to identify the LITE also includes a term that takes into account any period change.
Table~\ref{Tab-LITE-BIN} lists $\dot{P}$ (can be zero or not significant) for the stars that we tentatively identify 
to be in a binary system through the LITE, 
while Table~\ref{Tab-LITE-Pdot} lists additional stars with a significant $\dot{P}$. 
Table~\ref{Tab-LITE-Incon} lists a few stars that may show a period change.

The analysis of the LCs was performed primarily to find potential binaries, not to determine $\dot{P}$ per se.
We only analysed 133 of the 335 LCs. On the other hand, as explained in Sect.~\ref{Sect:Bin}, the others did not show any
unusual scatter in the rising branch of the published phased LC that would hint at (strong) changes in period. 
This implies that the period changes we publish here are likely to be larger than for the general population of T2Cs and ACs.

The evolutionary status of BLH, WVIR, and ACs was pioneered in \citet{Gingold_1976, Gingold_1985} and recently 
updated by \citet{Bono_2016} and discussed in \citet{Neilson_2016}. The period change in these stars can generally 
be described as the result of change in effective temperature as they are crossing the IS. Going from the blue to 
red region of the IS increases the period, while going from the red to blue decreases the pulsation period.

Period changes in T2Cs have been determined in about two dozen Galactic field and cluster objects and are summarised 
in \citet{Neilson_2016}. For periods below eight days the period changes are small and positive ($<$20 d/Myr) with one exception.
For longer periods very large changes are observed, from about $-400$ to $+400$ d/Myr, or $\pm 34$ s/yr in the units used 
in the present paper.

\citet{WehlauB82} used the models of \citet{Gingold_1976} to calculate the period changes expected for BLH objects, 
some of which may make three crossings, and find values of order $+20$, $-2$, $+1$ d/Myr. \citet{Neilson_2016} summarise 
that for periods below about six days the observed period changes are consistent with predictions from stellar evolution, 
but for longer periods these changes are apparently not consistent. It has been hypothesised that the WVir are AGB stars with envelope masses below 
about 0.02~\msol\ that move into the IS after a thermal pulse (and then back again if double-shell burning resumes). 
This would result in large positive and negative period changes, although no quantitative estimates appear to have been made.
RV Tauri stars are thought to be post-AGB stars, i.e. move from the AGB towards hotter temperatures. As the P-AGB phase is also 
a short phase (typically a few $10^3$ years; \citet{MB16}) one might expect large negative period changes.

That (very) large period changes are possible was noted in \cite{Soszynski_2011} for a T2C in the Galactic bulge, 
OGLE-BLG-T2CEP-059, although no quantitative estimate was given. The system is presented in Appendix~\ref{AppLITE} where 
we derive a period change of order 7000 d/Myr (606 s/yr).

In the case of six ACs that were examined with the LITE method, two show a small (both negative) period change. 
LMC-ACEP-024, -058, -070 and -083 have inconclusive results regarding the period change.

The period change in T2Cs is different. 
From the total of eight BLHs that were examined, four show no evidence for period change, three show a positive increase, and one, 
LMC-T2CEP-113, is decreasing in period, but its LC is very noisy and the OGLE-{\sc III} catalogue lists it as "uncertain".

\begin{figure}
\centering

\begin{minipage}{0.45\textwidth}
\resizebox{\hsize}{!}{\includegraphics[angle=-0]{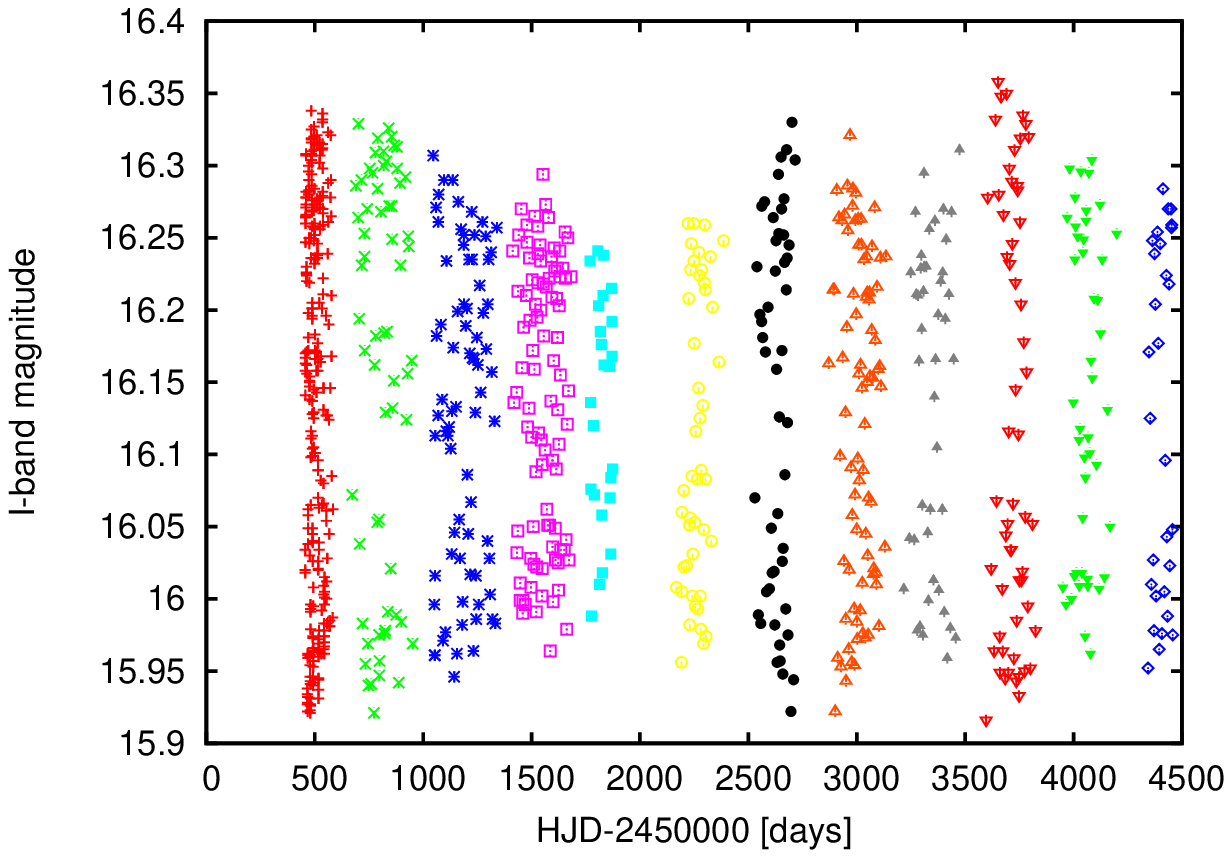}} 
\end{minipage}
\begin{minipage}{0.45\textwidth}
\resizebox{\hsize}{!}{\includegraphics[angle=-0]{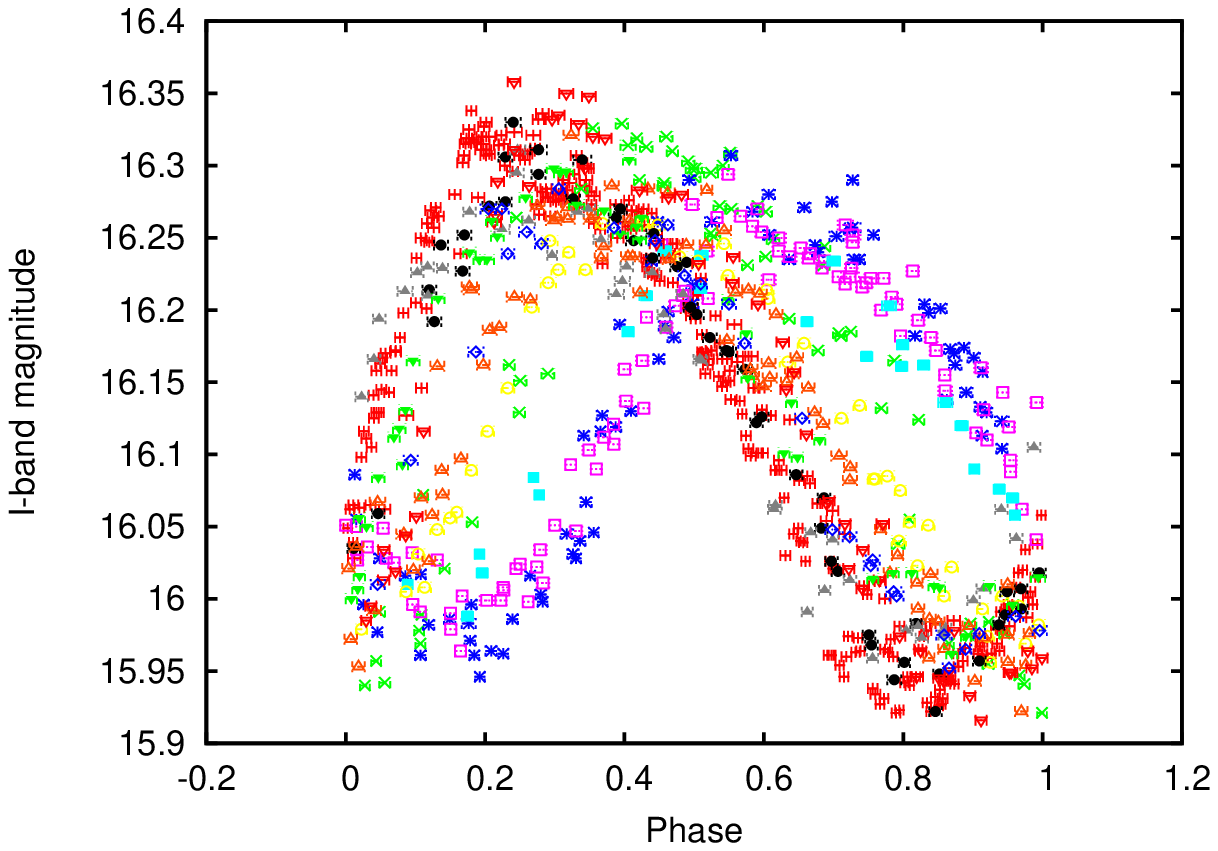}} 
\end{minipage}

\caption{
Segmented light curve of OGLE-LMC-T2CEP-127, showing the changes in the amplitude in the upper panel, 
and the phased light curve of OGLE-LMC-T2CEP-127, using a period of $P=12.6692$ days, in lower panel.
}

\label{figLMC127_LC}
\end{figure}

In the case of WVir stars things are more complicated. They should show increasing and decreasing period changes, 
since these stars are thought to undergo blue loops as they evolve from the AGB. From the stars included in
the LITE analysis 12 WVir and 4 pWVir were in a possible binary system, and show period changes 
in both directions (with the exception of LMC-T2CEP-087, LMC-T2CEP-040, and SMC-T2CEP-001, where the period
change was fixed to 0); see Tab.~\ref{Tab-LITE-BIN}. Even more interestingly in Table~\ref{Tab-LITE-Pdot} 4 pWVir 
and 22 WVir stars show a significant period change. A known feature of WVir stars is the amplitude variations 
in their LCs (\citet{Soszynski2008, Soszynski2010}, or in detail e.g. \citet{TempletonHenden_2007}, 
or more recently \citet{Plachy_2016}), while some of these changes can be interpreted as 
period doubling (see \cite{Moskalik_Buchler_1990, Moskalik_Buchler_1993, Smolec_Moskalik_2014, Smolec_2016}). 
In the case of a few of these stars there is a new feature in the behaviour of the LCs. 
One of the most prominent examples is OGLE-LMC-T2CEP-127, which is shown in Figure~\ref{figLMC127_LC}. While 
the different sections (shown with a different colour codes) in the LC have different amplitudes 
through the whole OGLE-{\sc III} data set, they can all be phased with a single period of $P=12.6692$ days. 
When phased, it becomes apparent that the LC is changing its shape, not only the amplitude. 
This feature is something new and we will be looking into this in more detail in the future.

Stars showing this kind of behaviour are detected by LITE and can be found in either 
Table~\ref{Tab-LITE-BIN} or Table~\ref{Tab-LITE-Pdot}: OGLE-LMC-T2CEP-026, 034, 044, 072, 100, 127, 
and OGLE-SMC-T2CEP-14, 32, 34. 

In the case of RVTs the increasingly erratic behaviour seen in their LCs, 
with alternating minima and maxima, and long-term amplitude changes on top of that, 
leads to detection of period change. The nature of the features seen in the 
infrared excess was discussed in Section~\ref{Sect:RVT}. 
While these stars are thought to be crossing from cool to hot temperatures and should be 
showing period decrease, this is not what we see in the LITE results.
Almost half of the RVT show a period increase.
The interpretation of this phenomena is beyond the scope of this paper.

\section{Summary and conclusions}
\label{Sect:Conclusion}

The SEDs of 335 Type II and anomalous Cepheids in the Small and Large Magellanic Clouds have been 
constructed using photometry from the literature and fitted with a dust radiative transfer code.
Luminosities and effective temperatures are derived from the fitting.

In a companion paper we use the derived luminosities and temperatures to discuss the period-luminosity and period-radius 
relation for T2Cs and ACs and to estimate the mass of these objects.

For $\sim60\%$ of the RVT and $\sim 10\%$ of the (p)WVir objects, an infrared excess is detected from the SED fitting.
For the RVT this is not unexpected as they are thought to evolve from the AGB to the PAGB phase.
The results of \cite{Kamath2016} are confirmed that stars exist with luminosities below that predicted from single-star 
stellar evolution with IR excess, and the shape of the IR excess points to the presence of a disc rather than an expanding shell 
structure, which in turn suggests a relation to binarity.

Based on the shape of the phased LCs about one-third of the sample was selected to look for the light-time effect signature 
of a binary system following the method of \cite{Hajdu2015}. Twenty-three systems appear to show the LITE, 
including a few known EBs. On the other hand LITE was not detected with any significance in seven known EBs. 

The analysis of the (O-C) diagram also allows us to detect period changes and values for $\dot{P}$ are given for about 40 stars.
Some of the period changes are much larger than predicted by standard evolution of single stars. Some are possibly related to the 
so-called binary evolutionary pulsators (BEP), the prototype of which, OGLE-BLG-RRLYR-02792, has a period change 
of $-8.4$ d/Myr ($-0.73$ s/yr) \citep{Pietrzynski2012}. Recently, \citet{Karczmarek2016} did extensive simulations to find 
contaminations of genuine RRL and classical Cepheids of  0.8 and 5\%, respectively, by BEP. They did not specify numbers for T2Cs, 
but using their data we find a median crossing time of the IS of 36 kyr for stellar systems with luminosities 
between $\log L = 2.4-2.9 $ , which is similar to the 28 kyr they quote for RRL imposters. However the crossing time of a single star T2C 
is typically 10 times less than the 10 Myr they quote for RRL, so that a contamination of several percent is plausible.

The position of the objects in the HRD is compared to evolutionary tracks.
In agreement with previous 
suggestions and predictions, the BLH can be explained by the evolution of $\sim$0.5-0.6~\msol\ stars (depending on metallicity) evolving 
off the ZAHB and the ACs can be explained by the evolution of $\sim$1.1-2.3~\msol\ stars.
The evolution of the WVir subclass is not clear. They are at higher luminosities than the ACs and indeed evolutionary 
tracks of $\sim$2.5-4~\msol\ stars cross this region in the HRD, but the periods of the WVir are longer than those of
the short period classical Cepheids at these luminosities, which points to a lower mass. 
Also the evolutionary timescale does not fit this picture.
It is shown that when a low-mass AGB star experiences a thermal pulse when the envelope mass is small, it can make a blue loop into 
the IS region of the WVir stars. But the timescale is extremely short, so this is also no explanation for the WVir as a class.
The connection to binarity might be at the origin of the WVir stars, which has already been explicitly suggested for the peculiar W Virginis stars.

\begin{acknowledgements}
M.I.J. acknowledges financial support from the Ministry of Education, Science and Technological Development of the 
Republic of Serbia through the project 176004, and the Hungarian National Research, 
Development and Innovation Office through NKFIH K-115709.
The authors sincerely thank Marcelo Miller Bertolami for providing his AGB and PAGB tracks.
This work has made use of BaSTI web tools.
This research has made use of the VizieR catalogue access tool, CDS, Strasbourg, France. 
The original description of the VizieR service was published in A\&AS 143, 23.
This research has made use of the NASA/IPAC Infrared Science Archive,
which is operated by the Jet Propulsion Laboratory, California Institute of Technology, 
under contract with the National Aeronautics and Space Administration.
This publication makes use of data products from the Wide-field Infrared Survey Explorer, 
which is a joint project of the University of California, Los Angeles, and the 
Jet Propulsion Laboratory/California Institute of Technology, funded by the 
National Aeronautics and Space Administration.

\end{acknowledgements}


\bibliographystyle{aa.bst}
        \bibliography{references.bib}

\begin{thebibliography}{82}
\expandafter\ifx\csname natexlab\endcsname\relax\def\natexlab#1{#1}\fi

\bibitem[{{Alcock} {et~al.}(1998){Alcock}, {Allsman}, {Alves}, {Axelrod},
  {Becker}, {Bennett}, {Cook}, {Freeman}, {Griest}, {Lawson}, {Lehner},
  {Marshall}, {Minniti}, {Peterson}, {Pollard}, {Pratt}, {Quinn}, {Rodgers},
  {Sutherland}, {Tomaney}, \& {Welch}}]{MACHO98}
{Alcock}, C., {Allsman}, R.~A., {Alves}, D.~R., {et~al.} 1998, \aj, 115, 1921

\bibitem[{{Alcock} {et~al.}(2002){Alcock}, {Allsman}, {Alves}, {Becker},
  {Bennett}, {Cook}, {Drake}, {Freeman}, {Griest}, {Hawley}, {Keller},
  {Lehner}, {Lepischak}, {Marshall}, {Minniti}, {Nelson}, {Peterson},
  {Popowski}, {Pratt}, {Quinn}, {Rodgers}, {Suntzeff}, {Sutherland},
  {Vandehei}, \& {Welch}}]{MACHO02}
{Alcock}, C., {Allsman}, R.~A., {Alves}, D.~R., {et~al.} 2002, \apj, 573, 338

\bibitem[{{Bono} {et~al.}(1997{\natexlab{a}}){Bono}, {Caputo}, \&
  {Santolamazza}}]{Bono_1997_T2CEP}
{Bono}, G., {Caputo}, F., \& {Santolamazza}, P. 1997{\natexlab{a}}, \aap, 317,
  171

\bibitem[{{Bono} {et~al.}(1997{\natexlab{b}}){Bono}, {Caputo}, {Santolamazza},
  {Cassisi}, \& {Piersimoni}}]{Bono_1997_ANCEP}
{Bono}, G., {Caputo}, F., {Santolamazza}, P., {Cassisi}, S., \& {Piersimoni},
  A. 1997{\natexlab{b}}, \aj, 113, 2209

\bibitem[{{Bono} {et~al.}(2000){Bono}, {Castellani}, \& {Marconi}}]{Bono_2000}
{Bono}, G., {Castellani}, V., \& {Marconi}, M. 2000, \apj, 529, 293

\bibitem[{{Bono} {et~al.}(2016){Bono}, {Pietrinferni}, {Marconi}, {Braga},
  {Fiorentino}, {Stetson}, {Buonanno}, {Castellani}, {Dall'Ora}, {Fabrizio},
  {Ferraro}, {Giuffrida}, {Iannicola}, {Marengo}, {Magurno},
  {Mart{\'{\i}}nez-V{\'a}zquez}, {Matsunaga}, {Monelli}, {Neeley}, {Rastello},
  {Salaris}, {Short}, \& {Stellingwerf}}]{Bono_2016}
{Bono}, G., {Pietrinferni}, A., {Marconi}, M., {et~al.} 2016, Commmunications
  of the Konkoly Observatory Hungary, 105, 149

\bibitem[{{Boyer} {et~al.}(2011){Boyer}, {Srinivasan}, {van Loon}, {McDonald},
  {Meixner}, {Zaritsky}, {Gordon}, {Kemper}, {Babler}, {Block}, {Bracker},
  {Engelbracht}, {Hora}, {Indebetouw}, {Meade}, {Misselt}, {Robitaille},
  {Sewi{\l}o}, {Shiao}, \& {Whitney}}]{Boyer_2011}
{Boyer}, M.~L., {Srinivasan}, S., {van Loon}, J.~T., {et~al.} 2011, \aj, 142,
  103

\bibitem[{{Bressan} {et~al.}(2012){Bressan}, {Marigo}, {Girardi}, {Salasnich},
  {Dal Cero}, {Rubele}, \& {Nanni}}]{Bressan2012}
{Bressan}, A., {Marigo}, P., {Girardi}, L., {et~al.} 2012, \mnras, 427, 127

\bibitem[{{Caputo} {et~al.}(2004){Caputo}, {Castellani}, {Degl'Innocenti},
  {Fiorentino}, \& {Marconi}}]{Caputo_2004}
{Caputo}, F., {Castellani}, V., {Degl'Innocenti}, S., {Fiorentino}, G., \&
  {Marconi}, M. 2004, \aap, 424, 927

\bibitem[{{Ciechanowska} {et~al.}(2010){Ciechanowska}, {Pietrzy{\'n}ski},
  {Szewczyk}, {Gieren}, \& {Soszy{\'n}ski}}]{Ciechanowska_2010}
{Ciechanowska}, A., {Pietrzy{\'n}ski}, G., {Szewczyk}, O., {Gieren}, W., \&
  {Soszy{\'n}ski}, I. 2010, \actaa, 60, 233

\bibitem[{{Cioni} {et~al.}(2011){Cioni}, {Clementini}, {Girardi}, {Guandalini},
  {Gullieuszik}, {Miszalski}, {Moretti}, {Ripepi}, {Rubele}, {Bagheri},
  {Bekki}, {Cross}, {de Blok}, {de Grijs}, {Emerson}, {Evans}, {Gibson},
  {Gonzales-Solares}, {Groenewegen}, {Irwin}, {Ivanov}, {Lewis}, {Marconi},
  {Marquette}, {Mastropietro}, {Moore}, {Napiwotzki}, {Naylor}, {Oliveira},
  {Read}, {Sutorius}, {van Loon}, {Wilkinson}, \& {Wood}}]{Cioni_VMC}
{Cioni}, M.-R.~L., {Clementini}, G., {Girardi}, L., {et~al.} 2011, \aap, 527,
  A116

\bibitem[{{Cutri} \& {et al.}(2014)}]{Cutri_Allwise}
{Cutri}, R.~M. \& {et al.} 2014, VizieR Online Data Catalog, 2328, 0

\bibitem[{{de Grijs} \& {Bono}(2015)}]{deGrijs2015}
{de Grijs}, R. \& {Bono}, G. 2015, \aj, 149, 179

\bibitem[{{de Grijs} {et~al.}(2014){de Grijs}, {Wicker}, \&
  {Bono}}]{deGrijs2014}
{de Grijs}, R., {Wicker}, J.~E., \& {Bono}, G. 2014, \aj, 147, 122

\bibitem[{{Di Criscienzo} {et~al.}(2007){Di Criscienzo}, {Caputo}, {Marconi},
  \& {Cassisi}}]{DiCriscienzo_2007}
{Di Criscienzo}, M., {Caputo}, F., {Marconi}, M., \& {Cassisi}, S. 2007, \aap,
  471, 893

\bibitem[{{Fiorentino} {et~al.}(2006){Fiorentino}, {Limongi}, {Caputo}, \&
  {Marconi}}]{Fiorentino2006}
{Fiorentino}, G., {Limongi}, M., {Caputo}, F., \& {Marconi}, M. 2006, \aap,
  460, 155

\bibitem[{{Fiorentino} \& {Monelli}(2012)}]{Fiorentino_Monelli_2012}
{Fiorentino}, G. \& {Monelli}, M. 2012, \aap, 540, A102

\bibitem[{{Gesicki} {et~al.}(2014){Gesicki}, {Zijlstra}, {Hajduk}, \&
  {Szyszka}}]{Gesicki_ea_2014}
{Gesicki}, K., {Zijlstra}, A.~A., {Hajduk}, M., \& {Szyszka}, C. 2014, \aap,
  566, A48

\bibitem[{{Gezer} {et~al.}(2015){Gezer}, {Van Winckel}, {Bozkurt}, {De Smedt},
  {Kamath}, {Hillen}, \& {Manick}}]{Gezer2015}
{Gezer}, I., {Van Winckel}, H., {Bozkurt}, Z., {et~al.} 2015, \mnras, 454, 804

\bibitem[{{Gielen} {et~al.}(2011){Gielen}, {Bouwman}, {van Winckel}, {Lloyd
  Evans}, {Woods}, {Kemper}, {Marengo}, {Meixner}, {Sloan}, \&
  {Tielens}}]{Gielen2011}
{Gielen}, C., {Bouwman}, J., {van Winckel}, H., {et~al.} 2011, \aap, 533, A99

\bibitem[{{Gielen} {et~al.}(2009){Gielen}, {van Winckel}, {Reyniers},
  {Zijlstra}, {Lloyd Evans}, {Gordon}, {Kemper}, {Indebetouw}, {Marengo},
  {Matsuura}, {Meixner}, {Sloan}, {Tielens}, \& {Woods}}]{Gielen2009}
{Gielen}, C., {van Winckel}, H., {Reyniers}, M., {et~al.} 2009, \aap, 508, 1391

\bibitem[{{Gingold}(1976)}]{Gingold_1976}
{Gingold}, R.~A. 1976, \apj, 204, 116

\bibitem[{{Gingold}(1985)}]{Gingold_1985}
{Gingold}, R.~A. 1985, \memsai, 56, 169

\bibitem[{{Groenewegen}(2004)}]{Groenewegen2004}
{Groenewegen}, M.~A.~T. 2004, \aap, 425, 595

\bibitem[{{Groenewegen}(2012)}]{Gr_MOD}
{Groenewegen}, M.~A.~T. 2012, \aap, 543, A36

\bibitem[{{Gustafsson} {et~al.}(2008){Gustafsson}, {Edvardsson}, {Eriksson},
  {J{\o}rgensen}, {Nordlund}, \& {Plez}}]{Gustafsson_MARCS}
{Gustafsson}, B., {Edvardsson}, B., {Eriksson}, K., {et~al.} 2008, \aap, 486,
  951

\bibitem[{{Hajdu} {et~al.}(2015){Hajdu}, {Catelan}, {Jurcsik},
  {D{\'e}k{\'a}ny}, {Drake}, \& {Marquette}}]{Hajdu2015}
{Hajdu}, G., {Catelan}, M., {Jurcsik}, J., {et~al.} 2015, \mnras, 449, L113

\bibitem[{{Haschke} {et~al.}(2012{\natexlab{a}}){Haschke}, {Grebel}, \&
  {Duffau}}]{HaschkeLMC_2012}
{Haschke}, R., {Grebel}, E.~K., \& {Duffau}, S. 2012{\natexlab{a}}, \aj, 144,
  106

\bibitem[{{Haschke} {et~al.}(2012{\natexlab{b}}){Haschke}, {Grebel}, \&
  {Duffau}}]{HaschkeSMC_2012}
{Haschke}, R., {Grebel}, E.~K., \& {Duffau}, S. 2012{\natexlab{b}}, \aj, 144,
  107

\bibitem[{{Hauschildt} {et~al.}(1999){Hauschildt}, {Allard}, \&
  {Baron}}]{Hauschildt1999}
{Hauschildt}, P.~H., {Allard}, F., \& {Baron}, E. 1999, \apj, 512, 377

\bibitem[{{Hillen} {et~al.}(2015){Hillen}, {de Vries}, {Menu}, {Van Winckel},
  {Min}, \& {Mulders}}]{Hillen_2015}
{Hillen}, M., {de Vries}, B.~L., {Menu}, J., {et~al.} 2015, \aap, 578, A40

\bibitem[{{Houck} {et~al.}(2004){Houck}, {Roellig}, {van Cleve}, {Forrest},
  {Herter}, {Lawrence}, {Matthews}, {Reitsema}, {Soifer}, {Watson}, {Weedman},
  {Huisjen}, {Troeltzsch}, {Barry}, {Bernard-Salas}, {Blacken}, {Brandl},
  {Charmandaris}, {Devost}, {Gull}, {Hall}, {Henderson}, {Higdon}, {Pirger},
  {Schoenwald}, {Sloan}, {Uchida}, {Appleton}, {Armus}, {Burgdorf},
  {Fajardo-Acosta}, {Grillmair}, {Ingalls}, {Morris}, \& {Teplitz}}]{Houck2004}
{Houck}, J.~R., {Roellig}, T.~L., {van Cleve}, J., {et~al.} 2004, \apjs, 154,
  18

\bibitem[{{Inno} {et~al.}(2016){Inno}, {Bono}, {Matsunaga}, {Fiorentino},
  {Marconi}, {Lemasle}, {da Silva}, {Soszy{\'n}ski}, {Udalski}, {Romaniello},
  \& {Rix}}]{Inno_2016}
{Inno}, L., {Bono}, G., {Matsunaga}, N., {et~al.} 2016, \apj, 832, 176

\bibitem[{{Irwin}(1952)}]{Irwin52}
{Irwin}, J.~B. 1952, \apj, 116, 211

\bibitem[{{Ita} {et~al.}(2010){Ita}, {Onaka}, {Tanab{\'e}}, {Matsunaga},
  {Matsuura}, {Yamamura}, {Nakada}, {Izumiura}, {Ueta}, {Mito}, {Fukushi}, \&
  {Kato}}]{Ita_AkariSMC}
{Ita}, Y., {Onaka}, T., {Tanab{\'e}}, T., {et~al.} 2010, \pasj, 62, 273

\bibitem[{{Ivezi{\'c}} {et~al.}(1999){Ivezi{\'c}}, {Nenkova}, \&
  {Elitzur}}]{Ivezic_D}
{Ivezi{\'c}}, {\v{Z}}., {Nenkova}, M., \& {Elitzur}, M. 1999, {DUSTY: Radiation
  transport in a dusty environment}, Astrophysics Source Code Library

\bibitem[{{Kamath} {et~al.}(2014){Kamath}, {Wood}, \& {Van
  Winckel}}]{Kamath2014}
{Kamath}, D., {Wood}, P.~R., \& {Van Winckel}, H. 2014, \mnras, 439, 2211

\bibitem[{{Kamath} {et~al.}(2015){Kamath}, {Wood}, \& {Van
  Winckel}}]{Kamath2015}
{Kamath}, D., {Wood}, P.~R., \& {Van Winckel}, H. 2015, \mnras, 454, 1468

\bibitem[{{Kamath} {et~al.}(2016){Kamath}, {Wood}, {Van Winckel}, \&
  {Nie}}]{Kamath2016}
{Kamath}, D., {Wood}, P.~R., {Van Winckel}, H., \& {Nie}, J.~D. 2016, \aap,
  586, L5

\bibitem[{{Karczmarek} {et~al.}(2017){Karczmarek}, {Wiktorowicz},
  {I{\l}kiewicz}, {Smolec}, {St{\c e}pie{\'n}}, {Pietrzy{\'n}ski}, {Gieren}, \&
  {Belczynski}}]{Karczmarek2016}
{Karczmarek}, P., {Wiktorowicz}, G., {I{\l}kiewicz}, K., {et~al.} 2017, \mnras,
  466, 2842

\bibitem[{{Kato} {et~al.}(2012){Kato}, {Ita}, {Onaka}, {Tanab{\'e}},
  {Shimonishi}, {Sakon}, {Kaneda}, {Kawamura}, {Wada}, {Usui}, {Koo},
  {Matsuura}, \& {Takahashi}}]{Kato_AkariLMC}
{Kato}, D., {Ita}, Y., {Onaka}, T., {et~al.} 2012, \aj, 144, 179

\bibitem[{{Kato} {et~al.}(2007){Kato}, {Nagashima}, {Nagayama}, {Kurita},
  {Koerwer}, {Kawai}, {Yamamuro}, {Zenno}, {Nishiyama}, {Baba}, {Kadowaki},
  {Haba}, {Hatano}, {Shimizu}, {Nishimura}, {Nagata}, {Sato}, {Murai},
  {Kawazu}, {Nakajima}, {Nakaya}, {Kandori}, {Kusakabe}, {Ishihara},
  {Kaneyasu}, {Hashimoto}, {Tamura}, {Tanab{\'e}}, {Ita}, {Matsunaga},
  {Nakada}, {Sugitani}, {Wakamatsu}, {Glass}, {Feast}, {Menzies}, {Whitelock},
  {Fourie}, {Stoffels}, {Evans}, \& {Hasegawa}}]{Kato_IRSF}
{Kato}, D., {Nagashima}, C., {Nagayama}, T., {et~al.} 2007, \pasj, 59, 615

\bibitem[{{Lenz} \& {Breger}(2005)}]{Lenz_Breger_2005}
{Lenz}, P. \& {Breger}, M. 2005, Communications in Asteroseismology, 146, 53

\bibitem[{{Macri} {et~al.}(2015){Macri}, {Ngeow}, {Kanbur}, {Mahzooni}, \&
  {Smitka}}]{Macri_2015}
{Macri}, L.~M., {Ngeow}, C.-C., {Kanbur}, S.~M., {Mahzooni}, S., \& {Smitka},
  M.~T. 2015, \aj, 149, 117

\bibitem[{{Marconi} {et~al.}(2015){Marconi}, {Coppola}, {Bono}, {Braga},
  {Pietrinferni}, {Buonanno}, {Castellani}, {Musella}, {Ripepi}, \&
  {Stellingwerf}}]{Marconi_2015}
{Marconi}, M., {Coppola}, G., {Bono}, G., {et~al.} 2015, \apj, 808, 50

\bibitem[{{Marconi} {et~al.}(2004){Marconi}, {Fiorentino}, \&
  {Caputo}}]{Marconi_2004}
{Marconi}, M., {Fiorentino}, G., \& {Caputo}, F. 2004, \aap, 417, 1101

\bibitem[{{Mart{\'{\i}}nez-V{\'a}zquez}
  {et~al.}(2016){Mart{\'{\i}}nez-V{\'a}zquez}, {Stetson}, {Monelli}, {Bernard},
  {Fiorentino}, {Gallart}, {Bono}, {Cassisi}, {Dall'Ora}, {Ferraro},
  {Iannicola}, \& {Walker}}]{MartinezVazques_2016}
{Mart{\'{\i}}nez-V{\'a}zquez}, C.~E., {Stetson}, P.~B., {Monelli}, M., {et~al.}
  2016, \mnras, 462, 4349

\bibitem[{{Massey}(2002)}]{Massey_2002}
{Massey}, P. 2002, \apjs, 141, 81

\bibitem[{{Miller Bertolami}(2016)}]{MB16}
{Miller Bertolami}, M.~M. 2016, \aap, 588, A25

\bibitem[{{Moskalik} \& {Buchler}(1990)}]{Moskalik_Buchler_1990}
{Moskalik}, P. \& {Buchler}, J.~R. 1990, \apj, 355, 590

\bibitem[{{Moskalik} \& {Buchler}(1993)}]{Moskalik_Buchler_1993}
{Moskalik}, P. \& {Buchler}, J.~R. 1993, \apj, 406, 190

\bibitem[{{Neilson} {et~al.}(2016){Neilson}, {Percy}, \&
  {Smith}}]{Neilson_2016}
{Neilson}, H.~R., {Percy}, J.~R., \& {Smith}, H.~A. 2016, Journal of the
  American Association of Variable Star Observers (JAAVSO), 44, 179

\bibitem[{{Pietrinferni} {et~al.}(2004){Pietrinferni}, {Cassisi}, {Salaris}, \&
  {Castelli}}]{Pietrinferni2004}
{Pietrinferni}, A., {Cassisi}, S., {Salaris}, M., \& {Castelli}, F. 2004, \apj,
  612, 168

\bibitem[{{Pietrzy{\'n}ski} {et~al.}(2012){Pietrzy{\'n}ski}, {Thompson},
  {Gieren}, {Graczyk}, {St{\c e}pie{\'n}}, {Bono}, {Moroni}, {Pilecki},
  {Udalski}, {Soszy{\'n}ski}, {Preston}, {Nardetto}, {McWilliam}, {Roederer},
  {G{\'o}rski}, {Konorski}, \& {Storm}}]{Pietrzynski2012}
{Pietrzy{\'n}ski}, G., {Thompson}, I.~B., {Gieren}, W., {et~al.} 2012, \nat,
  484, 75

\bibitem[{{Plachy} {et~al.}(2017){Plachy}, {Moln{\'a}r}, {Jurkovic}, {Smolec},
  {Moskalik}, {P{\'a}l}, {Szabados}, \& {Szab{\'o}}}]{Plachy_2016}
{Plachy}, E., {Moln{\'a}r}, L., {Jurkovic}, M.~I., {et~al.} 2017, \mnras, 465,
  173

\bibitem[{{Pollard} \& {Lloyd Evans}(2000)}]{Pollard2000}
{Pollard}, K.~R. \& {Lloyd Evans}, T. 2000, \aj, 120, 3098

\bibitem[{Press {et~al.}(1992)Press, Teukolsky, Vetterling, \&
  Flannery}]{Press1992}
Press, W., Teukolsky, S., Vetterling, W., \& Flannery, B. 1992, {Numerical
  Recipes in C} (Cambridge: Cambridge University Press)

\bibitem[{{Renzini} {et~al.}(1977){Renzini}, {Mengel}, \&
  {Sweigart}}]{Renzini_1977}
{Renzini}, A., {Mengel}, J.~G., \& {Sweigart}, A.~V. 1977, \aap, 56, 369

\bibitem[{{Reyniers} {et~al.}(2007){Reyniers}, {Abia}, {van Winckel}, {Lloyd
  Evans}, {Decin}, {Eriksson}, \& {Pollard}}]{Reyniers2007}
{Reyniers}, M., {Abia}, C., {van Winckel}, H., {et~al.} 2007, \aap, 461, 641

\bibitem[{{Reyniers} \& {van Winckel}(2007)}]{RW2007}
{Reyniers}, M. \& {van Winckel}, H. 2007, \aap, 463, L1

\bibitem[{{Ripepi} {et~al.}(2015){Ripepi}, {Moretti}, {Marconi}, {Clementini},
  {Cioni}, {de Grijs}, {Emerson}, {Groenewegen}, {Ivanov}, {Muraveva},
  {Piatti}, \& {Subramanian}}]{Ripepi_2015}
{Ripepi}, V., {Moretti}, M.~I., {Marconi}, M., {et~al.} 2015, \mnras, 446, 3034

\bibitem[{Schwarz(1978)}]{Schwarz1978}
Schwarz, G. 1978, Ann. Stat., 6, 461

\bibitem[{{Sebo} {et~al.}(2002){Sebo}, {Rawson}, {Mould}, {Madore}, {Putman},
  {Graham}, {Freedman}, {Gibson}, \& {Germany}}]{Sebo_2002}
{Sebo}, K.~M., {Rawson}, D., {Mould}, J., {et~al.} 2002, \apjs, 142, 71

\bibitem[{{Sloan} {et~al.}(2003){Sloan}, {Kraemer}, {Price}, \&
  {Shipman}}]{Sloan2003}
{Sloan}, G.~C., {Kraemer}, K.~E., {Price}, S.~D., \& {Shipman}, R.~F. 2003,
  \apjs, 147, 379

\bibitem[{{Smolec}(2016)}]{Smolec_2016}
{Smolec}, R. 2016, \mnras, 456, 3475

\bibitem[{{Smolec} \& {Moskalik}(2014)}]{Smolec_Moskalik_2014}
{Smolec}, R. \& {Moskalik}, P. 2014, \mnras, 441, 101

\bibitem[{{Soszy{\'n}ski} {et~al.}(2010{\natexlab{a}}){Soszy{\'n}ski},
  {Poleski}, {Udalski}, {Szyma{\'n}ski}, {Kubiak}, {Pietrzy{\'n}ski},
  {Wyrzykowski}, {Szewczyk}, \& {Ulaczyk}}]{Soszynski2010_ANC}
{Soszy{\'n}ski}, I., {Poleski}, R., {Udalski}, A., {et~al.} 2010{\natexlab{a}},
  \actaa, 60, 17

\bibitem[{{Soszy{\'n}ski} {et~al.}(2011){Soszy{\'n}ski}, {Udalski},
  {Pietrukowicz}, {Szyma{\'n}ski}, {Kubiak}, {Pietrzy{\'n}ski}, {Wyrzykowski},
  {Ulaczyk}, {Poleski}, \& {Koz{\l}owski}}]{Soszynski_2011}
{Soszy{\'n}ski}, I., {Udalski}, A., {Pietrukowicz}, P., {et~al.} 2011, \actaa,
  61, 285

\bibitem[{{Soszy{\'n}ski} {et~al.}(2010{\natexlab{b}}){Soszy{\'n}ski},
  {Udalski}, {Szyma{\'n}ski}, {Kubiak}, {Pietrzy{\~n}ski}, {Wyrzykowski},
  {Ulaczyk}, \& {Poleski}}]{Soszynski2010}
{Soszy{\'n}ski}, I., {Udalski}, A., {Szyma{\'n}ski}, M.~K., {et~al.}
  2010{\natexlab{b}}, \actaa, 60, 91

\bibitem[{{Soszy{\'n}ski} {et~al.}(2008){Soszy{\'n}ski}, {Udalski},
  {Szyma{\'n}ski}, {Kubiak}, {Pietrzy{\'n}ski}, {Wyrzykowski}, {Szewczyk},
  {Ulaczyk}, \& {Poleski}}]{Soszynski2008}
{Soszy{\'n}ski}, I., {Udalski}, A., {Szyma{\'n}ski}, M.~K., {et~al.} 2008,
  \actaa, 58, 293

\bibitem[{{Soszy{\'n}ski} {et~al.}(2014){Soszy{\'n}ski}, {Udalski},
  {Szyma{\'n}ski}, {Pietrukowicz}, {Mr{\'o}z}, {Skowron}, {Koz{\l}owski},
  {Poleski}, {Skowron}, {Pietrzy{\'n}ski}, {Wyrzykowski}, {Ulaczyk}, \&
  {Kubiak}}]{Soszynski_2014}
{Soszy{\'n}ski}, I., {Udalski}, A., {Szyma{\'n}ski}, M.~K., {et~al.} 2014,
  \actaa, 64, 177

\bibitem[{{Soszy{\'n}ski} {et~al.}(2015{\natexlab{a}}){Soszy{\'n}ski},
  {Udalski}, {Szyma{\'n}ski}, {Pietrzy{\'n}ski}, {Wyrzykowski}, {Ulaczyk},
  {Poleski}, {Pietrukowicz}, {Koz{\l}owski}, {Skowron}, {Skowron}, {Mr{\'o}z},
  \& {Pawlak}}]{Soszynski_2015}
{Soszy{\'n}ski}, I., {Udalski}, A., {Szyma{\'n}ski}, M.~K., {et~al.}
  2015{\natexlab{a}}, \actaa, 65, 233

\bibitem[{{Soszy{\'n}ski} {et~al.}(2015{\natexlab{b}}){Soszy{\'n}ski},
  {Udalski}, {Szyma{\'n}ski}, {Skowron}, {Pietrzy{\'n}ski}, {Poleski},
  {Pietrukowicz}, {Skowron}, {Mr{\'o}z}, {Koz{\l}owski}, {Wyrzykowski},
  {Ulaczyk}, \& {Pawlak}}]{Soszynski_2015Cep}
{Soszy{\'n}ski}, I., {Udalski}, A., {Szyma{\'n}ski}, M.~K., {et~al.}
  2015{\natexlab{b}}, \actaa, 65, 297

\bibitem[{{Sterken}(2005)}]{Sterken2005}
{Sterken}, C. 2005, in Astronomical Society of the Pacific Conference Series,
  Vol. 335, The Light-Time Effect in Astrophysics: Causes and cures of the O-C
  diagram, ed. C.~{Sterken}, 3

\bibitem[{{Templeton} \& {Henden}(2007)}]{TempletonHenden_2007}
{Templeton}, M.~R. \& {Henden}, A.~A. 2007, \aj, 134, 1999

\bibitem[{{van Aarle} {et~al.}(2011){van Aarle}, {van Winckel}, {Lloyd Evans},
  {Ueta}, {Wood}, \& {Ginsburg}}]{vanAarle11}
{van Aarle}, E., {van Winckel}, H., {Lloyd Evans}, T., {et~al.} 2011, \aap,
  530, A90

\bibitem[{{Vassiliadis} \& {Wood}(1993)}]{Vassiliadis_Wood_1993}
{Vassiliadis}, E. \& {Wood}, P.~R. 1993, \apj, 413, 641

\bibitem[{{Wallerstein}(2002)}]{Wallerstein_2002}
{Wallerstein}, G. 2002, \pasp, 114, 689

\bibitem[{{Wehlau} \& {Bohlender}(1982)}]{WehlauB82}
{Wehlau}, A. \& {Bohlender}, D. 1982, \aj, 87, 780

\bibitem[{{Welch}(2012)}]{Welch_2012}
{Welch}, D.~L. 2012, Journal of the American Association of Variable Star
  Observers (JAAVSO), 40, 492

\bibitem[{{Zaritsky} {et~al.}(2004){Zaritsky}, {Harris}, {Thompson}, \&
  {Grebel}}]{Zaritsky_2004}
{Zaritsky}, D., {Harris}, J., {Thompson}, I.~B., \& {Grebel}, E.~K. 2004, \aj,
  128, 1606

\bibitem[{{Zaritsky} {et~al.}(2002){Zaritsky}, {Harris}, {Thompson}, {Grebel},
  \& {Massey}}]{Zaritsky_2002}
{Zaritsky}, D., {Harris}, J., {Thompson}, I.~B., {Grebel}, E.~K., \& {Massey},
  P. 2002, \aj, 123, 855

\end{thebibliography}


\begin{appendix} 

\section{Results of the fitting of the SEDs}
\label{Sect:AppSED}

Table~\ref{table:params} lists in Cols. 1-3 the OGLE name, type, and subtype of the object.
Column 4 gives the pulsation period in days as reported by OGLE.
The remaining columns refer to the results of fitting the SEDs: luminosity with error, 
effective temperature with error, dust optical depth, dust temperature at the inner radius, 
and whether this parameter was fixed (fit=0) or fitted (fit=1).
The remarks column contains additional information, i.e. either regarding the pulsation period from \citet{MACHO98,MACHO02} 
or the possibility of (weak) IR excess emission not considered in the fitting.

\begin{table*}
\caption{Fit parameters}
\label{table:params}

\begin{tabular}{lrrrrrrrrlrrrrr}
\hline
Name               & Type  & Subtype & Period &   Luminosity  &    $T_{\rm eff}$  & $\tau$ & $T_{\rm c}$ & fit  & Remarks \\ 
                   &       &          & (d)    &   (\lsol)     &         (K)     &        &   (K)     &      &  \\
\hline
OGLE-LMC-ACEP-001 & ANCEP & F     &   0.850 &   78 $\pm$    3 & 6125 $\pm$  188 & 0.000 & 1000 & 0 &  \\ 
OGLE-LMC-ACEP-002 & ANCEP & F     &   0.977 &  119 $\pm$    5 & 6250 $\pm$  188 & 0.000 & 1000 & 0 &  \\ 
OGLE-LMC-ACEP-003 & ANCEP & 1O    &   0.382 &   66 $\pm$    2 & 6750 $\pm$  250 & 0.000 & 1000 & 0 &  \\ 
OGLE-LMC-ACEP-004 & ANCEP & F     &   1.862 &  200 $\pm$   17 & 6000 $\pm$  812 & 0.000 & 1000 & 0 &  \\ 
OGLE-LMC-ACEP-005 & ANCEP & F     &   0.932 &   92 $\pm$    1 & 6125 $\pm$  188 & 0.000 & 1000 & 0 &  \\ 
OGLE-LMC-ACEP-006 & ANCEP & 1O    &   0.850 &  200 $\pm$    5 & 6250 $\pm$  188 & 0.000 & 1000 & 0 &  \\ 
OGLE-LMC-ACEP-007 & ANCEP & F     &   0.896 &  114 $\pm$    4 & 6375 $\pm$  250 & 0.000 & 1000 & 0 &  \\ 
OGLE-LMC-ACEP-008 & ANCEP & 1O    &   0.749 &  168 $\pm$    5 & 6375 $\pm$  125 & 0.000 & 1000 & 0 &  \\ 
OGLE-LMC-ACEP-009 & ANCEP & 1O    &   0.800 &  153 $\pm$    3 & 6250 $\pm$  125 & 0.000 & 1000 & 0 &  \\ 
OGLE-LMC-ACEP-010 & ANCEP & F     &   0.834 &   81 $\pm$    3 & 6250 $\pm$  312 & 0.000 & 1000 & 0 &  \\ 
OGLE-LMC-ACEP-011 & ANCEP & F     &   0.999 &  118 $\pm$    7 & 5625 $\pm$  250 & 0.000 & 1000 & 0 &  \\ 
OGLE-LMC-ACEP-012 & ANCEP & F     &   0.829 &  107 $\pm$   12 & 5875 $\pm$ 1062 & 0.000 & 1000 & 0 &  \\ 
OGLE-LMC-ACEP-013 & ANCEP & 1O    &   0.501 &   77 $\pm$    3 & 6250 $\pm$  500 & 0.000 & 1000 & 0 &  \\ 
OGLE-LMC-ACEP-014 & ANCEP & F     &   2.291 &  299 $\pm$    7 & 6125 $\pm$  125 & 0.000 & 1000 & 0 &  \\ 
OGLE-LMC-ACEP-015 & ANCEP & 1O    &   1.181 &  218 $\pm$    4 & 6375 $\pm$  188 & 0.000 & 1000 & 0 &  \\ 
OGLE-LMC-ACEP-016 & ANCEP & F     &   1.546 &  214 $\pm$    9 & 6250 $\pm$  188 & 0.000 & 1000 & 0 &  \\ 
OGLE-LMC-ACEP-017 & ANCEP & F     &   0.930 &  127 $\pm$    3 & 6250 $\pm$  125 & 0.000 & 1000 & 0 &  \\ 
OGLE-LMC-ACEP-018 & ANCEP & F     &   1.019 &  124 $\pm$    4 & 6500 $\pm$  312 & 0.000 & 1000 & 0 &  \\ 
OGLE-LMC-ACEP-019 & ANCEP & F     &   0.909 &  159 $\pm$    3 & 6750 $\pm$  188 & 0.000 & 1000 & 0 &  \\ 
OGLE-LMC-ACEP-020 & ANCEP & 1O    &   0.382 &  100 $\pm$    4 & 7375 $\pm$  312 & 0.000 & 1000 & 0 &  \\ 
OGLE-LMC-ACEP-021 & ANCEP & F     &   1.296 &  159 $\pm$    7 & 6125 $\pm$  562 & 0.000 & 1000 & 0 &  \\ 
OGLE-LMC-ACEP-023 & ANCEP & 1O    &   0.723 &  190 $\pm$    9 & 6375 $\pm$  188 & 0.000 & 1000 & 0 &  \\ 
OGLE-LMC-ACEP-024 & ANCEP & F     &   0.794 &  139 $\pm$    3 & 7500 $\pm$  125 & 0.000 & 1000 & 0 &  \\ 
OGLE-LMC-ACEP-025 & ANCEP & 1O    &   0.474 &  100 $\pm$    6 & 7000 $\pm$  500 & 0.000 & 1000 & 0 &  \\ 
OGLE-LMC-ACEP-026 & ANCEP & F     &   1.739 &  228 $\pm$    3 & 5875 $\pm$  125 & 0.000 & 1000 & 0 &  \\ 
OGLE-LMC-ACEP-027 & ANCEP & F     &   1.267 &  222 $\pm$   10 & 6500 $\pm$  312 & 0.000 & 1000 & 0 &  \\ 
OGLE-LMC-ACEP-028 & ANCEP & 1O    &   0.599 &  111 $\pm$    5 & 5375 $\pm$  312 & 0.000 & 1000 & 0 &  \\ 
OGLE-LMC-ACEP-029 & ANCEP & F     &   0.802 &   77 $\pm$    2 & 6625 $\pm$  250 & 0.000 & 1000 & 0 &  \\ 
OGLE-LMC-ACEP-030 & ANCEP & 1O    &   0.667 &  173 $\pm$    6 & 6500 $\pm$  250 & 0.000 & 1000 & 0 &  \\ 
OGLE-LMC-ACEP-031 & ANCEP & 1O    &   0.840 &  167 $\pm$    7 & 6375 $\pm$  438 & 0.000 & 1000 & 0 &  \\ 
OGLE-LMC-ACEP-032 & ANCEP & F     &   1.316 &  174 $\pm$    4 & 6000 $\pm$  125 & 0.000 & 1000 & 0 &  \\ 
OGLE-LMC-ACEP-033 & ANCEP & F     &   2.347 &  294 $\pm$    5 & 6000 $\pm$  125 & 0.000 & 1000 & 0 &  \\ 
OGLE-LMC-ACEP-034 & ANCEP & F     &   0.734 &   96 $\pm$    8 & 6375 $\pm$  812 & 0.000 & 1000 & 0 &  \\ 
OGLE-LMC-ACEP-035 & ANCEP & 1O    &   0.446 &   76 $\pm$    1 & 6625 $\pm$  125 & 0.000 & 1000 & 0 &  \\ 
OGLE-LMC-ACEP-036 & ANCEP & F     &   1.258 &  170 $\pm$    3 & 6000 $\pm$  125 & 0.000 & 1000 & 0 &  \\ 
OGLE-LMC-ACEP-037 & ANCEP & F     &   1.258 &  258 $\pm$   17 & 5125 $\pm$  375 & 0.000 & 1000 & 0 &  \\ 
OGLE-LMC-ACEP-038 & ANCEP & F     &   1.335 &  164 $\pm$    4 & 6375 $\pm$  188 & 0.000 & 1000 & 0 &  \\ 
OGLE-LMC-ACEP-039 & ANCEP & F     &   0.992 &  115 $\pm$    1 & 6375 $\pm$   62 & 0.000 & 1000 & 0 &  \\ 
OGLE-LMC-ACEP-040 & ANCEP & F     &   0.961 &  136 $\pm$    5 & 6000 $\pm$  375 & 0.000 & 1000 & 0 &  \\ 
OGLE-LMC-ACEP-041 & ANCEP & F     &   0.878 &  101 $\pm$    7 & 6250 $\pm$  688 & 0.000 & 1000 & 0 &  \\ 
OGLE-LMC-ACEP-042 & ANCEP & F     &   1.079 &   82 $\pm$    3 & 5000 $\pm$  125 & 0.000 & 1000 & 0 &  \\ 
OGLE-LMC-ACEP-043 & ANCEP & 1O    &   0.506 &   85 $\pm$    1 & 6000 $\pm$  125 & 0.000 & 1000 & 0 &  \\ 
OGLE-LMC-ACEP-044 & ANCEP & F     &   1.309 &  229 $\pm$   11 & 6750 $\pm$  312 & 0.000 & 1000 & 0 &  \\ 
OGLE-LMC-ACEP-045 & ANCEP & F     &   0.678 &   59 $\pm$    1 & 6125 $\pm$  250 & 0.000 & 1000 & 0 &  \\ 
OGLE-LMC-ACEP-046 & ANCEP & F     &   1.264 &  179 $\pm$   11 & 6375 $\pm$  500 & 0.000 & 1000 & 0 &  \\ 
OGLE-LMC-ACEP-047 & ANCEP & F     &   2.178 &  237 $\pm$    8 & 6375 $\pm$  188 & 0.000 & 1000 & 0 &  \\ 
OGLE-LMC-ACEP-048 & ANCEP & F     &   1.546 &  230 $\pm$   12 & 6000 $\pm$  438 & 0.000 & 1000 & 0 &  \\ 
OGLE-LMC-ACEP-049 & ANCEP & F     &   0.645 &   89 $\pm$    6 & 6875 $\pm$  562 & 0.000 & 1000 & 0 &  \\ 
OGLE-LMC-ACEP-050 & ANCEP & 1O    &   1.045 &  335 $\pm$    9 & 6750 $\pm$  125 & 0.000 & 1000 & 0 &  \\ 
OGLE-LMC-ACEP-051 & ANCEP & F     &   0.709 &   68 $\pm$    1 & 6125 $\pm$   62 & 0.000 & 1000 & 0 &  \\ 
OGLE-LMC-ACEP-052 & ANCEP & F     &   1.263 &  191 $\pm$    9 & 6125 $\pm$  312 & 0.000 & 1000 & 0 &  \\ 
OGLE-LMC-ACEP-053 & ANCEP & F     &   1.888 &  267 $\pm$   12 & 5750 $\pm$  375 & 0.000 & 1000 & 0 &  \\ 
OGLE-LMC-ACEP-054 & ANCEP & F     &   0.980 &   82 $\pm$    2 & 5375 $\pm$  188 & 0.000 & 1000 & 0 &  \\ 
OGLE-LMC-ACEP-055 & ANCEP & F     &   1.607 &  188 $\pm$   10 & 5875 $\pm$  375 & 0.000 & 1000 & 0 &  \\ 
OGLE-LMC-ACEP-056 & ANCEP & F     &   1.124 &  144 $\pm$    7 & 6125 $\pm$  500 & 0.000 & 1000 & 0 &  \\ 
OGLE-LMC-ACEP-057 & ANCEP & F     &   1.710 &  250 $\pm$    6 & 6000 $\pm$  188 & 0.000 & 1000 & 0 &  \\ 
OGLE-LMC-ACEP-058 & ANCEP & 1O    &   0.485 &  142 $\pm$    9 & 7875 $\pm$  812 & 0.000 & 1000 & 0 &  \\ 
OGLE-LMC-ACEP-059 & ANCEP & F     &   0.835 &   42 $\pm$    2 & 4500 $\pm$  188 & 0.000 & 1000 & 0 &  \\ 
OGLE-LMC-ACEP-060 & ANCEP & F     &   1.276 &  221 $\pm$    4 & 6375 $\pm$  250 & 0.000 & 1000 & 0 &  \\ 

\hline
\end{tabular}
\end{table*}

\setcounter{table}{0}
\begin{table*}
\setlength{\tabcolsep}{1.0mm}
\caption{Continued}

\begin{tabular}{lrrrrrrrrlrrrrrr}
\hline
Name               & Type  & Subtype & Period &    Luminosity   &       $T_{\rm eff}$ & $\tau$ & $T_{\rm c}$ & fit  &  Remarks \\ 
                   &       &          & (d)    &    (\lsol)      &            (K)    &        &   (K)     &      &  \\
\hline
OGLE-LMC-ACEP-061 & ANCEP & F     &   0.848 &   99 $\pm$    2 & 6125 $\pm$  188 & 0.000 & 1000 & 0 &  \\ 
OGLE-LMC-ACEP-062 & ANCEP & F     &   1.059 &  216 $\pm$   14 & 7000 $\pm$  500 & 0.000 & 1000 & 0 &  \\ 
OGLE-LMC-ACEP-063 & ANCEP & F     &   0.893 &   70 $\pm$    2 & 5625 $\pm$  250 & 0.000 & 1000 & 0 &  \\ 
OGLE-LMC-ACEP-064 & ANCEP & F     &   1.357 &  203 $\pm$    4 & 6250 $\pm$  250 & 0.000 & 1000 & 0 &  \\ 
OGLE-LMC-ACEP-065 & ANCEP & F     &   1.322 &  195 $\pm$   10 & 6500 $\pm$  438 & 0.000 & 1000 & 0 &  \\ 
OGLE-LMC-ACEP-066 & ANCEP & F     &   1.040 &  124 $\pm$    3 & 6250 $\pm$  250 & 0.000 & 1000 & 0 &  \\ 
OGLE-LMC-ACEP-067 & ANCEP & F     &   0.821 &  101 $\pm$   11 & 5750 $\pm$ 1125 & 0.000 & 1000 & 0 &  \\ 
OGLE-LMC-ACEP-068 & ANCEP & F     &   0.626 &   65 $\pm$    5 & 6875 $\pm$  562 & 0.000 & 1000 & 0 &  IR excess? \\ 
OGLE-LMC-ACEP-069 & ANCEP & F     &   1.538 &  254 $\pm$    7 & 5875 $\pm$  250 & 0.000 & 1000 & 0 &  \\ 
OGLE-LMC-ACEP-070 & ANCEP & 1O    &   0.629 &   98 $\pm$    3 & 6500 $\pm$  312 & 0.000 & 1000 & 0 &  \\ 
OGLE-LMC-ACEP-071 & ANCEP & 1O    &   0.676 &  171 $\pm$   10 & 7125 $\pm$  375 & 0.000 & 1000 & 0 &  \\ 
OGLE-LMC-ACEP-072 & ANCEP & F     &   1.048 &  201 $\pm$    8 & 6500 $\pm$  375 & 0.000 & 1000 & 0 &  \\ 
OGLE-LMC-ACEP-073 & ANCEP & F     &   1.465 &  255 $\pm$   11 & 6375 $\pm$  438 & 0.000 & 1000 & 0 &  \\ 
OGLE-LMC-ACEP-074 & ANCEP & F     &   1.533 &  221 $\pm$    4 & 6250 $\pm$  188 & 0.000 & 1000 & 0 &  \\ 
OGLE-LMC-ACEP-075 & ANCEP & F     &   0.692 &   88 $\pm$    3 & 6500 $\pm$  312 & 0.000 & 1000 & 0 &  \\ 
OGLE-LMC-ACEP-076 & ANCEP & F     &   1.582 &  193 $\pm$    5 & 6250 $\pm$  312 & 0.000 & 1000 & 0 &  \\ 
OGLE-LMC-ACEP-077 & ANCEP & F     &   1.122 &  129 $\pm$    5 & 5875 $\pm$  125 & 0.000 & 1000 & 0 &  \\ 
OGLE-LMC-ACEP-078 & ANCEP & 1O    &   0.857 &  236 $\pm$    5 & 6875 $\pm$  188 & 0.000 & 1000 & 0 &  \\ 
OGLE-LMC-ACEP-079 & ANCEP & F     &   1.155 &  202 $\pm$   10 & 6625 $\pm$  250 & 0.000 & 1000 & 0 &  \\ 
OGLE-LMC-ACEP-080 & ANCEP & F     &   1.057 &  143 $\pm$    3 & 6375 $\pm$  250 & 0.000 & 1000 & 0 &  \\ 
OGLE-LMC-ACEP-081 & ANCEP & F     &   0.801 &   89 $\pm$    3 & 6375 $\pm$  188 & 0.000 & 1000 & 0 &  \\ 
OGLE-LMC-ACEP-082 & ANCEP & 1O    &   0.775 &  154 $\pm$    3 & 5625 $\pm$  125 & 0.000 & 1000 & 0 &  \\ 
OGLE-LMC-T2CEP-001 & T2CEP & BLHer &   1.814 &  101 $\pm$    1 & 6000 $\pm$   62 & 0.000 & 1000 & 0 &  \\ 
OGLE-LMC-T2CEP-002 & T2CEP & WVir  &  18.324 &  629 $\pm$   29 & 5250 $\pm$  312 & 0.000 & 1000 & 0 &  \\ 
OGLE-LMC-T2CEP-003 & T2CEP & RVTau &  35.660 & 4228 $\pm$  191 & 6000 $\pm$  438 & 1.382 & 1300 & 0 &  \\ 
OGLE-LMC-T2CEP-004 & T2CEP & BLHer &   1.916 &  143 $\pm$   12 & 5625 $\pm$  438 & 0.000 & 1000 & 0 &  \\ 
OGLE-LMC-T2CEP-005 & T2CEP & RVTau &  33.185 & 1277 $\pm$  136 & 4875 $\pm$  312 & 0.000 & 1000 & 0 &  \\ 
OGLE-LMC-T2CEP-006 & T2CEP & BLHer &   1.088 &   84 $\pm$    3 & 7000 $\pm$  312 & 0.000 & 1000 & 0 &  \\ 
OGLE-LMC-T2CEP-007 & T2CEP & BLHer &   1.243 &   87 $\pm$    3 & 6750 $\pm$  250 & 0.000 & 1000 & 0 &  \\ 
OGLE-LMC-T2CEP-008 & T2CEP & BLHer &   1.746 &   90 $\pm$    2 & 5875 $\pm$  125 & 0.000 & 1000 & 0 &  \\ 
OGLE-LMC-T2CEP-009 & T2CEP & BLHer &   1.761 &  104 $\pm$    2 & 6250 $\pm$  125 & 0.000 & 1000 & 0 &  \\ 
OGLE-LMC-T2CEP-010 & T2CEP & BLHer &   1.503 &   87 $\pm$    2 & 6500 $\pm$  250 & 0.000 & 1000 & 0 &  \\ 
OGLE-LMC-T2CEP-011 & T2CEP & RVTau &  39.257 & 2893 $\pm$   61 & 5875 $\pm$  125 & 0.045 &  906 & 1 &  \\ 
OGLE-LMC-T2CEP-012 & T2CEP & WVir  &  11.581 &  419 $\pm$   13 & 5250 $\pm$  250 & 0.000 & 1000 & 0 &  \\ 
OGLE-LMC-T2CEP-013 & T2CEP & WVir  &  11.545 &  388 $\pm$    8 & 5250 $\pm$  125 & 0.000 & 1000 & 0 &  \\ 
OGLE-LMC-T2CEP-014 & T2CEP & RVTau &  61.876 & 2325 $\pm$   52 & 5750 $\pm$  125 & 0.078 & 1100 & 0 &  \\ 
OGLE-LMC-T2CEP-015 & T2CEP & RVTau &  56.521 & 2910 $\pm$   53 & 5000 $\pm$  125 & 0.261 & 1200 & 0 & P=56.224 \citep{MACHO98} \\ 
OGLE-LMC-T2CEP-016 & T2CEP & RVTau &  20.296 & 1025 $\pm$   54 & 6750 $\pm$  312 & 0.122 &  600 & 0 &  \\ 
OGLE-LMC-T2CEP-017 & T2CEP & WVir  &  14.455 &  476 $\pm$   19 & 4875 $\pm$  188 & 0.000 & 1000 & 0 &  \\ 
OGLE-LMC-T2CEP-018 & T2CEP & BLHer &   1.380 &   88 $\pm$    1 & 6375 $\pm$  125 & 0.000 & 1000 & 0 &  \\ 
OGLE-LMC-T2CEP-019 & T2CEP & pWVir &   8.675 &  436 $\pm$   35 & 5000 $\pm$  750 & 0.000 & 1000 & 0 &  \\ 
OGLE-LMC-T2CEP-020 & T2CEP & BLHer &   1.108 &   93 $\pm$    4 & 6500 $\pm$  312 & 0.000 & 1000 & 0 &  \\ 
OGLE-LMC-T2CEP-021 & T2CEP & pWVir &   9.760 &  552 $\pm$   14 & 5750 $\pm$  125 & 0.000 & 1000 & 0 & EB \\ 
OGLE-LMC-T2CEP-022 & T2CEP & WVir  &  10.717 &  383 $\pm$    8 & 5250 $\pm$  188 & 0.000 & 1000 & 0 & W3 excess \\ 
OGLE-LMC-T2CEP-023 & T2CEP & pWVir &   5.235 &  837 $\pm$   26 & 6250 $\pm$  250 & 0.000 & 1000 & 0 & EB \\ 
OGLE-LMC-T2CEP-024 & T2CEP & BLHer &   1.247 &   75 $\pm$    3 & 6500 $\pm$  375 & 0.000 & 1000 & 0 &  \\ 
OGLE-LMC-T2CEP-025 & T2CEP & RVTau &  67.965 & 2911 $\pm$  179 & 4875 $\pm$  312 & 0.000 & 1000 & 0 & IR excess?  \\ 
OGLE-LMC-T2CEP-026 & T2CEP & WVir  &  13.578 &  443 $\pm$   10 & 5000 $\pm$  125 & 0.000 & 1000 & 0 &  \\ 
OGLE-LMC-T2CEP-027 & T2CEP & WVir  &  17.134 &  619 $\pm$   25 & 5500 $\pm$  375 & 0.000 & 1000 & 0 & P=17.127 \citep{MACHO98} \\ 
OGLE-LMC-T2CEP-028 & T2CEP & pWVir &   8.785 &  833 $\pm$   36 & 6375 $\pm$  375 & 0.000 & 1000 & 0 &  \\ 
OGLE-LMC-T2CEP-029 & T2CEP & RVTau &  31.245 & 2851 $\pm$   79 & 5750 $\pm$  188 & 1.347 &  746 & 1 & P=31.716 \citep{MACHO98} \\ 
OGLE-LMC-T2CEP-030 & T2CEP & BLHer &   3.935 &  202 $\pm$    5 & 5750 $\pm$  125 & 0.000 & 1000 & 0 &  \\ 
OGLE-LMC-T2CEP-031 & T2CEP & WVir  &   6.706 &  253 $\pm$    4 & 5375 $\pm$  125 & 0.000 & 1000 & 0 &  \\ 
OGLE-LMC-T2CEP-032 & T2CEP & RVTau &  44.561 & 3821 $\pm$  590 & 4625 $\pm$ 1000 & 1.763 &  800 & 0 &  \\ 
OGLE-LMC-T2CEP-033 & T2CEP & pWVir &   9.395 &  605 $\pm$   14 & 5875 $\pm$  250 & 0.000 & 1000 & 0 & P=9.387 \citep{MACHO98} \\ 
OGLE-LMC-T2CEP-034 & T2CEP & WVir  &  14.911 &  411 $\pm$   14 & 4750 $\pm$  125 & 0.000 & 1000 & 0 & P=14.906 \citep{MACHO98} \\ 
OGLE-LMC-T2CEP-035 & T2CEP & WVir  &   9.866 &  384 $\pm$   16 & 5000 $\pm$  250 & 0.000 & 1000 & 0 &  \\ 

\hline
\end{tabular}
\end{table*}

\setcounter{table}{0}
\begin{table*}
\setlength{\tabcolsep}{1.0mm}
\caption{Continued}

\begin{tabular}{lrrrrrrrrlrrrrr}
\hline
Name               & Type  & Subtype & Period &    Luminosity   &       $T_{\rm eff}$ & $\tau$ & $T_{\rm c}$ & fit  &  Remarks \\ 
                   &       &          & (d)    &    (\lsol)      &            (K)    &        &   (K)     &      &  \\
\hline
OGLE-LMC-T2CEP-036 & T2CEP & WVir  &  14.881 &  501 $\pm$   17 & 5625 $\pm$  188 & 0.000 & 1000 & 0 &  \\ 
OGLE-LMC-T2CEP-037 & T2CEP & WVir  &   6.897 &  266 $\pm$    4 & 5625 $\pm$  125 & 0.000 & 1000 & 0 &  \\ 
OGLE-LMC-T2CEP-038 & T2CEP & WVir  &   4.014 &  537 $\pm$   28 & 7250 $\pm$  375 & 0.000 & 1000 & 0 &  \\ 
OGLE-LMC-T2CEP-039 & T2CEP & WVir  &   8.716 &  367 $\pm$   13 & 5625 $\pm$  250 & 0.000 & 1000 & 0 & W3 excess \\ 
OGLE-LMC-T2CEP-040 & T2CEP & pWVir &   9.626 &  639 $\pm$   31 & 5000 $\pm$  375 & 0.000 & 1000 & 0 & P=9.622 \citep{MACHO98} \\ 
OGLE-LMC-T2CEP-041 & T2CEP & BLHer &   2.476 &  290 $\pm$   16 & 7250 $\pm$  562 & 0.000 & 1000 & 0 &  \\ 
OGLE-LMC-T2CEP-042 & T2CEP & pWVir &   4.923 &  384 $\pm$   23 & 6750 $\pm$  562 & 0.000 & 1000 & 0 &  \\ 
OGLE-LMC-T2CEP-043 & T2CEP & WVir  &   6.559 &  211 $\pm$    5 & 5375 $\pm$  188 & 0.000 & 1000 & 0 &  \\ 
OGLE-LMC-T2CEP-044 & T2CEP & WVir  &  13.270 &  439 $\pm$    9 & 5250 $\pm$  188 & 0.000 & 1000 & 0 & P=13.246 \citep{MACHO98} \\ 
OGLE-LMC-T2CEP-045 & T2CEP & RVTau &  63.386 & 3479 $\pm$  130 & 5125 $\pm$  125 & 0.000 & 1000 & 0 & IR excess? \\ 
OGLE-LMC-T2CEP-046 & T2CEP & WVir  &  14.744 &  879 $\pm$   21 & 5250 $\pm$  125 & 0.718 &  809 & 1 & P=14.752 \citep{MACHO98} \\ 
OGLE-LMC-T2CEP-047 & T2CEP & WVir  &   7.286 &  285 $\pm$    7 & 5500 $\pm$  250 & 0.000 & 1000 & 0 &  \\ 
OGLE-LMC-T2CEP-048 & T2CEP & BLHer &   1.445 &   92 $\pm$    8 & 6375 $\pm$  812 & 0.000 & 1000 & 0 &  \\ 
OGLE-LMC-T2CEP-049 & T2CEP & BLHer &   3.235 &  231 $\pm$   19 & 6375 $\pm$  812 & 0.000 & 1000 & 0 &  \\ 
OGLE-LMC-T2CEP-050 & T2CEP & RVTau &  34.748 & 1427 $\pm$   34 & 5875 $\pm$  125 & 0.117 & 1200 & 0 &  \\ 
OGLE-LMC-T2CEP-051 & T2CEP & RVTau &  40.606 & 1850 $\pm$   49 & 5500 $\pm$  188 & 0.000 & 1000 & 0 &  \\ 
OGLE-LMC-T2CEP-052 & T2CEP & pWVir &   4.688 &  448 $\pm$   29 & 7000 $\pm$  500 & 0.000 & 1000 & 0 & EB \\ 
OGLE-LMC-T2CEP-053 & T2CEP & BLHer &   1.043 &   81 $\pm$    2 & 6625 $\pm$  188 & 0.000 & 1000 & 0 &  \\ 
OGLE-LMC-T2CEP-054 & T2CEP & WVir  &   9.925 &  338 $\pm$    4 & 5125 $\pm$   62 & 0.000 & 1000 & 0 &  \\ 
OGLE-LMC-T2CEP-055 & T2CEP & RVTau &  41.005 & 2545 $\pm$   77 & 5750 $\pm$  188 & 0.080 & 1000 & 0 &  \\ 
OGLE-LMC-T2CEP-056 & T2CEP & WVir  &   7.290 &  246 $\pm$    3 & 5125 $\pm$   62 & 0.000 & 1000 & 0 &  \\ 
OGLE-LMC-T2CEP-057 & T2CEP & WVir  &  16.632 &  569 $\pm$   19 & 5125 $\pm$  312 & 0.000 & 1000 & 0 & P=16.602 \citep{MACHO98} \\ 
OGLE-LMC-T2CEP-058 & T2CEP & RVTau &  21.483 &  715 $\pm$   24 & 5125 $\pm$  312 & 0.000 & 1000 & 0 & P=21.486 \citep{MACHO98} \\   
OGLE-LMC-T2CEP-059 & T2CEP & WVir  &  16.736 &  720 $\pm$   26 & 5125 $\pm$  250 & 0.000 & 1000 & 0 & P=16.747 \citep{MACHO98} \\ 
OGLE-LMC-T2CEP-060 & T2CEP & BLHer &   1.237 &   73 $\pm$    4 & 6250 $\pm$  688 & 0.000 & 1000 & 0 &  \\ 
OGLE-LMC-T2CEP-061 & T2CEP & BLHer &   1.182 &   82 $\pm$    5 & 7125 $\pm$  562 & 0.000 & 1000 & 0 &  \\ 
OGLE-LMC-T2CEP-062 & T2CEP & WVir  &   6.047 &  190 $\pm$   19 & 4750 $\pm$  438 & 0.000 & 1000 & 0 &  \\ 
OGLE-LMC-T2CEP-063 & T2CEP & WVir  &   6.925 &  278 $\pm$    8 & 5625 $\pm$  250 & 0.000 & 1000 & 0 &  \\ 
OGLE-LMC-T2CEP-064 & T2CEP & BLHer &   2.128 &  121 $\pm$    8 & 6125 $\pm$  750 & 0.000 & 1000 & 0 &  \\ 
OGLE-LMC-T2CEP-065 & T2CEP & RVTau &  35.055 & 1563 $\pm$   20 & 5375 $\pm$  125 & 0.100 &  120 & 0 &  \\ 
OGLE-LMC-T2CEP-066 & T2CEP & WVir  &  13.109 &  412 $\pm$    6 & 5125 $\pm$   62 & 0.000 &  500 & 0 &  \\ 
OGLE-LMC-T2CEP-067 & T2CEP & RVTau &  48.232 & 6429 $\pm$  305 & 6125 $\pm$  500 & 1.743 & 1200 & 0 & P=48.539 \citep{MACHO98} \\ 
OGLE-LMC-T2CEP-068 & T2CEP & BLHer &   1.609 &  106 $\pm$    2 & 6500 $\pm$  250 & 0.000 & 1000 & 0 &  \\ 
OGLE-LMC-T2CEP-069 & T2CEP & BLHer &   1.021 &   93 $\pm$   14 & 6625 $\pm$ 1062 & 0.000 & 1000 & 0 &  \\ 
OGLE-LMC-T2CEP-070 & T2CEP & WVir  &  15.438 &  675 $\pm$   36 & 5875 $\pm$  312 & 0.000 & 1000 & 0 & W3 excess  \\ 
OGLE-LMC-T2CEP-071 & T2CEP & BLHer &   1.152 &   76 $\pm$    9 & 6375 $\pm$  875 & 0.000 & 1000 & 0 &  \\ 
OGLE-LMC-T2CEP-072 & T2CEP & WVir  &  14.514 &  534 $\pm$   11 & 5375 $\pm$  125 & 0.000 & 1000 & 0 &  \\ 
OGLE-LMC-T2CEP-073 & T2CEP & BLHer &   3.088 &  169 $\pm$    5 & 5875 $\pm$  250 & 0.000 & 1000 & 0 &  \\ 
OGLE-LMC-T2CEP-074 & T2CEP & WVir  &   8.988 &  456 $\pm$    9 & 5625 $\pm$  188 & 0.000 & 1000 & 0 &  \\ 
OGLE-LMC-T2CEP-075 & T2CEP & RVTau &  50.187 & 1868 $\pm$   68 & 5125 $\pm$  125 & 0.087 &  646 & 1 &  \\  
OGLE-LMC-T2CEP-076 & T2CEP & BLHer &   2.104 &   89 $\pm$    3 & 5500 $\pm$  312 & 0.000 & 1000 & 0 &  \\ 
OGLE-LMC-T2CEP-077 & T2CEP & BLHer &   1.214 &  132 $\pm$    5 & 7500 $\pm$  375 & 0.000 & 1000 & 0 & EB \\ 
OGLE-LMC-T2CEP-078 & T2CEP & pWVir &   6.716 &  404 $\pm$   14 & 4875 $\pm$  125 & 0.000 & 1000 & 0 &  \\ 
OGLE-LMC-T2CEP-079 & T2CEP & WVir  &  14.845 &  344 $\pm$   19 & 4875 $\pm$  188 & 0.000 & 1000 & 0 & P=14.855 \citep{MACHO98} \\ 
OGLE-LMC-T2CEP-080 & T2CEP & RVTau &  40.916 & 2395 $\pm$  104 & 5625 $\pm$  250 & 0.113 &  331 & 1 & P=41.118 \citep{MACHO98} \\ 
OGLE-LMC-T2CEP-081 & T2CEP & WVir  &   9.480 &  369 $\pm$    7 & 5375 $\pm$  125 & 0.000 & 1000 & 0 &  \\ 
OGLE-LMC-T2CEP-082 & T2CEP & RVTau &  35.124 & 1127 $\pm$   75 & 5125 $\pm$  312 & 0.000 & 1000 & 0 &  \\ 
OGLE-LMC-T2CEP-083 & T2CEP & pWVir &   5.968 &  284 $\pm$    6 & 5625 $\pm$  125 & 0.000 & 1000 & 0 &  \\ 
OGLE-LMC-T2CEP-084 & T2CEP & BLHer &   1.771 &  260 $\pm$   40 & 7750 $\pm$ 1562 & 0.000 & 1000 & 0 & EB  \\ 
OGLE-LMC-T2CEP-085 & T2CEP & BLHer &   3.405 &  177 $\pm$    6 & 6250 $\pm$  375 & 0.000 & 1000 & 0 &  \\ 
OGLE-LMC-T2CEP-086 & T2CEP & WVir  &  15.845 &  665 $\pm$   15 & 5500 $\pm$  188 & 0.000 & 1000 & 0 &  \\ 
OGLE-LMC-T2CEP-087 & T2CEP & WVir  &   5.185 &  213 $\pm$    9 & 5500 $\pm$  438 & 0.000 & 1000 & 0 &  \\ 
OGLE-LMC-T2CEP-088 & T2CEP & BLHer &   1.951 &  223 $\pm$   11 & 8000 $\pm$  375 & 0.000 & 1000 & 0 &  \\ 
OGLE-LMC-T2CEP-089 & T2CEP & BLHer &   1.167 &   88 $\pm$    2 & 6750 $\pm$  188 & 0.000 & 1000 & 0 &  \\ 
OGLE-LMC-T2CEP-090 & T2CEP & BLHer &   1.479 &   96 $\pm$    2 & 6250 $\pm$  250 & 0.000 & 1000 & 0 &  \\ 

\hline
\end{tabular}
\end{table*}

\setcounter{table}{0}
\begin{table*}
\setlength{\tabcolsep}{1.0mm}
\caption{Continued}

\begin{tabular}{lrrrrrrrrlrrrrr}
\hline
Name               & Type  & Subtype & Period &    Luminosity   &       $T_{\rm eff}$ & $\tau$ & $T_{\rm c}$ & fit  &  Remarks \\ 
                   &       &          & (d)    &    (\lsol)      &            (K)    &        &   (K)     &      &  \\
\hline
OGLE-LMC-T2CEP-091 & T2CEP & RVTau &  35.749 & 3880 $\pm$  319 & 6625 $\pm$  625 & 1.259 & 1100 & 0 &  \\ 
OGLE-LMC-T2CEP-092 & T2CEP & BLHer &   2.617 &  133 $\pm$    7 & 6000 $\pm$  625 & 0.000 & 1000 & 0 &  \\ 
OGLE-LMC-T2CEP-093 & T2CEP & WVir  &  17.593 & 1211 $\pm$   46 & 5875 $\pm$  250 & 0.000 & 1000 & 0 & EB, P=17.560 \citep{MACHO98} \\
                   &       &       &         &                 &                 &       &      &   & P=17.68586 \citep{MACHO02} \\ 
OGLE-LMC-T2CEP-094 & T2CEP & WVir  &   8.468 &  285 $\pm$    8 & 5000 $\pm$  125 & 0.000 & 1000 & 0 &  \\ 
OGLE-LMC-T2CEP-095 & T2CEP & WVir  &   5.000 &  187 $\pm$    3 & 5375 $\pm$   62 & 0.000 & 1000 & 0 &  \\ 
OGLE-LMC-T2CEP-096 & T2CEP & WVir  &  13.926 &  498 $\pm$   15 & 5375 $\pm$  312 & 0.000 & 1000 & 0 & P=13.925 \citep{MACHO98} \\ 
OGLE-LMC-T2CEP-097 & T2CEP & WVir  &  10.510 &  423 $\pm$   10 & 5500 $\pm$  250 & 0.000 & 1000 & 0 & P=10.509 \citep{MACHO98} \\ 
OGLE-LMC-T2CEP-098 & T2CEP & pWVir &   4.974 & 2857 $\pm$  169 & 7375 $\pm$  312 & 0.000 & 1000 & 0 & EB, P=4.97371 \citep{MACHO02} \\ 
OGLE-LMC-T2CEP-099 & T2CEP & WVir  &  15.487 &  516 $\pm$   17 & 4625 $\pm$  188 & 0.000 & 1000 & 0 &  \\ 
OGLE-LMC-T2CEP-100 & T2CEP & WVir  &   7.431 &  265 $\pm$    5 & 5875 $\pm$  188 & 0.000 & 1000 & 0 &  \\ 
OGLE-LMC-T2CEP-101 & T2CEP & WVir  &  11.419 &  499 $\pm$   17 & 5875 $\pm$  375 & 0.000 & 1000 & 0 & P=11.442 \citep{MACHO98} \\ 
OGLE-LMC-T2CEP-102 & T2CEP & BLHer &   1.266 &  115 $\pm$    5 & 6875 $\pm$  438 & 0.000 & 1000 & 0 &  \\ 
OGLE-LMC-T2CEP-103 & T2CEP & WVir  &  12.908 &  454 $\pm$   12 & 5375 $\pm$  250 & 0.000 & 1000 & 0 & P=12.902 \citep{MACHO98}  \\ 
OGLE-LMC-T2CEP-104 & T2CEP & RVTau &  24.880 & 1889 $\pm$   54 & 5500 $\pm$  438 & 1.168 & 1100 & 0 & P=24.848 \citep{MACHO98}  \\ 
OGLE-LMC-T2CEP-105 & T2CEP & BLHer &   1.489 &  117 $\pm$    6 & 6500 $\pm$  625 & 0.000 & 1000 & 0 &  \\ 
OGLE-LMC-T2CEP-106 & T2CEP & WVir  &   6.707 &  272 $\pm$    6 & 5500 $\pm$  188 & 0.000 & 1000 & 0 &  \\ 
OGLE-LMC-T2CEP-107 & T2CEP & BLHer &   1.209 &  105 $\pm$    5 & 5875 $\pm$  562 & 0.000 & 1000 & 0 &  \\ 
OGLE-LMC-T2CEP-108 & T2CEP & RVTau &  30.011 & 1654 $\pm$   37 & 5750 $\pm$  125 & 0.000 & 1000 & 0 &  \\ 
OGLE-LMC-T2CEP-109 & T2CEP & BLHer &   1.415 &   18 $\pm$    1 & 4125 $\pm$  112 & 0.000 & 1000 & 0 &  \\ 
OGLE-LMC-T2CEP-110 & T2CEP & WVir  &   7.078 &  242 $\pm$    7 & 5250 $\pm$  250 & 0.000 & 1000 & 0 &  \\ 
OGLE-LMC-T2CEP-111 & T2CEP & WVir  &   7.496 &  289 $\pm$    5 & 5500 $\pm$  125 & 0.000 & 1000 & 0 &  \\ 
OGLE-LMC-T2CEP-112 & T2CEP & RVTau &  39.398 & 3186 $\pm$  137 & 6000 $\pm$  188 & 0.091 & 1200 & 0 &  \\ 
OGLE-LMC-T2CEP-113 & T2CEP & BLHer &   3.085 &  267 $\pm$   26 & 6625 $\pm$  688 & 0.000 & 1000 & 0 &  \\ 
OGLE-LMC-T2CEP-114 & ANCEP & F     &   1.091 &   68 $\pm$    1 & 5250 $\pm$  188 & 0.000 & 1000 & 0 & \citet{Soszynski_2015} \\ 
OGLE-LMC-T2CEP-115 & T2CEP & RVTau &  24.967 &  768 $\pm$   29 & 5000 $\pm$  188 & 0.000 & 1000 & 0 & P=24.935 \citep{MACHO98} \\ 
OGLE-LMC-T2CEP-116 & T2CEP & BLHer &   1.967 &   79 $\pm$    3 & 5500 $\pm$  312 & 0.000 & 1000 & 0 &  \\ 
OGLE-LMC-T2CEP-117 & T2CEP & WVir  &   6.629 &  258 $\pm$    6 & 5500 $\pm$  125 & 0.000 & 1000 & 0 &  \\ 
OGLE-LMC-T2CEP-118 & T2CEP & WVir  &  12.699 &  428 $\pm$   12 & 5125 $\pm$  188 & 0.000 & 1000 & 0 & P=12.704 \citep{MACHO98} \\ 
OGLE-LMC-T2CEP-119 & T2CEP & RVTau &  33.825 & 3325 $\pm$  290 & 6250 $\pm$  625 & 1.440 & 1200 & 0 &  \\ 
OGLE-LMC-T2CEP-120 & T2CEP & WVir  &   4.559 &  185 $\pm$    4 & 5500 $\pm$  188 & 0.000 & 1000 & 0 &  \\ 
OGLE-LMC-T2CEP-121 & T2CEP & BLHer &   2.061 &  104 $\pm$    5 & 5875 $\pm$  562 & 0.000 & 1000 & 0 &  \\ 
OGLE-LMC-T2CEP-122 & T2CEP & BLHer &   1.539 &   63 $\pm$    2 & 5750 $\pm$  312 & 0.000 & 1000 & 0 &  \\ 
OGLE-LMC-T2CEP-123 & T2CEP & BLHer &   1.003 &   84 $\pm$    8 & 5000 $\pm$  438 & 0.000 & 1000 & 0 &  \\ 
OGLE-LMC-T2CEP-124 & T2CEP & BLHer &   1.735 &   83 $\pm$    4 & 6000 $\pm$  438 & 0.000 & 1000 & 0 &  \\ 
OGLE-LMC-T2CEP-125 & T2CEP & RVTau &  33.034 & 1208 $\pm$   53 & 5125 $\pm$  250 & 0.000 & 1000 & 0 & IR excess? \\ 
OGLE-LMC-T2CEP-126 & T2CEP & WVir  &  16.327 &  467 $\pm$   26 & 4750 $\pm$  125 & 0.981 &  313 & 1 &  \\ 
OGLE-LMC-T2CEP-127 & T2CEP & WVir  &  12.669 &  536 $\pm$   62 & 5500 $\pm$  500 & 0.305 &  363 & 1 &  \\ 
OGLE-LMC-T2CEP-128 & T2CEP & WVir  &  18.493 &  834 $\pm$   38 & 5125 $\pm$  250 & 0.000 & 1000 & 0 &  \\ 
OGLE-LMC-T2CEP-129 & T2CEP & RVTau &  62.509 & 3132 $\pm$   80 & 6000 $\pm$  125 & 0.091 &  700 & 0 &  \\ 
OGLE-LMC-T2CEP-130 & T2CEP & BLHer &   1.945 &  123 $\pm$    8 & 6375 $\pm$  688 & 0.000 & 1000 & 0 &  \\ 
OGLE-LMC-T2CEP-131 & T2CEP & BLHer &   1.413 &   65 $\pm$    1 & 6000 $\pm$  125 & 0.000 & 1000 & 0 &  \\ 
OGLE-LMC-T2CEP-132 & T2CEP & pWVir &  10.018 &  540 $\pm$   17 & 5625 $\pm$  312 & 0.000 & 1000 & 0 &  \\ 
OGLE-LMC-T2CEP-133 & T2CEP & WVir  &   6.282 &  269 $\pm$    7 & 5750 $\pm$  250 & 0.000 & 1000 & 0 &  \\ 
OGLE-LMC-T2CEP-134 & T2CEP & pWVir &   4.076 &  406 $\pm$    8 & 6125 $\pm$  250 & 0.000 & 1000 & 0 &  \\ 
OGLE-LMC-T2CEP-135 & T2CEP & RVTau &  26.522 & 1047 $\pm$   24 & 5000 $\pm$  188 & 0.000 & 1000 & 0 &  P=26.594 \citep{MACHO98} \\ 
OGLE-LMC-T2CEP-136 & T2CEP & BLHer &   1.323 &  163 $\pm$   21 & 5625 $\pm$ 1188 & 0.000 & 1000 & 0 &  \\ 
OGLE-LMC-T2CEP-137 & T2CEP & WVir  &   6.362 &  265 $\pm$    9 & 5625 $\pm$  375 & 0.000 & 1000 & 0 &  \\ 
OGLE-LMC-T2CEP-138 & T2CEP & BLHer &   1.394 &   79 $\pm$   12 & 5375 $\pm$ 1312 & 0.000 & 1000 & 0 &  \\ 
OGLE-LMC-T2CEP-139 & T2CEP & WVir  &  14.780 &  484 $\pm$   13 & 5000 $\pm$  125 & 0.000 & 1000 & 0 &  \\ 
OGLE-LMC-T2CEP-140 & T2CEP & BLHer &   1.841 &  103 $\pm$    3 & 6000 $\pm$  312 & 0.000 & 1000 & 0 &  \\ 
OGLE-LMC-T2CEP-141 & T2CEP & BLHer &   1.823 &   73 $\pm$    3 & 5875 $\pm$  438 & 0.000 & 1000 & 0 &  \\ 
OGLE-LMC-T2CEP-142 & T2CEP & BLHer &   1.761 &  108 $\pm$    2 & 5500 $\pm$   62 & 0.000 & 1000 & 0 &  \\ 
OGLE-LMC-T2CEP-143 & T2CEP & WVir  &  14.570 &  548 $\pm$   18 & 5750 $\pm$  312 & 0.000 & 1000 & 0 &  \\ 
OGLE-LMC-T2CEP-144 & T2CEP & BLHer &   1.937 &  103 $\pm$   26 & 5375 $\pm$ 1312 & 0.000 & 1000 & 0 &  \\ 
OGLE-LMC-T2CEP-145 & T2CEP & BLHer &   3.337 &  267 $\pm$   21 & 6500 $\pm$  688 & 0.000 & 1000 & 0 &  \\ 
OGLE-LMC-T2CEP-146 & T2CEP & WVir  &  10.080 &  298 $\pm$   11 & 5000 $\pm$  188 & 0.000 & 1000 & 0 &  \\ 
OGLE-LMC-T2CEP-147 & T2CEP & RVTau &  46.796 & 7160 $\pm$  259 & 6375 $\pm$  312 & 1.528 &  621 & 1 & P=46.542 \citep{MACHO98} \\ 
OGLE-LMC-T2CEP-148 & T2CEP & BLHer &   2.672 &  135 $\pm$    4 & 6250 $\pm$  250 & 0.000 & 1000 & 0 &  \\ 
OGLE-LMC-T2CEP-149 & T2CEP & RVTau &  42.481 & 2741 $\pm$  117 & 5750 $\pm$  250 & 0.000 & 1000 & 0 & P=42.079 \citep{MACHO98} \\ 
OGLE-LMC-T2CEP-150 & T2CEP & WVir  &   5.493 &  496 $\pm$    9 & 6500 $\pm$  125 & 0.152 & 1100 & 0 &  \\ 
\hline
\end{tabular}
\end{table*}

\setcounter{table}{0}
\begin{table*}
\setlength{\tabcolsep}{1.0mm}
\caption{Continued}

\begin{tabular}{lrrrrrrrrlrrrrr}
\hline
Name               & Type  & Subtype & Period &    Luminosity   &       $T_{\rm eff}$ & $\tau$ & $T_{\rm c}$ & fit  &  Remarks \\ 
                   &       &          & (d)    &    (\lsol)      &            (K)    &        &   (K)     &      &  \\
\hline
OGLE-LMC-T2CEP-151 & T2CEP & WVir  &   7.887 &  311 $\pm$    7 & 5500 $\pm$  250 & 0.000 & 1000 & 0 &  \\ 
OGLE-LMC-T2CEP-152 & T2CEP & WVir  &   9.315 &  356 $\pm$   12 & 5250 $\pm$  312 & 0.000 & 1000 & 0 & P=9.309 \citep{MACHO98} \\ 
OGLE-LMC-T2CEP-153 & T2CEP & BLHer &   1.175 &  465 $\pm$   19 & 8000 $\pm$  250 & 0.000 & 1000 & 0 &  \\ 
OGLE-LMC-T2CEP-154 & T2CEP & pWVir &   7.578 & 1071 $\pm$   28 & 6750 $\pm$  125 & 0.000 & 1000 & 0 &  \\ 
OGLE-LMC-T2CEP-155 & T2CEP & WVir  &   6.898 &  282 $\pm$   13 & 5000 $\pm$  438 & 0.000 & 1000 & 0 &  \\ 
OGLE-LMC-T2CEP-156 & T2CEP & WVir  &  15.387 &  581 $\pm$   79 & 4875 $\pm$  188 & 0.530 &  151 & 1 & P=15.391 \citep{MACHO98} \\ 
OGLE-LMC-T2CEP-157 & T2CEP & WVir  &  14.335 &  431 $\pm$   10 & 5000 $\pm$  125 & 0.000 & 1000 & 0 & P=14.337 \citep{MACHO98} \\ 
OGLE-LMC-T2CEP-158 & T2CEP & WVir  &   7.139 &  270 $\pm$    9 & 5500 $\pm$  188 & 0.150 &  300 & 0 &  \\ 
OGLE-LMC-T2CEP-159 & T2CEP & WVir  &   6.626 &  221 $\pm$    3 & 5125 $\pm$   62 & 0.000 & 1000 & 0 &  \\ 
OGLE-LMC-T2CEP-160 & T2CEP & BLHer &   1.757 &   90 $\pm$    3 & 5875 $\pm$  375 & 0.000 & 1000 & 0 &  \\ 
OGLE-LMC-T2CEP-161 & T2CEP & WVir  &   8.532 &  548 $\pm$   26 & 5125 $\pm$  312 & 0.000 & 1000 & 0 &  \\ 
OGLE-LMC-T2CEP-162 & T2CEP & RVTau &  30.394 & 1109 $\pm$   44 & 5000 $\pm$  188 & 0.183 & 1000 & 0 & P=30.408 \citep{MACHO98} \\ 
OGLE-LMC-T2CEP-163 & T2CEP & BLHer &   1.694 &  140 $\pm$   16 & 6250 $\pm$  875 & 0.000 & 1000 & 0 &  \\ 
OGLE-LMC-T2CEP-164 & T2CEP & pWVir &   8.495 &  550 $\pm$   14 & 5500 $\pm$  250 & 0.483 & 1300 & 0 &  \\ 
OGLE-LMC-T2CEP-165 & T2CEP & BLHer &   1.241 &   35 $\pm$    1 & 4875 $\pm$  125 & 0.000 & 1000 & 0 &  \\ 
OGLE-LMC-T2CEP-166 & T2CEP & BLHer &   2.111 &  211 $\pm$    7 & 5625 $\pm$  312 & 0.000 & 1000 & 0 &  \\ 
OGLE-LMC-T2CEP-167 & T2CEP & BLHer &   2.312 &   99 $\pm$    4 & 5375 $\pm$  438 & 0.000 & 1000 & 0 &  \\ 
OGLE-LMC-T2CEP-168 & T2CEP & WVir  &  15.698 &  554 $\pm$   15 & 5250 $\pm$  125 & 0.000 & 1000 & 0 &  \\ 
OGLE-LMC-T2CEP-169 & T2CEP & RVTau &  30.956 & 1893 $\pm$  379 & 6250 $\pm$  625 & 0.442 &  477 & 1 & P=31.127 \citep{MACHO98} \\ 
OGLE-LMC-T2CEP-170 & T2CEP & WVir  &   7.683 &  239 $\pm$    3 & 5125 $\pm$   62 & 0.000 & 1000 & 0 &  \\ 
OGLE-LMC-T2CEP-171 & T2CEP & BLHer &   1.555 &  109 $\pm$    6 & 6375 $\pm$  562 & 0.000 & 1000 & 0 &  \\ 
OGLE-LMC-T2CEP-172 & T2CEP & WVir  &  11.221 &  334 $\pm$   21 & 4875 $\pm$  250 & 0.000 & 1000 & 0 &  \\ 
OGLE-LMC-T2CEP-173 & T2CEP & WVir  &   4.148 &   55 $\pm$    3 & 3900 $\pm$  112 & 0.000 & 1000 & 0 &  \\ 
OGLE-LMC-T2CEP-174 & T2CEP & RVTau &  46.819 & 6549 $\pm$  342 & 6000 $\pm$  438 & 1.333 & 1100 & 0 & P=47.019 \citep{MACHO98} \\ 
OGLE-LMC-T2CEP-175 & T2CEP & WVir  &   9.326 &  331 $\pm$    6 & 5250 $\pm$  125 & 0.000 & 1000 & 0 &  \\ 
OGLE-LMC-T2CEP-176 & T2CEP & WVir  &   7.990 &  315 $\pm$    4 & 5375 $\pm$  125 & 0.000 & 1000 & 0 &  \\ 
OGLE-LMC-T2CEP-177 & T2CEP & WVir  &  15.036 &  440 $\pm$   14 & 5000 $\pm$  188 & 0.000 & 1000 & 0 &  \\ 
OGLE-LMC-T2CEP-178 & T2CEP & WVir  &  12.212 &  319 $\pm$   11 & 5000 $\pm$  188 & 0.000 & 1000 & 0 &  \\ 
OGLE-LMC-T2CEP-179 & T2CEP & WVir  &   8.050 &  230 $\pm$    4 & 5000 $\pm$   62 & 0.000 & 1000 & 0 &  \\ 
OGLE-LMC-T2CEP-180 & T2CEP & RVTau &  30.996 & 3139 $\pm$  182 & 5500 $\pm$  500 & 1.467 &  700 & 0 &  \\ 
OGLE-LMC-T2CEP-181 & T2CEP & pWVir &   7.213 &  427 $\pm$   24 & 5375 $\pm$  312 & 0.000 & 1000 & 0 &  \\ 
OGLE-LMC-T2CEP-182 & T2CEP & WVir  &   8.226 &  372 $\pm$    9 & 5250 $\pm$  125 & 0.000 & 1000 & 0 & P=8.240 \citep{MACHO98} \\ 
OGLE-LMC-T2CEP-183 & T2CEP & WVir  &   6.510 &  145 $\pm$    8 & 4625 $\pm$  125 & 0.000 & 1000 & 0 &  \\ 
OGLE-LMC-T2CEP-184 & T2CEP & WVir  &  14.840 &  309 $\pm$   16 & 4625 $\pm$  188 & 0.000 & 1000 & 0 &  \\ 
OGLE-LMC-T2CEP-185 & T2CEP & WVir  &  12.688 & 1842 $\pm$   33 & 4625 $\pm$   62 & 0.000 & 1000 & 0 &  \\ 
OGLE-LMC-T2CEP-186 & T2CEP & WVir  &  16.362 &  519 $\pm$   16 & 4875 $\pm$  125 & 0.000 & 1000 & 0 &  \\ 
OGLE-LMC-T2CEP-187 & T2CEP & BLHer &   2.404 &   96 $\pm$    4 & 5500 $\pm$  375 & 0.000 & 1000 & 0 &  \\ 
OGLE-LMC-T2CEP-188 & T2CEP & BLHer &   1.049 &   87 $\pm$    4 & 6125 $\pm$  375 & 0.000 & 1000 & 0 &  \\ 
OGLE-LMC-T2CEP-189 & T2CEP & BLHer &   1.308 &   79 $\pm$    2 & 6250 $\pm$  125 & 0.000 & 1000 & 0 &  \\ 
OGLE-LMC-T2CEP-190 & T2CEP & RVTau &  38.362 & 2450 $\pm$   56 & 5625 $\pm$  188 & 0.019 &  700 & 0 &  \\ 
OGLE-LMC-T2CEP-191 & T2CEP & RVTau &  34.345 & 3969 $\pm$  127 & 5750 $\pm$  250 & 1.251 &  700 & 0 &  \\ 
OGLE-LMC-T2CEP-192 & T2CEP & RVTau &  26.194 &  916 $\pm$   32 & 5375 $\pm$  250 & 0.000 & 1000 & 0 &  \\ 
OGLE-LMC-T2CEP-193 & T2CEP & WVir  &   7.005 &  273 $\pm$    4 & 5250 $\pm$   62 & 0.000 & 1000 & 0 &  \\ 
OGLE-LMC-T2CEP-194 & T2CEP & BLHer &   1.314 &  101 $\pm$    5 & 6375 $\pm$  250 & 0.000 & 1000 & 0 &  \\ 
OGLE-LMC-T2CEP-195 & T2CEP & BLHer &   2.753 &  141 $\pm$    5 & 5875 $\pm$  312 & 0.000 & 1000 & 0 &  \\ 
OGLE-LMC-T2CEP-196 & T2CEP & WVir  &  14.958 &  668 $\pm$   20 & 5125 $\pm$  188 & 0.000 & 1000 & 0 &  \\ 
OGLE-LMC-T2CEP-197 & T2CEP & BLHer &   1.224 &   96 $\pm$    7 & 6375 $\pm$  438 & 0.000 & 1000 & 0 &  \\ 
OGLE-LMC-T2CEP-198 & T2CEP & RVTau &  38.274 &  908 $\pm$   32 & 4500 $\pm$  125 & 0.000 & 1000 & 0 &  \\ 
OGLE-LMC-T2CEP-199 & T2CEP & RVTau &  37.204 & 7090 $\pm$  338 & 8600 $\pm$  300 & 0.024 & 1000 & 0 &  \\ 
OGLE-LMC-T2CEP-200 & T2CEP & RVTau &  34.917 & 1695 $\pm$  114 & 4875 $\pm$  375 & 1.158 &  700 & 0 &  \\ 
OGLE-LMC-T2CEP-201 & T2CEP & pWVir &  11.007 & 2172 $\pm$   84 & 6500 $\pm$  250 & 0.150 &  180 & 0 &  \\ 
OGLE-LMC-T2CEP-202 & T2CEP & RVTau &  38.136 & 1028 $\pm$   24 & 4625 $\pm$  125 & 0.000 & 1000 & 0 &  \\ 
OGLE-LMC-T2CEP-203 & T2CEP & RVTau &  37.127 &  887 $\pm$   36 & 4625 $\pm$  188 & 0.000 & 1000 & 0 &  \\ 
\hline
\end{tabular}
\end{table*}

\setcounter{table}{0}
\begin{table*}
\caption{Continued}

\begin{tabular}{lrrrrrrrrlrrrrrr}
\hline
Name               & Type  & Subtype & Period &    Luminosity   &       $T_{\rm eff}$ & $\tau$ & $T_{\rm c}$ & fit  &  Remarks \\ 
                   &       &          & (d)    &    (\lsol)      &            (K)    &        &   (K)     &      &  \\
\hline
OGLE-SMC-ACEP-01   & ANCEP & 1O    &   0.621 &  118 $\pm$    3 & 7250 $\pm$  188 & 0.000 & 1000 & 0 &  \\ 
OGLE-SMC-ACEP-02   & ANCEP & F     &   0.828 &   78 $\pm$    1 & 6125 $\pm$   62 & 0.000 & 1000 & 0 &  \\ 
OGLE-SMC-ACEP-03   & ANCEP & 1O    &   0.570 &  103 $\pm$    3 & 7125 $\pm$  250 & 0.000 & 1000 & 0 &  \\ 
OGLE-SMC-ACEP-04   & ANCEP & F     &   0.830 &  100 $\pm$    1 & 6375 $\pm$  125 & 0.000 & 1000 & 0 &  \\ 
OGLE-SMC-ACEP-05   & ANCEP & 1O    &   0.521 &   99 $\pm$    2 & 7000 $\pm$  125 & 0.000 & 1000 & 0 &  \\ 
OGLE-SMC-ACEP-06   & ANCEP & F     &   1.256 &  167 $\pm$   10 & 6875 $\pm$  250 & 0.000 & 1000 & 0 &  \\ 
OGLE-SMC-T2CEP-001 & T2CEP & pWVir &  11.869 & 2570 $\pm$  102 & 6250 $\pm$  125 & 0.000 & 1000 & 0 &  \\ 
OGLE-SMC-T2CEP-002 & T2CEP & BLHer &   1.372 &  109 $\pm$    4 & 6750 $\pm$  312 & 0.000 & 1000 & 0 &  \\ 
OGLE-SMC-T2CEP-003 & T2CEP & WVir  &   4.360 &  173 $\pm$    7 & 5875 $\pm$  312 & 0.000 & 1000 & 0 &  \\ 
OGLE-SMC-T2CEP-004 & T2CEP & WVir  &   6.533 &  299 $\pm$   14 & 5375 $\pm$  375 & 0.000 & 1000 & 0 &  \\ 
OGLE-SMC-T2CEP-005 & T2CEP & WVir  &   8.206 &  282 $\pm$    3 & 5375 $\pm$   62 & 0.000 & 1000 & 0 &  \\ 
OGLE-SMC-T2CEP-006 & T2CEP & BLHer &   1.236 &   80 $\pm$    1 & 6375 $\pm$   62 & 0.000 & 1000 & 0 &  \\ 
OGLE-SMC-T2CEP-007 & T2CEP & RVTau &  30.961 & 7560 $\pm$ 1970 & 6125 $\pm$  750 & 0.000 & 1000 & 0 &  \\  
OGLE-SMC-T2CEP-008 & T2CEP & BLHer &   1.490 &  117 $\pm$    2 & 5875 $\pm$  125 & 0.000 & 1000 & 0 &  \\ 
OGLE-SMC-T2CEP-009 & T2CEP & BLHer &   2.971 &  151 $\pm$    3 & 5625 $\pm$  125 & 0.000 & 1000 & 0 &  \\ 
OGLE-SMC-T2CEP-010 & T2CEP & pWVir &  17.481 & 4264 $\pm$  319 & 6000 $\pm$  312 & 0.000 & 1000 & 0 &  \\ 
OGLE-SMC-T2CEP-011 & T2CEP & pWVir &   9.925 & 3758 $\pm$  136 & 7250 $\pm$  250 & 0.028 & 1000 & 0 &  \\ 
OGLE-SMC-T2CEP-012 & T2CEP & RVTau &  29.219 & 1301 $\pm$   41 & 5375 $\pm$   62 & 0.000 & 1000 & 0 &  \\ 
OGLE-SMC-T2CEP-013 & T2CEP & WVir  &  13.810 &  478 $\pm$   13 & 5375 $\pm$   62 & 0.000 & 1000 & 0 &  \\ 
OGLE-SMC-T2CEP-014 & T2CEP & WVir  &  13.878 &  426 $\pm$    8 & 5375 $\pm$   62 & 0.000 & 1000 & 0 &  \\ 
OGLE-SMC-T2CEP-015 & T2CEP & BLHer &   2.570 &  458 $\pm$   21 & 7500 $\pm$  312 & 0.000 & 1000 & 0 &  \\ 
OGLE-SMC-T2CEP-016 & T2CEP & BLHer &   2.113 &  122 $\pm$    3 & 6000 $\pm$  188 & 0.000 & 1000 & 0 &  \\ 
OGLE-SMC-T2CEP-017 & T2CEP & BLHer &   1.299 &  115 $\pm$    5 & 6375 $\pm$  375 & 0.000 & 1000 & 0 &  \\ 
OGLE-SMC-T2CEP-018 & T2CEP & RVTau &  39.519 & 3539 $\pm$  166 & 5875 $\pm$  375 & 1.616 & 1200 & 0 &  \\ 
OGLE-SMC-T2CEP-019 & T2CEP & RVTau &  40.912 & 3481 $\pm$  101 & 6250 $\pm$  125 & 0.037 &  598 & 1 &  \\ 
OGLE-SMC-T2CEP-020 & T2CEP & RVTau &  50.623 & 1885 $\pm$   84 & 5375 $\pm$  188 & 0.000 & 1000 & 0 &  \\  
OGLE-SMC-T2CEP-021 & T2CEP & BLHer &   2.313 &   94 $\pm$    1 & 6000 $\pm$  125 & 0.000 & 1000 & 0 &  \\ 
OGLE-SMC-T2CEP-022 & T2CEP & BLHer &   1.471 &   61 $\pm$    2 & 6000 $\pm$  250 & 0.000 & 1000 & 0 &  \\ 
OGLE-SMC-T2CEP-023 & T2CEP & pWVir &  17.675 & 1508 $\pm$   28 & 5750 $\pm$  125 & 0.000 & 1000 & 0 & EB \\ 
OGLE-SMC-T2CEP-024 & T2CEP & RVTau &  43.961 & 3075 $\pm$  126 & 6000 $\pm$  188 & 0.077 &  700 & 0 &  \\ 
OGLE-SMC-T2CEP-025 & T2CEP & pWVir &  14.171 &  941 $\pm$   34 & 5750 $\pm$  125 & 0.000 & 1000 & 0 &  \\ 
OGLE-SMC-T2CEP-026 & T2CEP & BLHer &   1.705 &  153 $\pm$    2 & 6375 $\pm$  125 & 0.000 & 1000 & 0 &  \\ 
OGLE-SMC-T2CEP-027 & T2CEP & BLHer &   1.542 &   81 $\pm$    2 & 6250 $\pm$  125 & 0.000 & 1000 & 0 &  \\ 
OGLE-SMC-T2CEP-028 & T2CEP & pWVir &  15.264 & 1854 $\pm$   53 & 5375 $\pm$   62 & 0.000 & 1000 & 0 & EB \\ 
OGLE-SMC-T2CEP-029 & T2CEP & RVTau &  33.676 & 6273 $\pm$  157 & 5375 $\pm$   62 & 0.000 & 1000 & 0 & EB \\ 
OGLE-SMC-T2CEP-030 & T2CEP & BLHer &   3.389 &  418 $\pm$    9 & 6750 $\pm$  188 & 0.000 & 1000 & 0 &  \\ 
OGLE-SMC-T2CEP-031 & T2CEP & WVir  &   7.895 &  323 $\pm$   21 & 5375 $\pm$  312 & 0.000 & 1000 & 0 &  \\ 
OGLE-SMC-T2CEP-032 & T2CEP & WVir  &  14.247 &  693 $\pm$   30 & 5750 $\pm$  188 & 0.086 &  600 & 0 &  \\ 
OGLE-SMC-T2CEP-033 & T2CEP & BLHer &   1.878 &  198 $\pm$    6 & 6375 $\pm$  188 & 0.000 & 1000 & 0 &  \\ 
OGLE-SMC-T2CEP-034 & T2CEP & WVir  &  20.121 &  994 $\pm$   39 & 5375 $\pm$   62 & 0.000 & 1000 & 0 &  \\ 
OGLE-SMC-T2CEP-035 & T2CEP & WVir  &  17.181 &  730 $\pm$   31 & 5500 $\pm$  312 & 0.025 & 1100 & 0 &  \\ 
OGLE-SMC-T2CEP-036 & T2CEP & BLHer &   1.092 &  141 $\pm$    2 & 6500 $\pm$  125 & 0.000 & 1000 & 0 &  \\ 
OGLE-SMC-T2CEP-037 & T2CEP & BLHer &   1.559 &  127 $\pm$    3 & 6125 $\pm$  125 & 0.000 & 1000 & 0 &  \\ 
OGLE-SMC-T2CEP-038 & T2CEP & pWVir &   4.444 &  734 $\pm$   32 & 6500 $\pm$  250 & 0.000 & 1000 & 0 &  \\ 
OGLE-SMC-T2CEP-039 & T2CEP & BLHer &   1.888 &  142 $\pm$    1 & 5875 $\pm$   62 & 0.000 & 1000 & 0 &  \\ 
OGLE-SMC-T2CEP-040 & T2CEP & WVir  &  16.111 &  662 $\pm$   64 & 5375 $\pm$  312 & 0.000 & 1000 & 0 &  \\ 
OGLE-SMC-T2CEP-041 & T2CEP & RVTau &  29.118 & 1393 $\pm$   49 & 5750 $\pm$  125 & 0.000 & 1000 & 0 &  \\ 
OGLE-SMC-T2CEP-042 & T2CEP & BLHer &   1.487 &  100 $\pm$    2 & 6250 $\pm$   62 & 0.000 & 1000 & 0 &  \\ 
OGLE-SMC-T2CEP-043 & T2CEP & RVTau &  23.743 & 1285 $\pm$   57 & 5375 $\pm$  125 & 0.000 & 1000 & 0 & W3 excess \\

\hline
\end{tabular}
\end{table*}

\clearpage

\section{The modelled SEDs}
\label{Sect:App_modelSED}

Figure~\ref{Fig_SEDs} shows the fits to the photometry.
 
\begin{figure*}
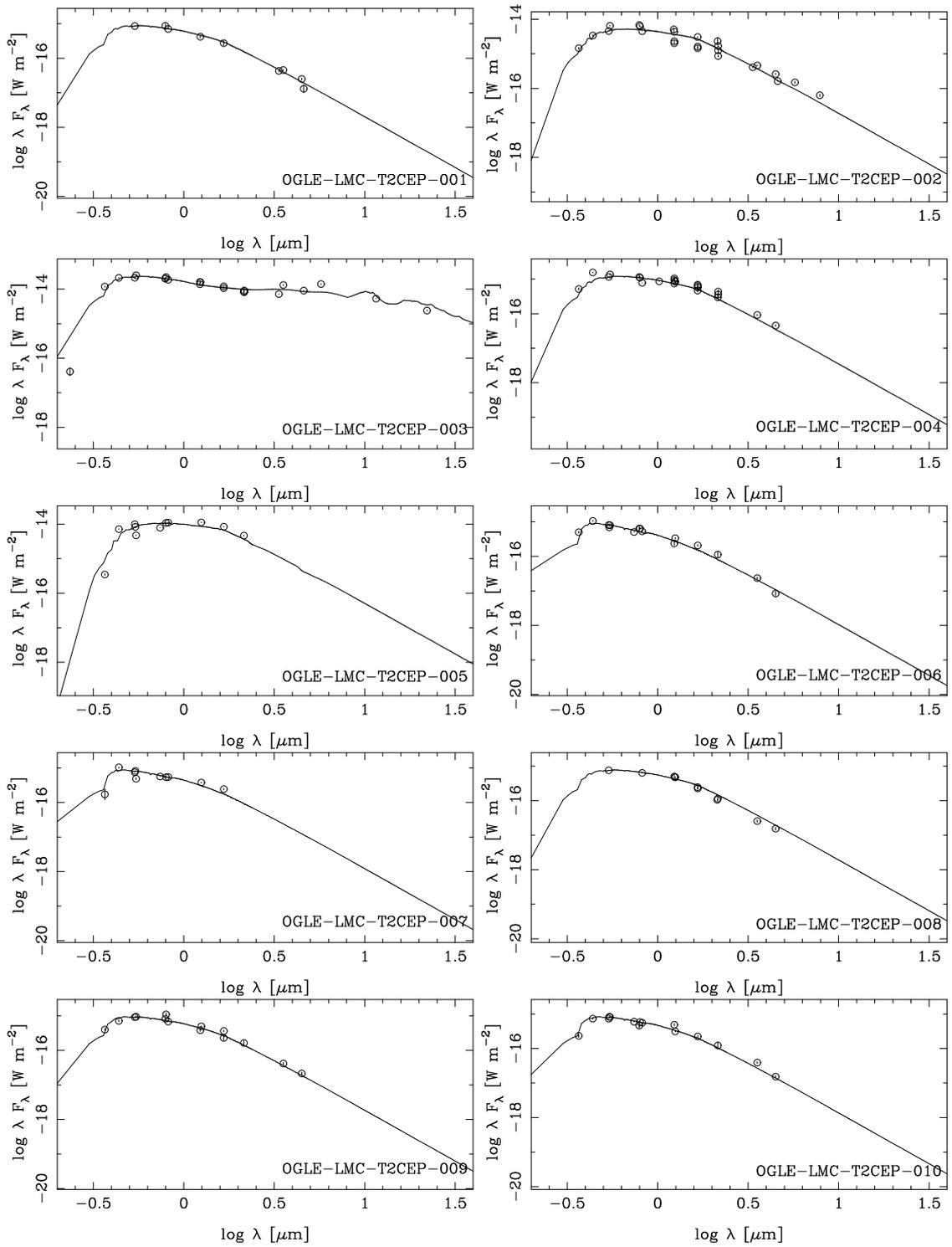

\centering

\begin{minipage}{0.4\textwidth}
\resizebox{\hsize}{!}{\includegraphics[angle=-0]{SEDall/OGLE-LMC-T2CEP-001_sed.ps}} 
\end{minipage}
\begin{minipage}{0.4\textwidth}
\resizebox{\hsize}{!}{\includegraphics[angle=-0]{SEDall/OGLE-LMC-T2CEP-002_sed.ps}} 
\end{minipage}
 
\begin{minipage}{0.4\textwidth}
\resizebox{\hsize}{!}{\includegraphics[angle=-0]{SEDall/OGLE-LMC-T2CEP-003_sed.ps}} 
\end{minipage}
\begin{minipage}{0.4\textwidth}
\resizebox{\hsize}{!}{\includegraphics[angle=-0]{SEDall/OGLE-LMC-T2CEP-004_sed.ps}} 
\end{minipage}
 
\begin{minipage}{0.4\textwidth}
\resizebox{\hsize}{!}{\includegraphics[angle=-0]{SEDall/OGLE-LMC-T2CEP-005_sed.ps}} 
\end{minipage}
\begin{minipage}{0.4\textwidth}
\resizebox{\hsize}{!}{\includegraphics[angle=-0]{SEDall/OGLE-LMC-T2CEP-006_sed.ps}} 
\end{minipage}
 
\begin{minipage}{0.4\textwidth}
\resizebox{\hsize}{!}{\includegraphics[angle=-0]{SEDall/OGLE-LMC-T2CEP-007_sed.ps}} 
\end{minipage}
\begin{minipage}{0.4\textwidth}
\resizebox{\hsize}{!}{\includegraphics[angle=-0]{SEDall/OGLE-LMC-T2CEP-008_sed.ps}} 
\end{minipage}

\begin{minipage}{0.4\textwidth}
\resizebox{\hsize}{!}{\includegraphics[angle=-0]{SEDall/OGLE-LMC-T2CEP-009_sed.ps}} 
\end{minipage}
\begin{minipage}{0.4\textwidth}
\resizebox{\hsize}{!}{\includegraphics[angle=-0]{SEDall/OGLE-LMC-T2CEP-010_sed.ps}} 
\end{minipage}

\caption{Fits to the SEDs.}
\label{Fig_SEDs}

\end{figure*}

\setcounter{figure}{0}
\begin{figure*}
\centering

\begin{minipage}{0.4\textwidth}
\resizebox{\hsize}{!}{\includegraphics[angle=-0]{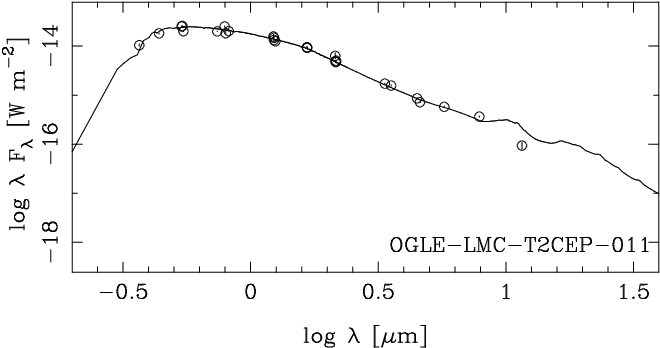}} 
\end{minipage}
\begin{minipage}{0.4\textwidth}
\resizebox{\hsize}{!}{\includegraphics[angle=-0]{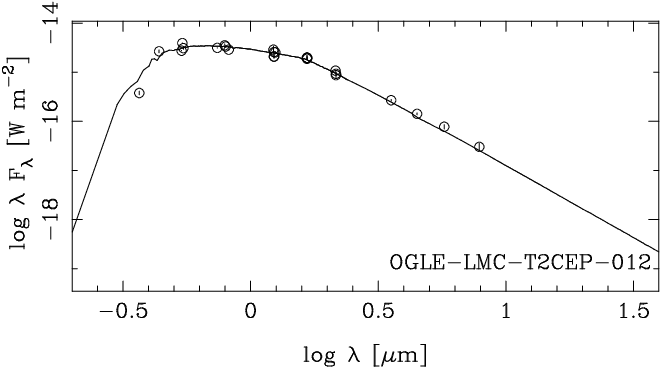}} 
\end{minipage}
 
\begin{minipage}{0.4\textwidth}
\resizebox{\hsize}{!}{\includegraphics[angle=-0]{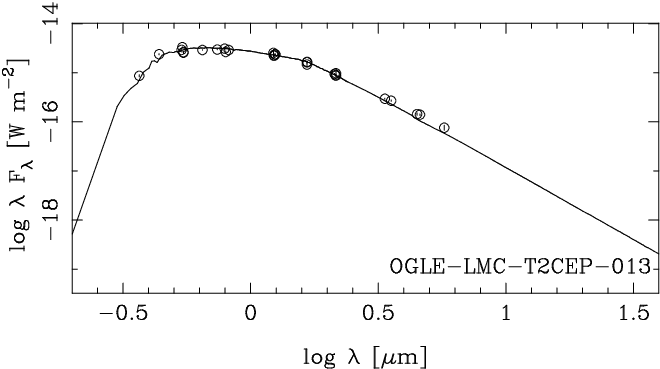}} 
\end{minipage}
\begin{minipage}{0.4\textwidth}
\resizebox{\hsize}{!}{\includegraphics[angle=-0]{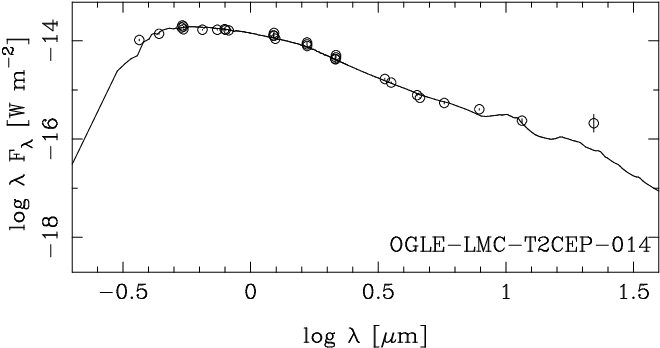}} 
\end{minipage}

\begin{minipage}{0.4\textwidth}
\resizebox{\hsize}{!}{\includegraphics[angle=-0]{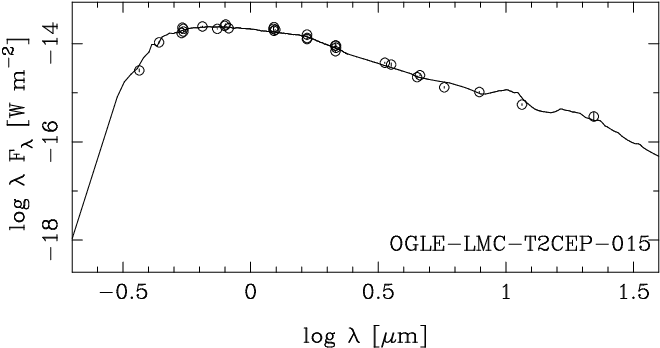}} 
\end{minipage}
\begin{minipage}{0.4\textwidth}
\resizebox{\hsize}{!}{\includegraphics[angle=-0]{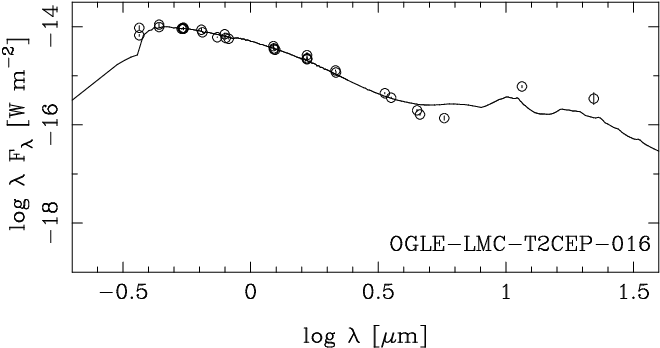}} 
\end{minipage}
 
\begin{minipage}{0.4\textwidth}
\resizebox{\hsize}{!}{\includegraphics[angle=-0]{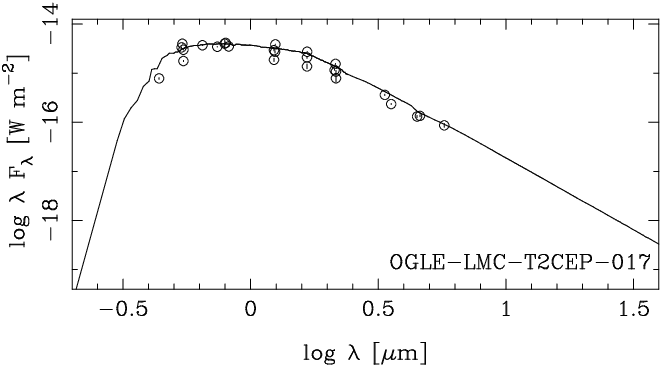}} 
\end{minipage}
\begin{minipage}{0.4\textwidth}
\resizebox{\hsize}{!}{\includegraphics[angle=-0]{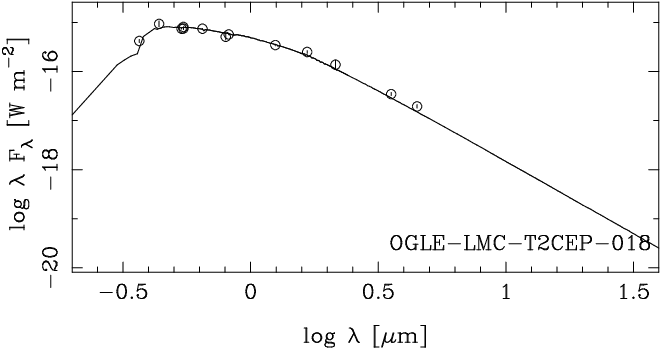}} 
\end{minipage}
 
\begin{minipage}{0.4\textwidth}
\resizebox{\hsize}{!}{\includegraphics[angle=-0]{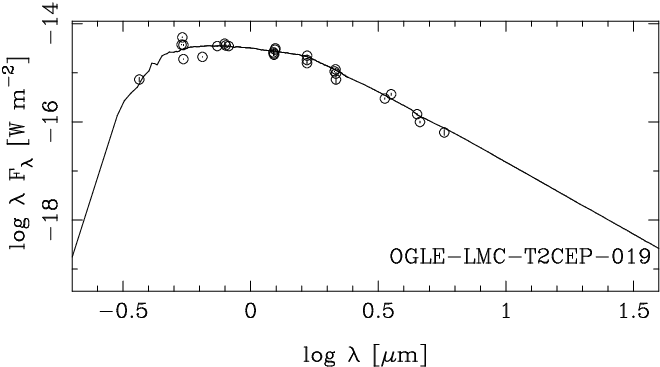}} 
\end{minipage}
\begin{minipage}{0.4\textwidth}
\resizebox{\hsize}{!}{\includegraphics[angle=-0]{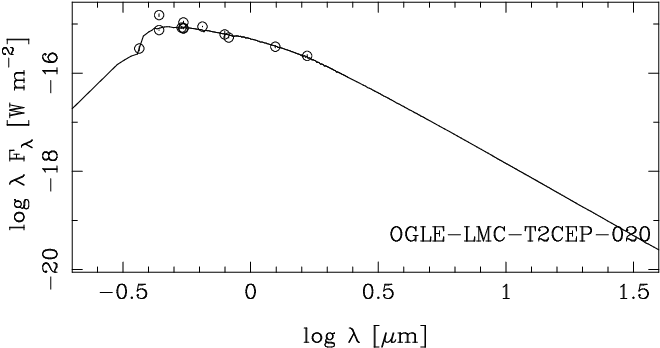}} 
\end{minipage}

\caption{Continued}
\end{figure*}

\setcounter{figure}{0}
\begin{figure*}
\centering

\begin{minipage}{0.4\textwidth}
\resizebox{\hsize}{!}{\includegraphics[angle=-0]{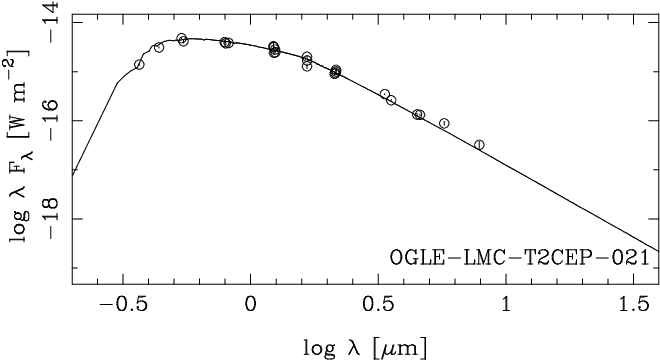}} 
\end{minipage}
\begin{minipage}{0.4\textwidth}
\resizebox{\hsize}{!}{\includegraphics[angle=-0]{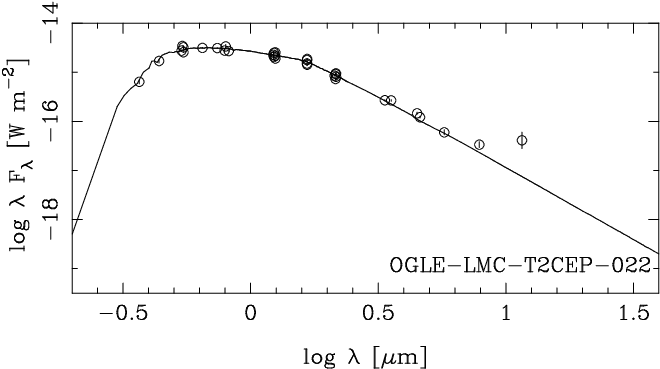}} 
\end{minipage}
 
\begin{minipage}{0.4\textwidth}
\resizebox{\hsize}{!}{\includegraphics[angle=-0]{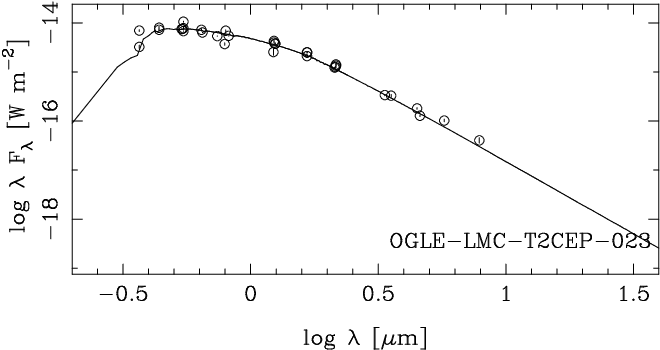}} 
\end{minipage}
\begin{minipage}{0.4\textwidth}
\resizebox{\hsize}{!}{\includegraphics[angle=-0]{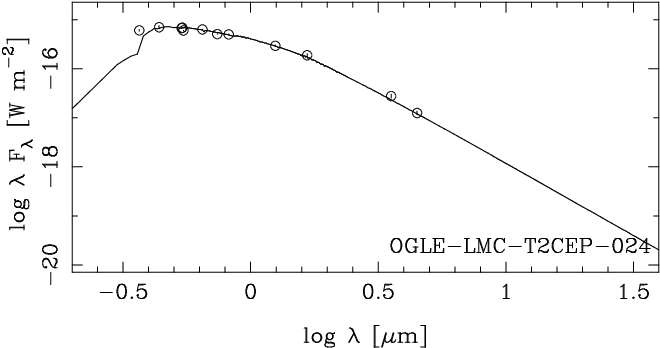}} 
\end{minipage}
 
\begin{minipage}{0.4\textwidth}
\resizebox{\hsize}{!}{\includegraphics[angle=-0]{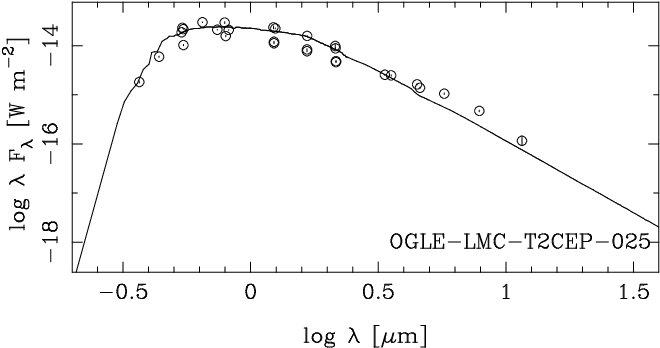}} 
\end{minipage}
\begin{minipage}{0.4\textwidth}
\resizebox{\hsize}{!}{\includegraphics[angle=-0]{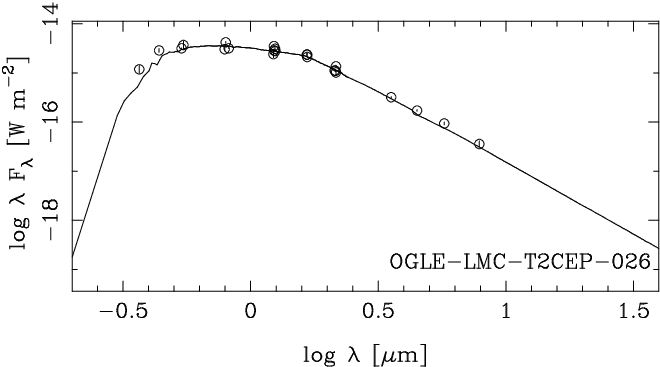}} 
\end{minipage}
 
\begin{minipage}{0.4\textwidth}
\resizebox{\hsize}{!}{\includegraphics[angle=-0]{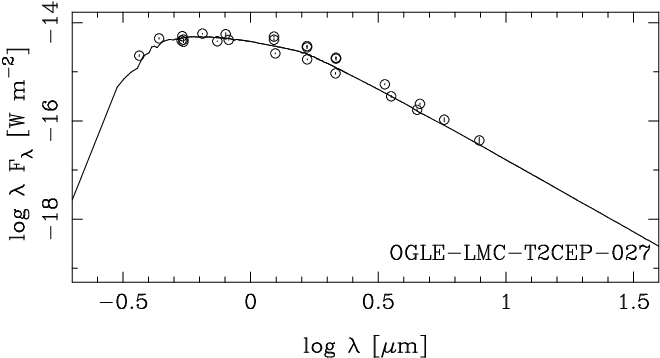}} 
\end{minipage}
\begin{minipage}{0.4\textwidth}
\resizebox{\hsize}{!}{\includegraphics[angle=-0]{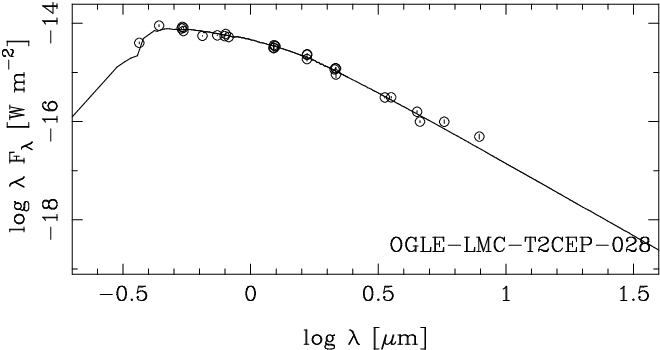}} 
\end{minipage}
 
\begin{minipage}{0.4\textwidth}
\resizebox{\hsize}{!}{\includegraphics[angle=-0]{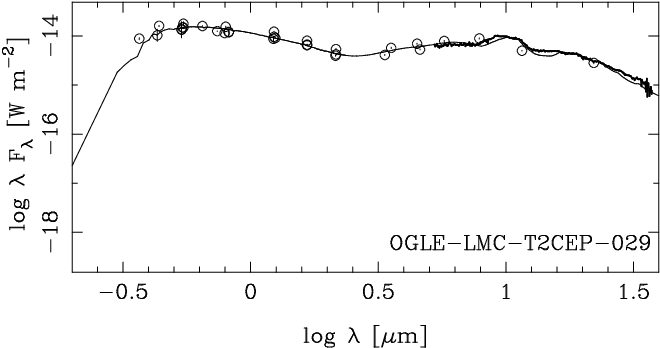}} 
\end{minipage}
\begin{minipage}{0.4\textwidth}
\resizebox{\hsize}{!}{\includegraphics[angle=-0]{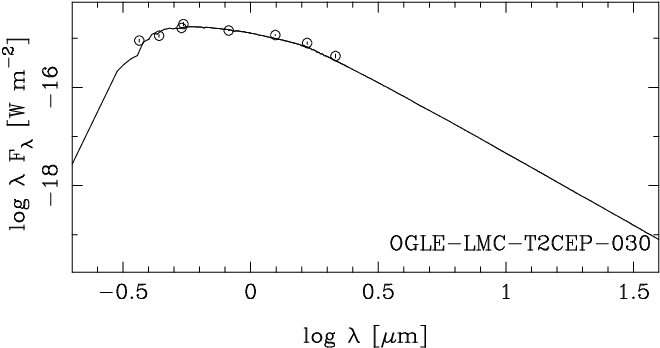}} 
\end{minipage}
\caption{Continued}
\end{figure*}

\setcounter{figure}{0}
\begin{figure*}
\centering

\begin{minipage}{0.4\textwidth}
\resizebox{\hsize}{!}{\includegraphics[angle=-0]{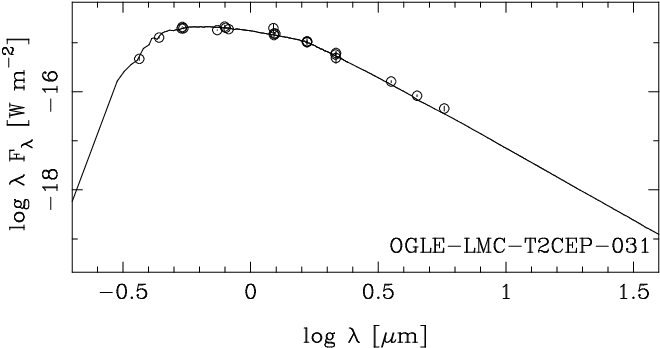}} 
\end{minipage}
\begin{minipage}{0.4\textwidth}
\resizebox{\hsize}{!}{\includegraphics[angle=-0]{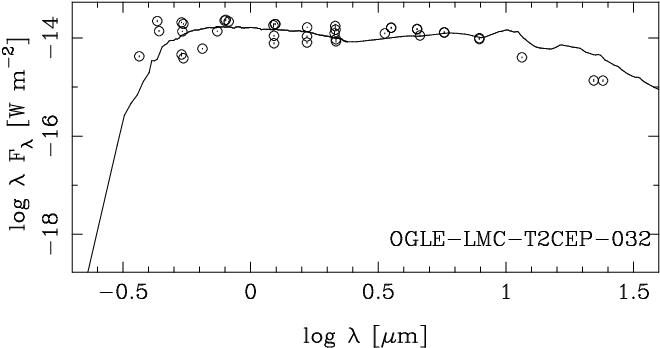}} 
\end{minipage}
 
\begin{minipage}{0.4\textwidth}
\resizebox{\hsize}{!}{\includegraphics[angle=-0]{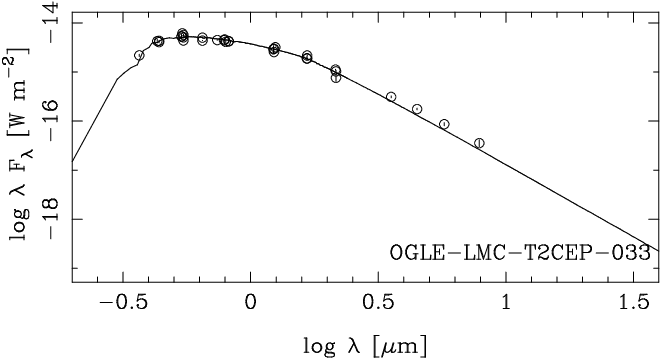}} 
\end{minipage}
\begin{minipage}{0.4\textwidth}
\resizebox{\hsize}{!}{\includegraphics[angle=-0]{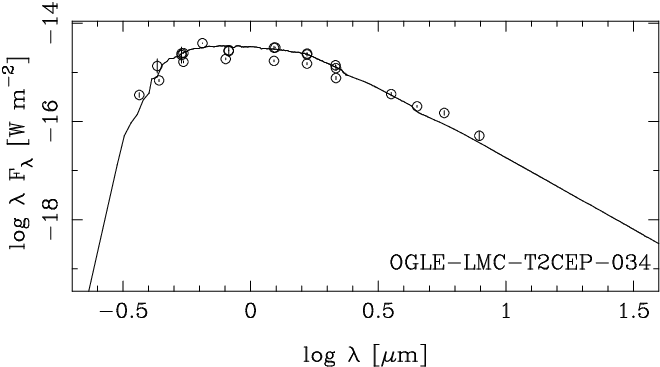}} 
\end{minipage}
 
\begin{minipage}{0.4\textwidth}
\resizebox{\hsize}{!}{\includegraphics[angle=-0]{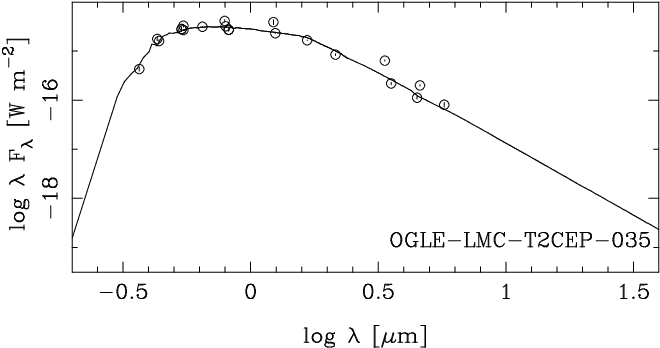}} 
\end{minipage}
\begin{minipage}{0.4\textwidth}
\resizebox{\hsize}{!}{\includegraphics[angle=-0]{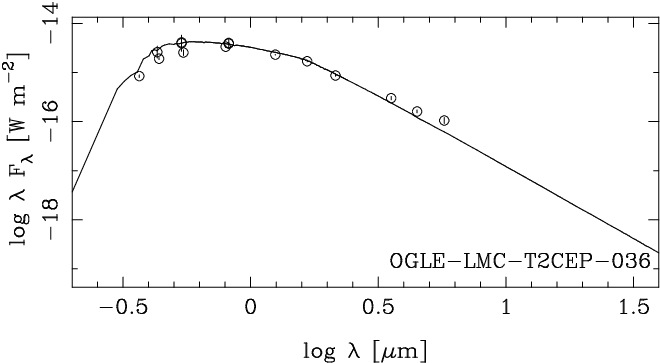}} 
\end{minipage}
 
\begin{minipage}{0.4\textwidth}
\resizebox{\hsize}{!}{\includegraphics[angle=-0]{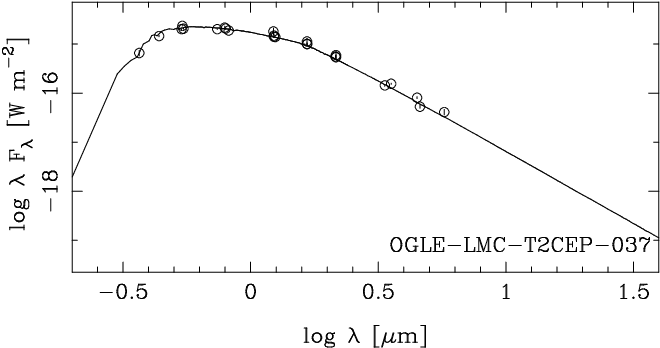}} 
\end{minipage}
\begin{minipage}{0.4\textwidth}
\resizebox{\hsize}{!}{\includegraphics[angle=-0]{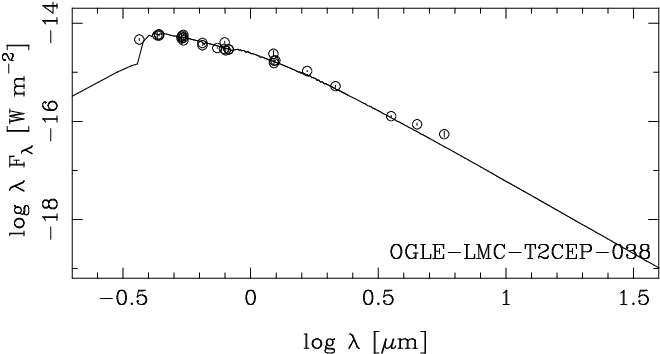}} 
\end{minipage}
 
\begin{minipage}{0.4\textwidth}
\resizebox{\hsize}{!}{\includegraphics[angle=-0]{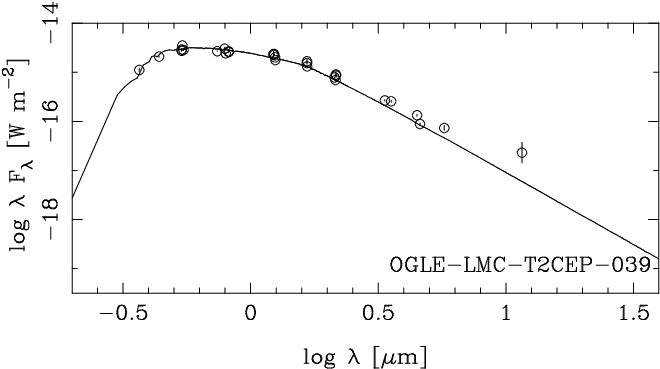}} 
\end{minipage}
\begin{minipage}{0.4\textwidth}
\resizebox{\hsize}{!}{\includegraphics[angle=-0]{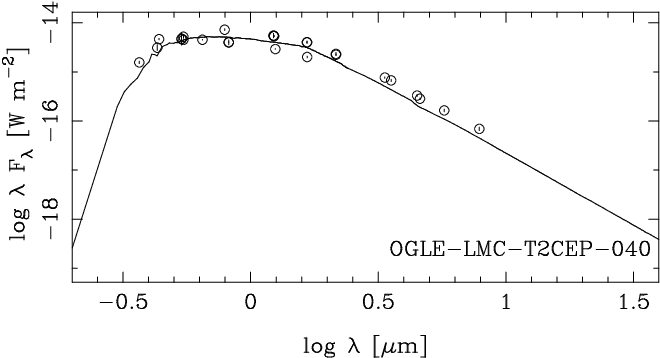}} 
\end{minipage}
\caption{Continued}
\end{figure*}

\setcounter{figure}{0}
\begin{figure*}
\centering

\begin{minipage}{0.4\textwidth}
\resizebox{\hsize}{!}{\includegraphics[angle=-0]{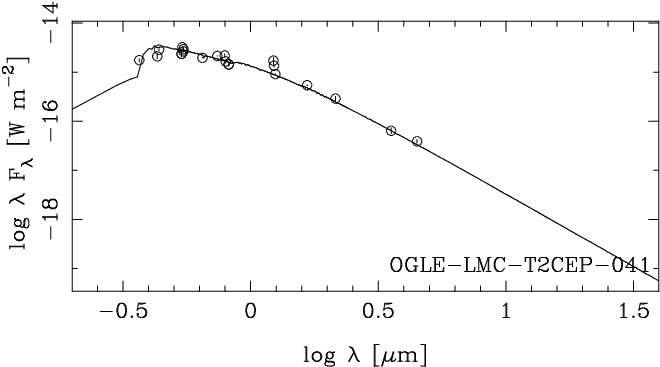}} 
\end{minipage}
\begin{minipage}{0.4\textwidth}
\resizebox{\hsize}{!}{\includegraphics[angle=-0]{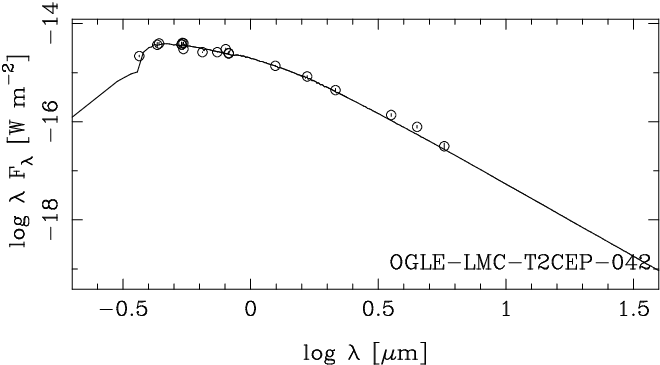}} 
\end{minipage}
 
\begin{minipage}{0.4\textwidth}
\resizebox{\hsize}{!}{\includegraphics[angle=-0]{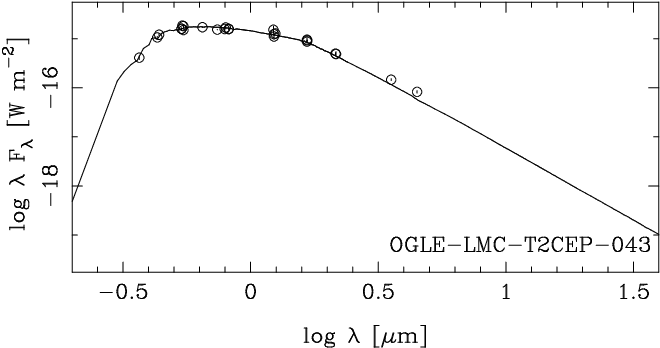}} 
\end{minipage}
\begin{minipage}{0.4\textwidth}
\resizebox{\hsize}{!}{\includegraphics[angle=-0]{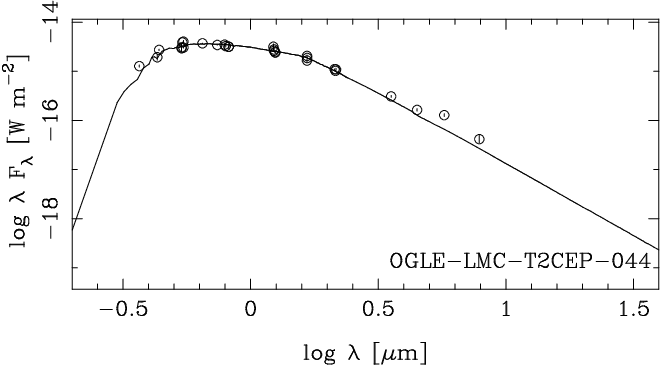}} 
\end{minipage}
 
\begin{minipage}{0.4\textwidth}
\resizebox{\hsize}{!}{\includegraphics[angle=-0]{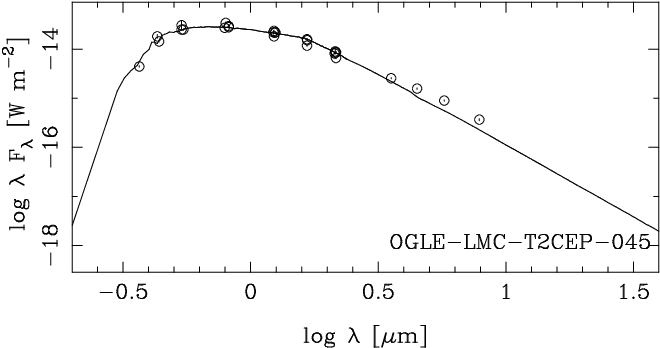}} 
\end{minipage}
\begin{minipage}{0.4\textwidth}
\resizebox{\hsize}{!}{\includegraphics[angle=-0]{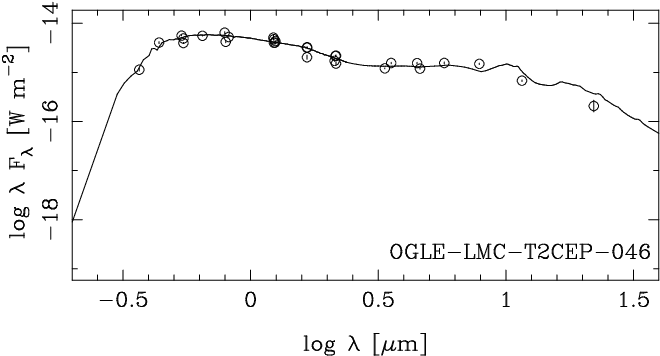}} 
\end{minipage}
 
\begin{minipage}{0.4\textwidth}
\resizebox{\hsize}{!}{\includegraphics[angle=-0]{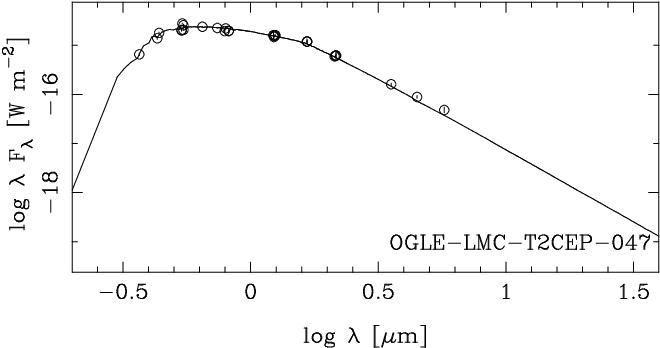}} 
\end{minipage}
\begin{minipage}{0.4\textwidth}
\resizebox{\hsize}{!}{\includegraphics[angle=-0]{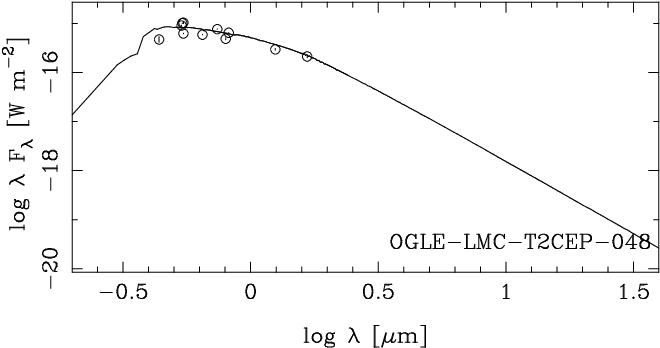}} 
\end{minipage}
 
\begin{minipage}{0.4\textwidth}
\resizebox{\hsize}{!}{\includegraphics[angle=-0]{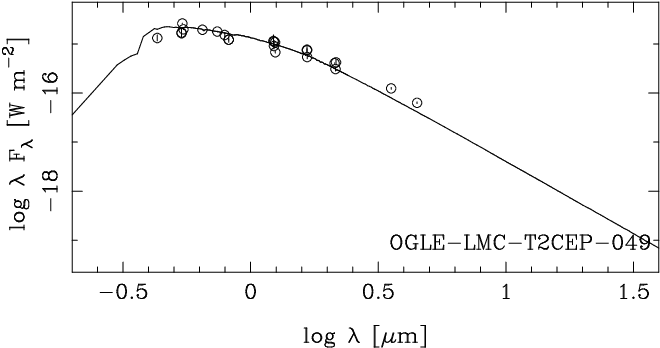}} 
\end{minipage}
\begin{minipage}{0.4\textwidth}
\resizebox{\hsize}{!}{\includegraphics[angle=-0]{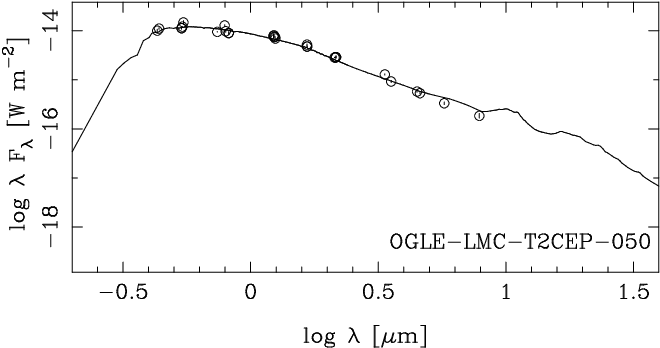}} 
\end{minipage}

\caption{Continued}
\end{figure*}

\setcounter{figure}{0}
\begin{figure*}
\centering

\begin{minipage}{0.4\textwidth}
\resizebox{\hsize}{!}{\includegraphics[angle=-0]{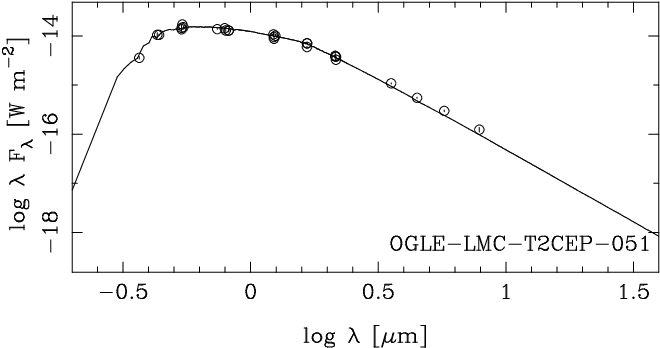}} 
\end{minipage}
\begin{minipage}{0.4\textwidth}
\resizebox{\hsize}{!}{\includegraphics[angle=-0]{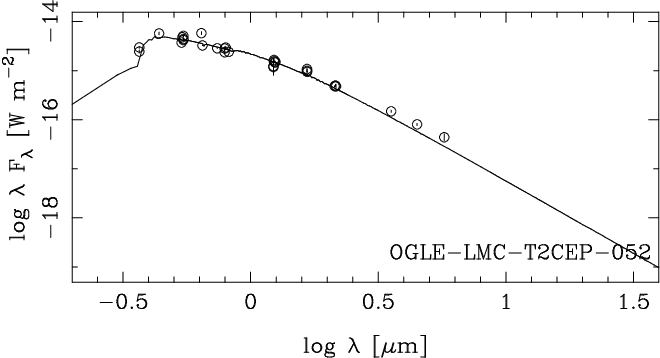}} 
\end{minipage}
 
\begin{minipage}{0.4\textwidth}
\resizebox{\hsize}{!}{\includegraphics[angle=-0]{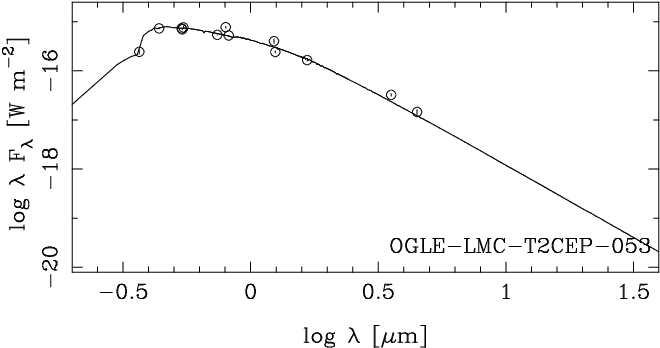}} 
\end{minipage}
\begin{minipage}{0.4\textwidth}
\resizebox{\hsize}{!}{\includegraphics[angle=-0]{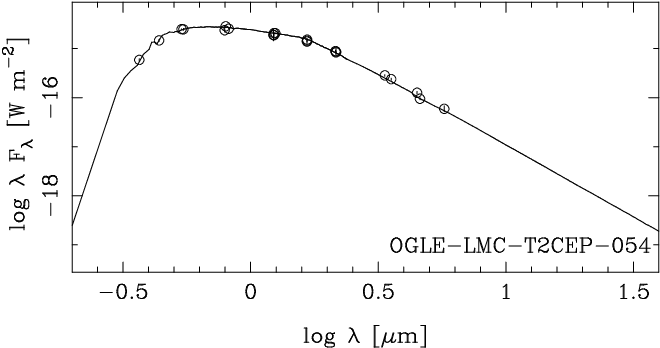}} 
\end{minipage}
 
\begin{minipage}{0.4\textwidth}
\resizebox{\hsize}{!}{\includegraphics[angle=-0]{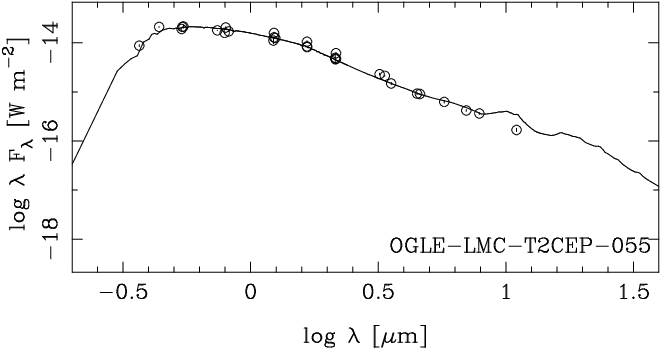}} 
\end{minipage}
\begin{minipage}{0.4\textwidth}
\resizebox{\hsize}{!}{\includegraphics[angle=-0]{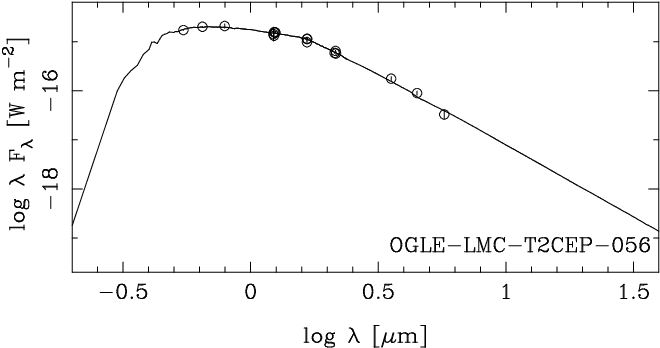}} 
\end{minipage}
 
\begin{minipage}{0.4\textwidth}
\resizebox{\hsize}{!}{\includegraphics[angle=-0]{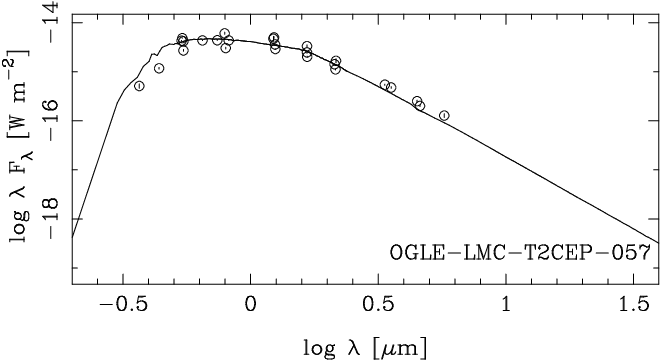}} 
\end{minipage}
\begin{minipage}{0.4\textwidth}
\resizebox{\hsize}{!}{\includegraphics[angle=-0]{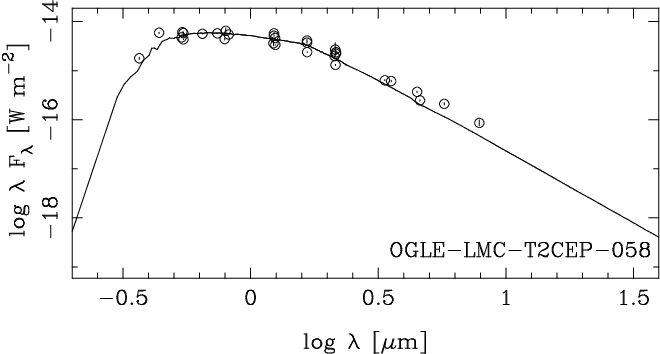}} 
\end{minipage}
 
\begin{minipage}{0.4\textwidth}
\resizebox{\hsize}{!}{\includegraphics[angle=-0]{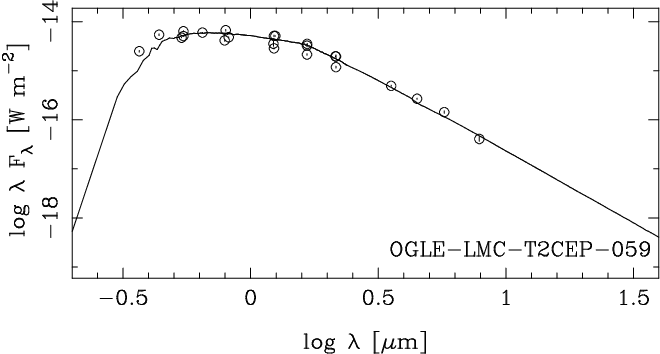}} 
\end{minipage}
\begin{minipage}{0.4\textwidth}
\resizebox{\hsize}{!}{\includegraphics[angle=-0]{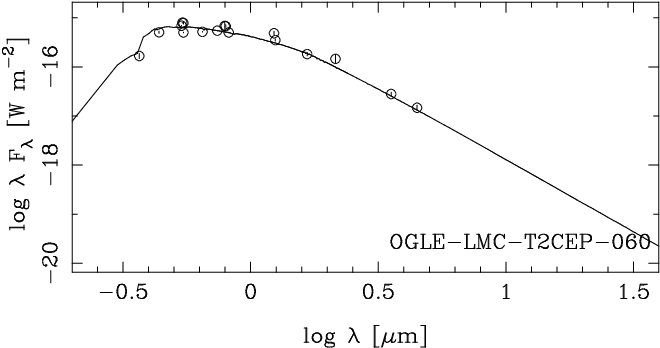}} 
\end{minipage}
 
\caption{Continued}
\end{figure*}

\setcounter{figure}{0}
\begin{figure*}
\centering

\begin{minipage}{0.4\textwidth}
\resizebox{\hsize}{!}{\includegraphics[angle=-0]{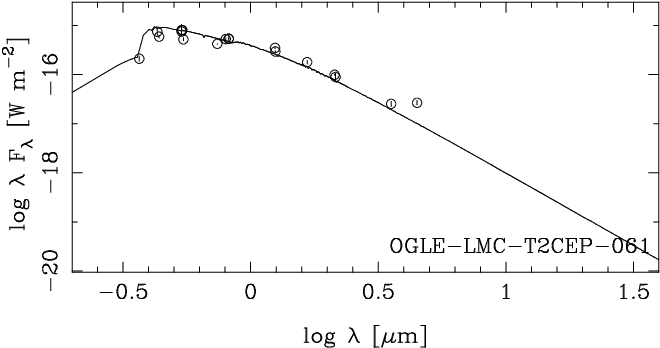}} 
\end{minipage}
\begin{minipage}{0.4\textwidth}
\resizebox{\hsize}{!}{\includegraphics[angle=-0]{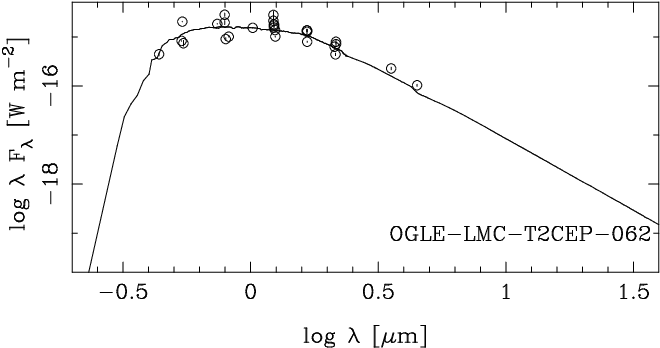}} 
\end{minipage}
 
\begin{minipage}{0.4\textwidth}
\resizebox{\hsize}{!}{\includegraphics[angle=-0]{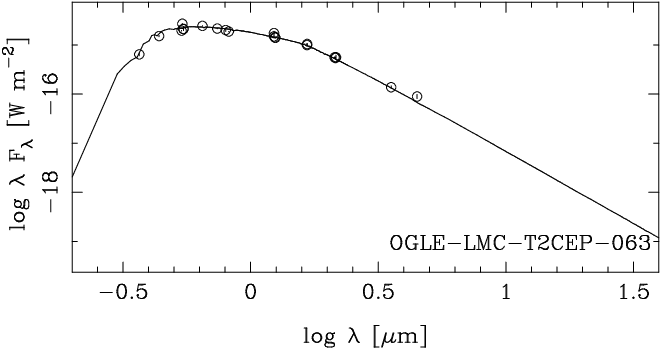}} 
\end{minipage}
\begin{minipage}{0.4\textwidth}
\resizebox{\hsize}{!}{\includegraphics[angle=-0]{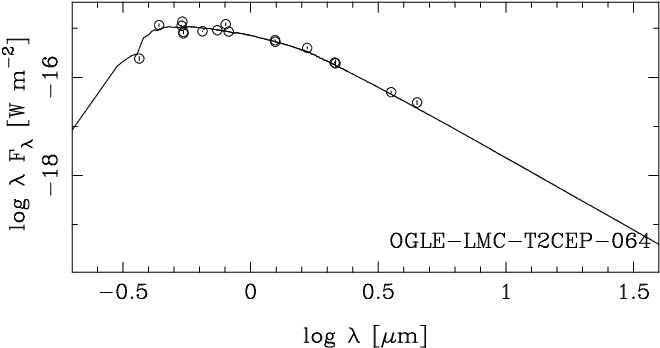}} 
\end{minipage}
 
\begin{minipage}{0.4\textwidth}
\resizebox{\hsize}{!}{\includegraphics[angle=-0]{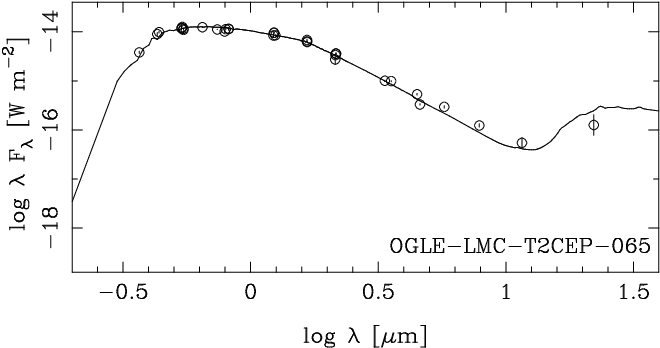}} 
\end{minipage}
\begin{minipage}{0.4\textwidth}
\resizebox{\hsize}{!}{\includegraphics[angle=-0]{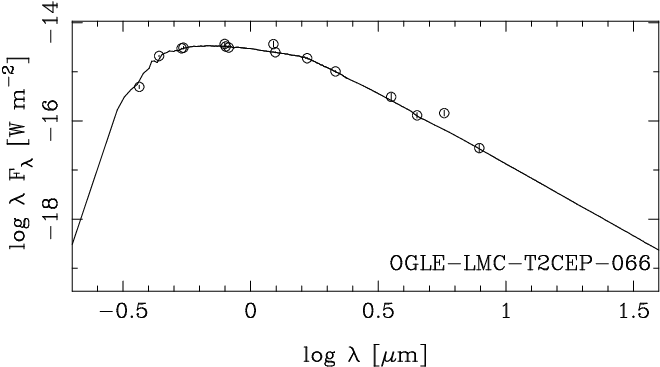}} 
\end{minipage}
 
\begin{minipage}{0.4\textwidth}
\resizebox{\hsize}{!}{\includegraphics[angle=-0]{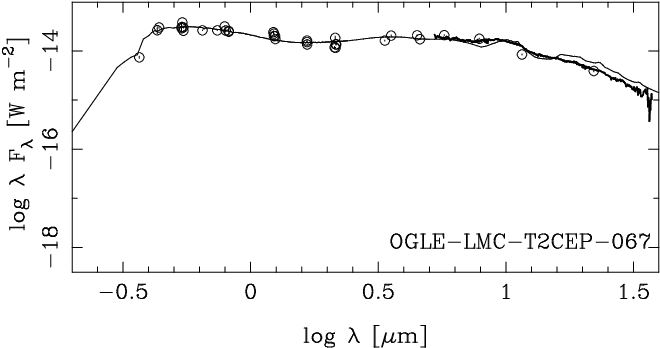}} 
\end{minipage}
\begin{minipage}{0.4\textwidth}
\resizebox{\hsize}{!}{\includegraphics[angle=-0]{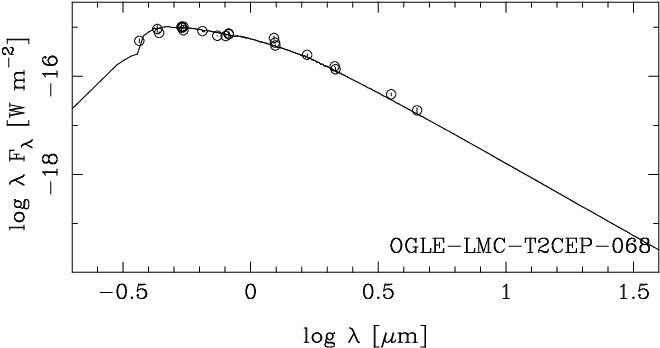}} 
\end{minipage}
 
\begin{minipage}{0.4\textwidth}
\resizebox{\hsize}{!}{\includegraphics[angle=-0]{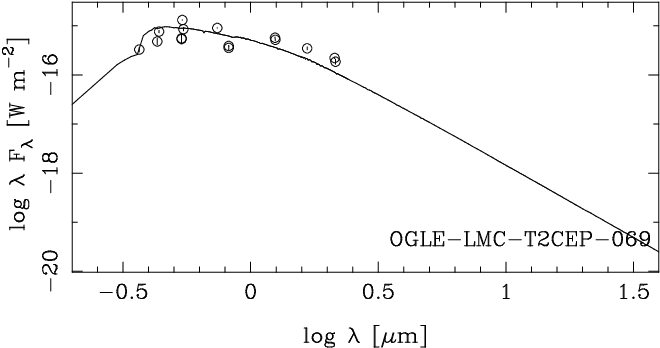}} 
\end{minipage}
\begin{minipage}{0.4\textwidth}
\resizebox{\hsize}{!}{\includegraphics[angle=-0]{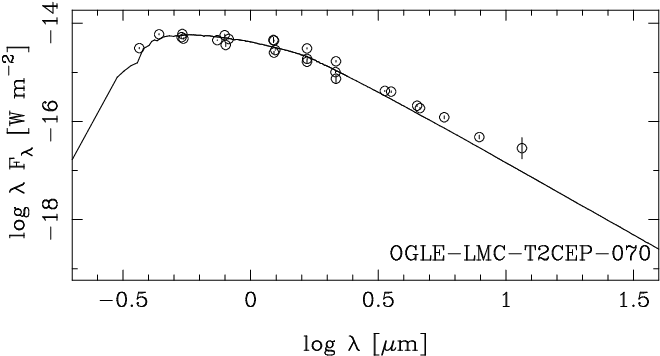}} 
\end{minipage}

\caption{Continued}
\end{figure*}

\setcounter{figure}{0}
\begin{figure*}
\centering

\begin{minipage}{0.4\textwidth}
\resizebox{\hsize}{!}{\includegraphics[angle=-0]{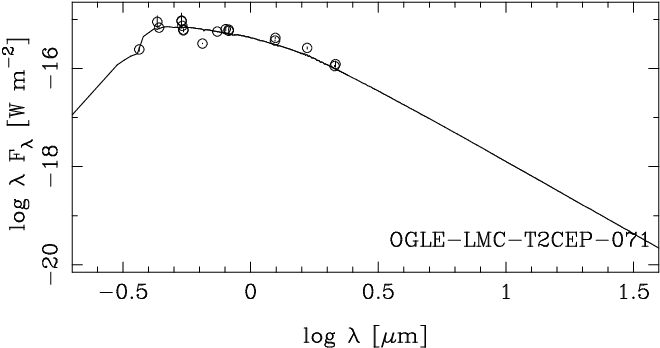}} 
\end{minipage}
\begin{minipage}{0.4\textwidth}
\resizebox{\hsize}{!}{\includegraphics[angle=-0]{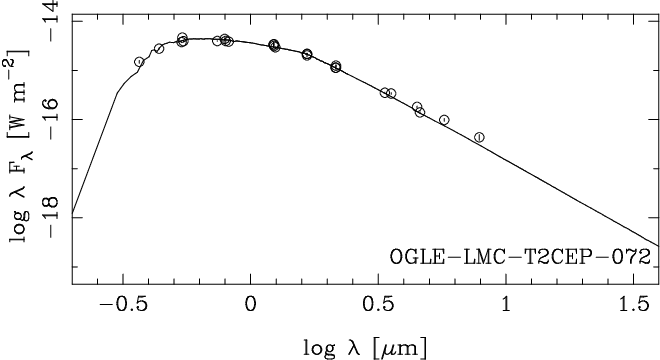}} 
\end{minipage}
 
\begin{minipage}{0.4\textwidth}
\resizebox{\hsize}{!}{\includegraphics[angle=-0]{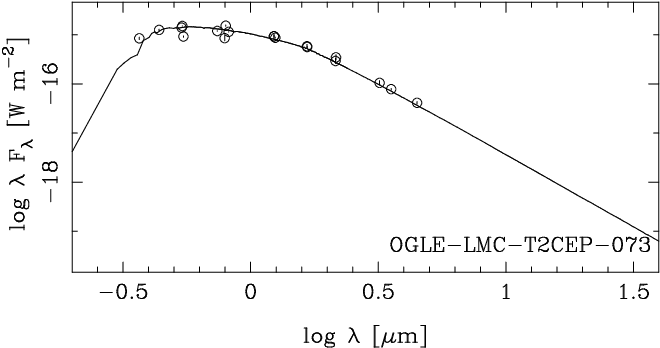}} 
\end{minipage}
\begin{minipage}{0.4\textwidth}
\resizebox{\hsize}{!}{\includegraphics[angle=-0]{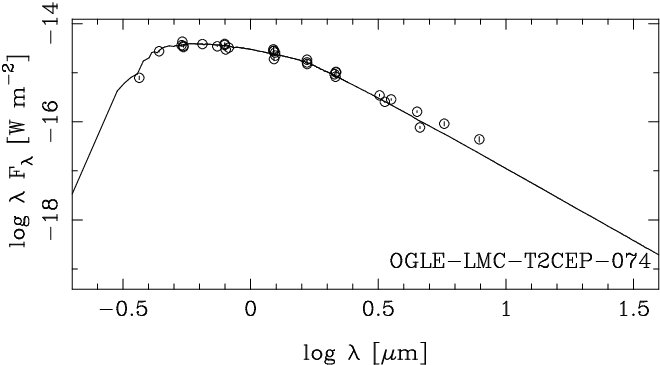}} 
\end{minipage}
 
\begin{minipage}{0.4\textwidth}
\resizebox{\hsize}{!}{\includegraphics[angle=-0]{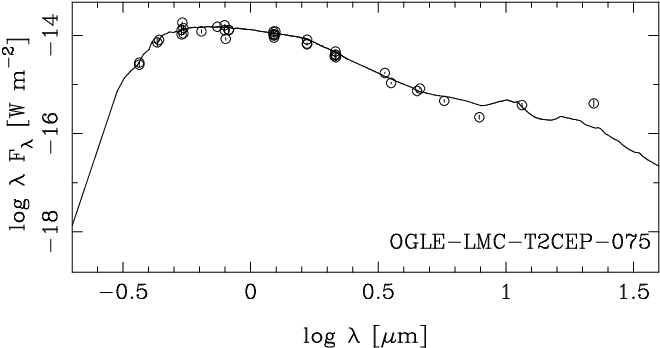}} 
\end{minipage}
\begin{minipage}{0.4\textwidth}
\resizebox{\hsize}{!}{\includegraphics[angle=-0]{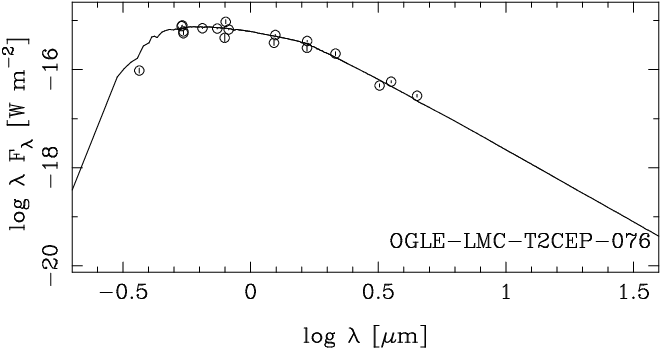}} 
\end{minipage}
 
\begin{minipage}{0.4\textwidth}
\resizebox{\hsize}{!}{\includegraphics[angle=-0]{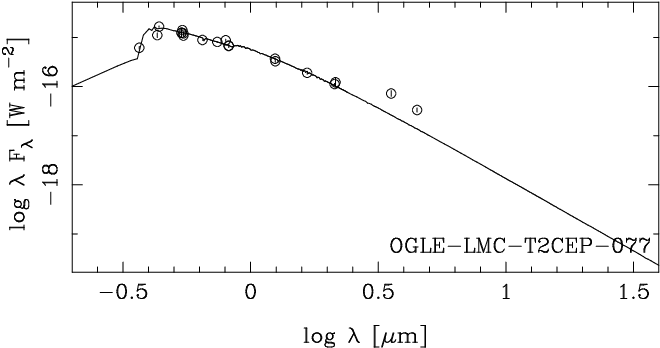}} 
\end{minipage}
\begin{minipage}{0.4\textwidth}
\resizebox{\hsize}{!}{\includegraphics[angle=-0]{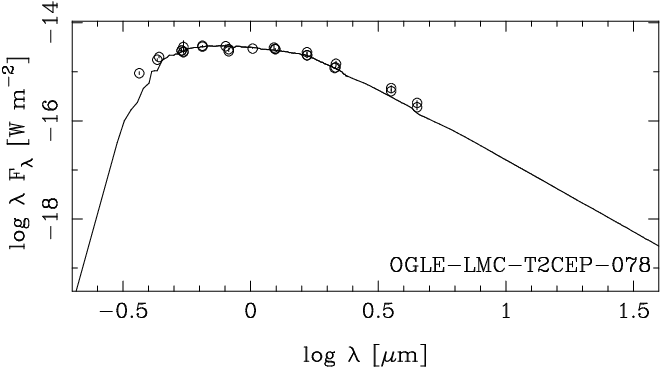}} 
\end{minipage}
 
\begin{minipage}{0.4\textwidth}
\resizebox{\hsize}{!}{\includegraphics[angle=-0]{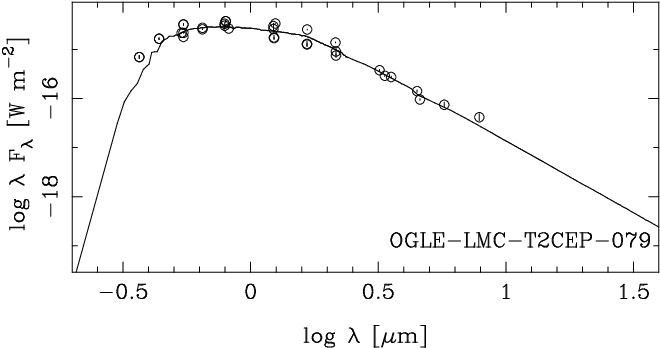}} 
\end{minipage}
\begin{minipage}{0.4\textwidth}
\resizebox{\hsize}{!}{\includegraphics[angle=-0]{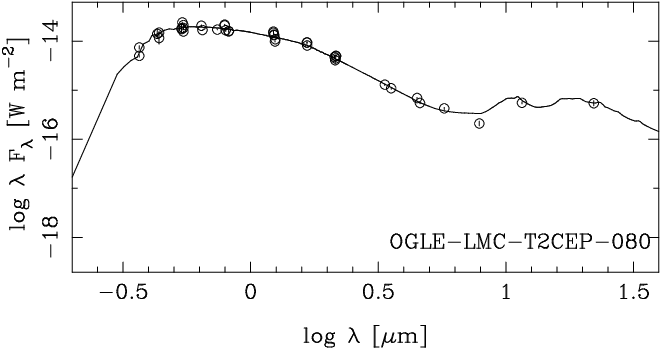}} 
\end{minipage}

\caption{Continued}
\end{figure*}

\setcounter{figure}{0}
\begin{figure*}
\centering

\begin{minipage}{0.4\textwidth}
\resizebox{\hsize}{!}{\includegraphics[angle=-0]{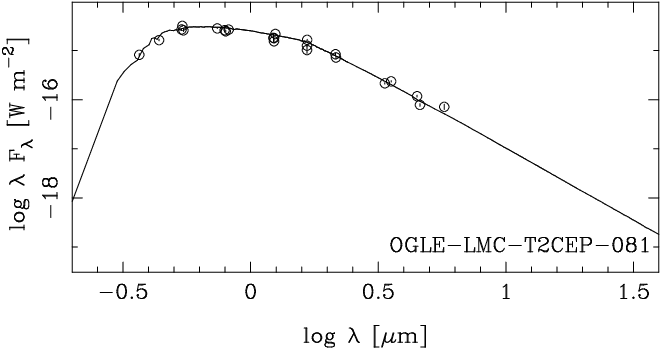}} 
\end{minipage}
\begin{minipage}{0.4\textwidth}
\resizebox{\hsize}{!}{\includegraphics[angle=-0]{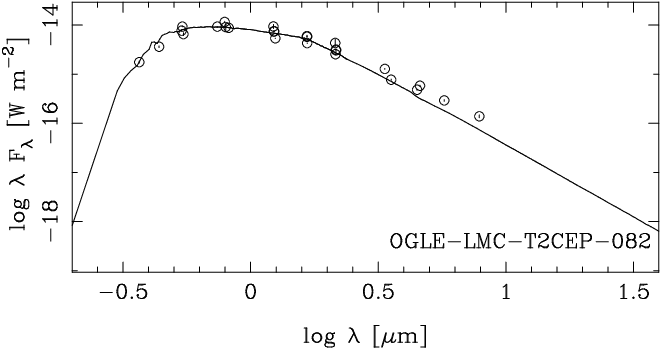}} 
\end{minipage}
 
\begin{minipage}{0.4\textwidth}
\resizebox{\hsize}{!}{\includegraphics[angle=-0]{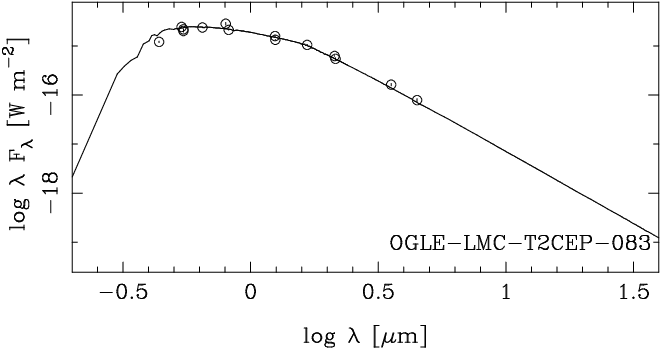}} 
\end{minipage}
\begin{minipage}{0.4\textwidth}
\resizebox{\hsize}{!}{\includegraphics[angle=-0]{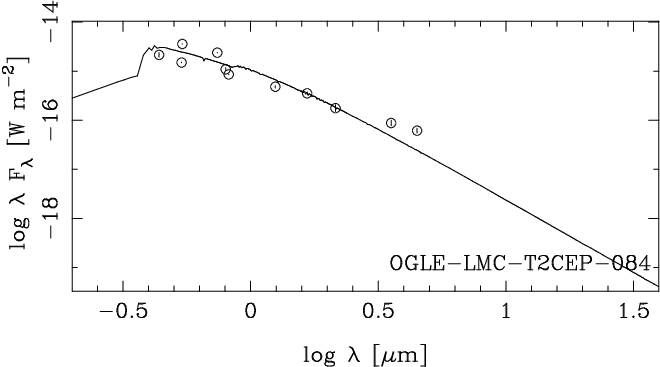}} 
\end{minipage}
 
\begin{minipage}{0.4\textwidth}
\resizebox{\hsize}{!}{\includegraphics[angle=-0]{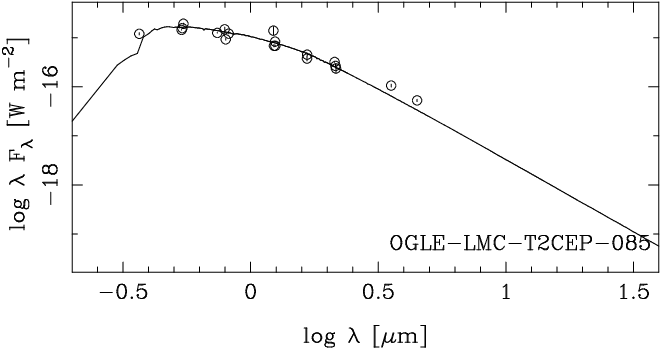}} 
\end{minipage}
\begin{minipage}{0.4\textwidth}
\resizebox{\hsize}{!}{\includegraphics[angle=-0]{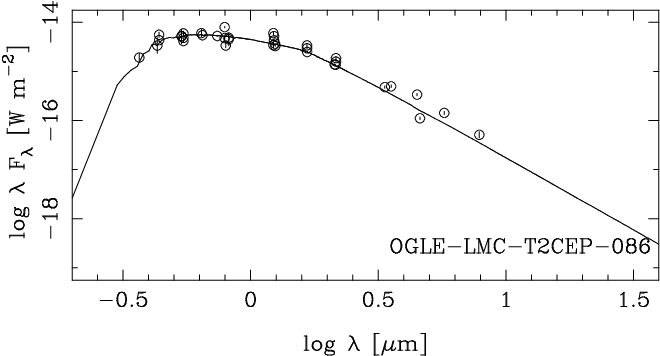}} 
\end{minipage}
 
\begin{minipage}{0.4\textwidth}
\resizebox{\hsize}{!}{\includegraphics[angle=-0]{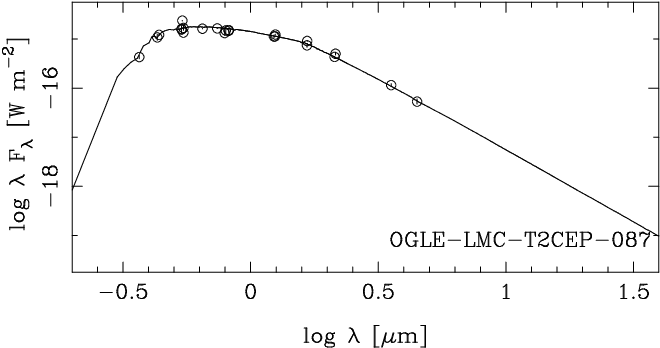}} 
\end{minipage}
\begin{minipage}{0.4\textwidth}
\resizebox{\hsize}{!}{\includegraphics[angle=-0]{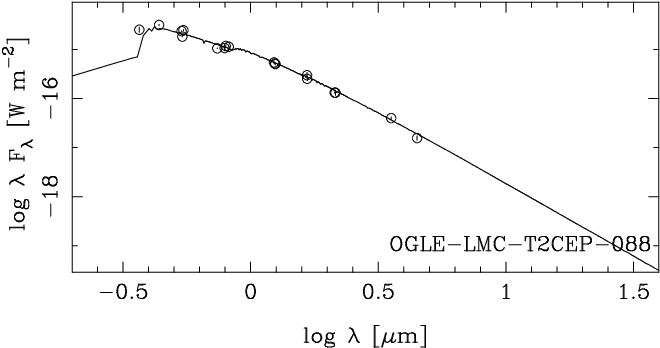}} 
\end{minipage}
 
\begin{minipage}{0.4\textwidth}
\resizebox{\hsize}{!}{\includegraphics[angle=-0]{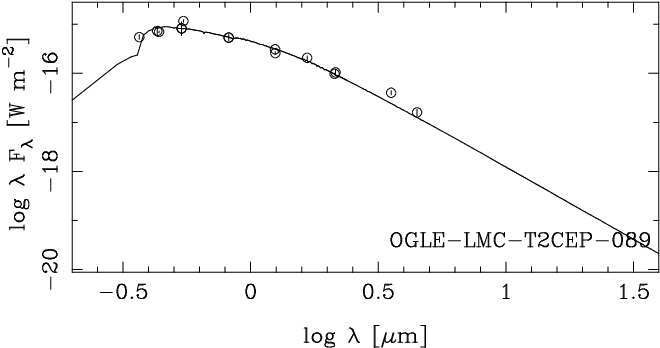}} 
\end{minipage}
\begin{minipage}{0.4\textwidth}
\resizebox{\hsize}{!}{\includegraphics[angle=-0]{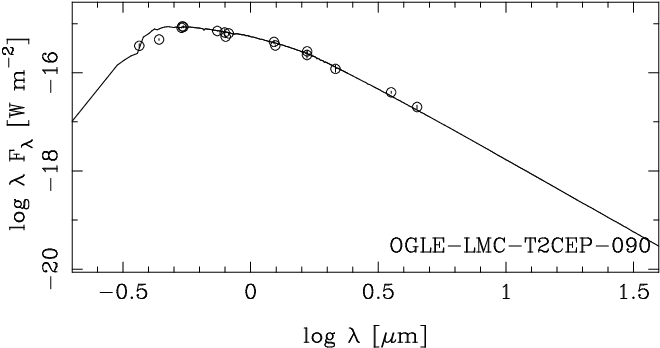}} 
\end{minipage}

\caption{Continued}
\end{figure*}

\setcounter{figure}{0}
\begin{figure*}
\centering

\begin{minipage}{0.4\textwidth}
\resizebox{\hsize}{!}{\includegraphics[angle=-0]{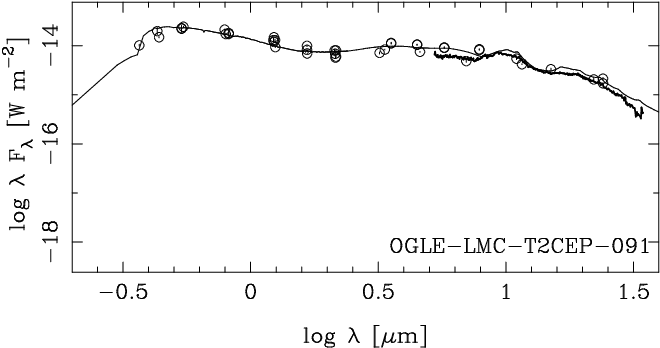}} 
\end{minipage}
\begin{minipage}{0.4\textwidth}
\resizebox{\hsize}{!}{\includegraphics[angle=-0]{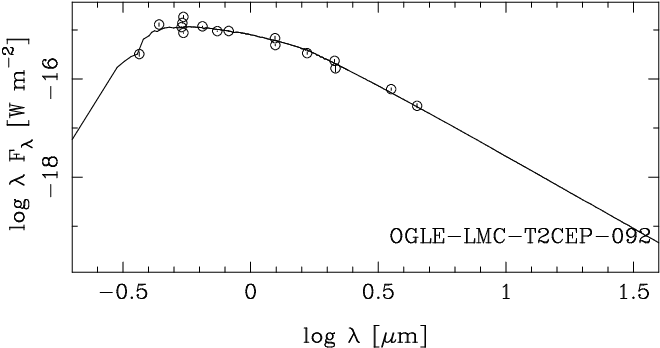}} 
\end{minipage}
 
\begin{minipage}{0.4\textwidth}
\resizebox{\hsize}{!}{\includegraphics[angle=-0]{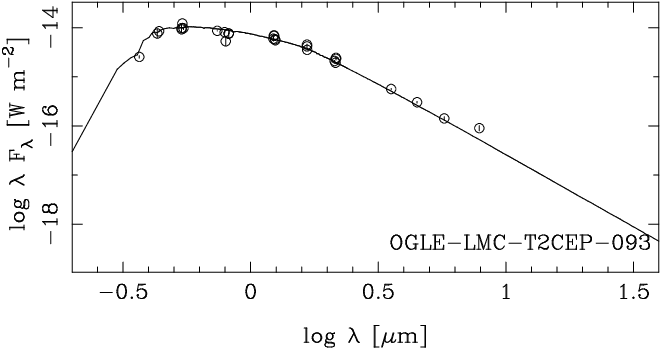}} 
\end{minipage}
\begin{minipage}{0.4\textwidth}
\resizebox{\hsize}{!}{\includegraphics[angle=-0]{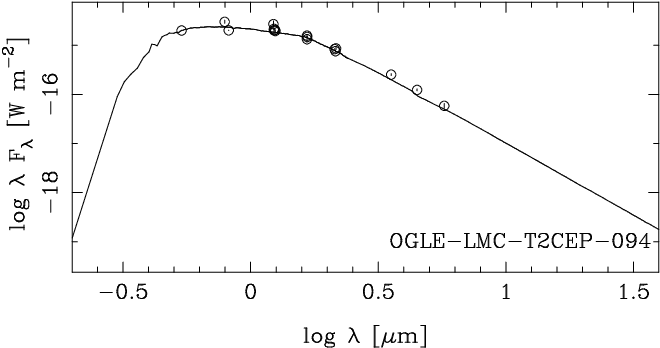}} 
\end{minipage}
 
\begin{minipage}{0.4\textwidth}
\resizebox{\hsize}{!}{\includegraphics[angle=-0]{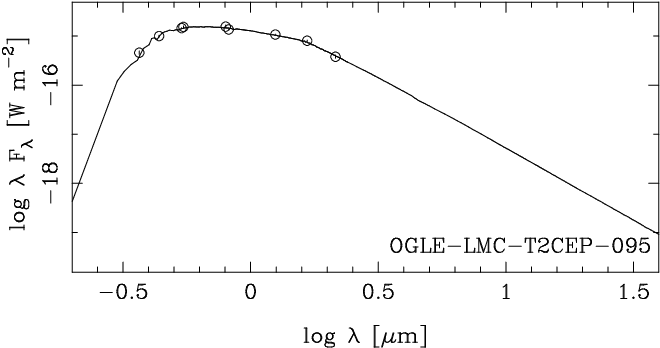}} 
\end{minipage}
\begin{minipage}{0.4\textwidth}
\resizebox{\hsize}{!}{\includegraphics[angle=-0]{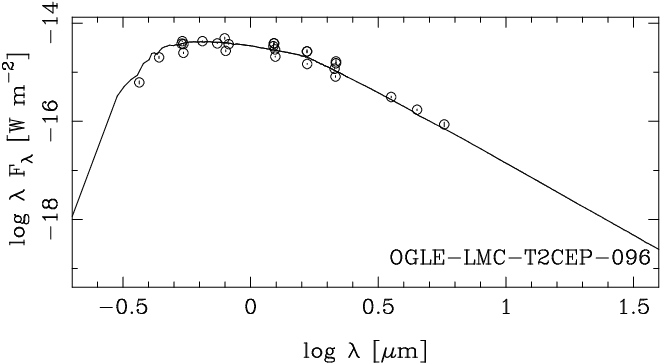}} 
\end{minipage}
 
\begin{minipage}{0.4\textwidth}
\resizebox{\hsize}{!}{\includegraphics[angle=-0]{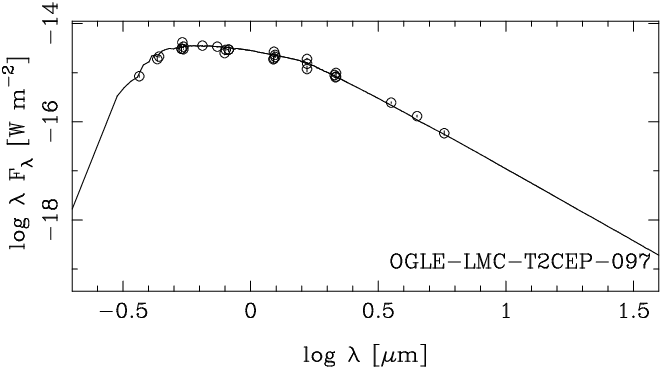}} 
\end{minipage}
\begin{minipage}{0.4\textwidth}
\resizebox{\hsize}{!}{\includegraphics[angle=-0]{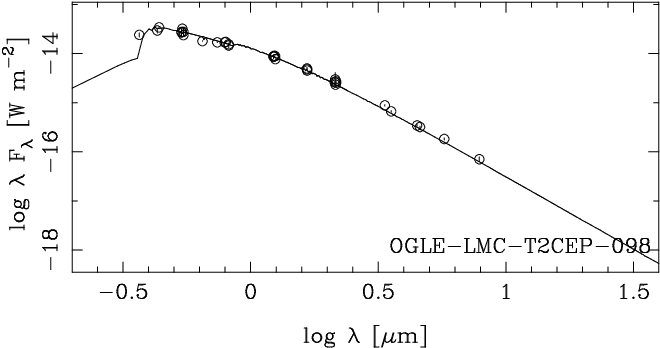}} 
\end{minipage}
 
\begin{minipage}{0.4\textwidth}
\resizebox{\hsize}{!}{\includegraphics[angle=-0]{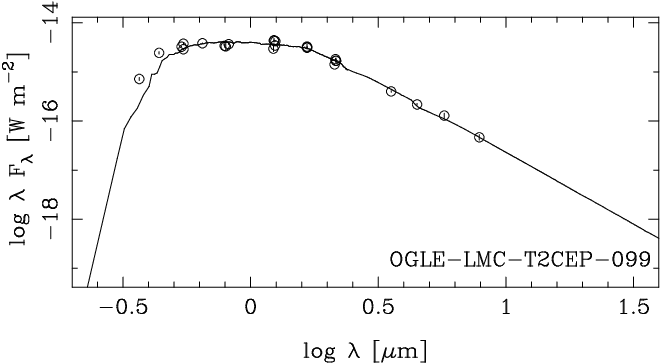}} 
\end{minipage}
\begin{minipage}{0.4\textwidth}
\resizebox{\hsize}{!}{\includegraphics[angle=-0]{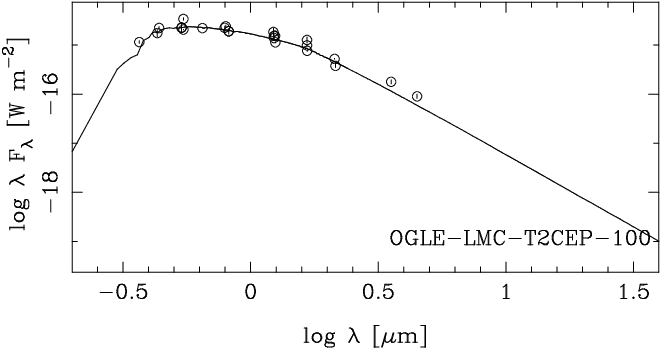}} 
\end{minipage}
 
\caption{Continued}
\end{figure*}
 
\setcounter{figure}{0}
\begin{figure*}
\centering

\begin{minipage}{0.4\textwidth}
\resizebox{\hsize}{!}{\includegraphics[angle=-0]{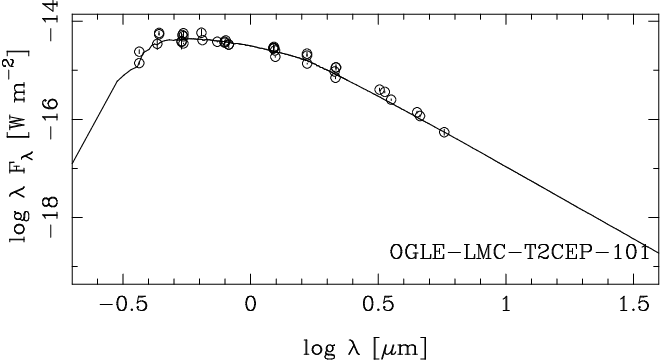}} 
\end{minipage}
\begin{minipage}{0.4\textwidth}
\resizebox{\hsize}{!}{\includegraphics[angle=-0]{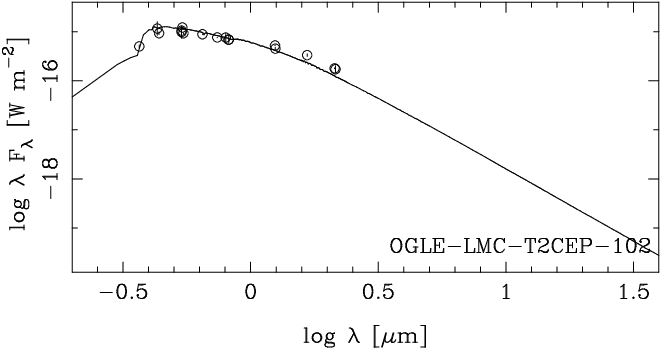}} 
\end{minipage}
 
\begin{minipage}{0.4\textwidth}
\resizebox{\hsize}{!}{\includegraphics[angle=-0]{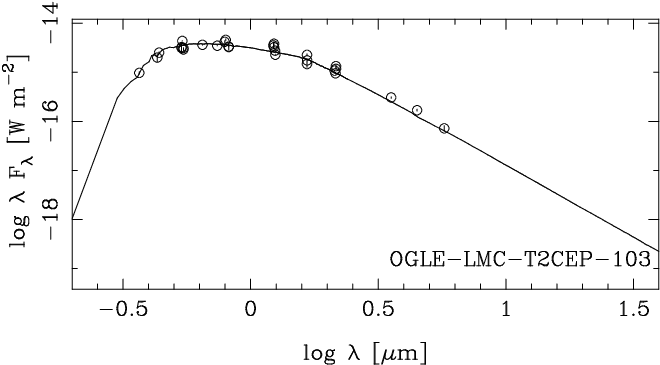}} 
\end{minipage}
\begin{minipage}{0.4\textwidth}
\resizebox{\hsize}{!}{\includegraphics[angle=-0]{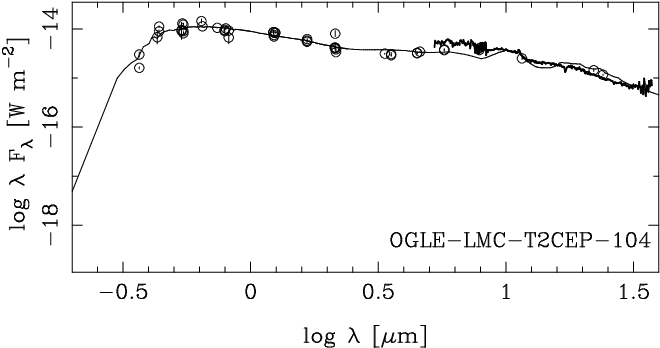}} 
\end{minipage}
 
\begin{minipage}{0.4\textwidth}
\resizebox{\hsize}{!}{\includegraphics[angle=-0]{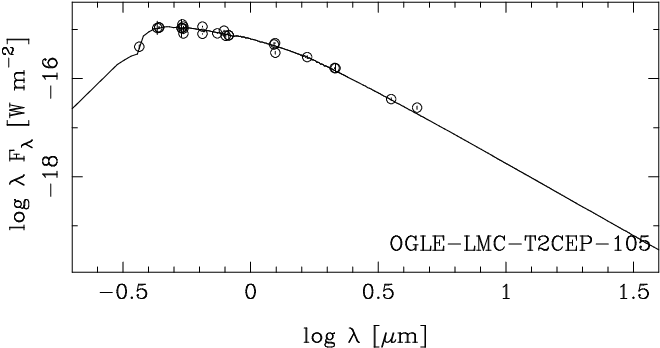}} 
\end{minipage}
\begin{minipage}{0.4\textwidth}
\resizebox{\hsize}{!}{\includegraphics[angle=-0]{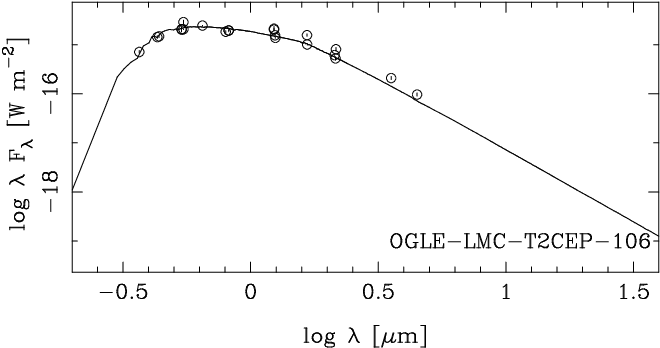}} 
\end{minipage}
 
\begin{minipage}{0.4\textwidth}
\resizebox{\hsize}{!}{\includegraphics[angle=-0]{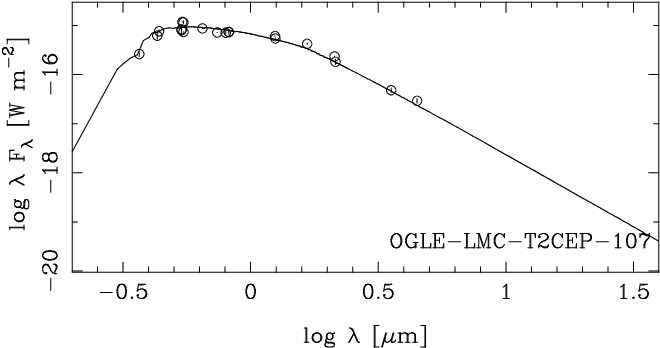}} 
\end{minipage}
\begin{minipage}{0.4\textwidth}
\resizebox{\hsize}{!}{\includegraphics[angle=-0]{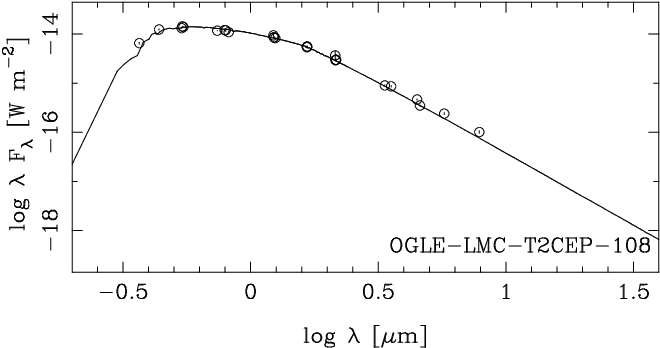}} 
\end{minipage}
 
 \begin{minipage}{0.4\textwidth}
\resizebox{\hsize}{!}{\includegraphics[angle=-0]{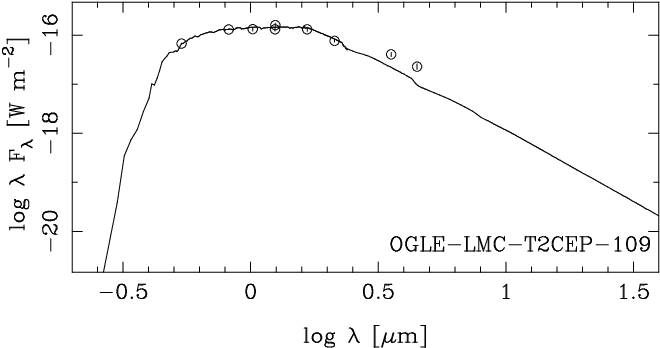}} 
\end{minipage}
\begin{minipage}{0.4\textwidth}
\resizebox{\hsize}{!}{\includegraphics[angle=-0]{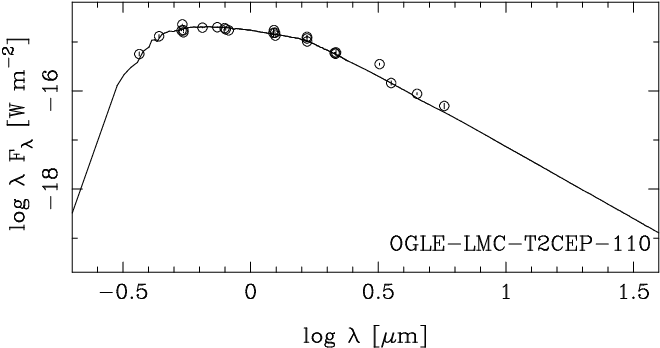}} 
\end{minipage}

\caption{Continued}
\end{figure*}
 
\setcounter{figure}{0}
\begin{figure*}
\centering

\begin{minipage}{0.4\textwidth}
\resizebox{\hsize}{!}{\includegraphics[angle=-0]{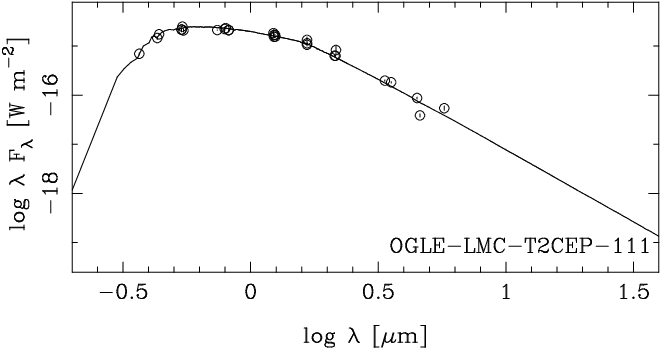}} 
\end{minipage}
\begin{minipage}{0.4\textwidth}
\resizebox{\hsize}{!}{\includegraphics[angle=-0]{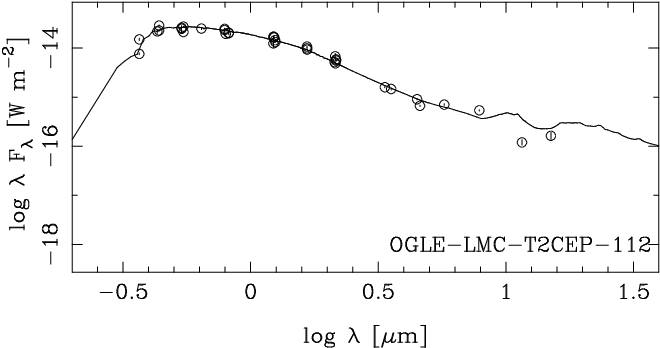}} 
\end{minipage}
 
\begin{minipage}{0.4\textwidth}
\resizebox{\hsize}{!}{\includegraphics[angle=-0]{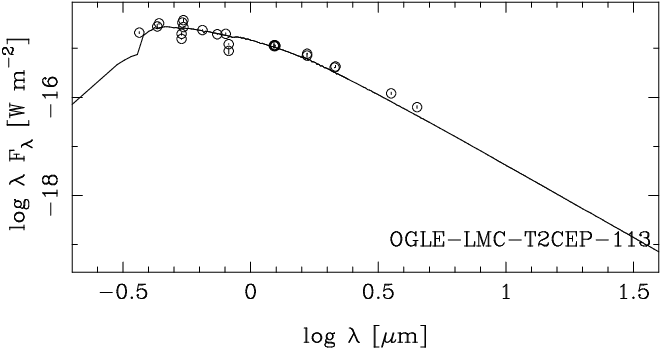}} 
\end{minipage}
\begin{minipage}{0.4\textwidth}
\resizebox{\hsize}{!}{\includegraphics[angle=-0]{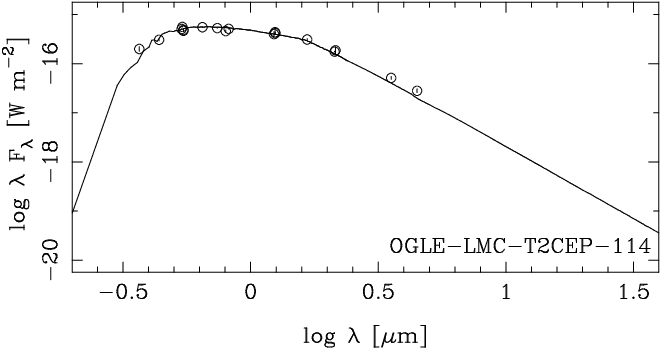}} 
\end{minipage}
 
\begin{minipage}{0.4\textwidth}
\resizebox{\hsize}{!}{\includegraphics[angle=-0]{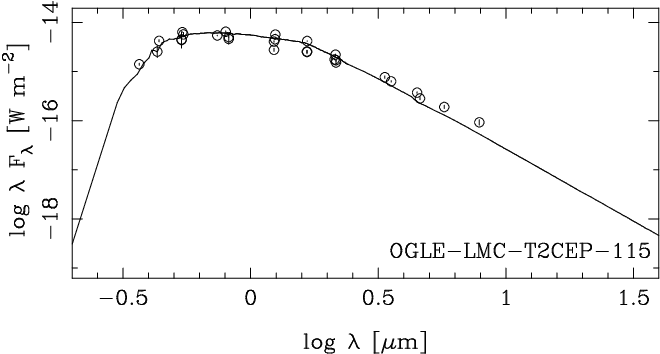}} 
\end{minipage}
\begin{minipage}{0.4\textwidth}
\resizebox{\hsize}{!}{\includegraphics[angle=-0]{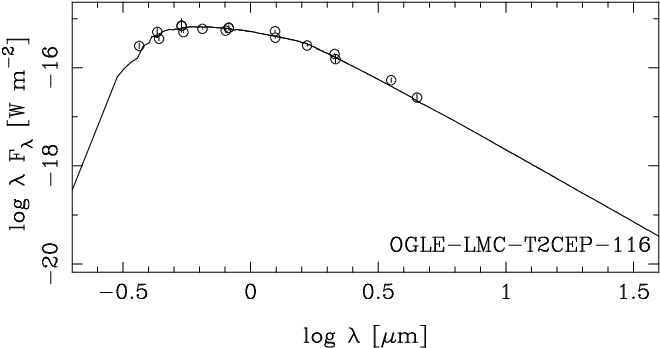}} 
\end{minipage}
 
\begin{minipage}{0.4\textwidth}
\resizebox{\hsize}{!}{\includegraphics[angle=-0]{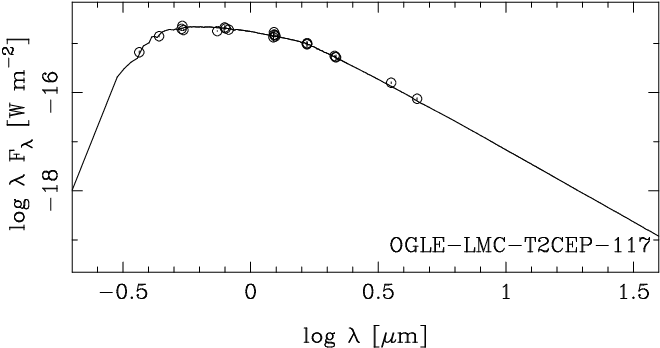}} 
\end{minipage}
\begin{minipage}{0.4\textwidth}
\resizebox{\hsize}{!}{\includegraphics[angle=-0]{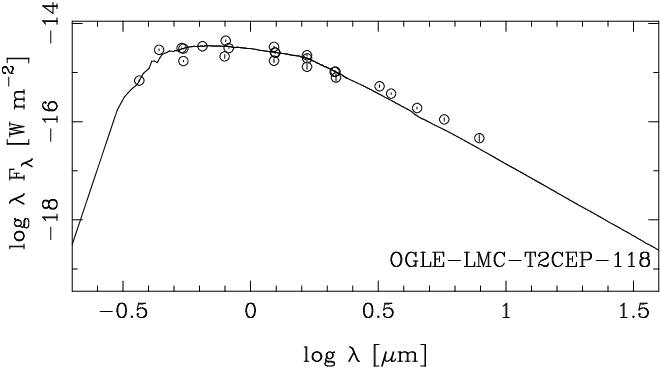}} 
\end{minipage}
 
\begin{minipage}{0.4\textwidth}
\resizebox{\hsize}{!}{\includegraphics[angle=-0]{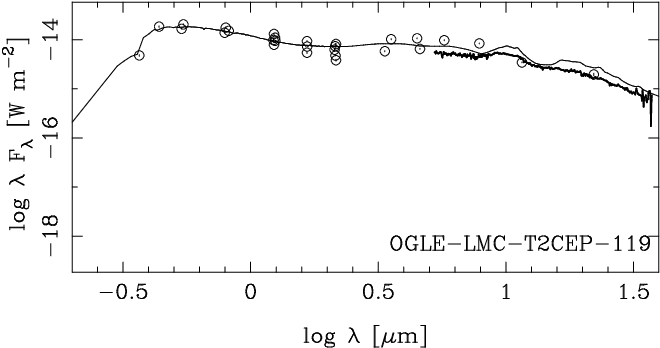}} 
\end{minipage}
\begin{minipage}{0.4\textwidth}
\resizebox{\hsize}{!}{\includegraphics[angle=-0]{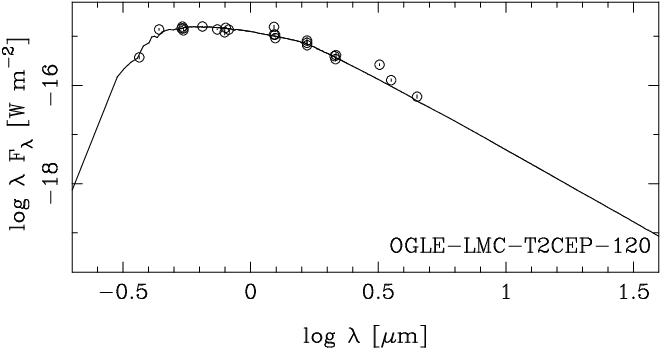}} 
\end{minipage}
 
\caption{Continued}
\end{figure*}

\setcounter{figure}{0}
\begin{figure*}
\centering

\begin{minipage}{0.4\textwidth}
\resizebox{\hsize}{!}{\includegraphics[angle=-0]{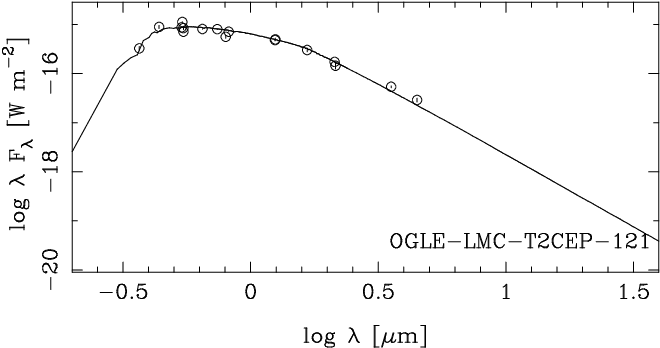}} 
\end{minipage}
\begin{minipage}{0.4\textwidth}
\resizebox{\hsize}{!}{\includegraphics[angle=-0]{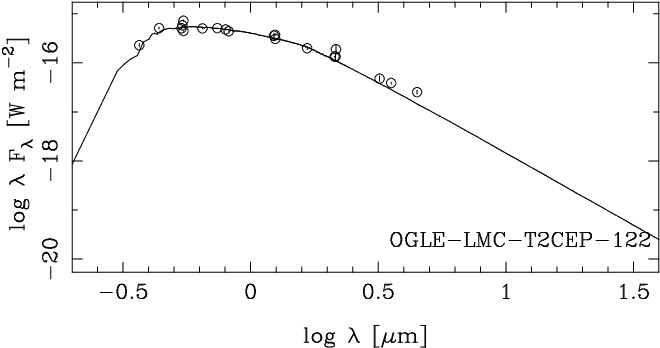}} 
\end{minipage}
 
\begin{minipage}{0.4\textwidth}
\resizebox{\hsize}{!}{\includegraphics[angle=-0]{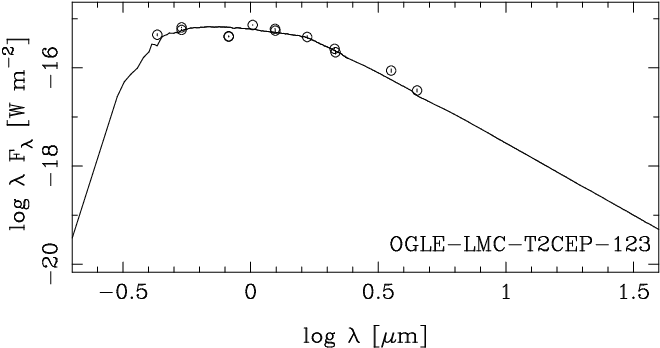}} 
\end{minipage}
\begin{minipage}{0.4\textwidth}
\resizebox{\hsize}{!}{\includegraphics[angle=-0]{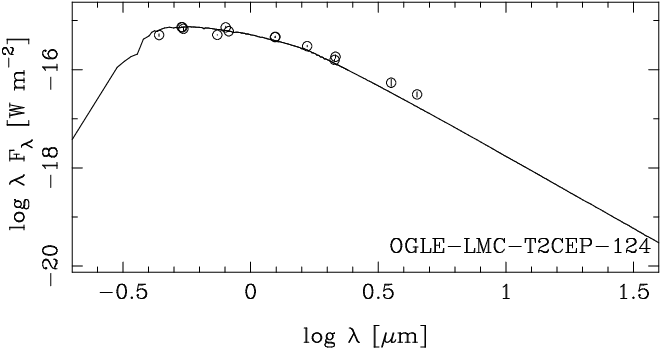}} 
\end{minipage}
 
\begin{minipage}{0.4\textwidth}
\resizebox{\hsize}{!}{\includegraphics[angle=-0]{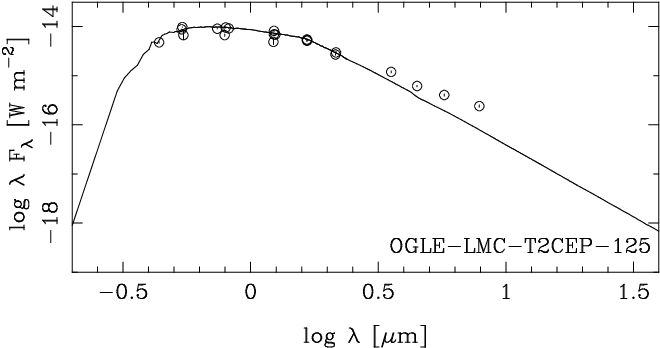}} 
\end{minipage}
\begin{minipage}{0.4\textwidth}
\resizebox{\hsize}{!}{\includegraphics[angle=-0]{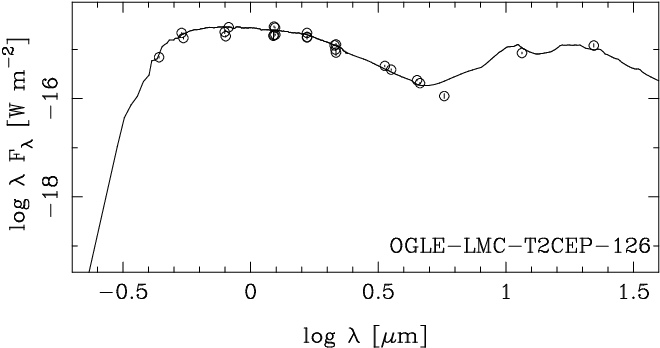}} 
\end{minipage}
 
\begin{minipage}{0.4\textwidth}
\resizebox{\hsize}{!}{\includegraphics[angle=-0]{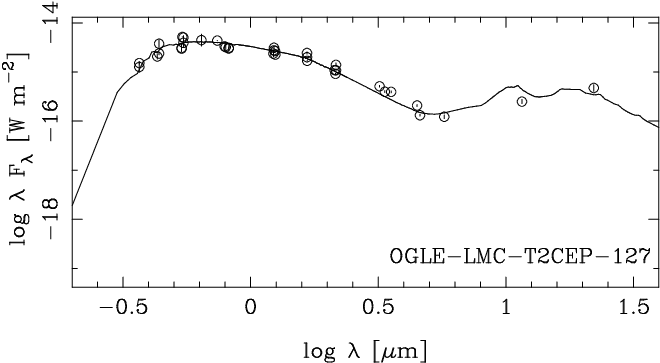}} 
\end{minipage}
\begin{minipage}{0.4\textwidth}
\resizebox{\hsize}{!}{\includegraphics[angle=-0]{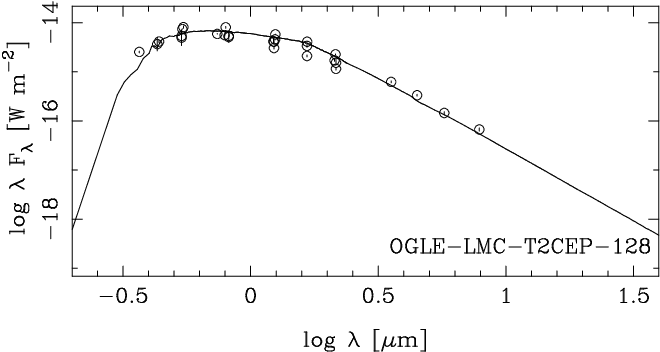}} 
\end{minipage}
 
\begin{minipage}{0.4\textwidth}
\resizebox{\hsize}{!}{\includegraphics[angle=-0]{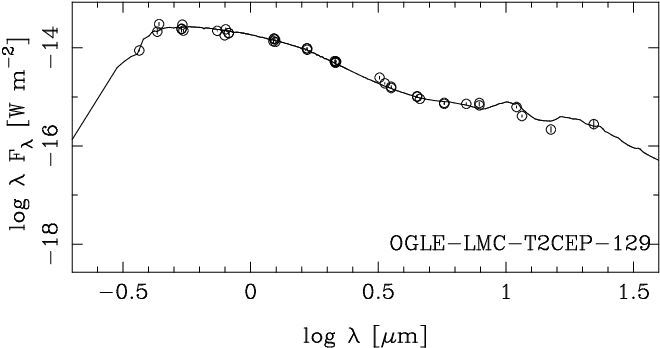}} 
\end{minipage}
\begin{minipage}{0.4\textwidth}
\resizebox{\hsize}{!}{\includegraphics[angle=-0]{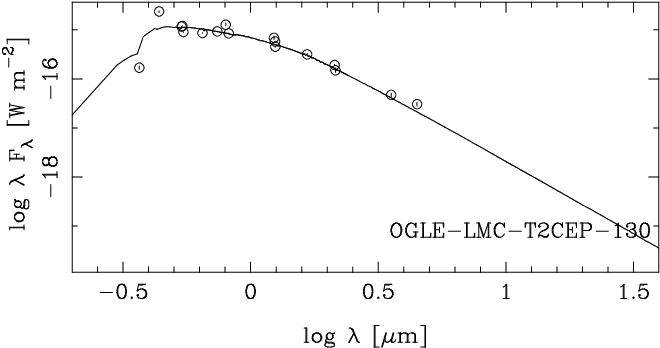}} 
\end{minipage}

\caption{Continued}
\end{figure*}
 
\setcounter{figure}{0}
\begin{figure*}
\centering

\begin{minipage}{0.4\textwidth}
\resizebox{\hsize}{!}{\includegraphics[angle=-0]{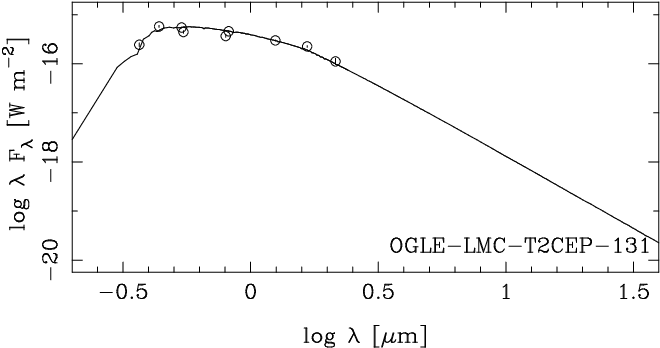}} 
\end{minipage}
\begin{minipage}{0.4\textwidth}
\resizebox{\hsize}{!}{\includegraphics[angle=-0]{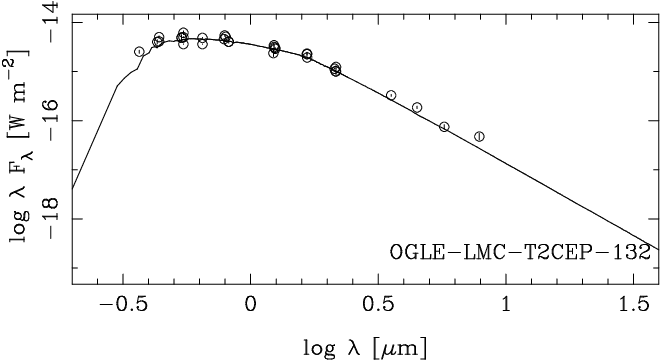}} 
\end{minipage}

\begin{minipage}{0.4\textwidth}
\resizebox{\hsize}{!}{\includegraphics[angle=-0]{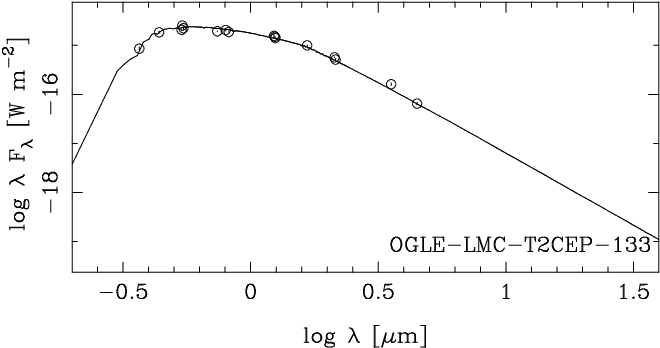}} 
\end{minipage}
\begin{minipage}{0.4\textwidth}
\resizebox{\hsize}{!}{\includegraphics[angle=-0]{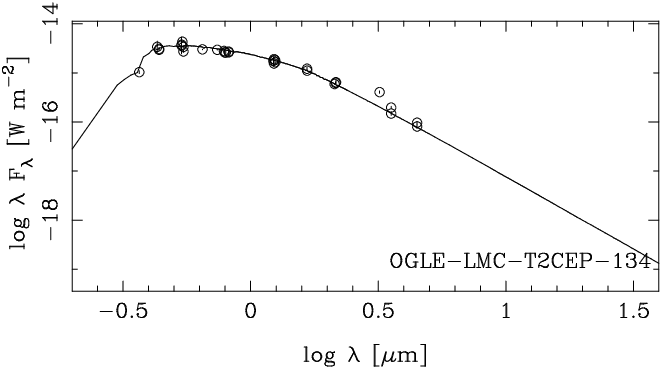}} 
\end{minipage}
 
\begin{minipage}{0.4\textwidth}
\resizebox{\hsize}{!}{\includegraphics[angle=-0]{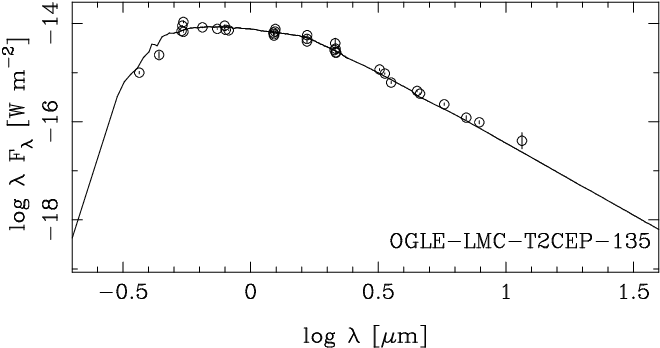}} 
\end{minipage}
\begin{minipage}{0.4\textwidth}
\resizebox{\hsize}{!}{\includegraphics[angle=-0]{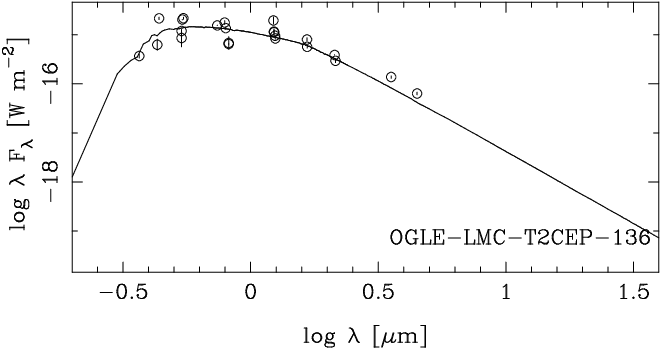}} 
\end{minipage}
 
\begin{minipage}{0.4\textwidth}
\resizebox{\hsize}{!}{\includegraphics[angle=-0]{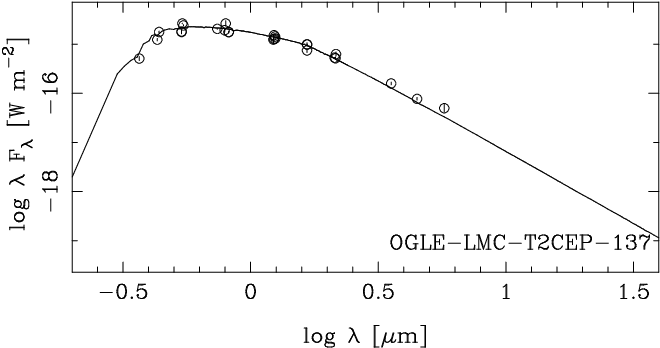}} 
\end{minipage}
\begin{minipage}{0.4\textwidth}
\resizebox{\hsize}{!}{\includegraphics[angle=-0]{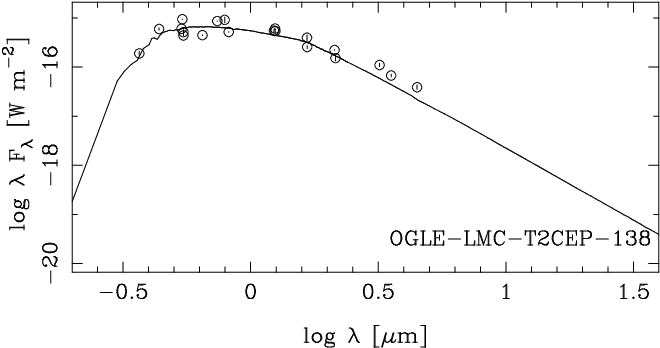}} 
\end{minipage}
 
\begin{minipage}{0.4\textwidth}
\resizebox{\hsize}{!}{\includegraphics[angle=-0]{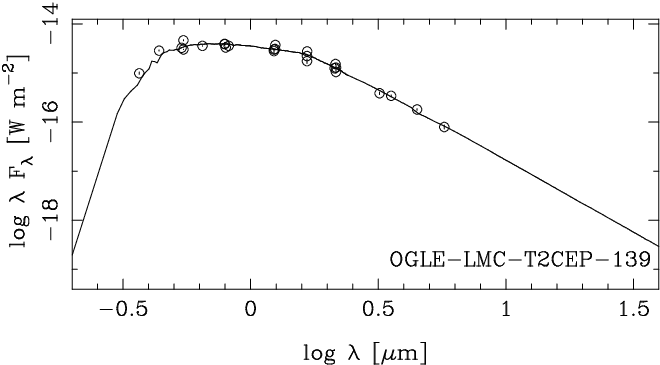}} 
\end{minipage}
\begin{minipage}{0.4\textwidth}
\resizebox{\hsize}{!}{\includegraphics[angle=-0]{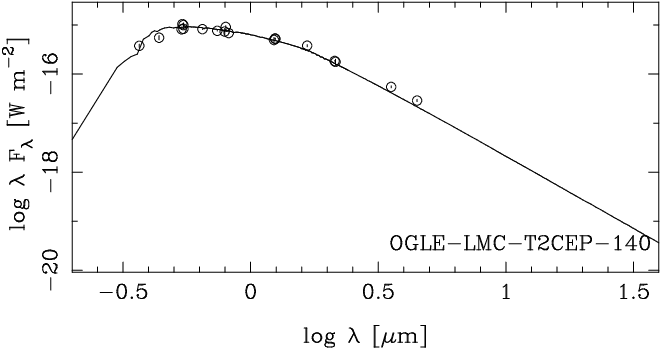}} 
\end{minipage}

\caption{Continued}
\end{figure*}
 
\setcounter{figure}{0}
\begin{figure*}
\centering

\begin{minipage}{0.4\textwidth}
\resizebox{\hsize}{!}{\includegraphics[angle=-0]{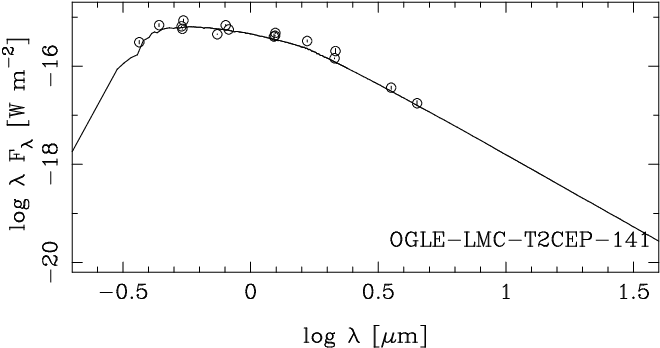}} 
\end{minipage}
\begin{minipage}{0.4\textwidth}
\resizebox{\hsize}{!}{\includegraphics[angle=-0]{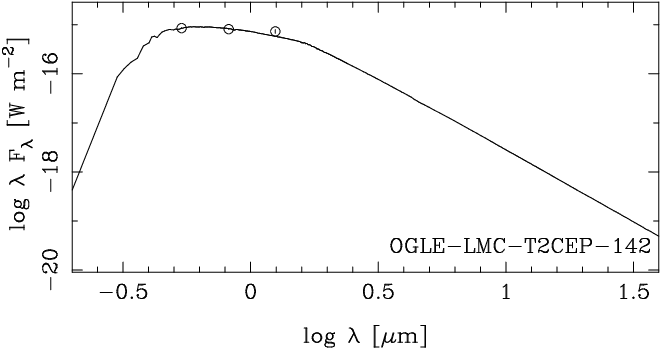}} 
\end{minipage}
 
\begin{minipage}{0.4\textwidth}
\resizebox{\hsize}{!}{\includegraphics[angle=-0]{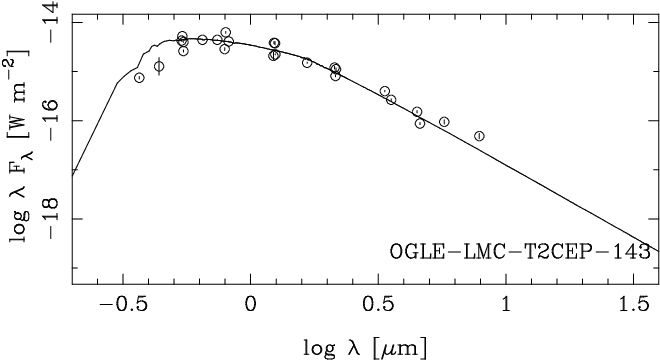}} 
\end{minipage}
\begin{minipage}{0.4\textwidth}
\resizebox{\hsize}{!}{\includegraphics[angle=-0]{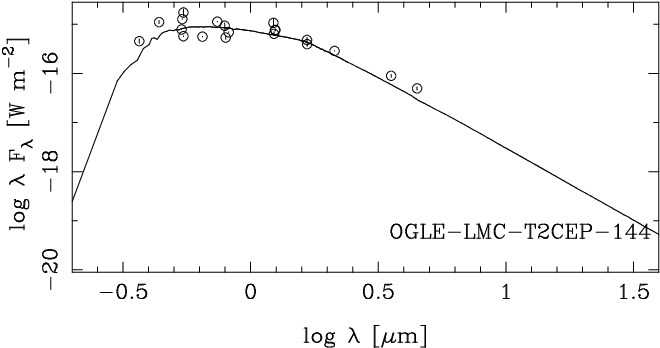}} 
\end{minipage}
 
\begin{minipage}{0.4\textwidth}
\resizebox{\hsize}{!}{\includegraphics[angle=-0]{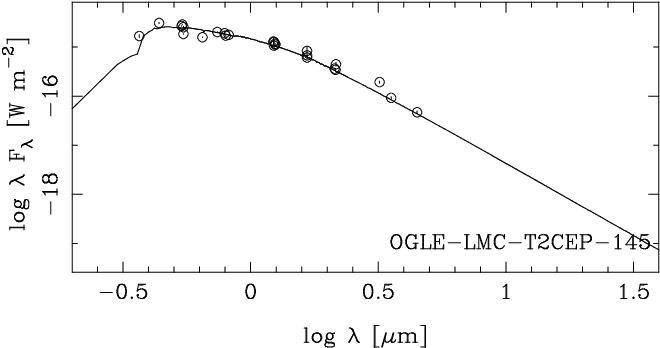}} 
\end{minipage}
\begin{minipage}{0.4\textwidth}
\resizebox{\hsize}{!}{\includegraphics[angle=-0]{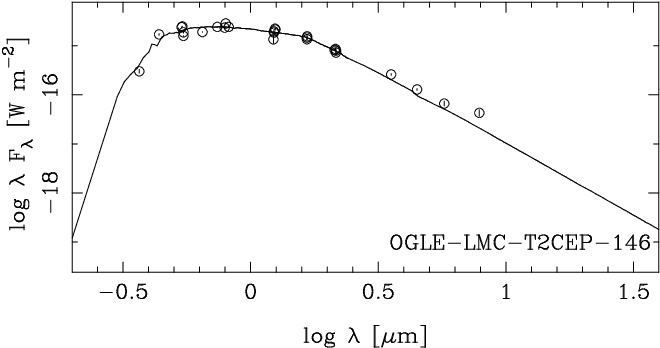}} 
\end{minipage}
 
\begin{minipage}{0.4\textwidth}
\resizebox{\hsize}{!}{\includegraphics[angle=-0]{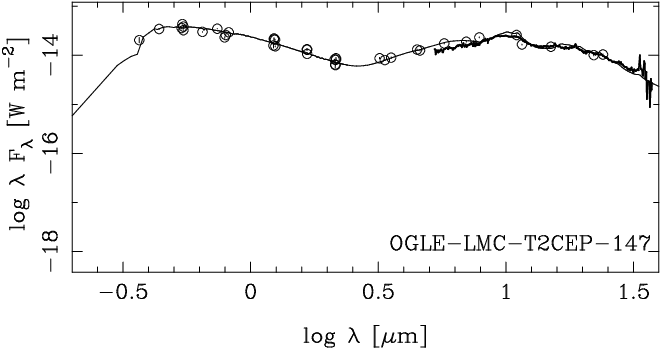}} 
\end{minipage}
\begin{minipage}{0.4\textwidth}
\resizebox{\hsize}{!}{\includegraphics[angle=-0]{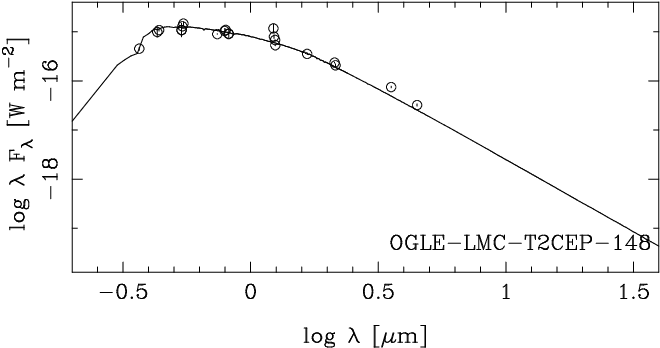}} 
\end{minipage}
 
\begin{minipage}{0.4\textwidth}
\resizebox{\hsize}{!}{\includegraphics[angle=-0]{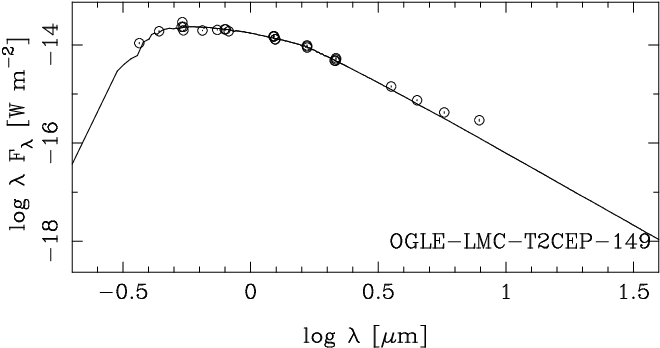}} 
\end{minipage}
\begin{minipage}{0.4\textwidth}
\resizebox{\hsize}{!}{\includegraphics[angle=-0]{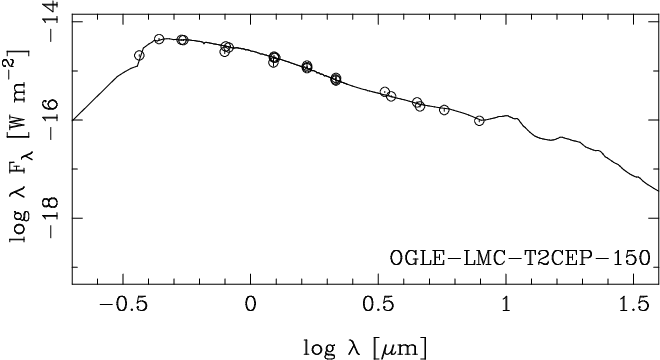}} 
\end{minipage}

\caption{Continued}
\end{figure*}

\clearpage

\setcounter{figure}{0}
\begin{figure*}
\centering

\begin{minipage}{0.4\textwidth}
\resizebox{\hsize}{!}{\includegraphics[angle=-0]{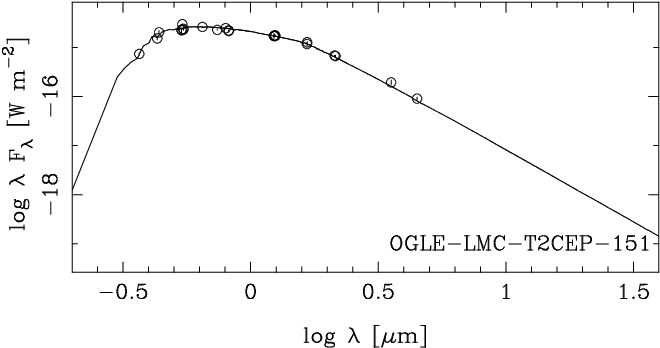}} 
\end{minipage}
\begin{minipage}{0.4\textwidth}
\resizebox{\hsize}{!}{\includegraphics[angle=-0]{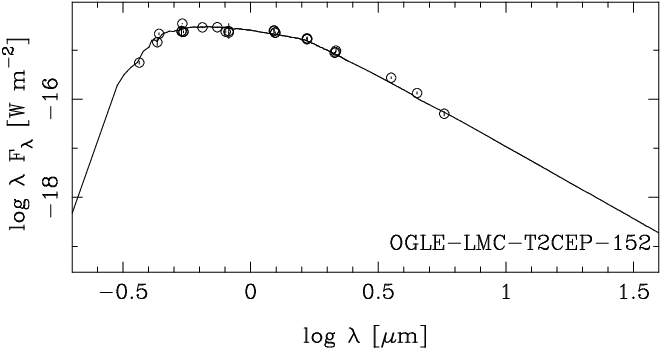}} 
\end{minipage}
 
\begin{minipage}{0.4\textwidth}
\resizebox{\hsize}{!}{\includegraphics[angle=-0]{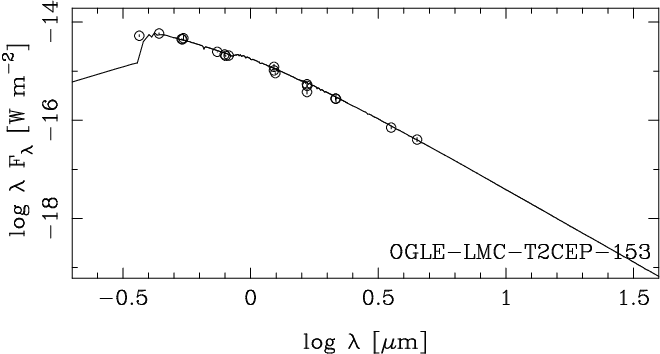}} 
\end{minipage}
\begin{minipage}{0.4\textwidth}
\resizebox{\hsize}{!}{\includegraphics[angle=-0]{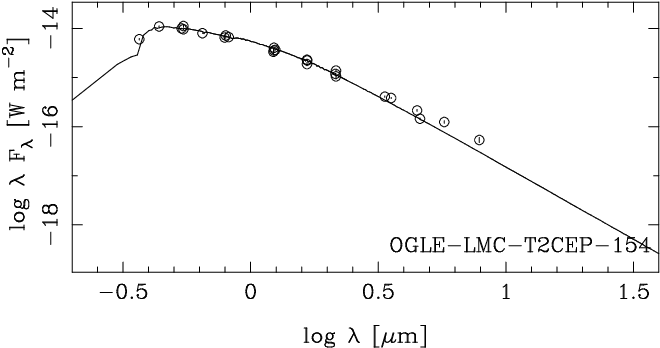}} 
\end{minipage}
 
\begin{minipage}{0.4\textwidth}
\resizebox{\hsize}{!}{\includegraphics[angle=-0]{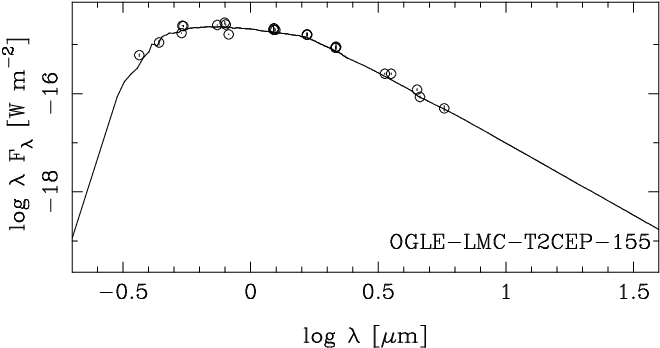}} 
\end{minipage}
\begin{minipage}{0.4\textwidth}
\resizebox{\hsize}{!}{\includegraphics[angle=-0]{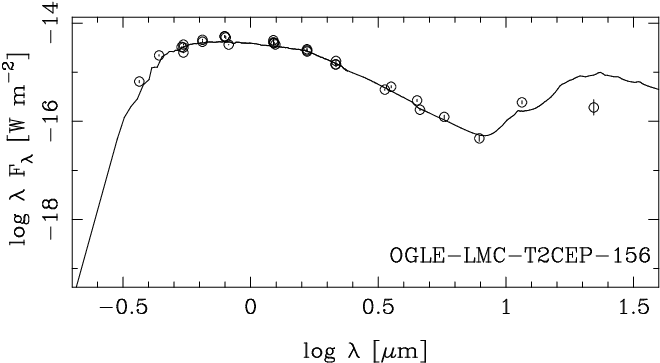}} 
\end{minipage}

\begin{minipage}{0.4\textwidth}
\resizebox{\hsize}{!}{\includegraphics[angle=-0]{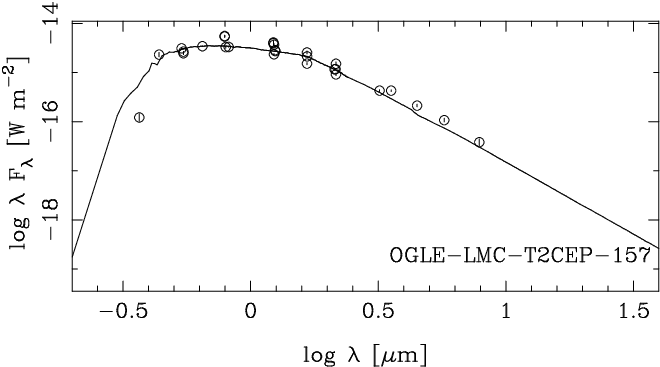}} 
\end{minipage}
\begin{minipage}{0.4\textwidth}
\resizebox{\hsize}{!}{\includegraphics[angle=-0]{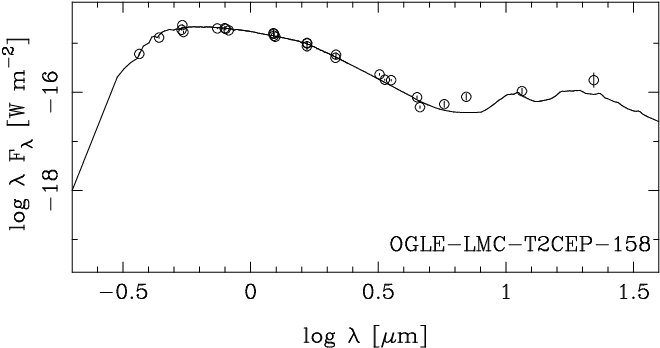}} 
\end{minipage}
 
\begin{minipage}{0.4\textwidth}
\resizebox{\hsize}{!}{\includegraphics[angle=-0]{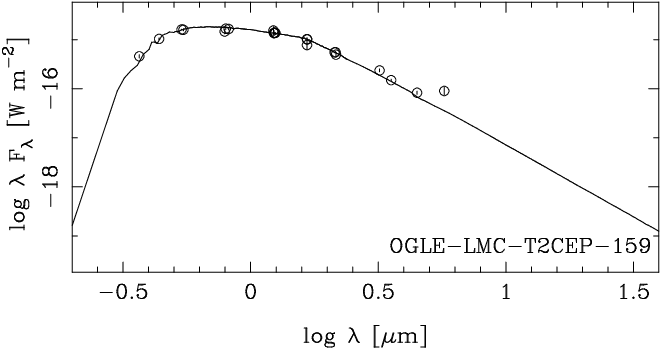}} 
\end{minipage}
\begin{minipage}{0.4\textwidth}
\resizebox{\hsize}{!}{\includegraphics[angle=-0]{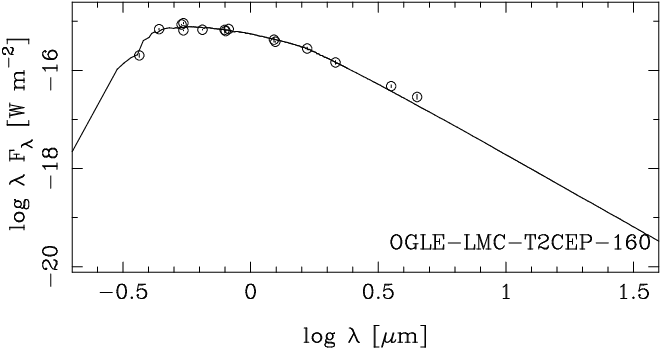}} 
\end{minipage}

\caption{Continued}
\end{figure*}
 
\setcounter{figure}{0}
\begin{figure*}
\centering

\begin{minipage}{0.4\textwidth}
\resizebox{\hsize}{!}{\includegraphics[angle=-0]{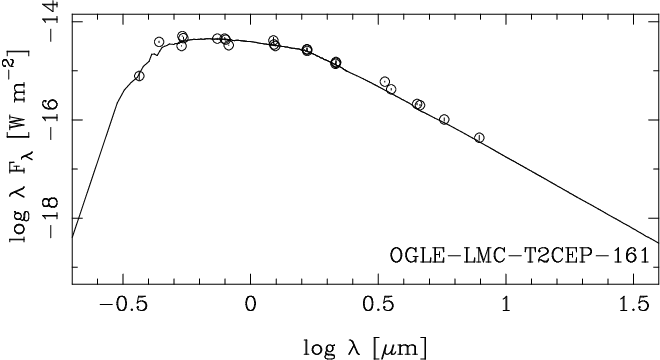}} 
\end{minipage}
\begin{minipage}{0.4\textwidth}
\resizebox{\hsize}{!}{\includegraphics[angle=-0]{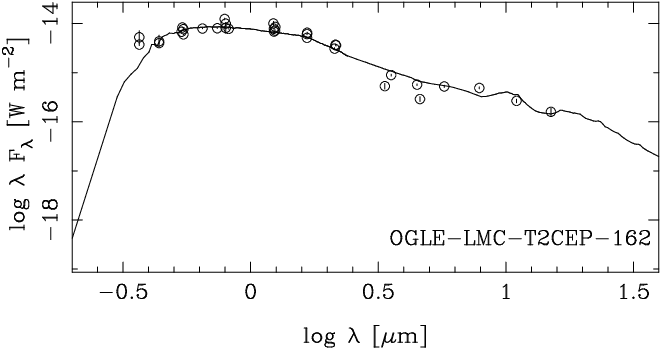}} 
\end{minipage}
 
\begin{minipage}{0.4\textwidth}
\resizebox{\hsize}{!}{\includegraphics[angle=-0]{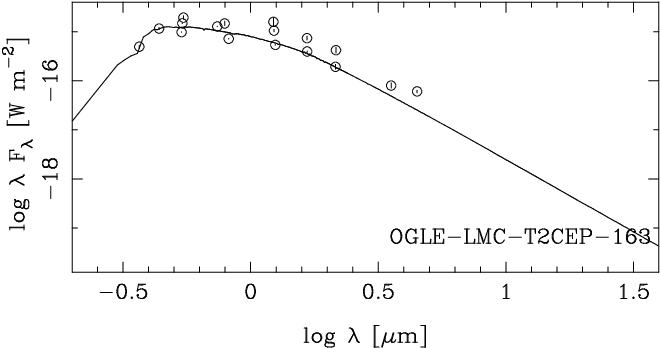}} 
\end{minipage}
\begin{minipage}{0.4\textwidth}
\resizebox{\hsize}{!}{\includegraphics[angle=-0]{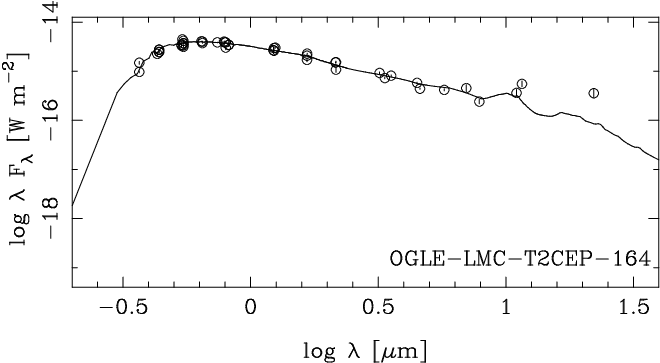}} 
\end{minipage}
 
\begin{minipage}{0.4\textwidth}
\resizebox{\hsize}{!}{\includegraphics[angle=-0]{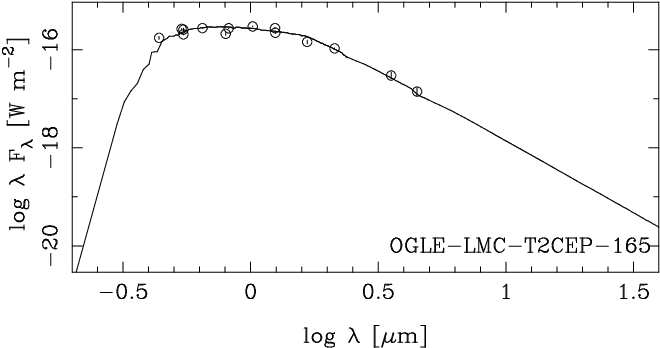}} 
\end{minipage}
\begin{minipage}{0.4\textwidth}
\resizebox{\hsize}{!}{\includegraphics[angle=-0]{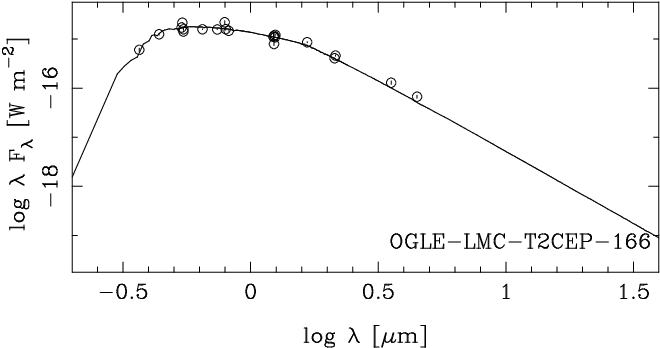}} 
\end{minipage}
 
\begin{minipage}{0.4\textwidth}
\resizebox{\hsize}{!}{\includegraphics[angle=-0]{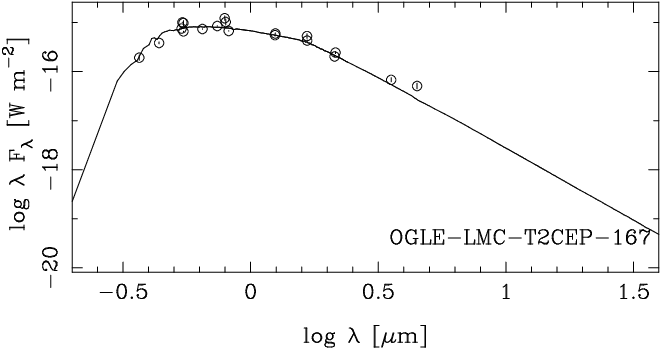}} 
\end{minipage}
\begin{minipage}{0.4\textwidth}
\resizebox{\hsize}{!}{\includegraphics[angle=-0]{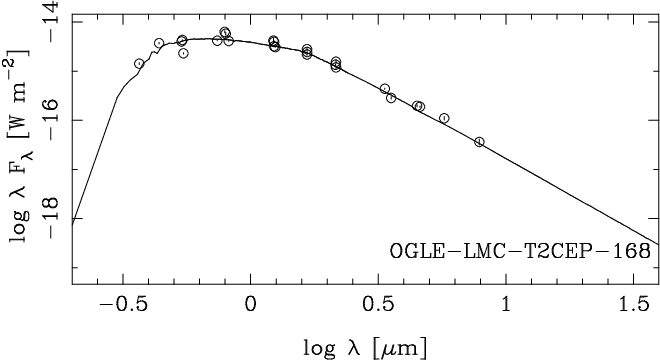}} 
\end{minipage}

\begin{minipage}{0.4\textwidth}
\resizebox{\hsize}{!}{\includegraphics[angle=-0]{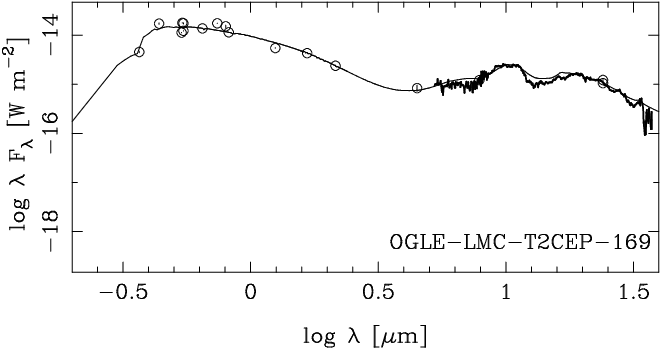}} 
\end{minipage}
\begin{minipage}{0.4\textwidth}
\resizebox{\hsize}{!}{\includegraphics[angle=-0]{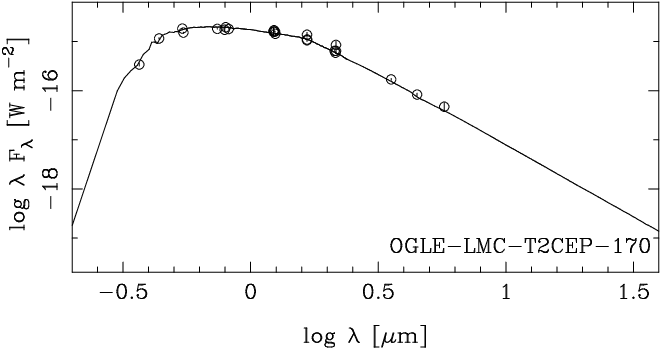}} 
\end{minipage}

\caption{Continued}
\end{figure*}

\setcounter{figure}{0}
\begin{figure*}
\centering

\begin{minipage}{0.4\textwidth}
\resizebox{\hsize}{!}{\includegraphics[angle=-0]{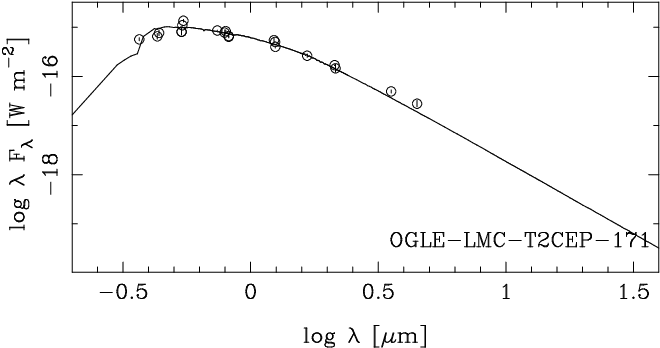}} 
\end{minipage}
\begin{minipage}{0.4\textwidth}
\resizebox{\hsize}{!}{\includegraphics[angle=-0]{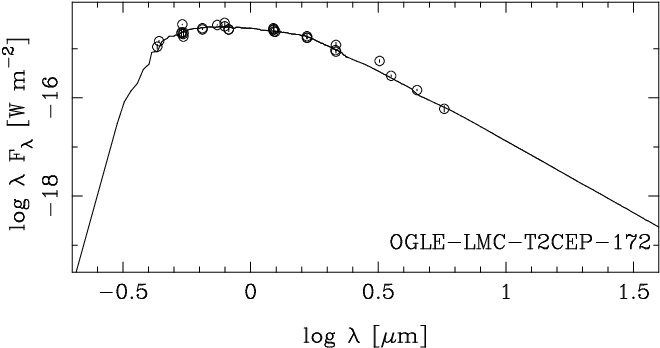}} 
\end{minipage}
 
\begin{minipage}{0.4\textwidth}
\resizebox{\hsize}{!}{\includegraphics[angle=-0]{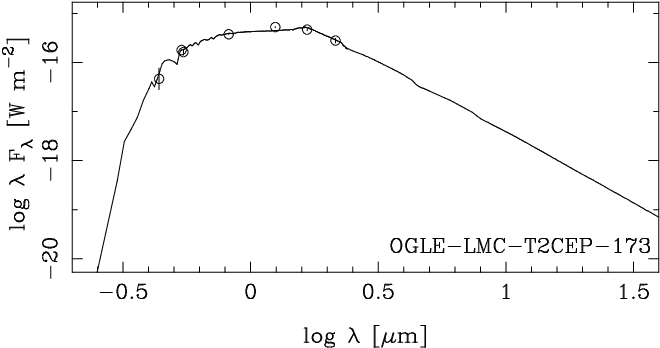}} 
\end{minipage}
\begin{minipage}{0.4\textwidth}
\resizebox{\hsize}{!}{\includegraphics[angle=-0]{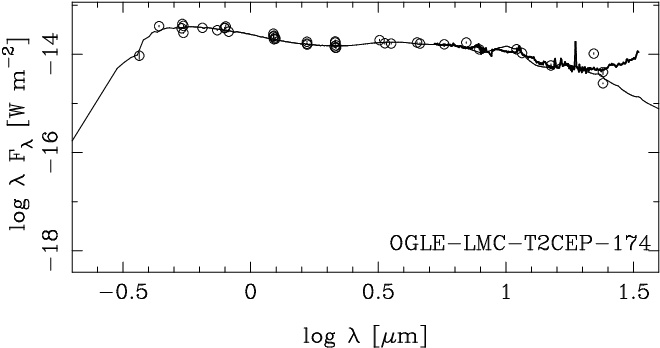}} 
\end{minipage}
 
\begin{minipage}{0.4\textwidth}
\resizebox{\hsize}{!}{\includegraphics[angle=-0]{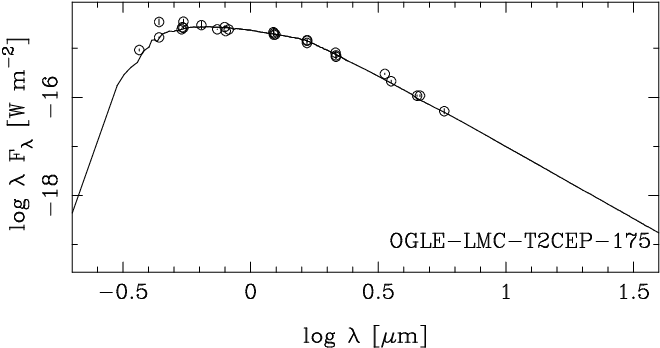}} 
\end{minipage}
\begin{minipage}{0.4\textwidth}
\resizebox{\hsize}{!}{\includegraphics[angle=-0]{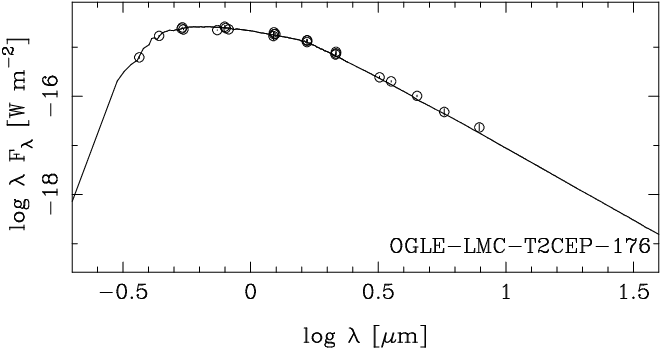}} 
\end{minipage}
 
\begin{minipage}{0.4\textwidth}
\resizebox{\hsize}{!}{\includegraphics[angle=-0]{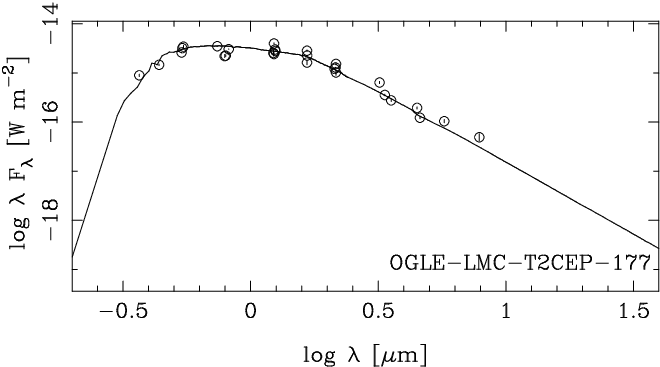}} 
\end{minipage}
\begin{minipage}{0.4\textwidth}
\resizebox{\hsize}{!}{\includegraphics[angle=-0]{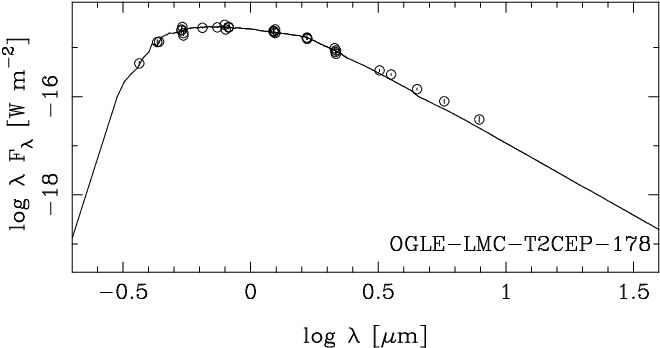}} 
\end{minipage}
 
\begin{minipage}{0.4\textwidth}
\resizebox{\hsize}{!}{\includegraphics[angle=-0]{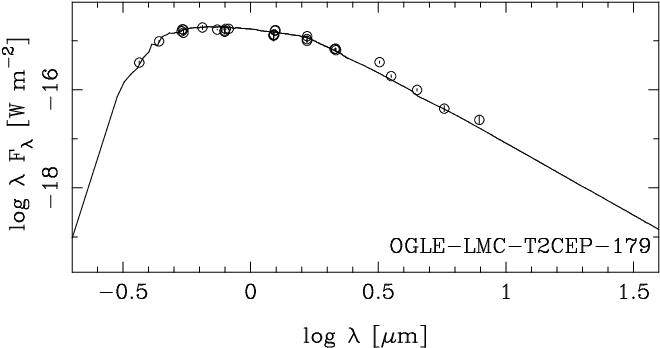}} 
\end{minipage}
\begin{minipage}{0.4\textwidth}
\resizebox{\hsize}{!}{\includegraphics[angle=-0]{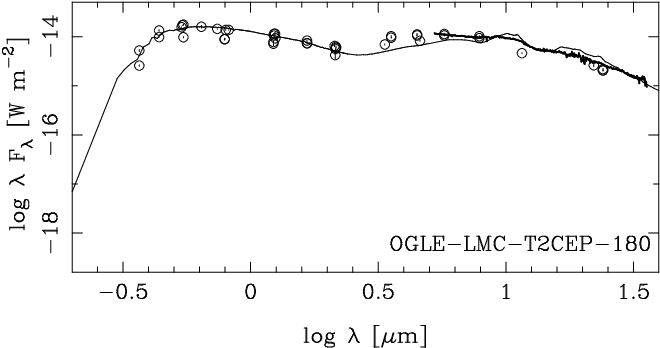}} 
\end{minipage}
 
\caption{Continued}
\end{figure*}

\setcounter{figure}{0}
\begin{figure*}
\centering

\begin{minipage}{0.4\textwidth}
\resizebox{\hsize}{!}{\includegraphics[angle=-0]{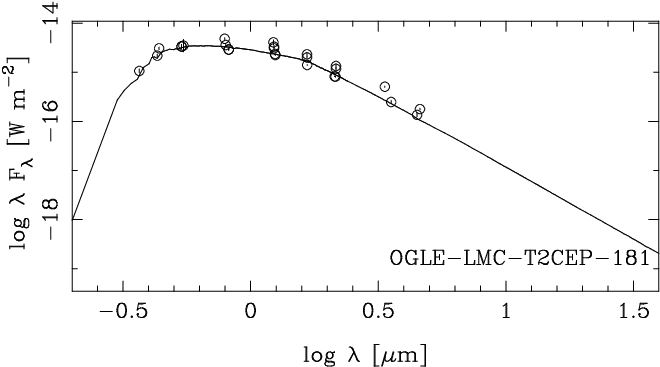}} 
\end{minipage}
\begin{minipage}{0.4\textwidth}
\resizebox{\hsize}{!}{\includegraphics[angle=-0]{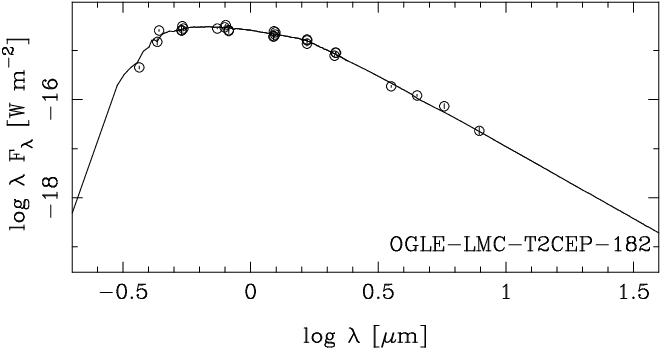}} 
\end{minipage}
 
\begin{minipage}{0.4\textwidth}
\resizebox{\hsize}{!}{\includegraphics[angle=-0]{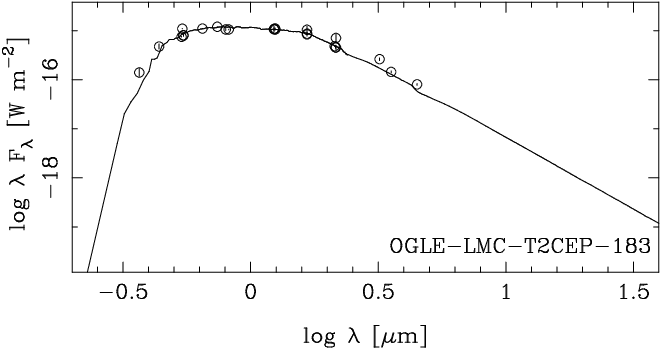}} 
\end{minipage}
\begin{minipage}{0.4\textwidth}
\resizebox{\hsize}{!}{\includegraphics[angle=-0]{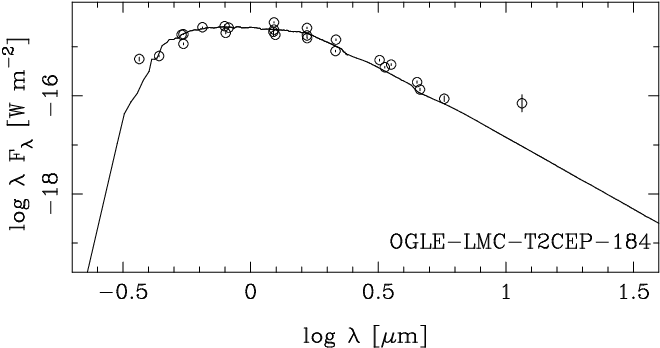}} 
\end{minipage}
 
\begin{minipage}{0.4\textwidth}
\resizebox{\hsize}{!}{\includegraphics[angle=-0]{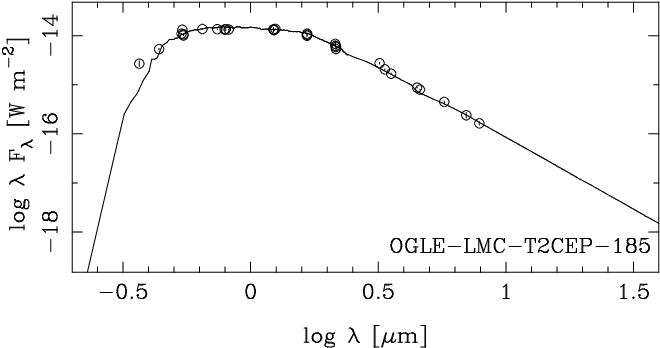}} 
\end{minipage}
\begin{minipage}{0.4\textwidth}
\resizebox{\hsize}{!}{\includegraphics[angle=-0]{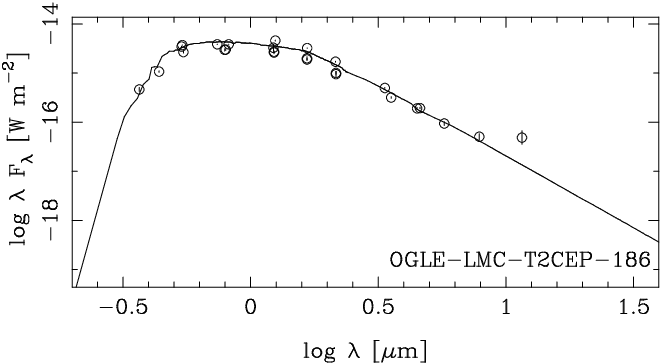}} 
\end{minipage}
 
\begin{minipage}{0.4\textwidth}
\resizebox{\hsize}{!}{\includegraphics[angle=-0]{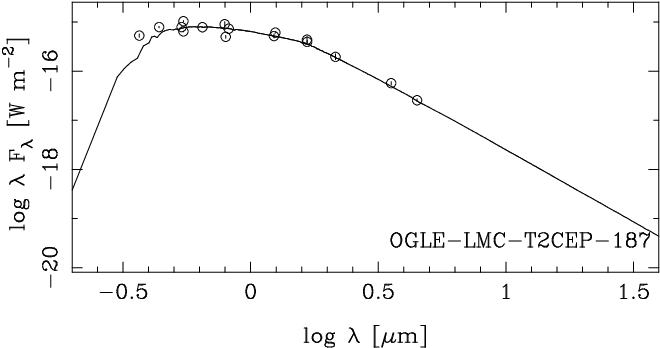}} 
\end{minipage}
\begin{minipage}{0.4\textwidth}
\resizebox{\hsize}{!}{\includegraphics[angle=-0]{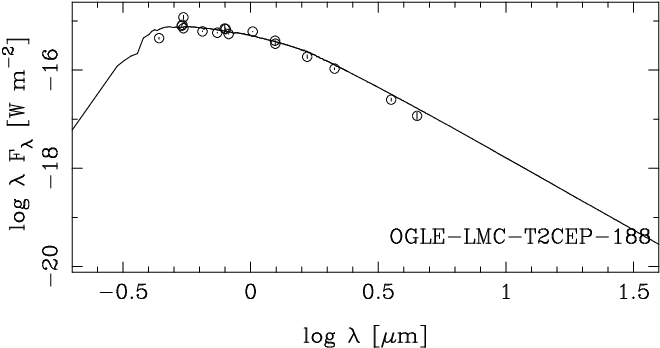}} 
\end{minipage}
 
\begin{minipage}{0.4\textwidth}
\resizebox{\hsize}{!}{\includegraphics[angle=-0]{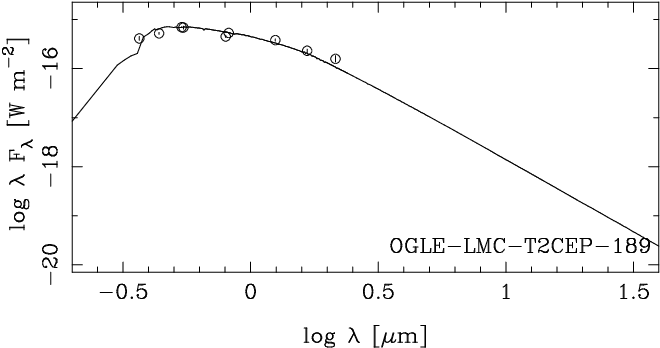}} 
\end{minipage}
\begin{minipage}{0.4\textwidth}
\resizebox{\hsize}{!}{\includegraphics[angle=-0]{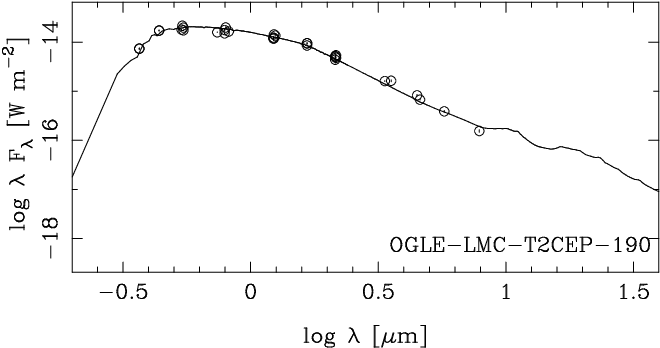}} 
\end{minipage}

\caption{Continued}
\end{figure*}

\setcounter{figure}{0}
\begin{figure*}
\centering

\begin{minipage}{0.4\textwidth}
\resizebox{\hsize}{!}{\includegraphics[angle=-0]{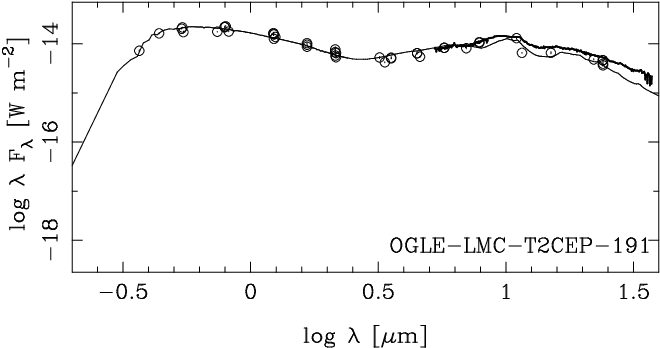}} 
\end{minipage}
\begin{minipage}{0.4\textwidth}
\resizebox{\hsize}{!}{\includegraphics[angle=-0]{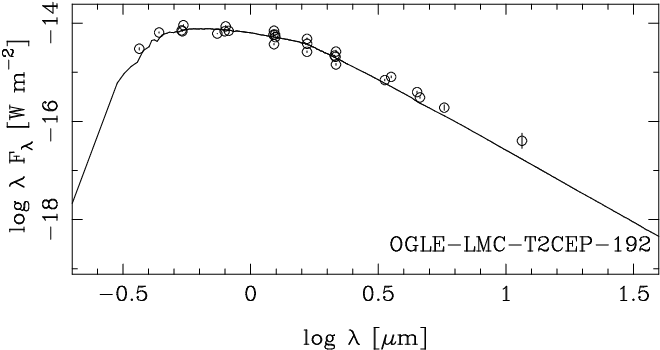}} 
\end{minipage}
 
\begin{minipage}{0.4\textwidth}
\resizebox{\hsize}{!}{\includegraphics[angle=-0]{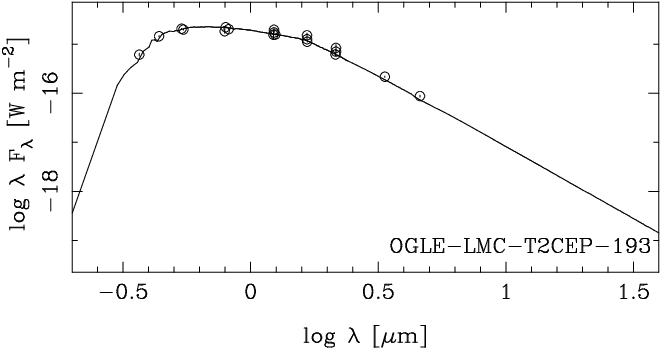}} 
\end{minipage}
\begin{minipage}{0.4\textwidth}
\resizebox{\hsize}{!}{\includegraphics[angle=-0]{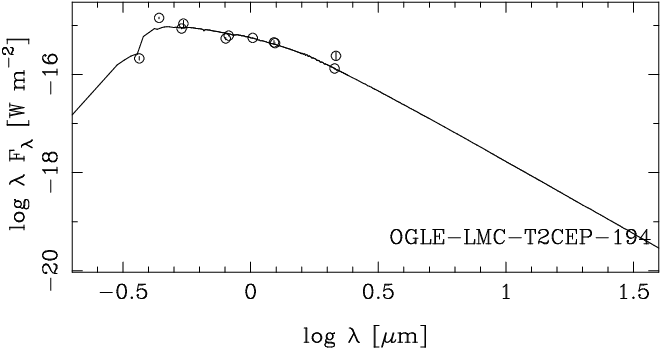}} 
\end{minipage}
 
\begin{minipage}{0.4\textwidth}
\resizebox{\hsize}{!}{\includegraphics[angle=-0]{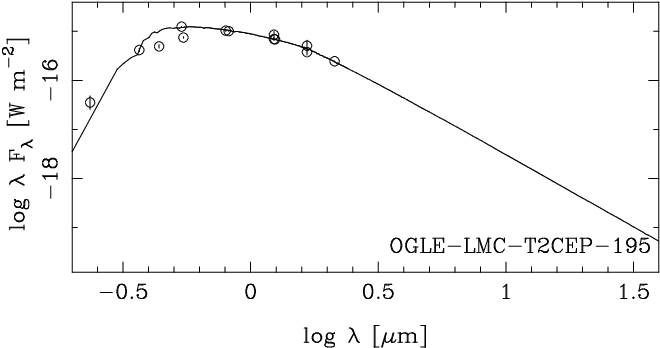}} 
\end{minipage}
\begin{minipage}{0.4\textwidth}
\resizebox{\hsize}{!}{\includegraphics[angle=-0]{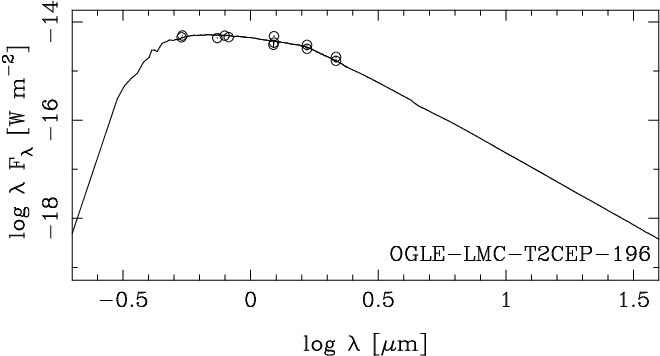}} 
\end{minipage}
 
\begin{minipage}{0.4\textwidth}
\resizebox{\hsize}{!}{\includegraphics[angle=-0]{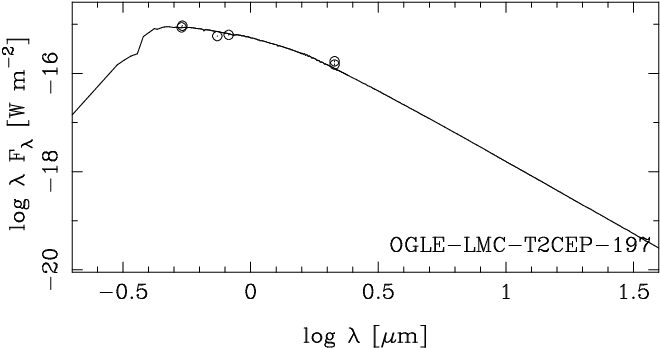}} 
\end{minipage}
\begin{minipage}{0.4\textwidth}
\resizebox{\hsize}{!}{\includegraphics[angle=-0]{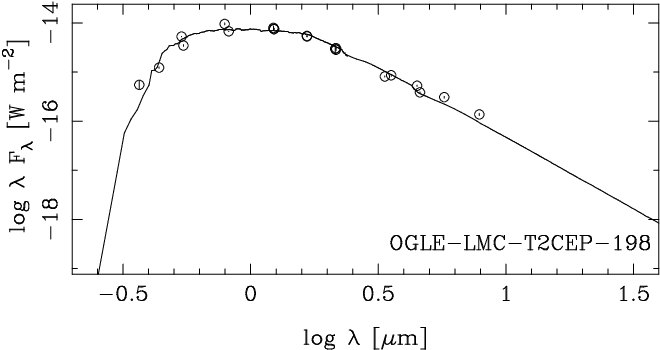}} 
\end{minipage}
 
\begin{minipage}{0.4\textwidth}
\resizebox{\hsize}{!}{\includegraphics[angle=-0]{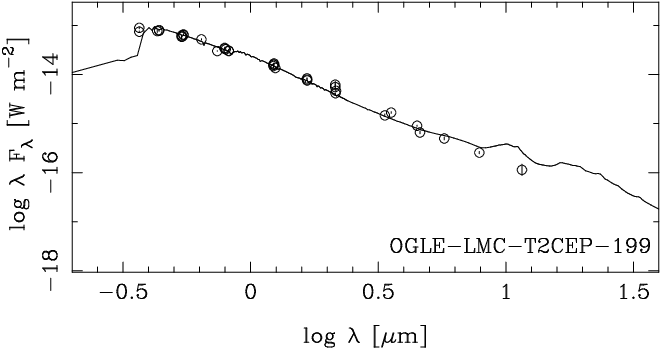}} 
\end{minipage}
\begin{minipage}{0.4\textwidth}
\resizebox{\hsize}{!}{\includegraphics[angle=-0]{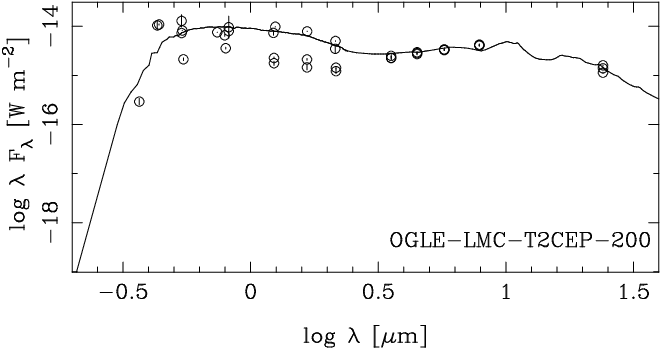}} 
\end{minipage}

\caption{Continued}
\end{figure*}

\setcounter{figure}{0}
\begin{figure*}
\centering

\begin{minipage}{0.4\textwidth}
\resizebox{\hsize}{!}{\includegraphics[angle=-0]{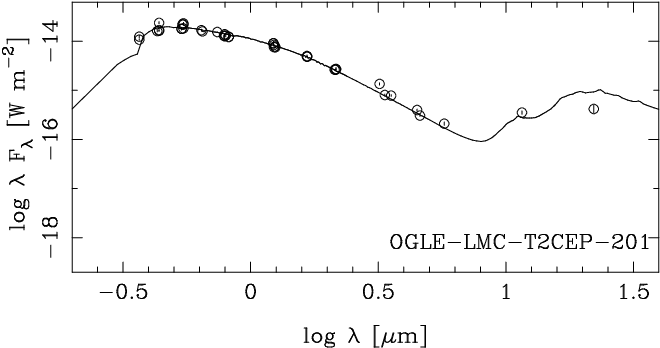}} 
\end{minipage}
\begin{minipage}{0.4\textwidth}
\resizebox{\hsize}{!}{\includegraphics[angle=-0]{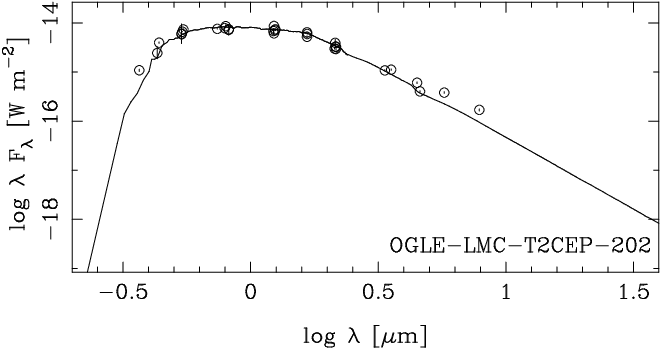}} 
\end{minipage}
 
\begin{minipage}{0.4\textwidth}
\resizebox{\hsize}{!}{\includegraphics[angle=-0]{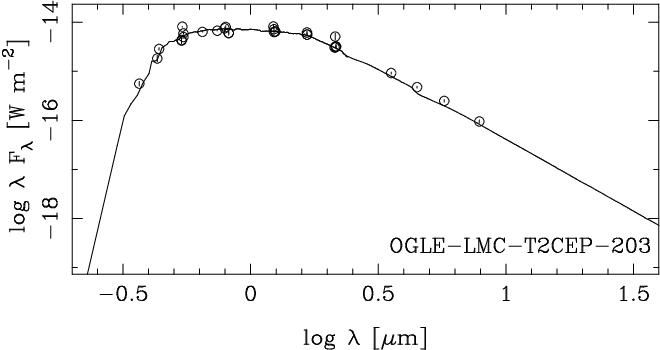}} 
\end{minipage}
\begin{minipage}{0.4\textwidth}
\resizebox{\hsize}{!}{\includegraphics[angle=-0]{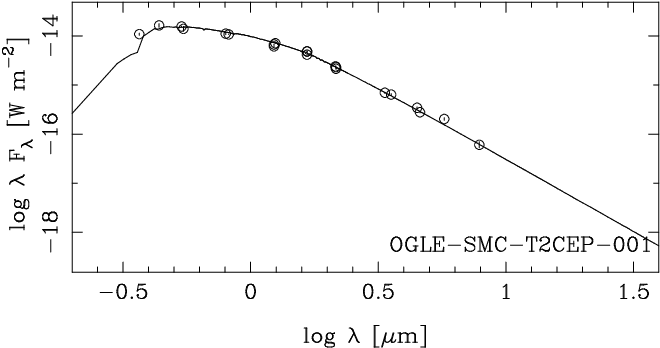}} 
\end{minipage}
 
\begin{minipage}{0.4\textwidth}
\resizebox{\hsize}{!}{\includegraphics[angle=-0]{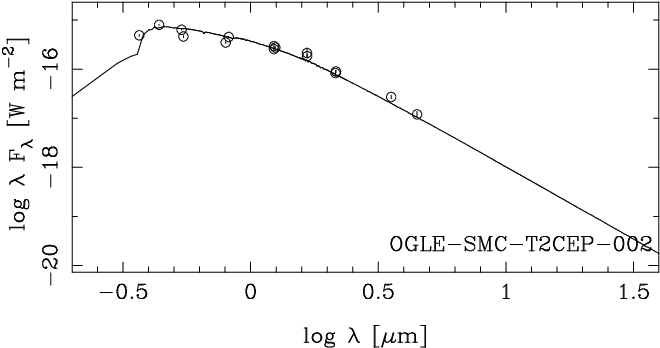}} 
\end{minipage}
\begin{minipage}{0.4\textwidth}
\resizebox{\hsize}{!}{\includegraphics[angle=-0]{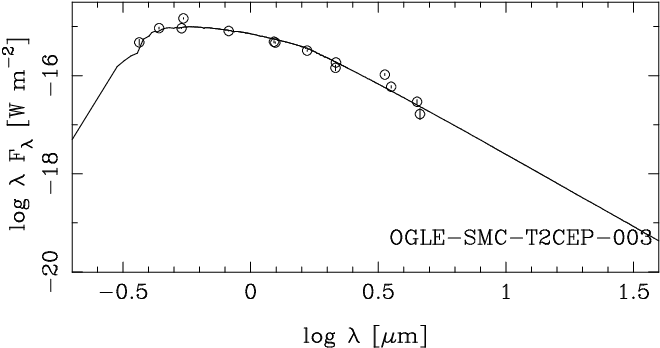}} 
\end{minipage}
 
\begin{minipage}{0.4\textwidth}
\resizebox{\hsize}{!}{\includegraphics[angle=-0]{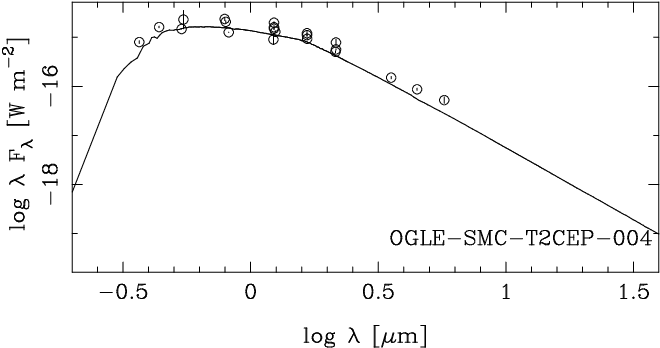}} 
\end{minipage}
\begin{minipage}{0.4\textwidth}
\resizebox{\hsize}{!}{\includegraphics[angle=-0]{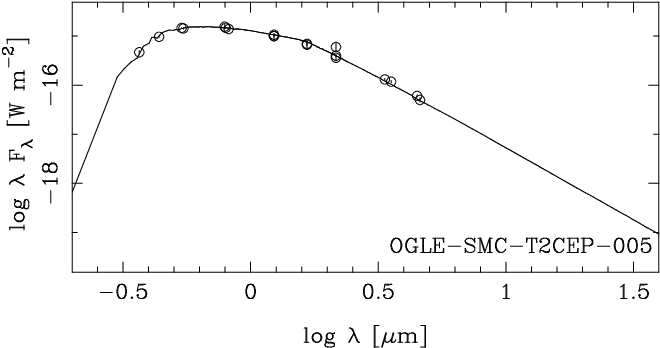}} 
\end{minipage}
 
\begin{minipage}{0.4\textwidth}
\resizebox{\hsize}{!}{\includegraphics[angle=-0]{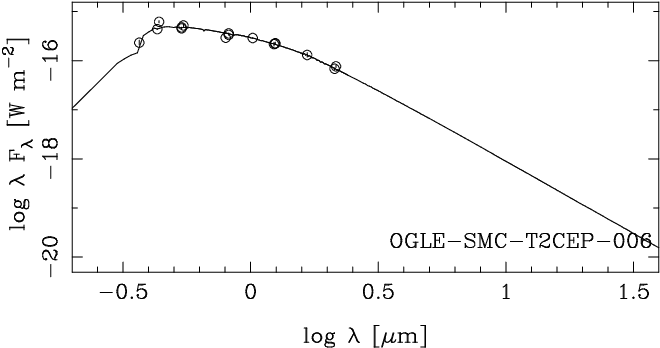}} 
\end{minipage}
\begin{minipage}{0.4\textwidth}
\resizebox{\hsize}{!}{\includegraphics[angle=-0]{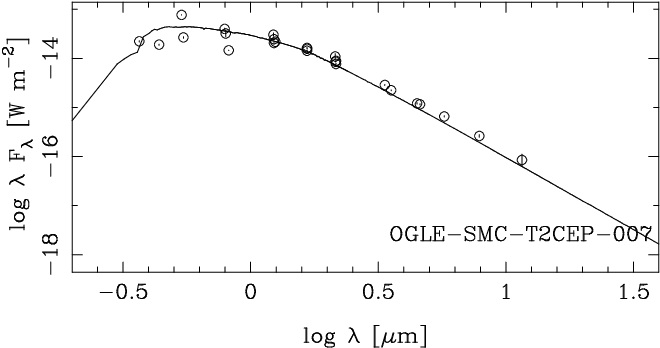}} 
\end{minipage}

\caption{Continued}
\end{figure*}

\setcounter{figure}{0}
\begin{figure*}
\centering

\begin{minipage}{0.4\textwidth}
\resizebox{\hsize}{!}{\includegraphics[angle=-0]{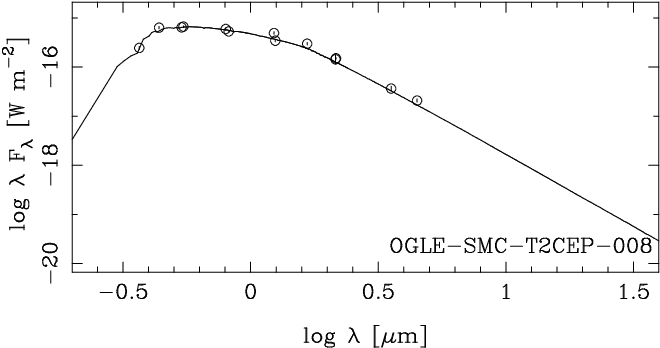}} 
\end{minipage}
\begin{minipage}{0.4\textwidth}
\resizebox{\hsize}{!}{\includegraphics[angle=-0]{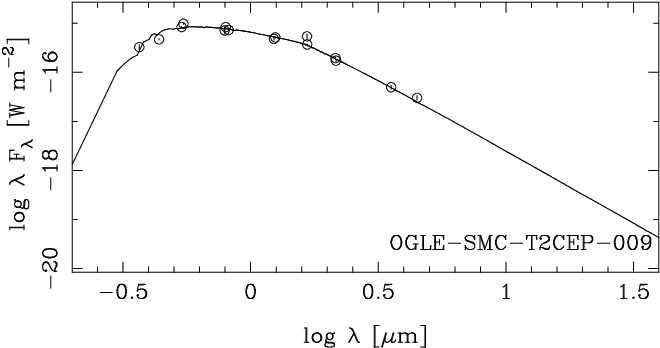}} 
\end{minipage}
 
\begin{minipage}{0.4\textwidth}
\resizebox{\hsize}{!}{\includegraphics[angle=-0]{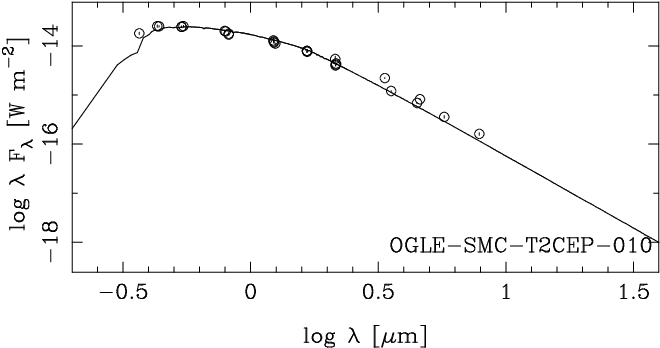}} 
\end{minipage}
\begin{minipage}{0.4\textwidth}
\resizebox{\hsize}{!}{\includegraphics[angle=-0]{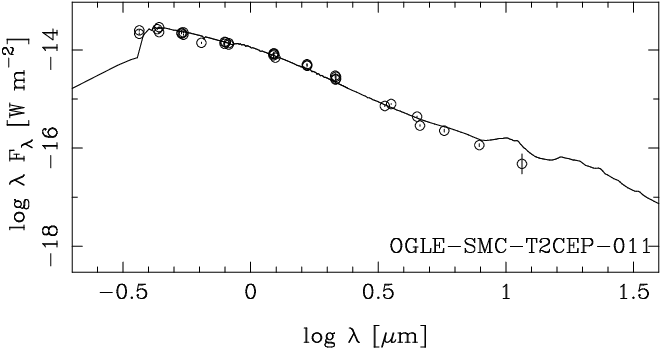}} 
\end{minipage}
 
\begin{minipage}{0.4\textwidth}
\resizebox{\hsize}{!}{\includegraphics[angle=-0]{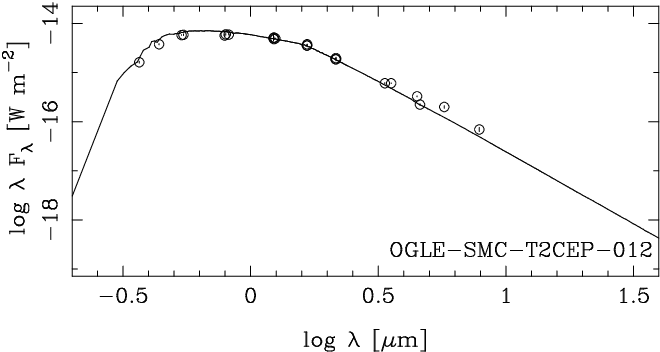}} 
\end{minipage}
\begin{minipage}{0.4\textwidth}
\resizebox{\hsize}{!}{\includegraphics[angle=-0]{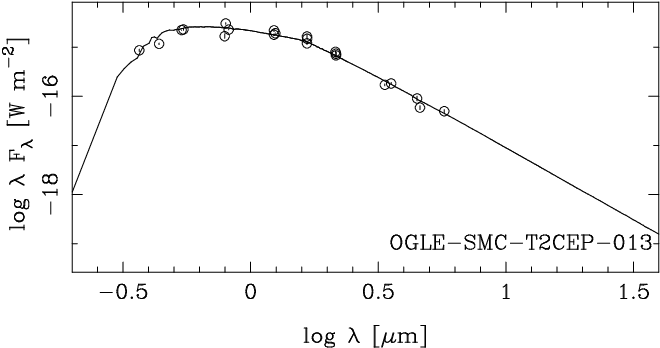}} 
\end{minipage}

\begin{minipage}{0.4\textwidth}
\resizebox{\hsize}{!}{\includegraphics[angle=-0]{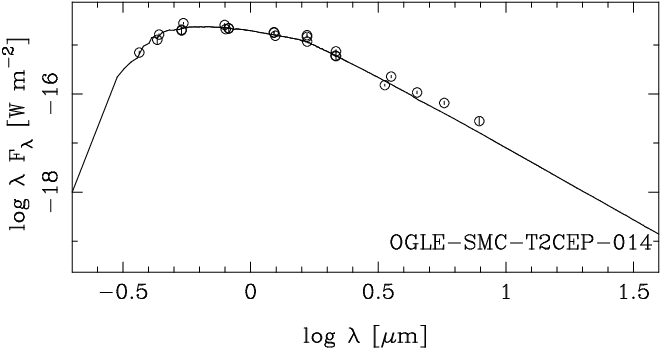}} 
\end{minipage}
\begin{minipage}{0.4\textwidth}
\resizebox{\hsize}{!}{\includegraphics[angle=-0]{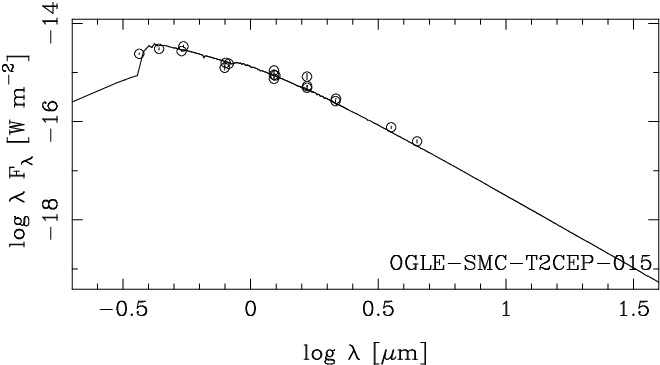}} 
\end{minipage}
 
\begin{minipage}{0.4\textwidth}
\resizebox{\hsize}{!}{\includegraphics[angle=-0]{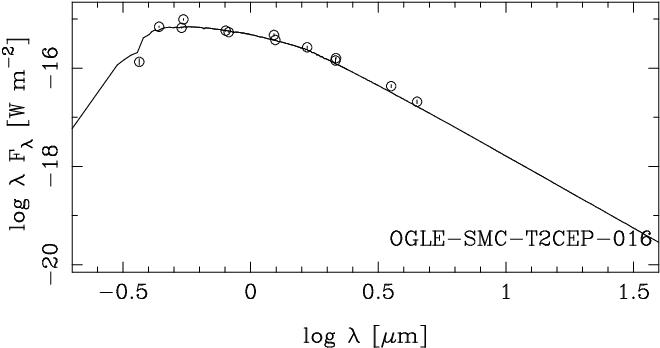}} 
\end{minipage}
\begin{minipage}{0.4\textwidth}
\resizebox{\hsize}{!}{\includegraphics[angle=-0]{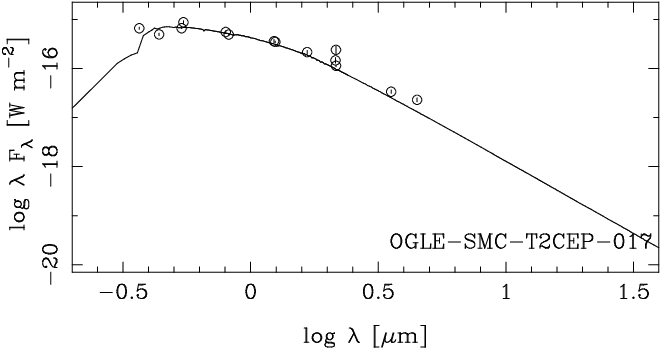}} 
\end{minipage}

\caption{Continued}
\end{figure*}

\setcounter{figure}{0}
\begin{figure*}
\centering

\begin{minipage}{0.4\textwidth}
\resizebox{\hsize}{!}{\includegraphics[angle=-0]{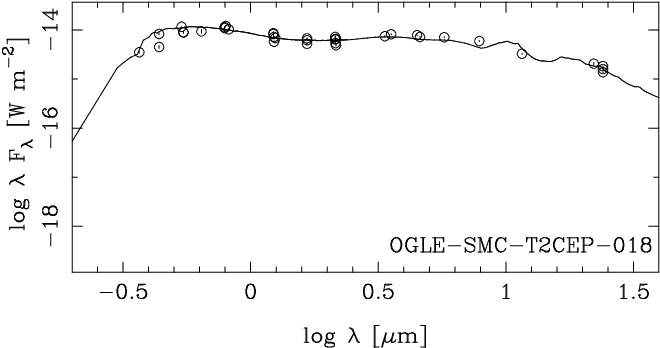}} 
\end{minipage}
\begin{minipage}{0.4\textwidth}
\resizebox{\hsize}{!}{\includegraphics[angle=-0]{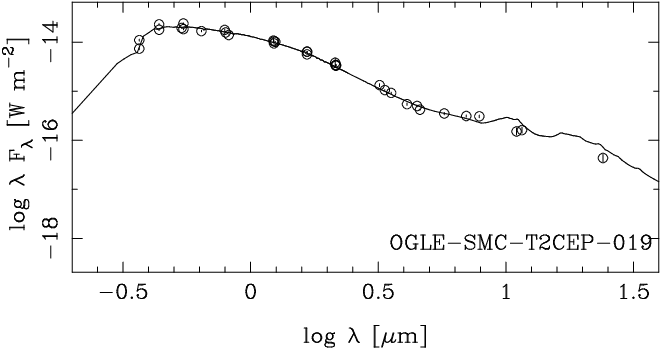}} 
\end{minipage}
 
\begin{minipage}{0.4\textwidth}
\resizebox{\hsize}{!}{\includegraphics[angle=-0]{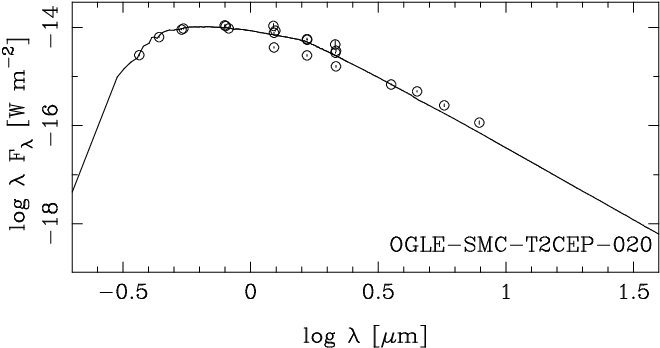}} 
\end{minipage}
\begin{minipage}{0.4\textwidth}
\resizebox{\hsize}{!}{\includegraphics[angle=-0]{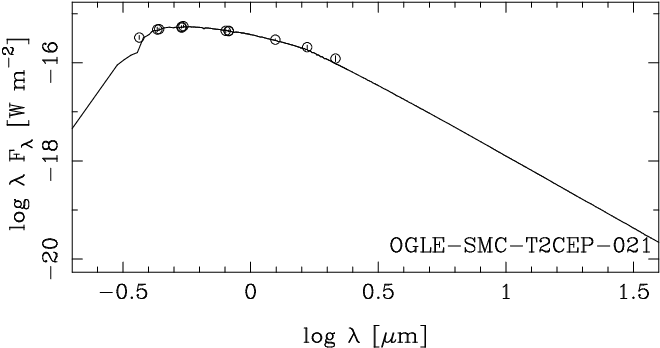}} 
\end{minipage}
 
\begin{minipage}{0.4\textwidth}
\resizebox{\hsize}{!}{\includegraphics[angle=-0]{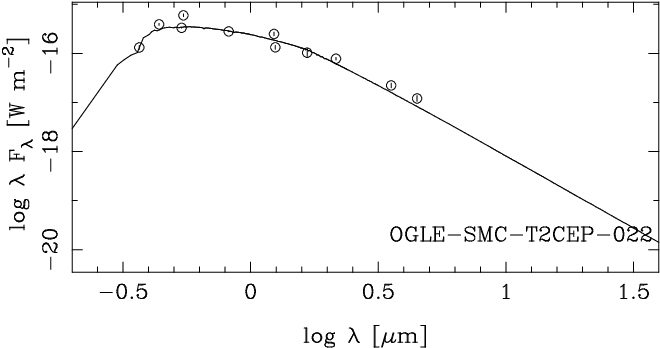}} 
\end{minipage}
\begin{minipage}{0.4\textwidth}
\resizebox{\hsize}{!}{\includegraphics[angle=-0]{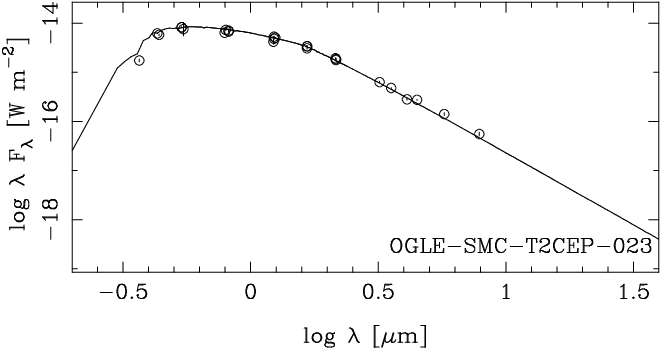}} 
\end{minipage}
 
\begin{minipage}{0.4\textwidth}
\resizebox{\hsize}{!}{\includegraphics[angle=-0]{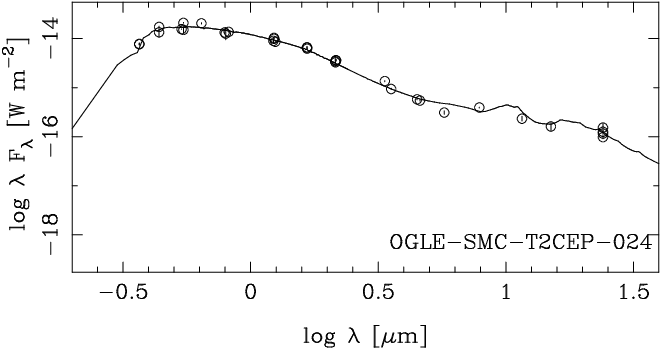}} 
\end{minipage}
\begin{minipage}{0.4\textwidth}
\resizebox{\hsize}{!}{\includegraphics[angle=-0]{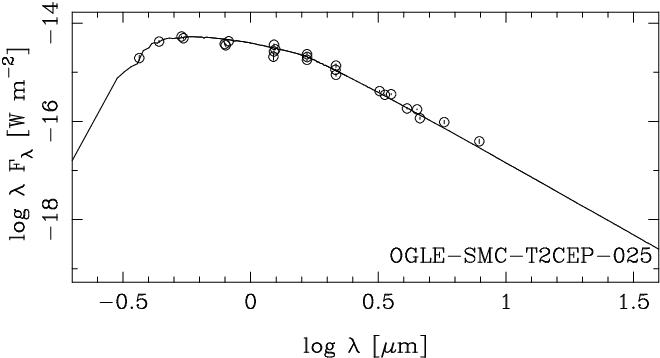}} 
\end{minipage}

\begin{minipage}{0.4\textwidth}
\resizebox{\hsize}{!}{\includegraphics[angle=-0]{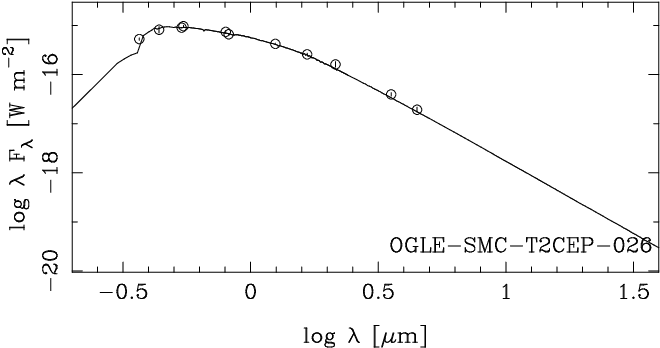}} 
\end{minipage}
\begin{minipage}{0.4\textwidth}
\resizebox{\hsize}{!}{\includegraphics[angle=-0]{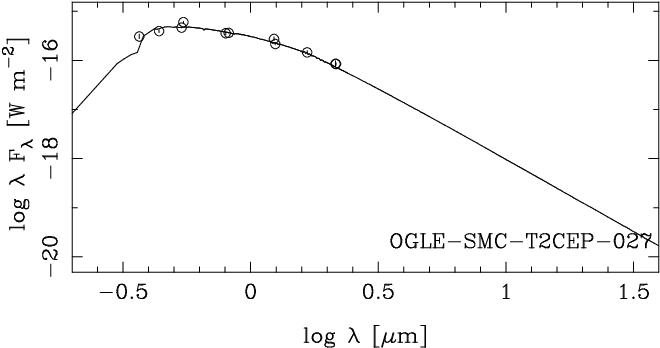}} 
\end{minipage}

\caption{Continued}
\end{figure*}
 
\setcounter{figure}{0}
\begin{figure*}
\centering

\begin{minipage}{0.4\textwidth}
\resizebox{\hsize}{!}{\includegraphics[angle=-0]{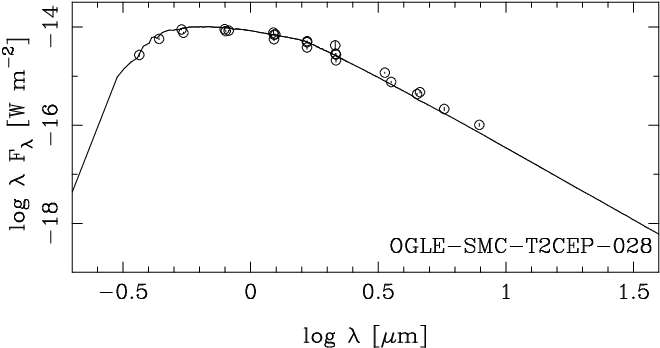}} 
\end{minipage}
\begin{minipage}{0.4\textwidth}
\resizebox{\hsize}{!}{\includegraphics[angle=-0]{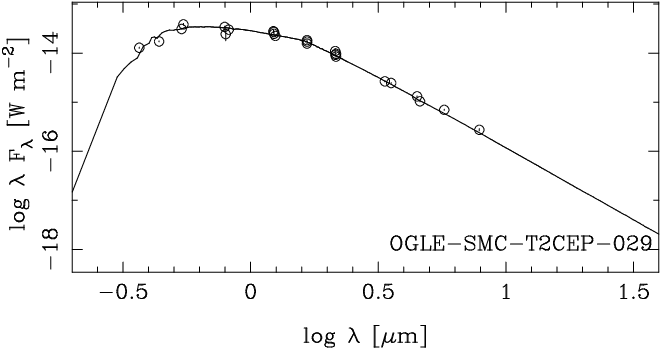}} 
\end{minipage}
 
\begin{minipage}{0.4\textwidth}
\resizebox{\hsize}{!}{\includegraphics[angle=-0]{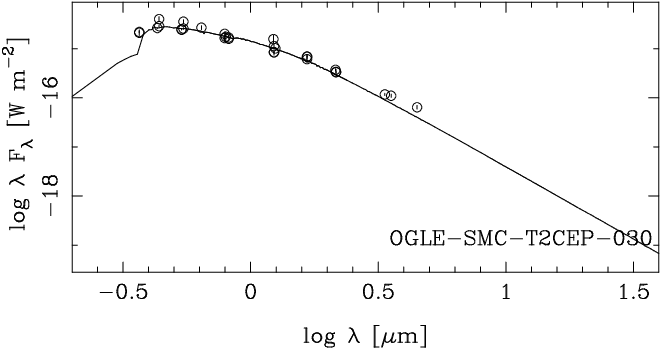}} 
\end{minipage}
\begin{minipage}{0.4\textwidth}
\resizebox{\hsize}{!}{\includegraphics[angle=-0]{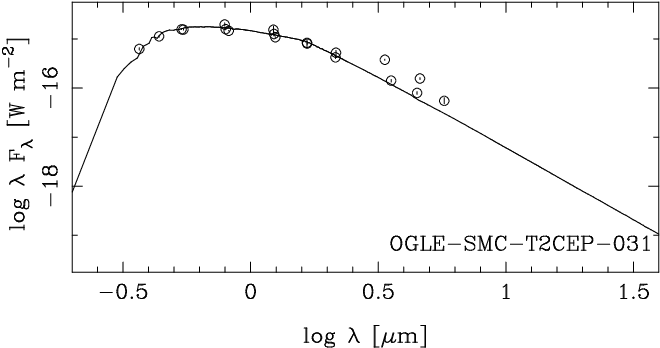}} 
\end{minipage}
 
\begin{minipage}{0.4\textwidth}
\resizebox{\hsize}{!}{\includegraphics[angle=-0]{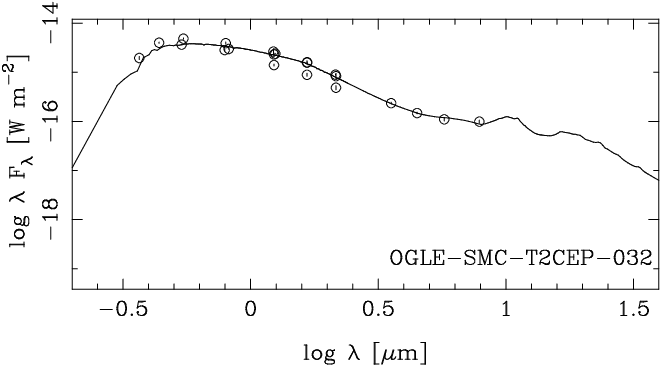}} 
\end{minipage}
\begin{minipage}{0.4\textwidth}
\resizebox{\hsize}{!}{\includegraphics[angle=-0]{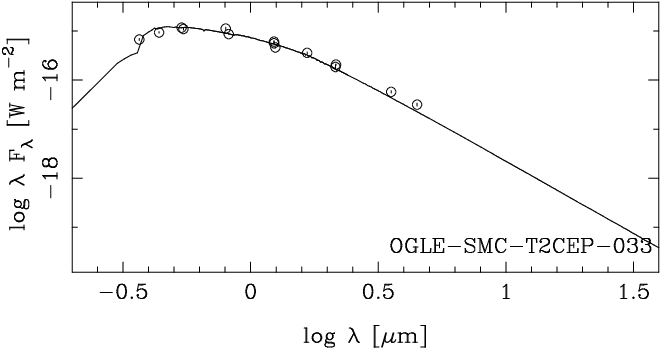}} 
\end{minipage}
 
\begin{minipage}{0.4\textwidth}
\resizebox{\hsize}{!}{\includegraphics[angle=-0]{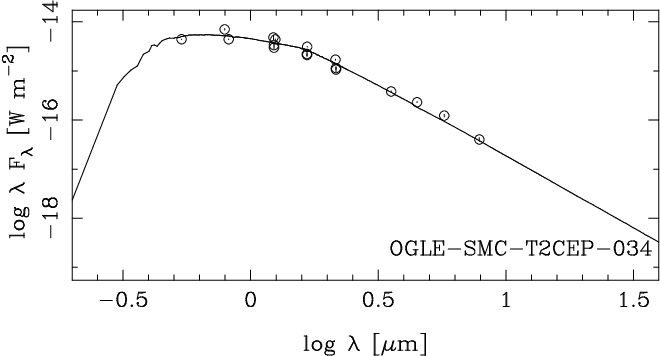}} 
\end{minipage}
\begin{minipage}{0.4\textwidth}
\resizebox{\hsize}{!}{\includegraphics[angle=-0]{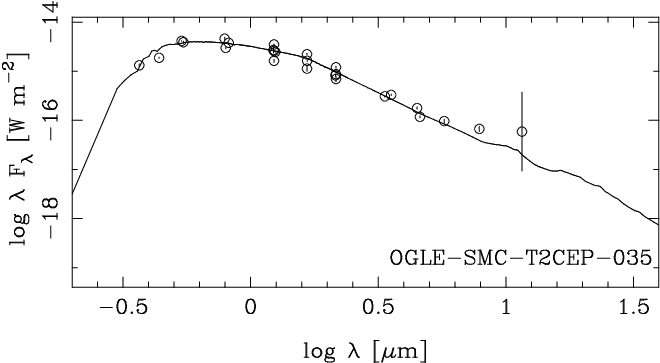}} 
\end{minipage}
 
\begin{minipage}{0.4\textwidth}
\resizebox{\hsize}{!}{\includegraphics[angle=-0]{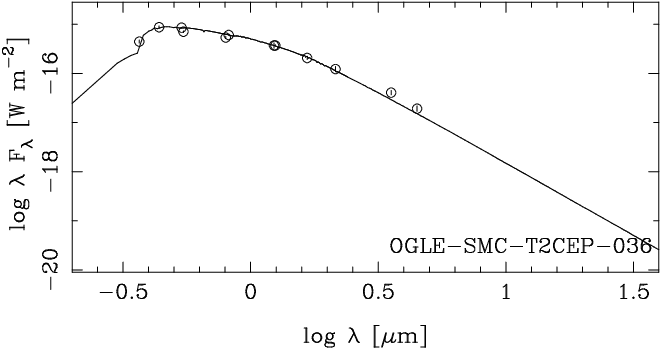}} 
\end{minipage}
\begin{minipage}{0.4\textwidth}
\resizebox{\hsize}{!}{\includegraphics[angle=-0]{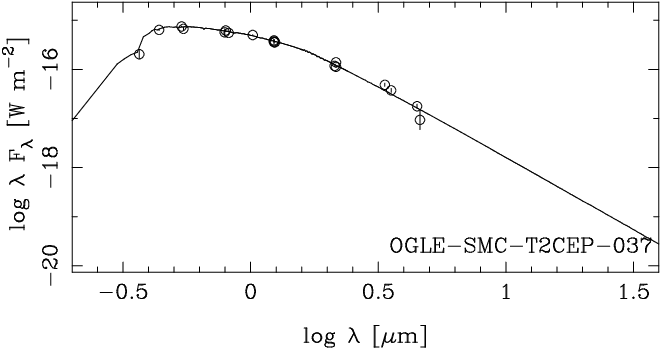}} 
\end{minipage}
 
\caption{Continued}
\end{figure*}

\setcounter{figure}{0}
\begin{figure*}
\centering

\begin{minipage}{0.4\textwidth}
\resizebox{\hsize}{!}{\includegraphics[angle=-0]{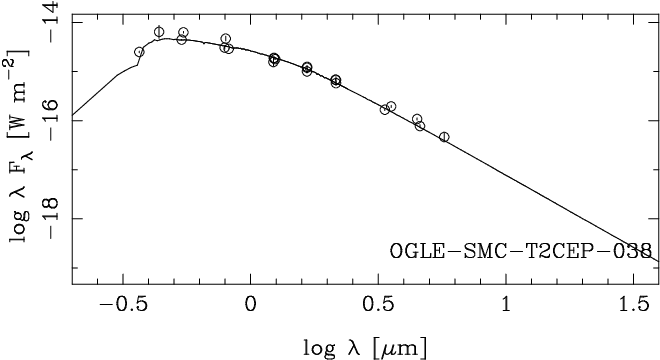}} 
\end{minipage}
\begin{minipage}{0.4\textwidth}
\resizebox{\hsize}{!}{\includegraphics[angle=-0]{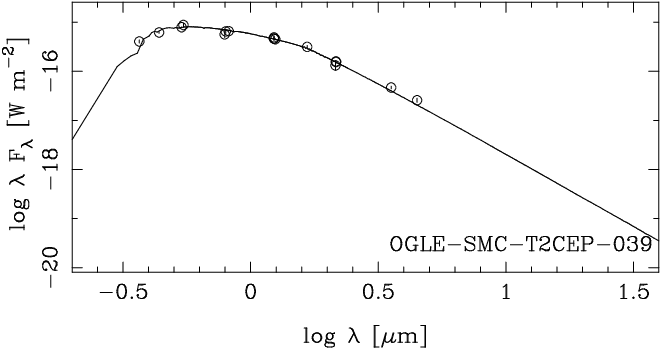}} 
\end{minipage}
 
\begin{minipage}{0.4\textwidth}
\resizebox{\hsize}{!}{\includegraphics[angle=-0]{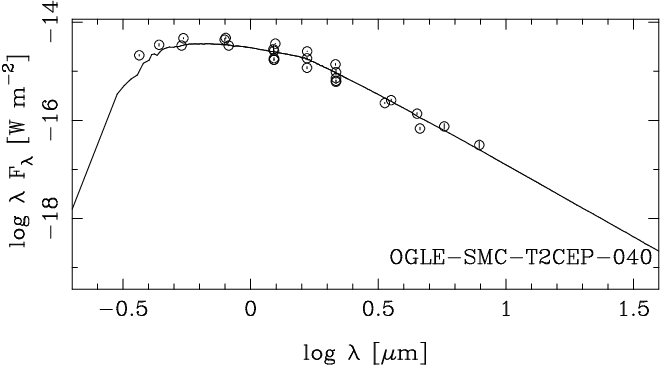}} 
\end{minipage}
\begin{minipage}{0.4\textwidth}
\resizebox{\hsize}{!}{\includegraphics[angle=-0]{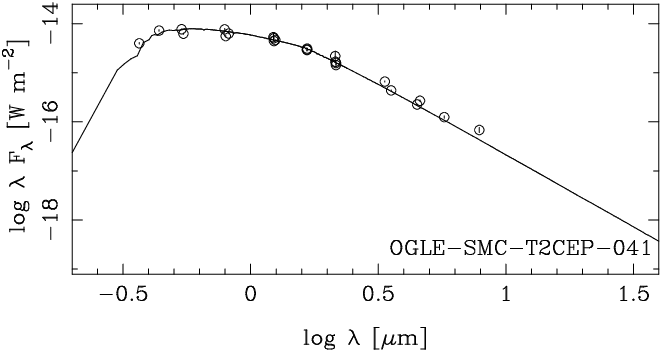}} 
\end{minipage}
 
\begin{minipage}{0.4\textwidth}
\resizebox{\hsize}{!}{\includegraphics[angle=-0]{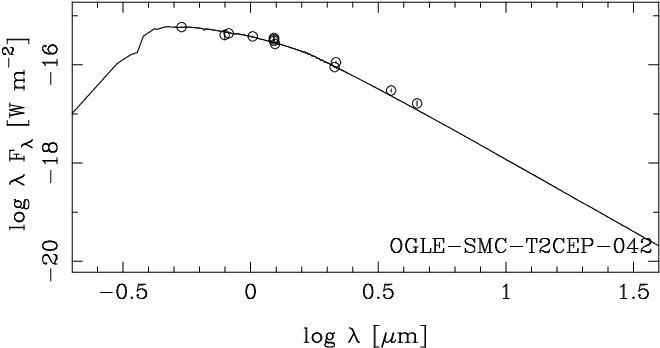}} 
\end{minipage}
\begin{minipage}{0.4\textwidth}
\resizebox{\hsize}{!}{\includegraphics[angle=-0]{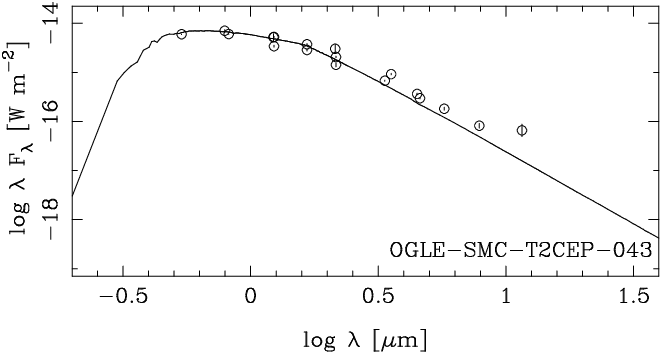}} 
\end{minipage}
 
\begin{minipage}{0.4\textwidth}
\resizebox{\hsize}{!}{\includegraphics[angle=-0]{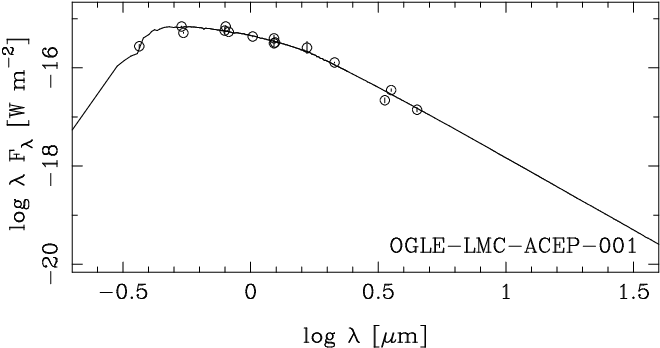}} 
\end{minipage}
\begin{minipage}{0.4\textwidth}
\resizebox{\hsize}{!}{\includegraphics[angle=-0]{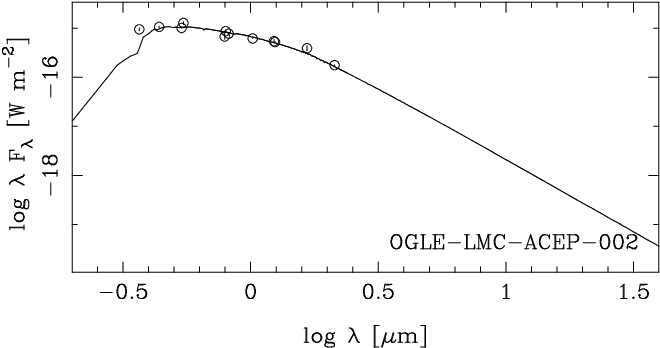}} 
\end{minipage}
 
\begin{minipage}{0.4\textwidth}
\resizebox{\hsize}{!}{\includegraphics[angle=-0]{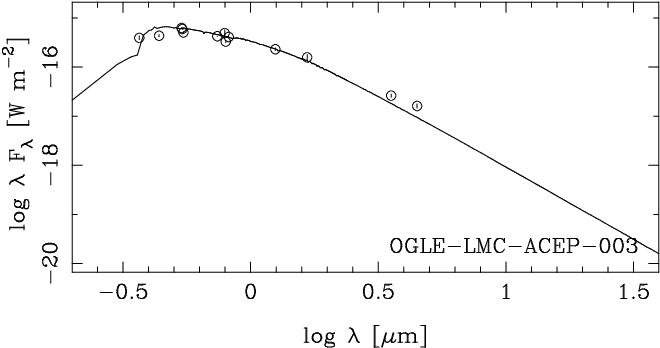}} 
\end{minipage}
\begin{minipage}{0.4\textwidth}
\resizebox{\hsize}{!}{\includegraphics[angle=-0]{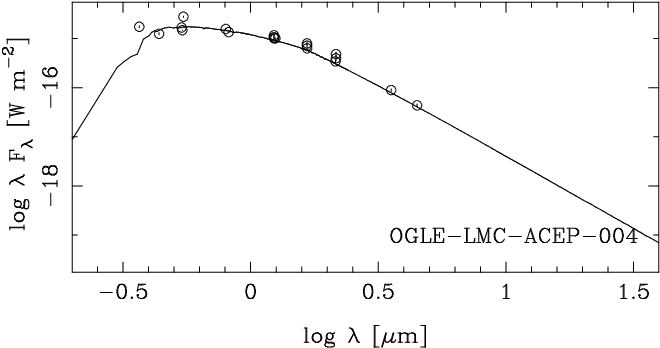}} 
\end{minipage}

\caption{Continued}
\end{figure*}

\setcounter{figure}{0}
\begin{figure*}
\centering

\begin{minipage}{0.4\textwidth}
\resizebox{\hsize}{!}{\includegraphics[angle=-0]{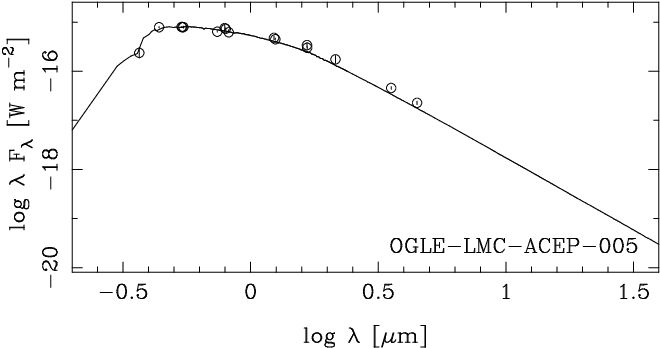}} 
\end{minipage}
\begin{minipage}{0.4\textwidth}
\resizebox{\hsize}{!}{\includegraphics[angle=-0]{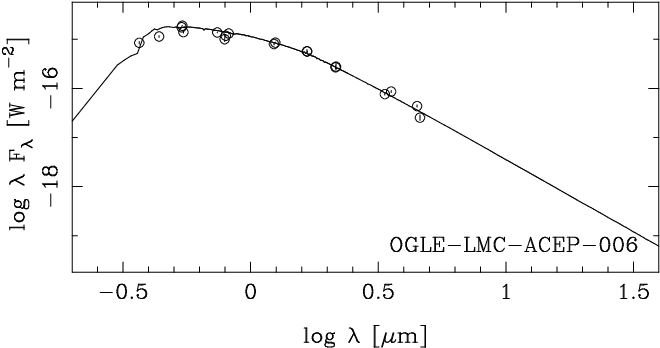}} 
\end{minipage}
 
\begin{minipage}{0.4\textwidth}
\resizebox{\hsize}{!}{\includegraphics[angle=-0]{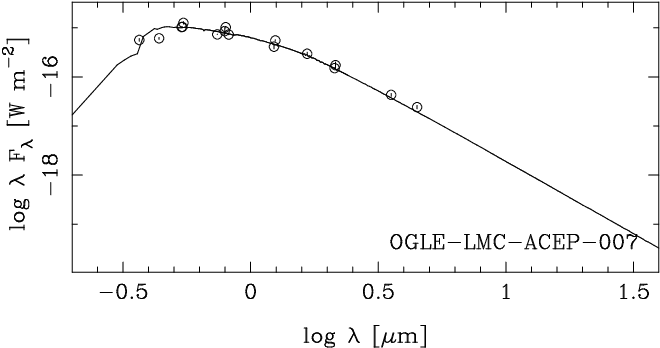}} 
\end{minipage}
\begin{minipage}{0.4\textwidth}
\resizebox{\hsize}{!}{\includegraphics[angle=-0]{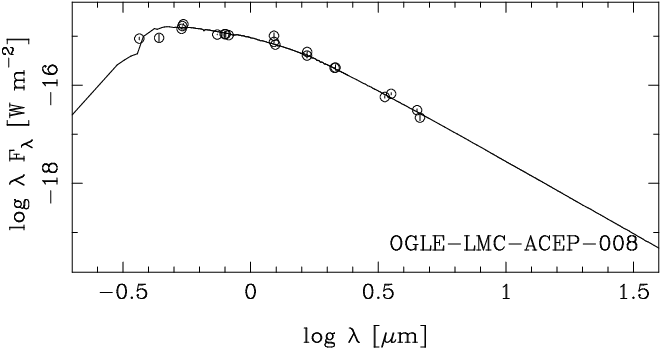}} 
\end{minipage}
 
\begin{minipage}{0.4\textwidth}
\resizebox{\hsize}{!}{\includegraphics[angle=-0]{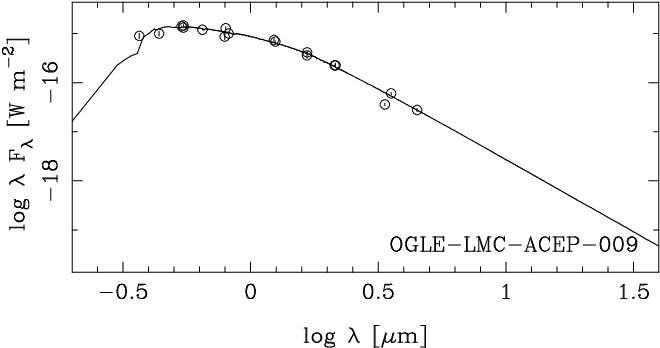}} 
\end{minipage}
\begin{minipage}{0.4\textwidth}
\resizebox{\hsize}{!}{\includegraphics[angle=-0]{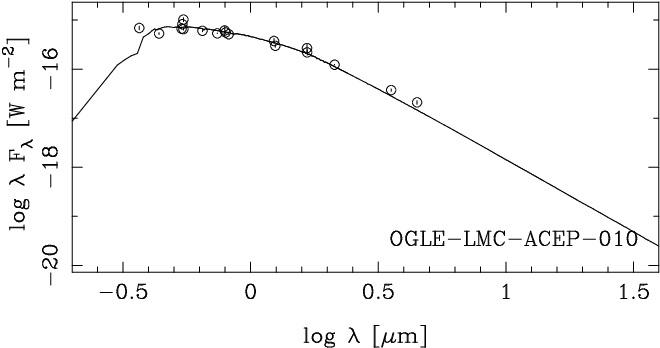}} 
\end{minipage}
 
\begin{minipage}{0.4\textwidth}
\resizebox{\hsize}{!}{\includegraphics[angle=-0]{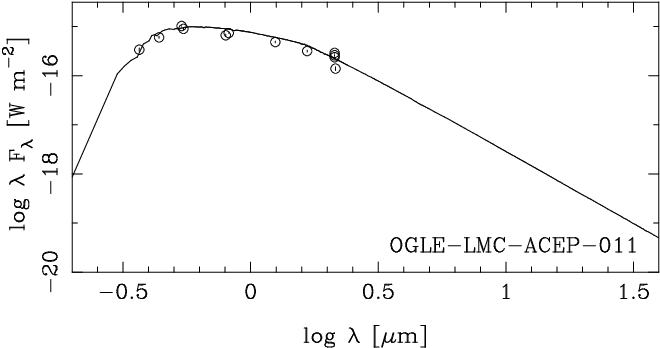}} 
\end{minipage}
\begin{minipage}{0.4\textwidth}
\resizebox{\hsize}{!}{\includegraphics[angle=-0]{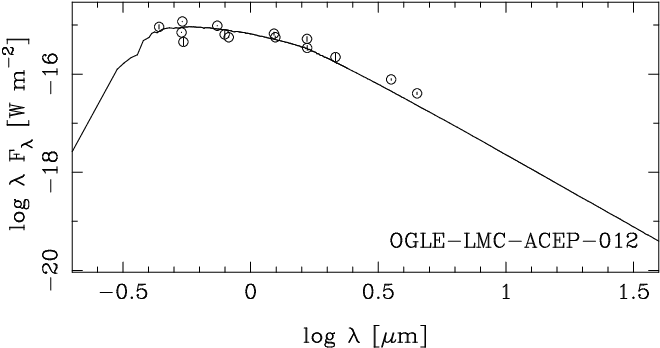}} 
\end{minipage}
 
\begin{minipage}{0.4\textwidth}
\resizebox{\hsize}{!}{\includegraphics[angle=-0]{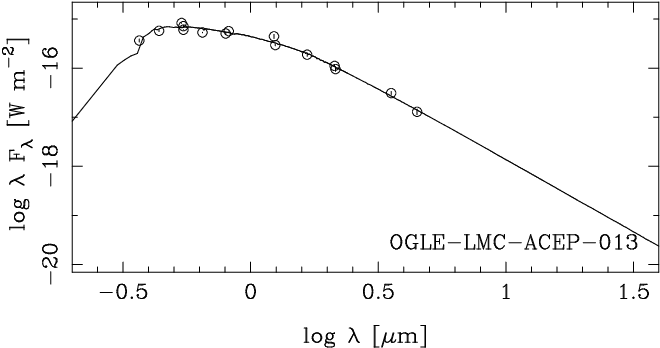}} 
\end{minipage}
\begin{minipage}{0.4\textwidth}
\resizebox{\hsize}{!}{\includegraphics[angle=-0]{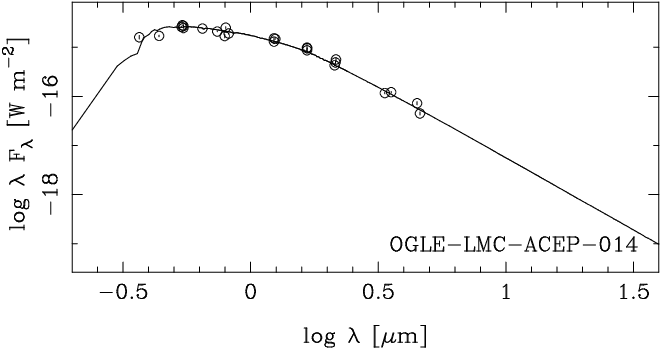}} 
\end{minipage}
 
\begin{minipage}{0.4\textwidth}
\resizebox{\hsize}{!}{\includegraphics[angle=-0]{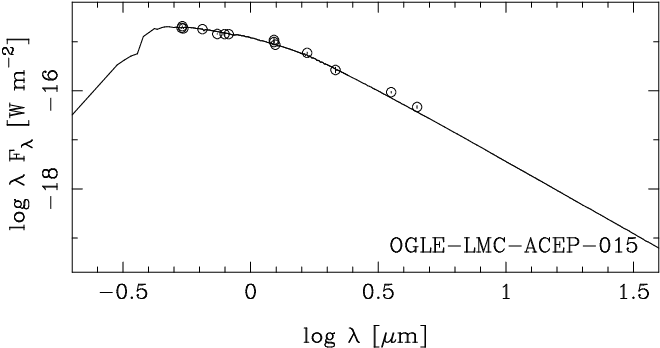}} 
\end{minipage}
\begin{minipage}{0.4\textwidth}
\resizebox{\hsize}{!}{\includegraphics[angle=-0]{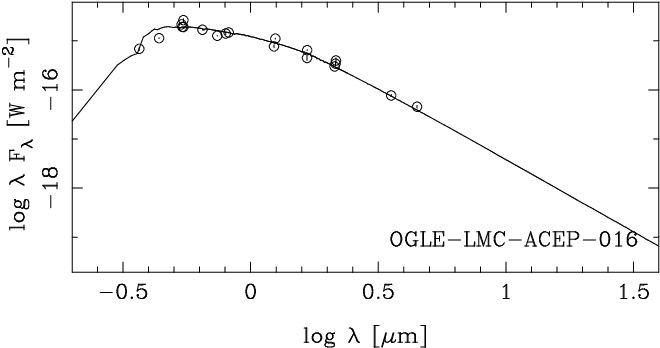}} 
\end{minipage}

\caption{Continued}
\end{figure*}

\setcounter{figure}{0}
\begin{figure*}
\centering

\begin{minipage}{0.4\textwidth}
\resizebox{\hsize}{!}{\includegraphics[angle=-0]{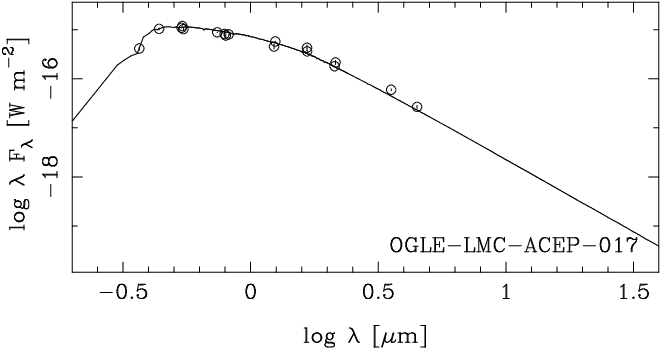}} 
\end{minipage}
\begin{minipage}{0.4\textwidth}
\resizebox{\hsize}{!}{\includegraphics[angle=-0]{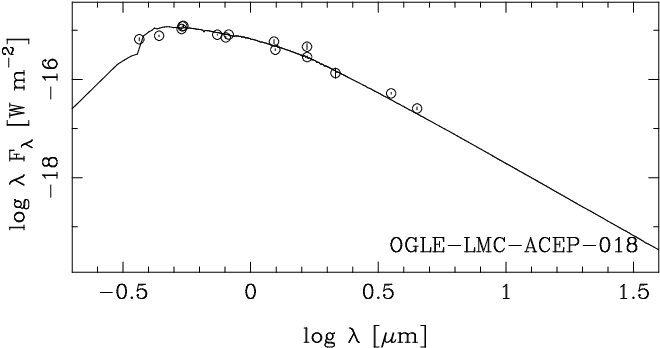}} 
\end{minipage}
 
\begin{minipage}{0.4\textwidth}
\resizebox{\hsize}{!}{\includegraphics[angle=-0]{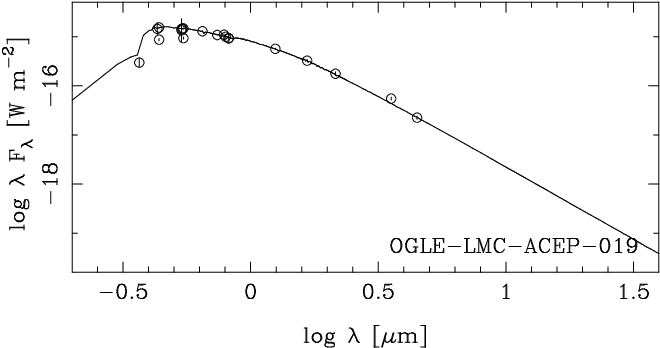}} 
\end{minipage}
\begin{minipage}{0.4\textwidth}
\resizebox{\hsize}{!}{\includegraphics[angle=-0]{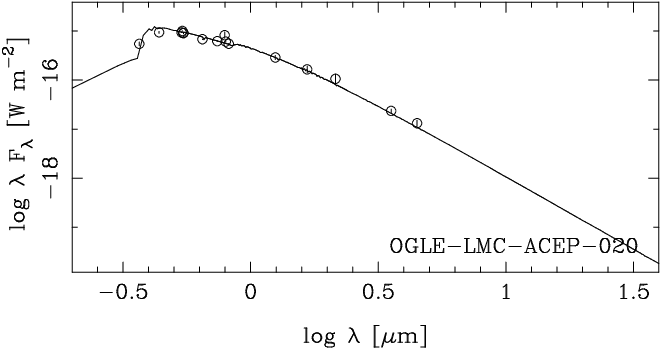}} 
\end{minipage}
 
\begin{minipage}{0.4\textwidth}
\resizebox{\hsize}{!}{\includegraphics[angle=-0]{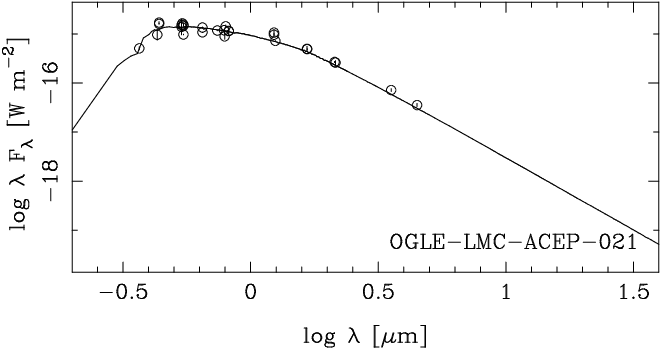}} 
\end{minipage}
\begin{minipage}{0.4\textwidth}
\resizebox{\hsize}{!}{\includegraphics[angle=-0]{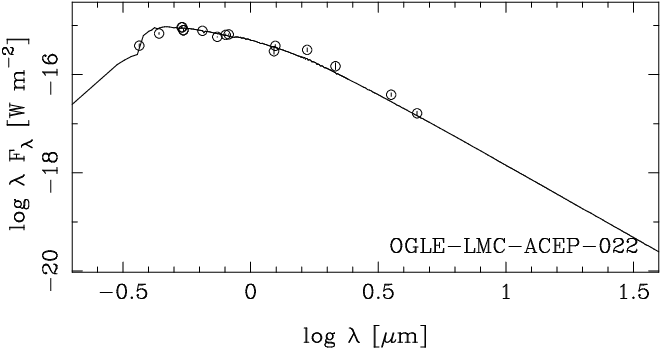}} 
\end{minipage}
 
\begin{minipage}{0.4\textwidth}
\resizebox{\hsize}{!}{\includegraphics[angle=-0]{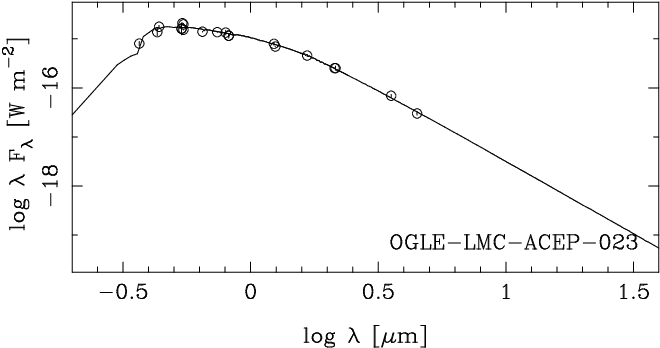}} 
\end{minipage}
\begin{minipage}{0.4\textwidth}
\resizebox{\hsize}{!}{\includegraphics[angle=-0]{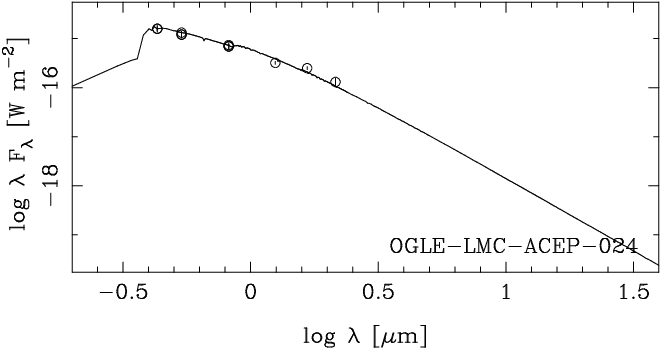}} 
\end{minipage}
 
\begin{minipage}{0.4\textwidth}
\resizebox{\hsize}{!}{\includegraphics[angle=-0]{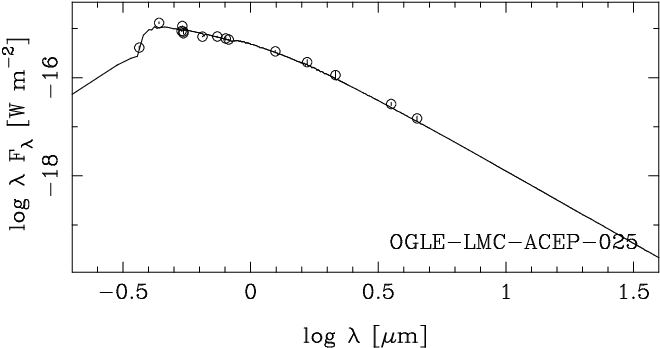}} 
\end{minipage}
\begin{minipage}{0.4\textwidth}
\resizebox{\hsize}{!}{\includegraphics[angle=-0]{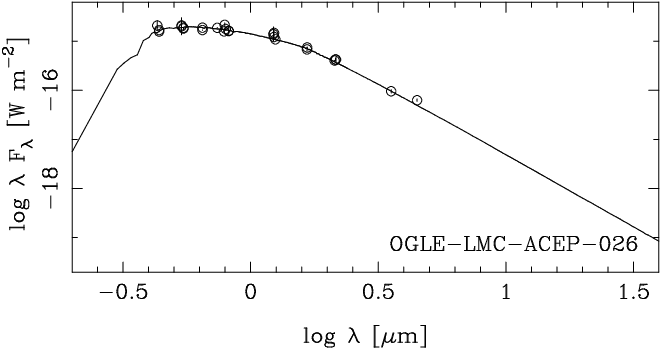}} 
\end{minipage}

\caption{Continued}
\end{figure*}

\setcounter{figure}{0}
\begin{figure*}
\centering

\begin{minipage}{0.4\textwidth}
\resizebox{\hsize}{!}{\includegraphics[angle=-0]{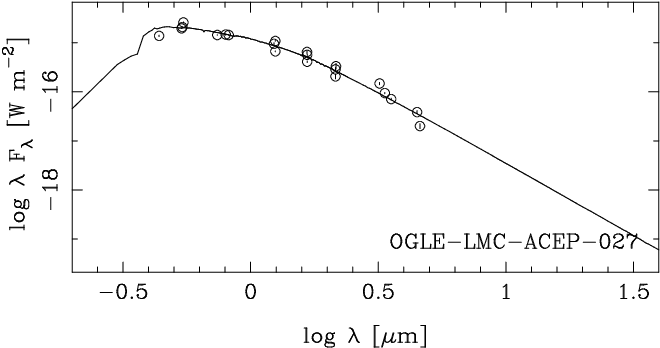}} 
\end{minipage}
\begin{minipage}{0.4\textwidth}
\resizebox{\hsize}{!}{\includegraphics[angle=-0]{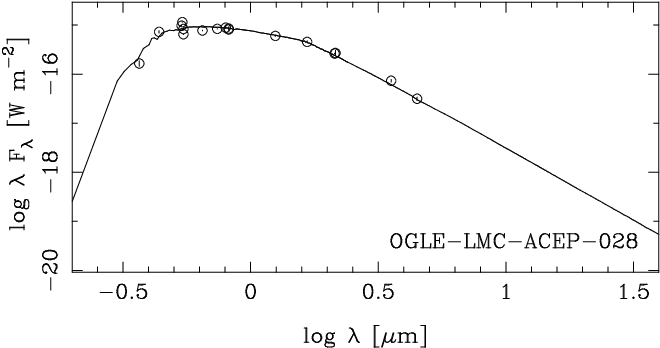}} 
\end{minipage}
 
\begin{minipage}{0.4\textwidth}
\resizebox{\hsize}{!}{\includegraphics[angle=-0]{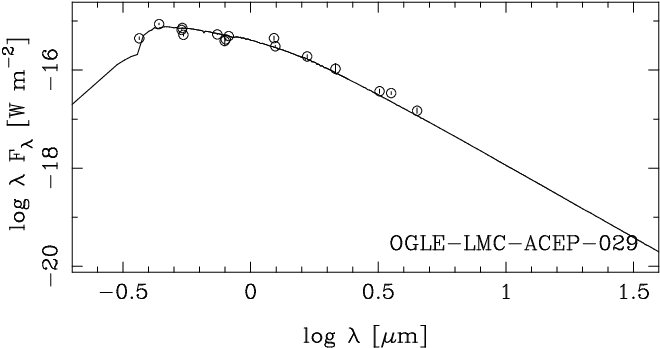}} 
\end{minipage}
\begin{minipage}{0.4\textwidth}
\resizebox{\hsize}{!}{\includegraphics[angle=-0]{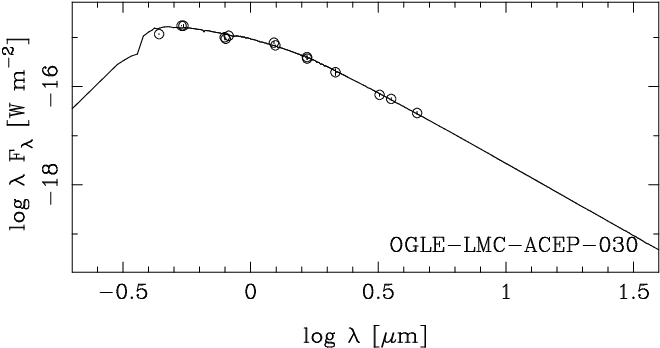}} 
\end{minipage}

\begin{minipage}{0.4\textwidth}
\resizebox{\hsize}{!}{\includegraphics[angle=-0]{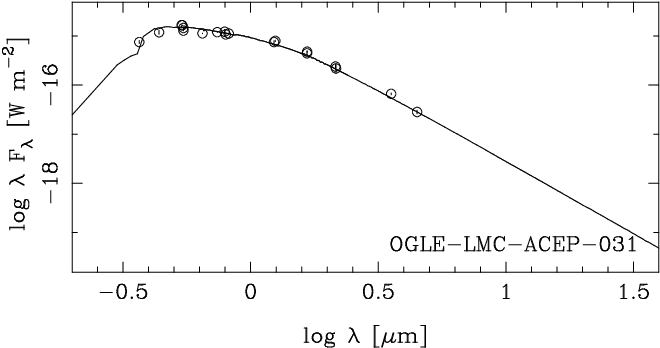}} 
\end{minipage}
\begin{minipage}{0.4\textwidth}
\resizebox{\hsize}{!}{\includegraphics[angle=-0]{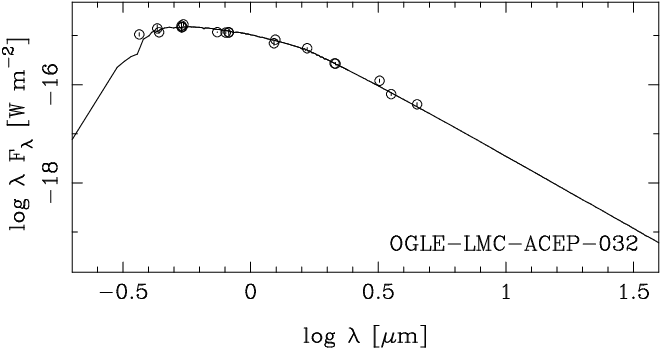}} 
\end{minipage}
 
\begin{minipage}{0.4\textwidth}
\resizebox{\hsize}{!}{\includegraphics[angle=-0]{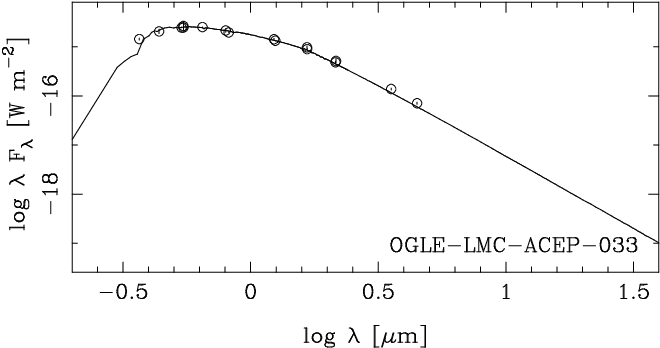}} 
\end{minipage}
\begin{minipage}{0.4\textwidth}
\resizebox{\hsize}{!}{\includegraphics[angle=-0]{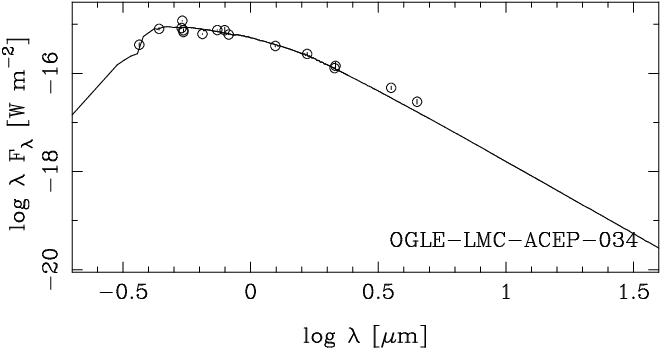}} 
\end{minipage}
 
\begin{minipage}{0.4\textwidth}
\resizebox{\hsize}{!}{\includegraphics[angle=-0]{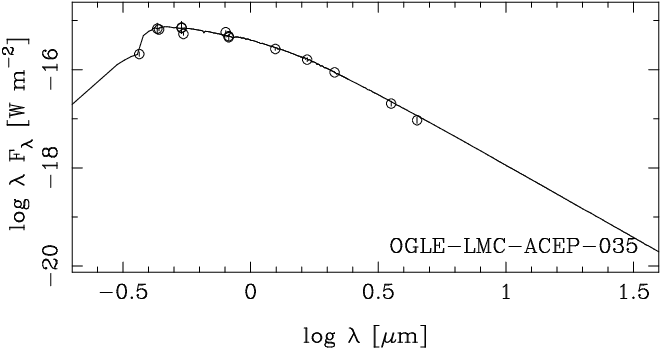}} 
\end{minipage}
\begin{minipage}{0.4\textwidth}
\resizebox{\hsize}{!}{\includegraphics[angle=-0]{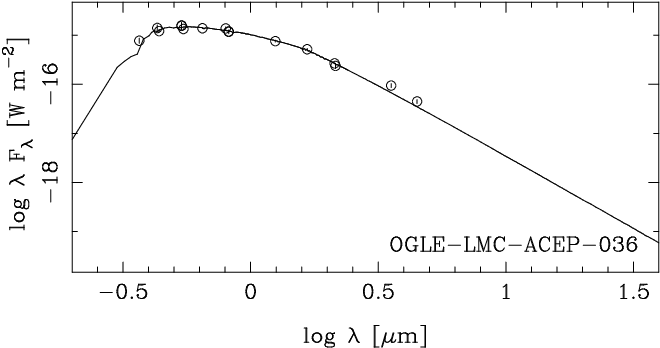}} 
\end{minipage}

\caption{Continued}
\end{figure*}

\setcounter{figure}{0}
\begin{figure*}
\centering

\begin{minipage}{0.4\textwidth}
\resizebox{\hsize}{!}{\includegraphics[angle=-0]{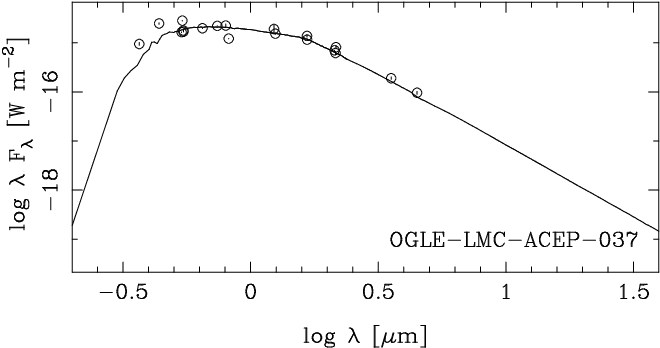}} 
\end{minipage}
\begin{minipage}{0.4\textwidth}
\resizebox{\hsize}{!}{\includegraphics[angle=-0]{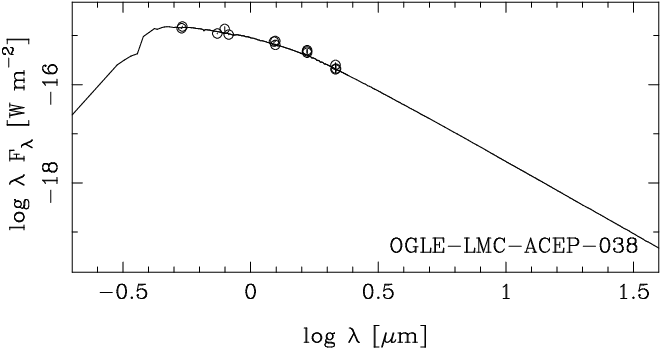}} 
\end{minipage}
 
\begin{minipage}{0.4\textwidth}
\resizebox{\hsize}{!}{\includegraphics[angle=-0]{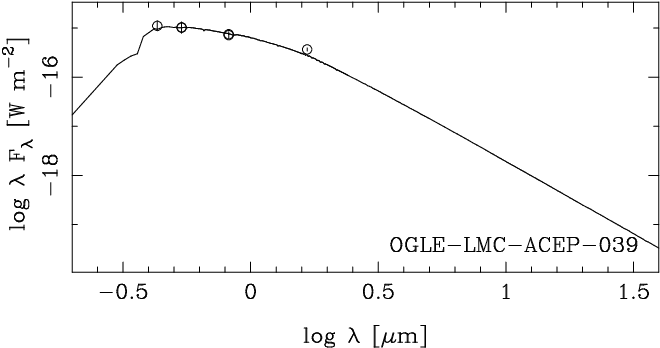}} 
\end{minipage}
\begin{minipage}{0.4\textwidth}
\resizebox{\hsize}{!}{\includegraphics[angle=-0]{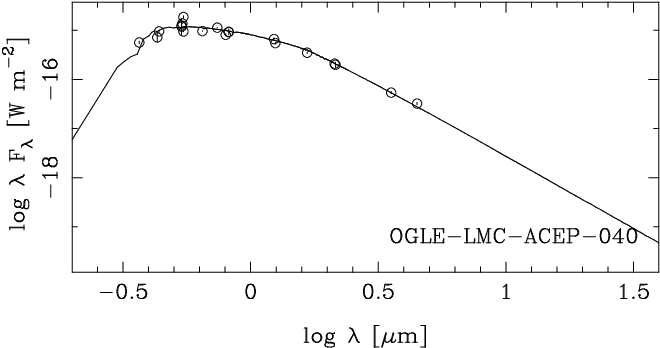}} 
\end{minipage}
 
\begin{minipage}{0.4\textwidth}
\resizebox{\hsize}{!}{\includegraphics[angle=-0]{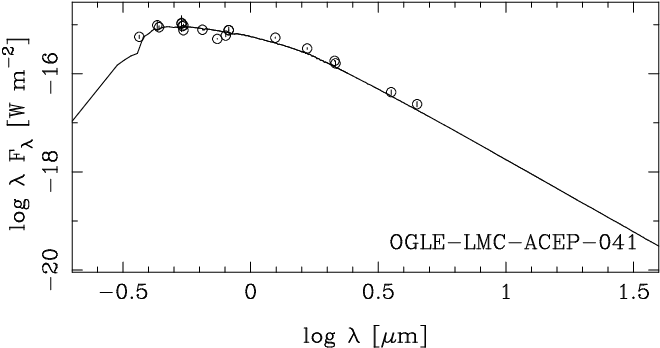}} 
\end{minipage}
\begin{minipage}{0.4\textwidth}
\resizebox{\hsize}{!}{\includegraphics[angle=-0]{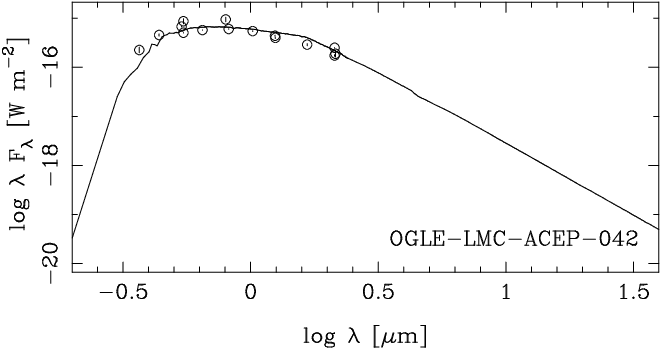}} 
\end{minipage}

\begin{minipage}{0.4\textwidth}
\resizebox{\hsize}{!}{\includegraphics[angle=-0]{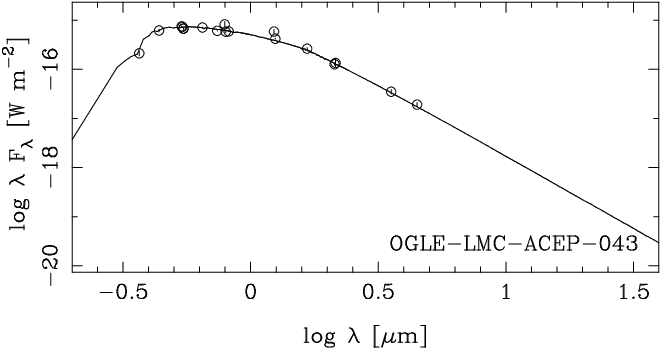}} 
\end{minipage}
\begin{minipage}{0.4\textwidth}
\resizebox{\hsize}{!}{\includegraphics[angle=-0]{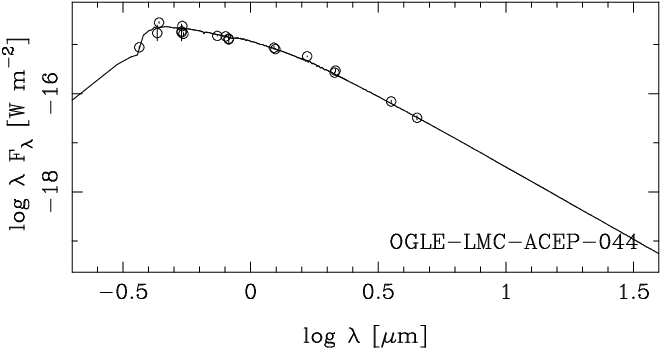}} 
\end{minipage}
 
\begin{minipage}{0.4\textwidth}
\resizebox{\hsize}{!}{\includegraphics[angle=-0]{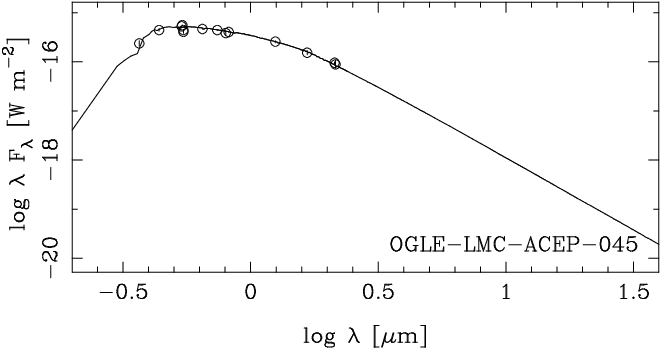}} 
\end{minipage}
\begin{minipage}{0.4\textwidth}
\resizebox{\hsize}{!}{\includegraphics[angle=-0]{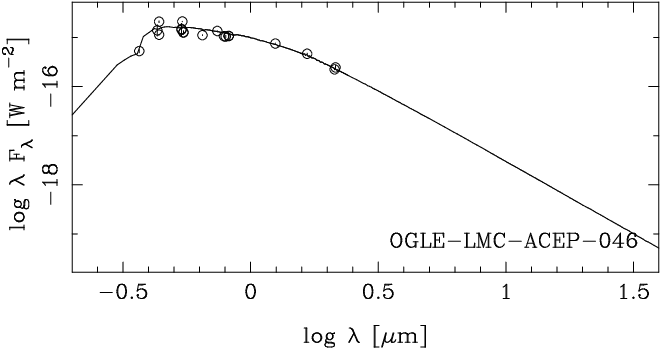}} 
\end{minipage}

\caption{Continued}
\end{figure*}

\setcounter{figure}{0}
\begin{figure*}
\centering

\begin{minipage}{0.4\textwidth}
\resizebox{\hsize}{!}{\includegraphics[angle=-0]{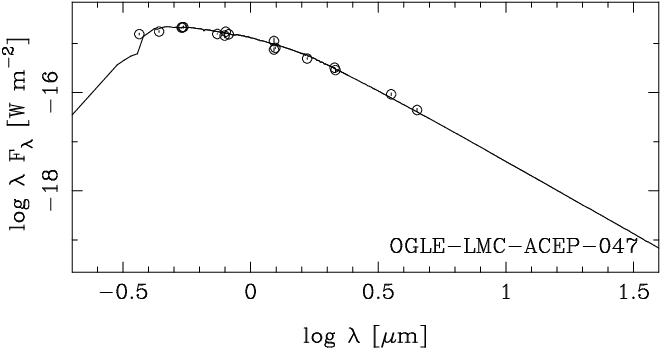}} 
\end{minipage}
\begin{minipage}{0.4\textwidth}
\resizebox{\hsize}{!}{\includegraphics[angle=-0]{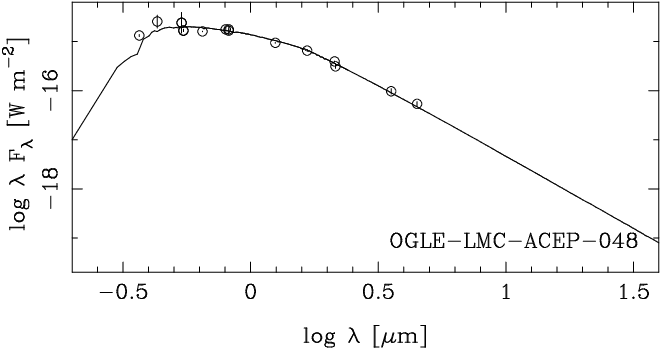}} 
\end{minipage}
 
\begin{minipage}{0.4\textwidth}
\resizebox{\hsize}{!}{\includegraphics[angle=-0]{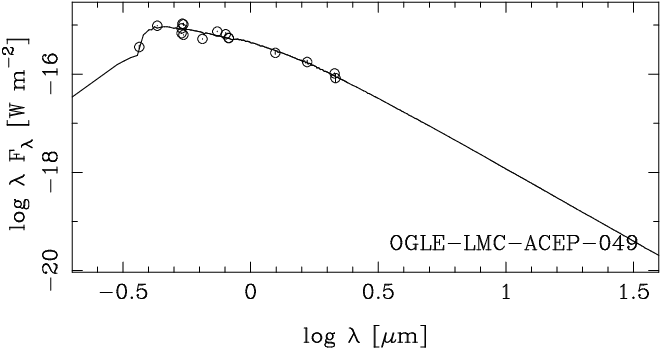}} 
\end{minipage}
\begin{minipage}{0.4\textwidth}
\resizebox{\hsize}{!}{\includegraphics[angle=-0]{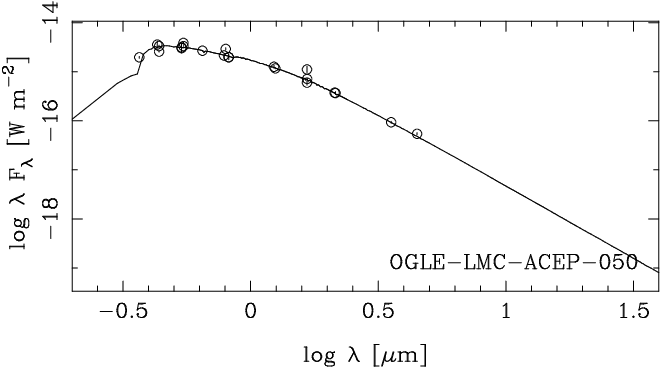}} 
\end{minipage}
 
\begin{minipage}{0.4\textwidth}
\resizebox{\hsize}{!}{\includegraphics[angle=-0]{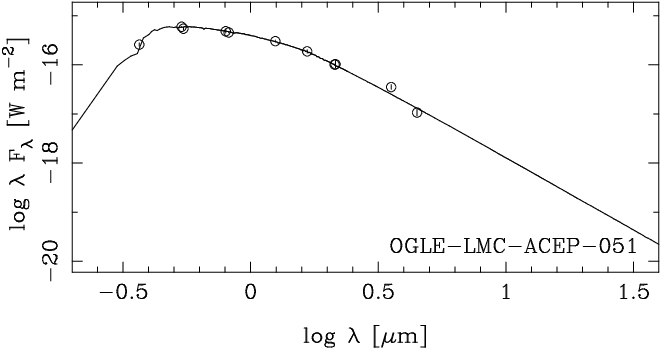}} 
\end{minipage}
\begin{minipage}{0.4\textwidth}
\resizebox{\hsize}{!}{\includegraphics[angle=-0]{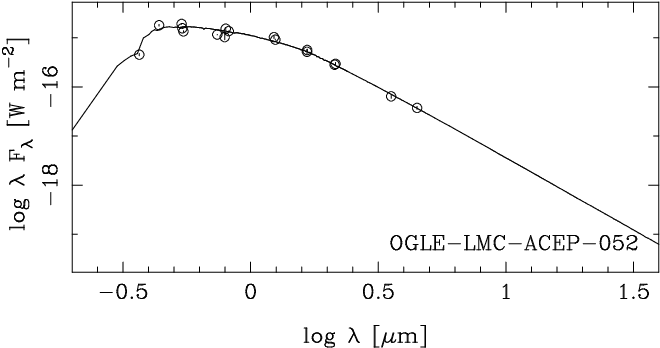}} 
\end{minipage}
 
\begin{minipage}{0.4\textwidth}
\resizebox{\hsize}{!}{\includegraphics[angle=-0]{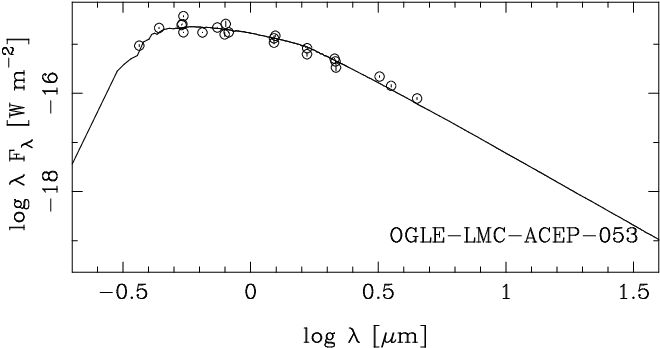}} 
\end{minipage}
\begin{minipage}{0.4\textwidth}
\resizebox{\hsize}{!}{\includegraphics[angle=-0]{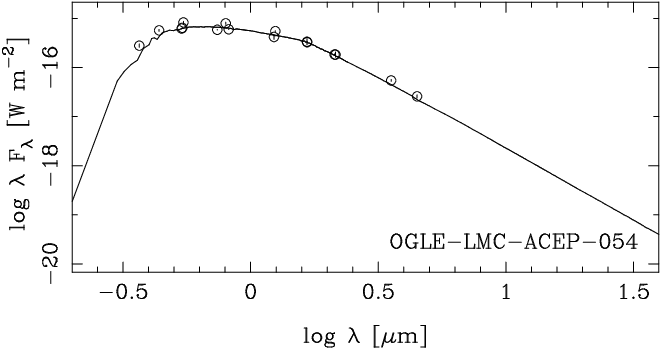}} 
\end{minipage}

\begin{minipage}{0.4\textwidth}
\resizebox{\hsize}{!}{\includegraphics[angle=-0]{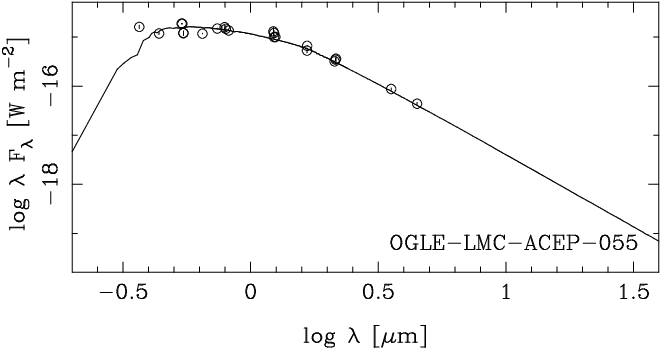}} 
\end{minipage}
\begin{minipage}{0.4\textwidth}
\resizebox{\hsize}{!}{\includegraphics[angle=-0]{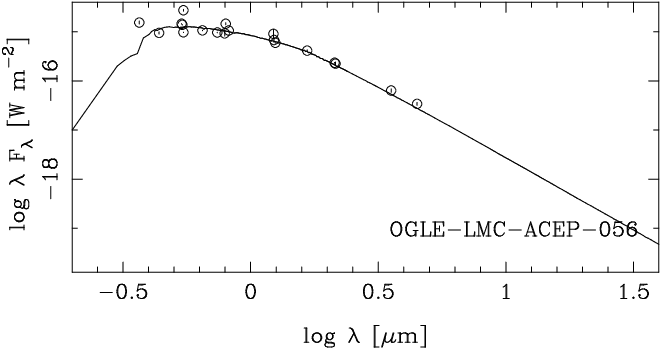}} 
\end{minipage}

\caption{Continued}
\end{figure*}

\setcounter{figure}{0}
\begin{figure*}
\centering

\begin{minipage}{0.4\textwidth}
\resizebox{\hsize}{!}{\includegraphics[angle=-0]{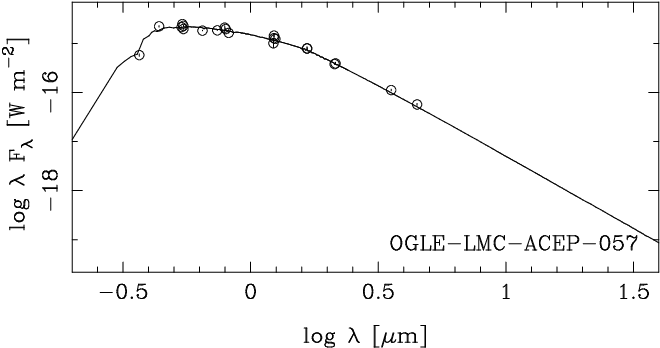}} 
\end{minipage}
\begin{minipage}{0.4\textwidth}
\resizebox{\hsize}{!}{\includegraphics[angle=-0]{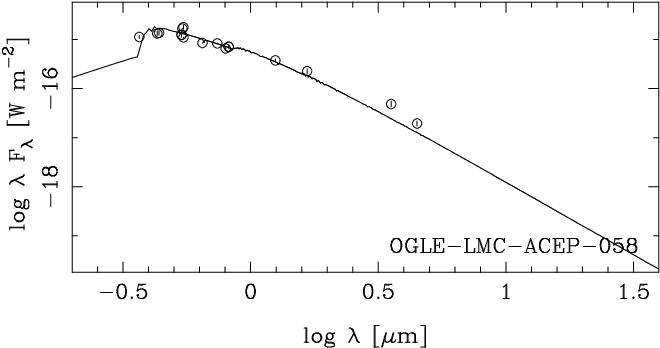}} 
\end{minipage}
 
\begin{minipage}{0.4\textwidth}
\resizebox{\hsize}{!}{\includegraphics[angle=-0]{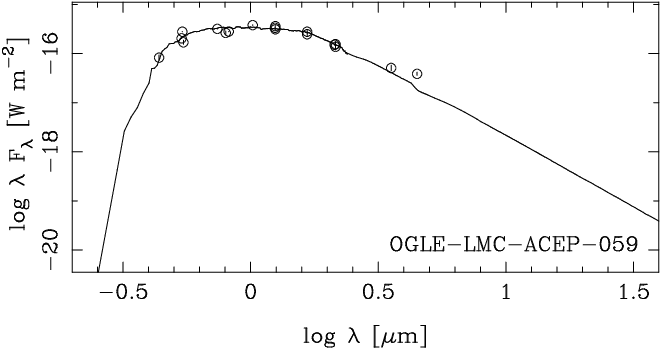}} 
\end{minipage}
\begin{minipage}{0.4\textwidth}
\resizebox{\hsize}{!}{\includegraphics[angle=-0]{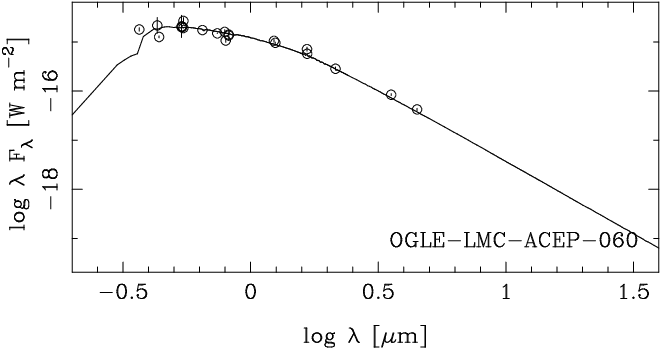}} 
\end{minipage}
 
\begin{minipage}{0.4\textwidth}
\resizebox{\hsize}{!}{\includegraphics[angle=-0]{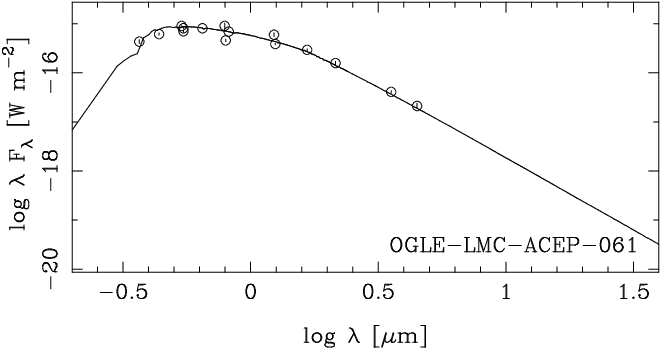}} 
\end{minipage}
\begin{minipage}{0.4\textwidth}
\resizebox{\hsize}{!}{\includegraphics[angle=-0]{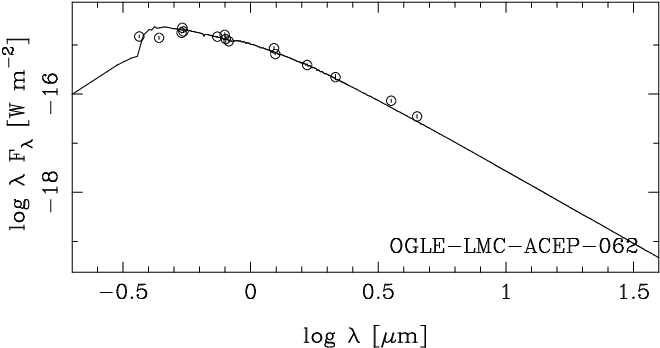}} 
\end{minipage}
 
\begin{minipage}{0.4\textwidth}
\resizebox{\hsize}{!}{\includegraphics[angle=-0]{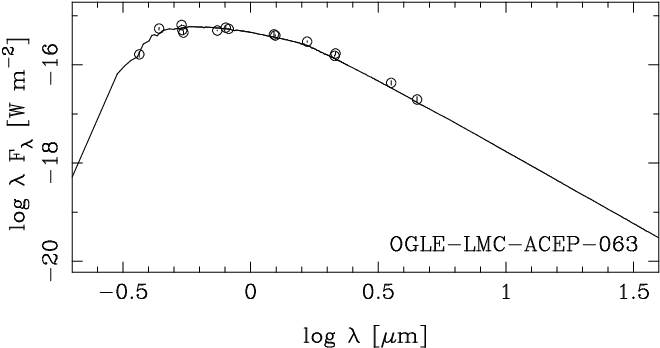}} 
\end{minipage}
\begin{minipage}{0.4\textwidth}
\resizebox{\hsize}{!}{\includegraphics[angle=-0]{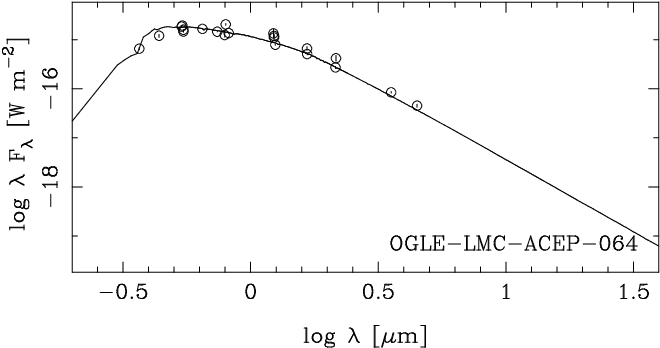}} 
\end{minipage}
 
\begin{minipage}{0.4\textwidth}
\resizebox{\hsize}{!}{\includegraphics[angle=-0]{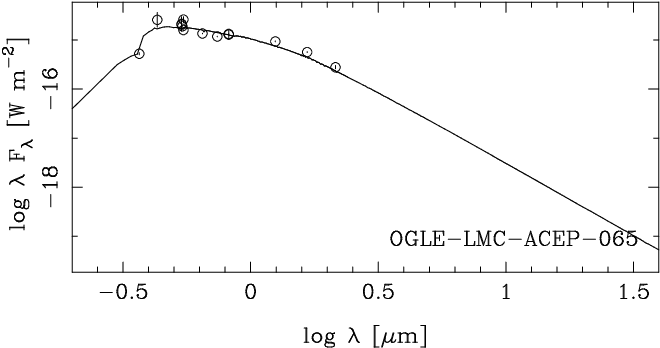}} 
\end{minipage}
\begin{minipage}{0.4\textwidth}
\resizebox{\hsize}{!}{\includegraphics[angle=-0]{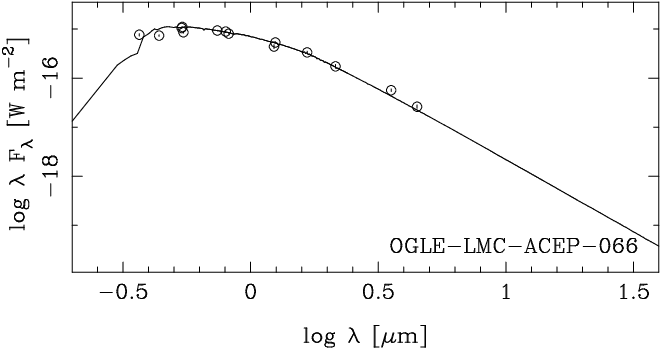}} 
\end{minipage}
 
\caption{Continued}
\end{figure*}

\clearpage
 
\setcounter{figure}{0}
\begin{figure*}
\centering

\begin{minipage}{0.4\textwidth}
\resizebox{\hsize}{!}{\includegraphics[angle=-0]{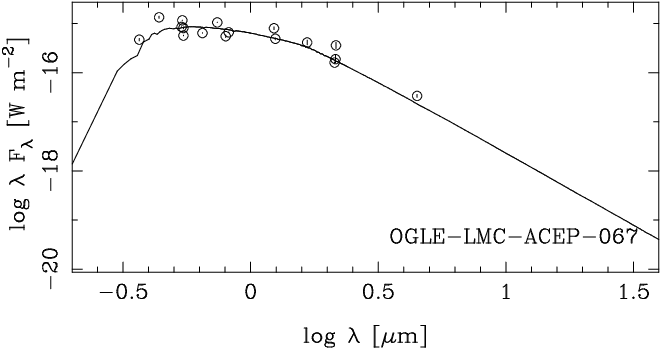}} 
\end{minipage}
\begin{minipage}{0.4\textwidth}
\resizebox{\hsize}{!}{\includegraphics[angle=-0]{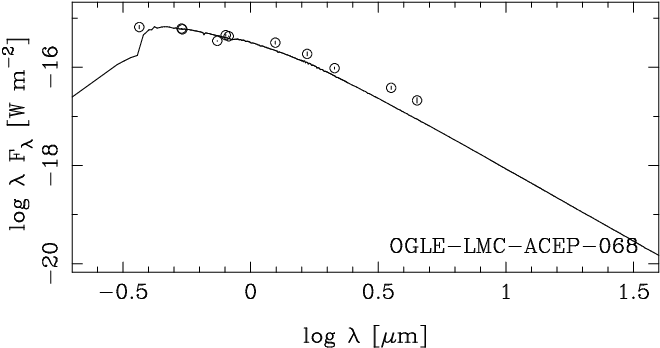}} 
\end{minipage}
 
\begin{minipage}{0.4\textwidth}
\resizebox{\hsize}{!}{\includegraphics[angle=-0]{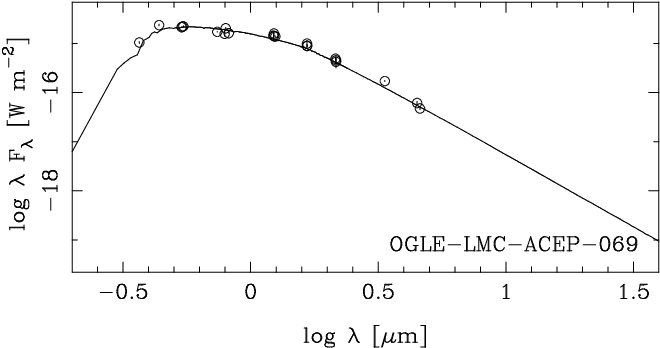}} 
\end{minipage}
\begin{minipage}{0.4\textwidth}
\resizebox{\hsize}{!}{\includegraphics[angle=-0]{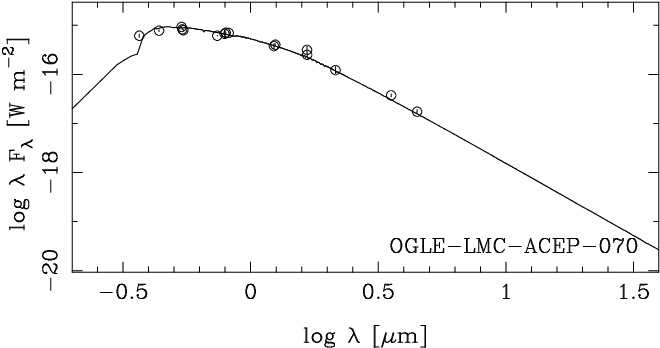}} 
\end{minipage}
 
\begin{minipage}{0.4\textwidth}
\resizebox{\hsize}{!}{\includegraphics[angle=-0]{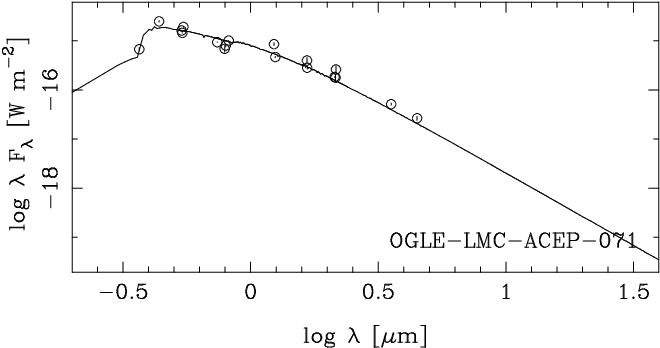}} 
\end{minipage}
\begin{minipage}{0.4\textwidth}
\resizebox{\hsize}{!}{\includegraphics[angle=-0]{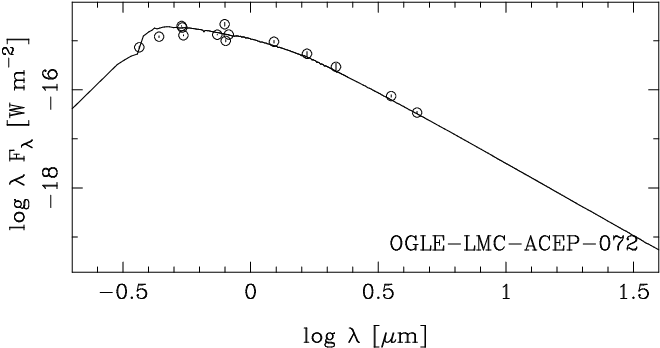}} 
\end{minipage}
 
\begin{minipage}{0.4\textwidth}
\resizebox{\hsize}{!}{\includegraphics[angle=-0]{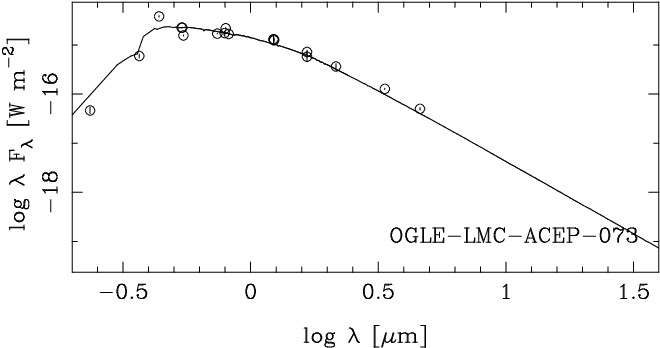}} 
\end{minipage}
\begin{minipage}{0.4\textwidth}
\resizebox{\hsize}{!}{\includegraphics[angle=-0]{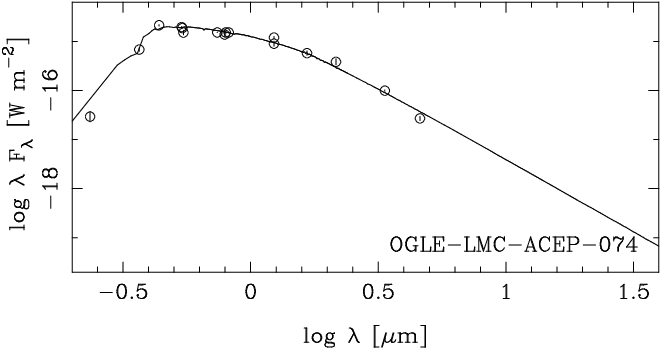}} 
\end{minipage}
 
\begin{minipage}{0.4\textwidth}
\resizebox{\hsize}{!}{\includegraphics[angle=-0]{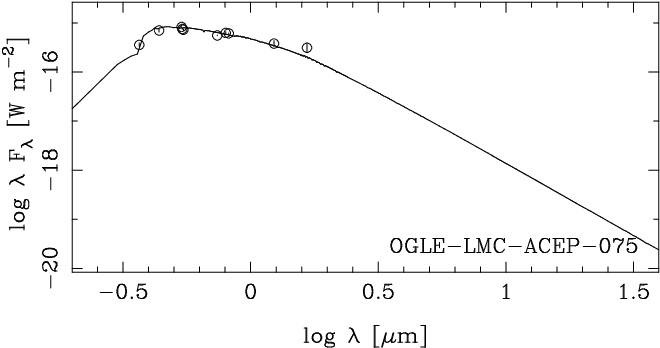}} 
\end{minipage}
\begin{minipage}{0.4\textwidth}
\resizebox{\hsize}{!}{\includegraphics[angle=-0]{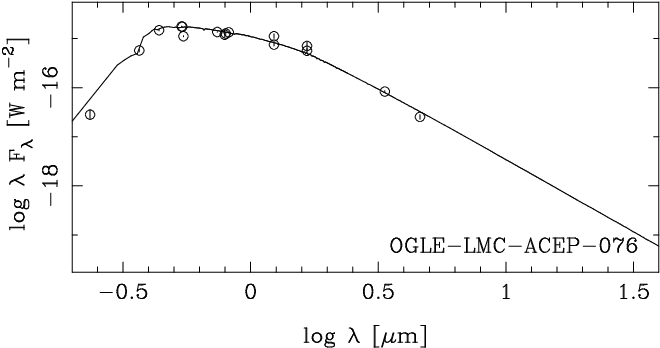}} 
\end{minipage}

\caption{Continued}
\end{figure*}

\setcounter{figure}{0}
\begin{figure*}
\centering

\begin{minipage}{0.4\textwidth}
\resizebox{\hsize}{!}{\includegraphics[angle=-0]{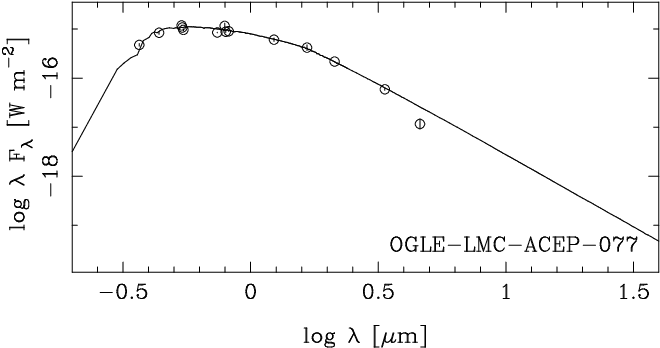}} 
\end{minipage}
\begin{minipage}{0.4\textwidth}
\resizebox{\hsize}{!}{\includegraphics[angle=-0]{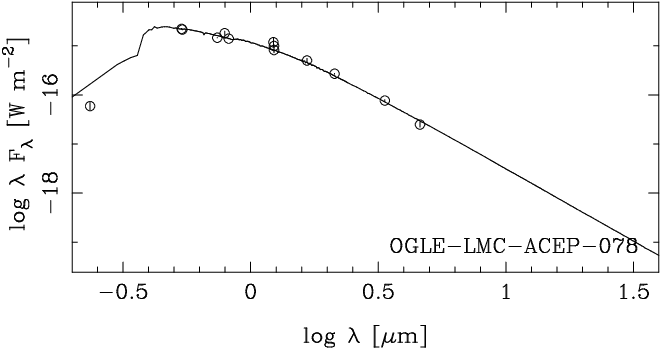}} 
\end{minipage}
 
\begin{minipage}{0.4\textwidth}
\resizebox{\hsize}{!}{\includegraphics[angle=-0]{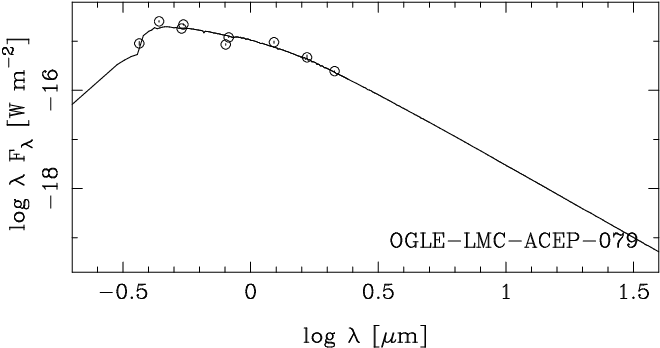}} 
\end{minipage}
\begin{minipage}{0.4\textwidth}
\resizebox{\hsize}{!}{\includegraphics[angle=-0]{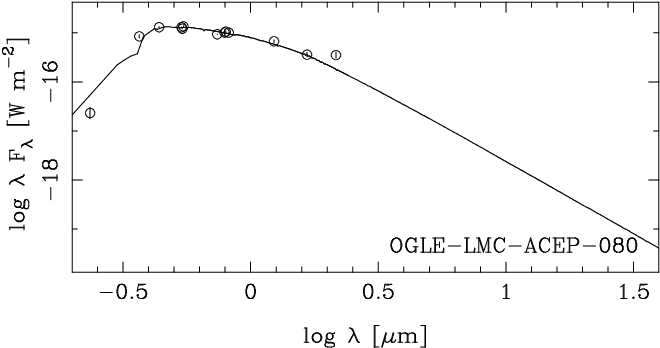}} 
\end{minipage}
 
\begin{minipage}{0.4\textwidth}
\resizebox{\hsize}{!}{\includegraphics[angle=-0]{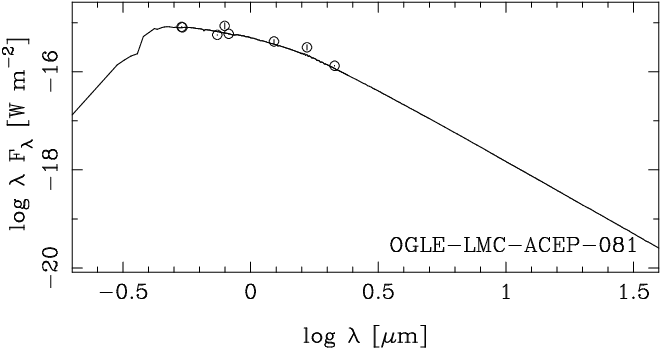}} 
\end{minipage}
\begin{minipage}{0.4\textwidth}
\resizebox{\hsize}{!}{\includegraphics[angle=-0]{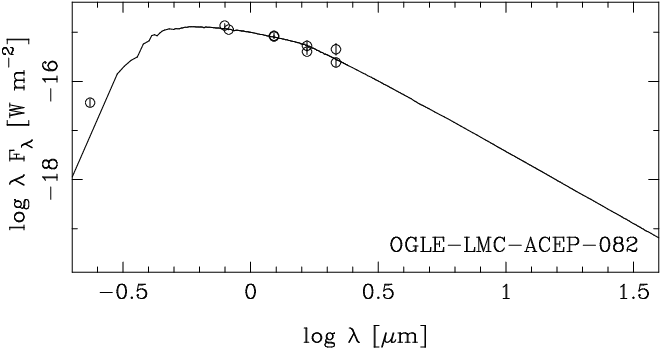}} 
\end{minipage}

\caption{Continued}
\end{figure*}

\setcounter{figure}{0}
\begin{figure*}
\centering

\begin{minipage}{0.4\textwidth}
\resizebox{\hsize}{!}{\includegraphics[angle=-0]{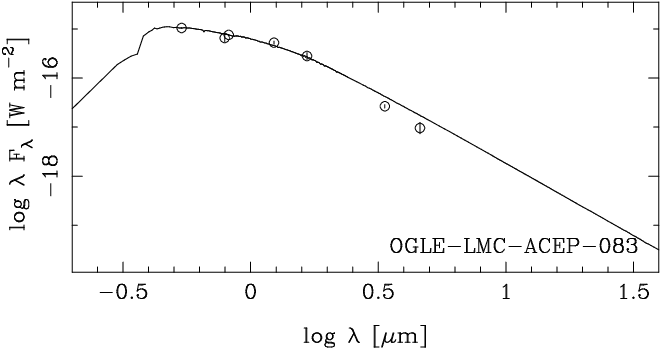}} 
\end{minipage}
\begin{minipage}{0.4\textwidth}
\resizebox{\hsize}{!}{\includegraphics[angle=-0]{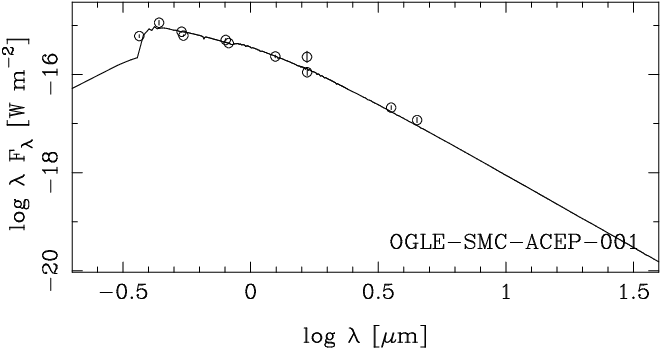}} 
\end{minipage}
 
\begin{minipage}{0.4\textwidth}
\resizebox{\hsize}{!}{\includegraphics[angle=-0]{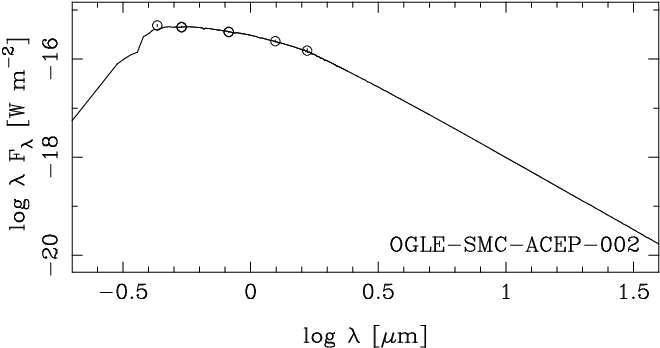}} 
\end{minipage}
\begin{minipage}{0.4\textwidth}
\resizebox{\hsize}{!}{\includegraphics[angle=-0]{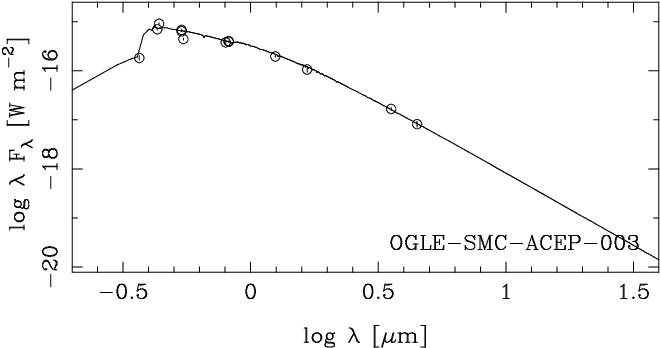}} 
\end{minipage}
 
\begin{minipage}{0.4\textwidth}
\resizebox{\hsize}{!}{\includegraphics[angle=-0]{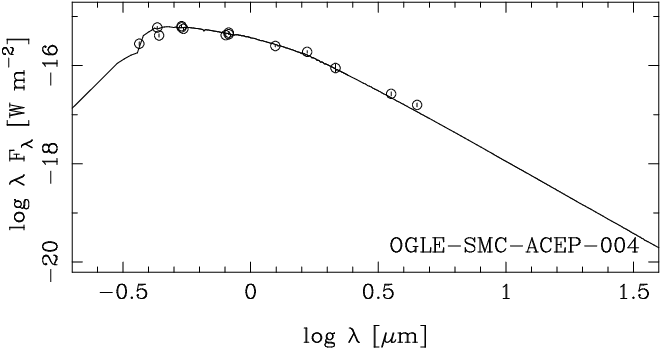}} 
\end{minipage}
\begin{minipage}{0.4\textwidth}
\resizebox{\hsize}{!}{\includegraphics[angle=-0]{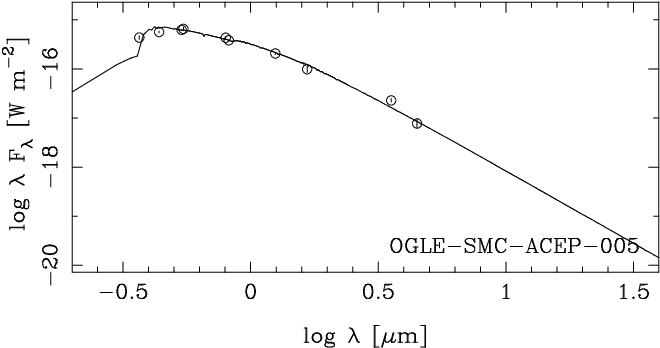}} 
\end{minipage}
 
\begin{minipage}{0.4\textwidth}
\resizebox{\hsize}{!}{\includegraphics[angle=-0]{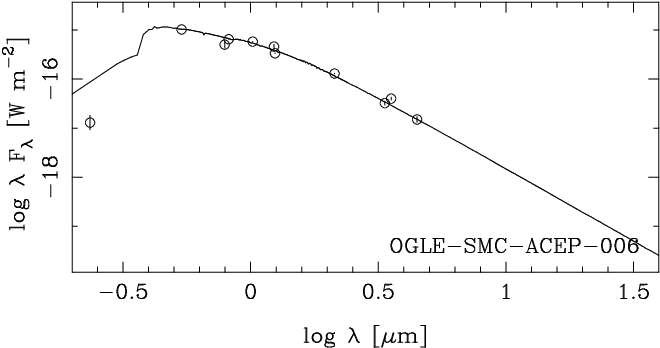}} 
\end{minipage}
 
\caption{Ccontinued}
\end{figure*}

\clearpage

\section{The light-time travel effect}
\label{AppLITE}

In this Appendix we describe the implementation of the method outlined in \citet{Hajdu2015} to construct O-C diagrams and investigate
the presence of period changes or binarity based on the LITE.
The code described below is written in Fortran77. 
The programme reads in an input file containing the name of the source, an estimate of a time of maximum light 
(typically near the midpoint of the available time series) and pulsation period, 
an estimate of the mean magnitude, the number of harmonics ($N$) to be fitted to the phased LC, and a 
file containing estimates of the Fourier coefficients of the LC. All these parameters are determined externally to the code, 
and specifically we use the publicly available code {\it Period04} for this part \citep{Lenz_Breger_2005}.

These parameters are read in the code, and in a first step, an $N$-term Fourier series 
(specifically in the form of Eq.~1 in Appendix~A in \cite{Groenewegen2004}) is fitted to the entire time series using the
{\it mrqmin} routine from Numerical Recipes \citep{Press1992}.
The LC and phased LC are inspected, and at this point some data may be removed from the time series,
either photometric outliers, or, specifically in the case of EBs, data taken at the time of eclipses.

Then the code is run on a limited stretch of data to determine a well-defined template LC and its Fourier parameters are saved.

The next step is the determination of the O-C diagram.
The DT consecutive data points in time are selected; DT is typically 50. The "time" associated with this chunk of data is the 
median value of the individual observing times. 
Forty trial shifts in phase are examined over a total phase range ($\delta \phi$), which varies from star to star depending on the 
O-C variation relative to the period (see below).
In the extreme case, 40 points increasing by 0.025 in phase would allow us to probe O-C variations from $-0.5P$ to $+0.5P$.
For each trial shift, $j$, the template LC is compared to the actual LC and a reduced $\chi_{\rm r}^2$ determined.
The key assumption in the entire method is that the template LC remains constant in time.
This is a reasonable assumption for classical variables, but T2Cs often display amplitude modulations as well.
To allow for this simplest of modification of a strictly time-independent template LC, a scaling of the 
Fourier amplitudes is applied in determining the $\chi_{\rm r}^2$.
The index $j_{\rm min}$ of the lowest $\chi_{\rm r}^2$ is found and a parabola is fitted to the $\chi_{\rm r}^2$s at indices $j_{\rm min}-1$, 
$j_{\rm min}$, and $j_{\rm min}+1$ to determine the final best-fitting phase shift, hence O-C value, and final $\chi_{\rm r, min}^2$.
This is repeated for the next DT number of points, etc. 
An error in O-C is determined by finding the roots of the parabolic equation at a value higher than $\chi_{\rm r, min}^2$. 
If $\chi^2_{\rm r, min}$ were of order unity this value would be $\chi^2_{\rm r, min} +1$ but this is rarely the case here.
Practically, the errors are determined by finding the range in phase shifts at a level $(F \cdot \chi^2_{\rm r, min})$, where 
$F$ is typically 1.15 and ranges between 1.01 and 1.5 (see below).
Also at this point the range in the values of $j_{\rm min}$ for the individual chuncks of data are known, 
and $\delta \phi$ can be changed so that these values cover roughly the range between 1 and 40.

The O-C diagram is plotted and a model can be fitted to the data. The model is a combination of a changing period 
($\dot{P}$, $\ddot{P}$), and a binary model
\begin{equation}
(O - C) (t)= c_0 + c_1 t + c_2 t^2 + c_3 t^3 + ...                                       \label{parabola1}
\end{equation}
\begin{equation}
     \hspace{30mm} ... + (a \sin i) \, \frac{1-e^2}{1+e\cos(\nu)} \, \sin(\nu+\omega),   \label{lteffect1}
\end{equation}
\noindent where $a$ is the semi-major axis, $i$ is the inclination, $e$ is the eccentricity, and $\omega$ is the
argument of the periastron. The true anomaly $\nu$ is a function of the time $t$,
the orbital period $P_{\rm orb}$, the time of periastron passage $T_{\rm peri}$, and $e$.
All parameters of the model ($c_0, c_1, c_2, c_3$, $P_{\rm orb}$, $T_{\rm peri}$, $a \sin i$, $e$, $\omega$) can be fitted or fixed.
Starting values for $c_0, c_1, c_2, c_3$ are zero, if a binary model is used the initial values for 
$P_{\rm orb}$, $T_{\rm peri}$, $a \sin i$ are estimated from the O-C diagram. Some trial and error starting values are 
sometimes recommended, especially for non-zero eccentric orbits.
The fitting is done again with the {\it mrqmin} routine. By changing the factor $F$ in the determination of the error 
in the individual O-C values the reduced $\chi^2$ of the model fit is tuned to be approximately unity.
This then provides the model parameters and their errors.

\begin{figure}
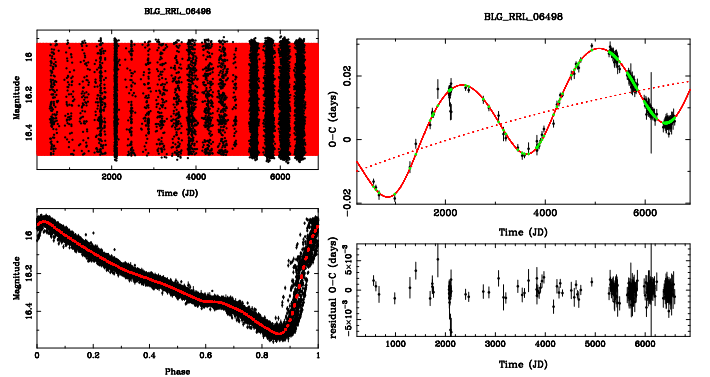


\centering
\begin{minipage}{0.22\textwidth}
\resizebox{\hsize}{!}{\includegraphics[angle=-0]{LITE/BLG_RRL_06498_OLC_new.ps}} 
\end{minipage}
\begin{minipage}{0.26\textwidth}
\resizebox{\hsize}{!}{\includegraphics[angle=-0]{LITE/BLG_RRL_06498_OmC_new.ps}} 
\end{minipage}

\caption{Light curves and O-C diagrams of BLG-RRL-06498.
The left-hand panel shows the time series (top) and phased light curve (bottom).
The right-hand panel shows the O-C diagram (top) and the residual after subtracting the model (bottom).
The full model is represented by the full line. The contribution due to a changing period is indicated by the dotted line.
}

\label{Fig-LITE-TEST}
\end{figure}

The code was tested against the showcase example of BLG-RRL-06498 in \citet{Hajdu2015}. 
The parameters are compared in Table~\ref{Tab-LITE-TEST} and the (phased) LC and O-C diagram 
and model fit are shown in Figure~\ref{Fig-LITE-TEST}; cf. Fig.~2 in \citet{Hajdu2015}. 
Initially only OGLE-III was available to us but in the course of this project the OGLE-IV data also became public via \citet{Soszynski_2014},
which were already available to \citet{Hajdu2015}.
Both solutions are given in the table illustrating the power of the increased time span of the observations.
The derived parameters agree within the error.

Table~\ref{Tab-LITE-BIN} in the main text lists the systems where this investigation suggested a possible binary system based 
on the LITE (it includes known EBs). The (phased) LC and the O-C diagrams with the model fits are shown in Figure~\ref{Fig-LITE-BIN}.

Table~\ref{Tab-LITE-Pdot} lists the systems for which a significant period change is suggested; it includes known EBs for 
which the LITE could not be established. 
The (phased) LC and the O-C diagrams with the model fits are shown in Figure~\ref{Fig-LITE-Pdot}.

Table~\ref{Tab-LITE-Incon} lists the systems that were investigated but with inconclusive results regarding the LITE 
or significant period changes. Additional information is available from the first author.

\begin{table*}
\setlength{\tabcolsep}{1.2mm}
\caption{Comparison of results for BLG-RRL-06498}

\begin{tabular}{ccccccl}
\hline
$P_{\rm bin}$  &   $T_{\rm peri}$  &        $e$          &  $\omega$   &  $a \sin i$    &     $\dot{P}$  & Remarks  \\  
    (d)      &   (JD-2450000)  &                     &     (deg)    &    (AU)        &     (d/Myr)    &   \\
\hline
2789 $\pm$ 18 & 8137 $\pm$ 134 & 0.12 $\pm$ 0.04       & $-82 \pm 16$ & 2.35 $\pm$ 0.05 & $-0.11~\pm$ 0.03 & \citet{Hajdu2015}, based on OGLE-III + IV data \\
2804 $\pm$ 30 & 7917 $\pm$ 116 & 0.17 $^{+0.05}_{-0.04}$ & $-85 \pm 15$ & 2.45 $\pm$ 0.05 & $+0.06~\pm$ 0.07 & present paper, based on OGLE-III data \\
2766 $\pm$ 10 & 7931 $\pm$  52 & 0.16 $\pm$ 0.02       & $-73 \pm  6$ & 2.42 $\pm$ 0.03 & $-0.10~\pm$ 0.02 & present paper, based on OGLE-III +IV data \\
\\
3470 $\pm$ 160 & 7850 $\pm$ 64 & 0 (fixed)             &   --      &  276  $\pm$ 13 & $ +7000~\pm$ 350 & OGLE-BLG-T2CEP-059 \\
  --           &    --         &  --                   &   --      &       --       & $+11200~\pm$ 200 & OGLE-BLG-T2CEP-059, only period change \\
\hline
\end{tabular}

\label{Tab-LITE-TEST}
\end{table*}

\begin{figure}

\centering
\begin{minipage}{0.24\textwidth}
\resizebox{\hsize}{!}{\includegraphics[angle=-90]{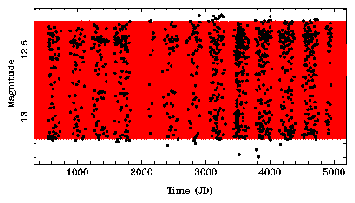}} 
\end{minipage}
\begin{minipage}{0.24\textwidth}
\resizebox{\hsize}{!}{\includegraphics[angle=-90]{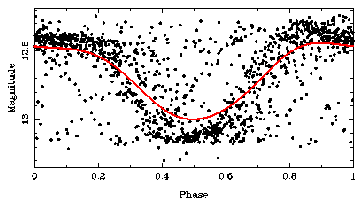}} 
\end{minipage}

\begin{minipage}{0.24\textwidth}
\resizebox{\hsize}{!}{\includegraphics[angle=-0]{BLG-T2CEP-059_OmC_PdotONLY.ps}} 
\end{minipage}
\begin{minipage}{0.24\textwidth}
\resizebox{\hsize}{!}{\includegraphics[angle=-0]{BLG-T2CEP-059_OmC_PdotBin.ps}} 
\end{minipage}

\caption{Light curves and O-C diagrams of BLG-T2CEP-059. 
The top left-hand and right-hand panels show the time series (top left), and phased light curve (top right) for a period of 12.637 days.
The bottom panels show the O-C diagram for a model with only a period change (bottom left), and a binary + $\dot{P}$ model.
}

\label{Fig-LITE-BLG}
\end{figure}

\medskip
Figure~\ref{Fig-LITE-BLG} and Table~\ref{Tab-LITE-TEST} show the analysis of the system OGLE-BLG-T2CEP-059 that is 
mentioned in the main text and was noted by \cite{Soszynski_2011} to have a large period change.
The bottom left panel shows the result when only a period change is considered. A value of 11200 d/Myr is found but 
the plot with the residuals shows a pattern.
Within the framework of the model that is used here a binary model would be proposed and the fits are shown in the bottom right panel.
The residuals are flat within the errors. The period change is 7700 d/Myr in this model.

\clearpage

\begin{figure*}
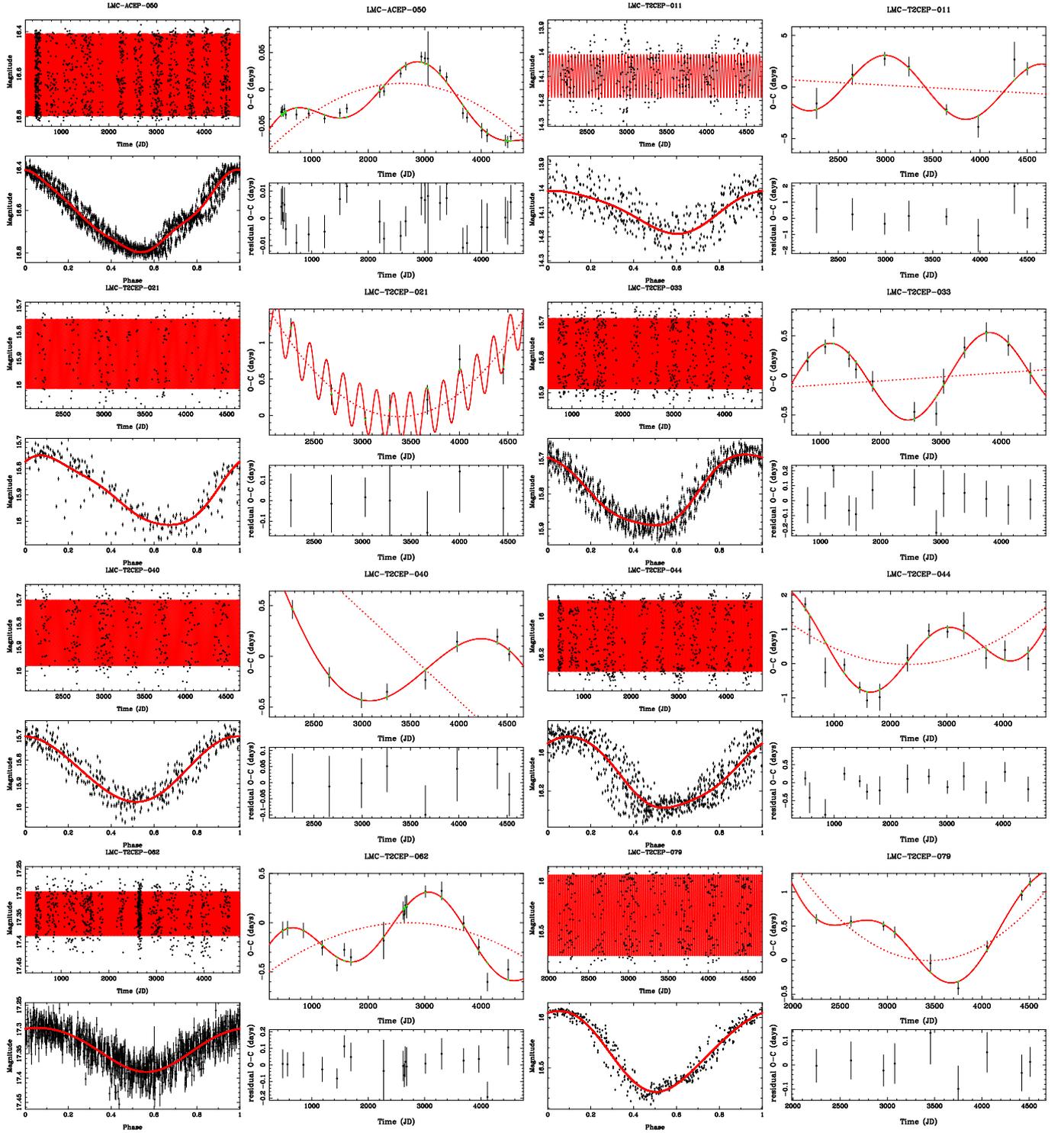


\centering

\begin{minipage}{0.22\textwidth}
\resizebox{\hsize}{!}{\includegraphics[angle=-0]{LITE/LMC-ACEP-050_OLC.ps}} 
\end{minipage}
\begin{minipage}{0.26\textwidth}
\resizebox{\hsize}{!}{\includegraphics[angle=-0]{LITE/LMC-ACEP-050_OmC.ps}} 
\end{minipage}
\begin{minipage}{0.22\textwidth}
\resizebox{\hsize}{!}{\includegraphics[angle=-0]{LITE/LMC-T2CEP-011_OLC.ps}} 
\end{minipage}
\begin{minipage}{0.26\textwidth}
\resizebox{\hsize}{!}{\includegraphics[angle=-0]{LITE/LMC-T2CEP-011_OmC.ps}} 
\end{minipage}
 
\begin{minipage}{0.22\textwidth}
\resizebox{\hsize}{!}{\includegraphics[angle=-0]{LITE/LMC-T2CEP-021_OLC.ps}} 
\end{minipage}
\begin{minipage}{0.26\textwidth}
\resizebox{\hsize}{!}{\includegraphics[angle=-0]{LITE/LMC-T2CEP-021_OmC.ps}} 
\end{minipage}
\begin{minipage}{0.22\textwidth}
\resizebox{\hsize}{!}{\includegraphics[angle=-0]{LITE/LMC-T2CEP-033_OLC.ps}} 
\end{minipage}
\begin{minipage}{0.26\textwidth}
\resizebox{\hsize}{!}{\includegraphics[angle=-0]{LITE/LMC-T2CEP-033_OmC.ps}} 
\end{minipage}

\begin{minipage}{0.22\textwidth}
\resizebox{\hsize}{!}{\includegraphics[angle=-0]{LITE/LMC-T2CEP-040_OLC.ps}} 
\end{minipage}
\begin{minipage}{0.26\textwidth}
\resizebox{\hsize}{!}{\includegraphics[angle=-0]{LITE/LMC-T2CEP-040_OmC.ps}} 
\end{minipage}
\begin{minipage}{0.22\textwidth}
\resizebox{\hsize}{!}{\includegraphics[angle=-0]{LITE/LMC-T2CEP-044_OLC.ps}} 
\end{minipage}
\begin{minipage}{0.26\textwidth}
\resizebox{\hsize}{!}{\includegraphics[angle=-0]{LITE/LMC-T2CEP-044_OmC.ps}} 
\end{minipage}

\begin{minipage}{0.22\textwidth}
\resizebox{\hsize}{!}{\includegraphics[angle=-0]{LITE/LMC-T2CEP-062_OLC.ps}} 
\end{minipage}
\begin{minipage}{0.26\textwidth}
\resizebox{\hsize}{!}{\includegraphics[angle=-0]{LITE/LMC-T2CEP-062_OmC.ps}} 
\end{minipage}
\begin{minipage}{0.22\textwidth}
\resizebox{\hsize}{!}{\includegraphics[angle=-0]{LITE/LMC-T2CEP-079_OLC.ps}} 
\end{minipage}
\begin{minipage}{0.26\textwidth}
\resizebox{\hsize}{!}{\includegraphics[angle=-0]{LITE/LMC-T2CEP-079_OmC.ps}} 
\end{minipage}

\caption{Light curves and O-C diagrams of candidate binary stars based on the LITE.
The left-hand panel shows the time series (top) and phased light curve (bottom).
The right-hand panel shows the O-C diagram (top) and the residual after subtracting the model (bottom).
The full model is represented by the full line. The contribution due to a changing period is indicated by the dotted line.
}

\label{Fig-LITE-BIN}
\end{figure*}

\setcounter{figure}{2}
\begin{figure*}
\centering

\begin{minipage}{0.22\textwidth}
\resizebox{\hsize}{!}{\includegraphics[angle=-0]{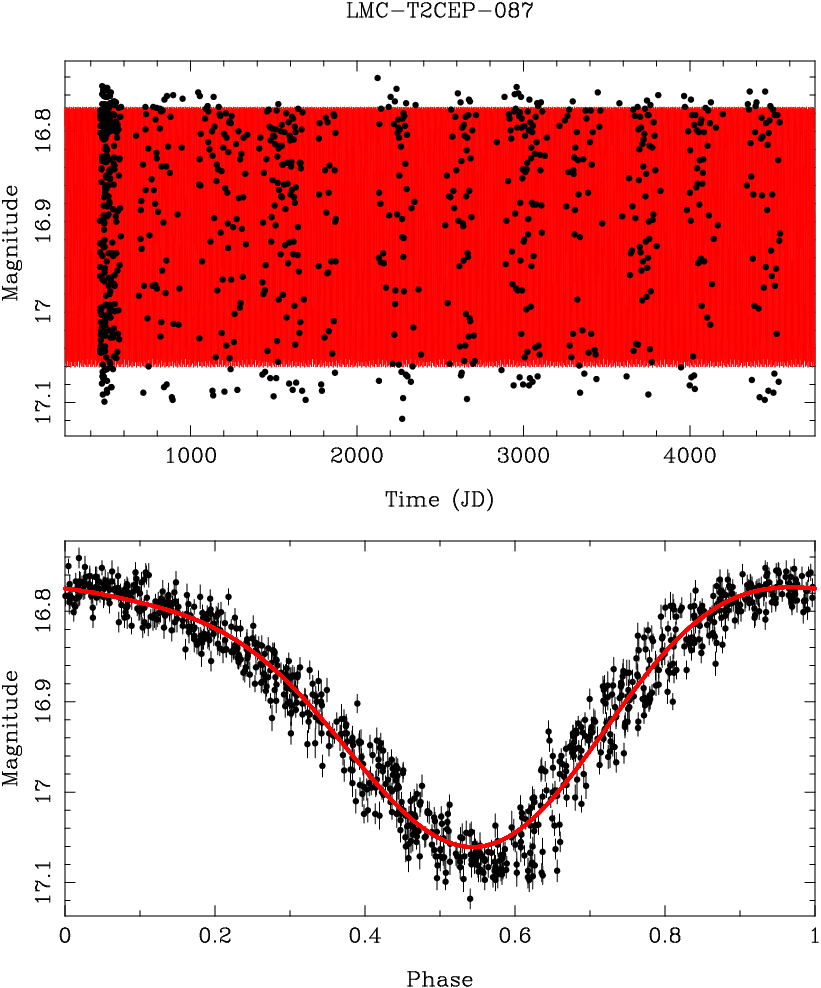}} 
\end{minipage}
\begin{minipage}{0.26\textwidth}
\resizebox{\hsize}{!}{\includegraphics[angle=-0]{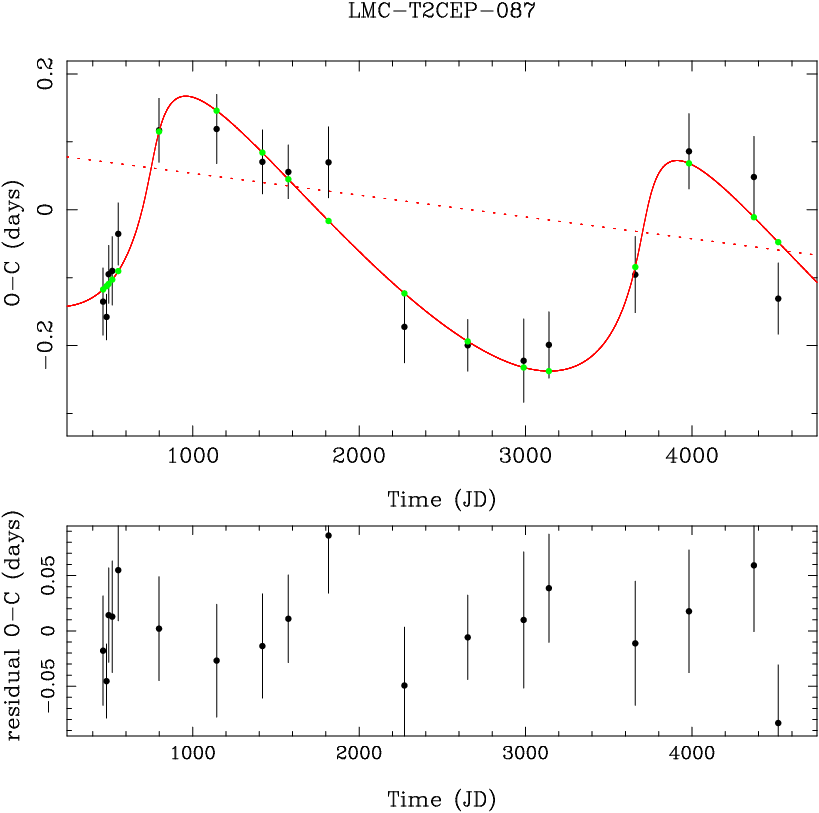}} 
\end{minipage}
\begin{minipage}{0.22\textwidth}
\resizebox{\hsize}{!}{\includegraphics[angle=-0]{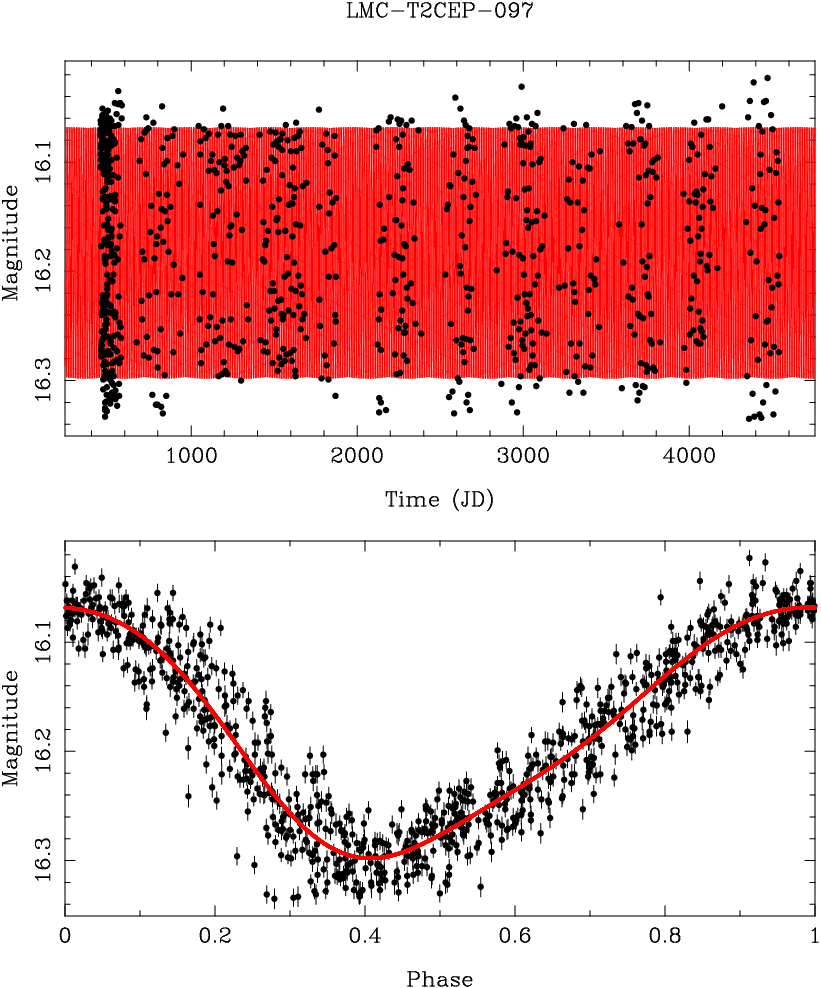}} 
\end{minipage}
\begin{minipage}{0.26\textwidth}
\resizebox{\hsize}{!}{\includegraphics[angle=-0]{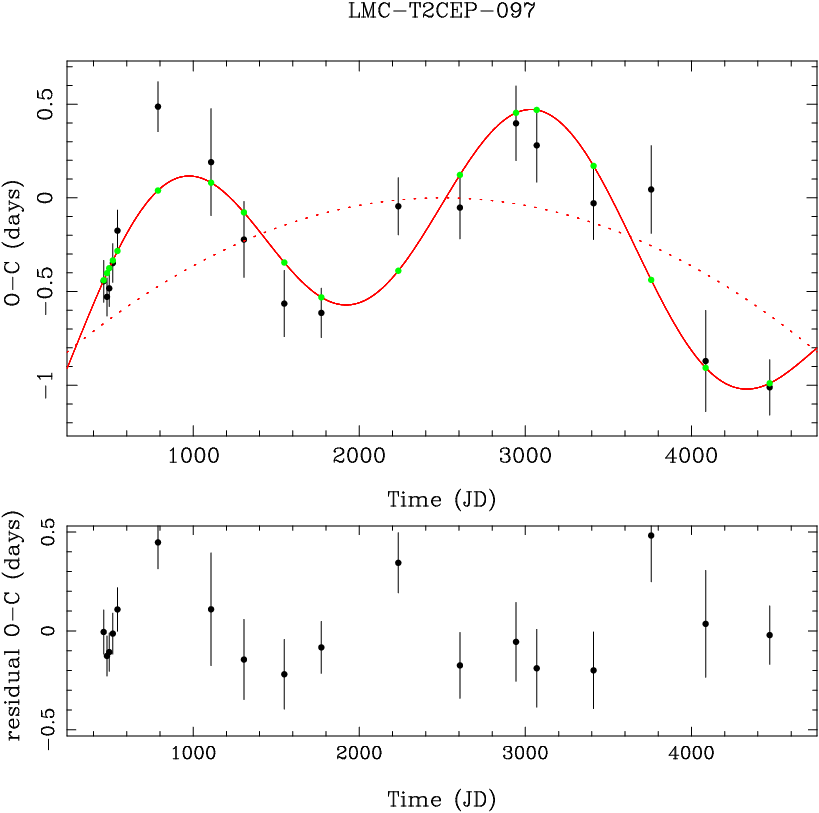}} 
\end{minipage}

\begin{minipage}{0.22\textwidth}
\resizebox{\hsize}{!}{\includegraphics[angle=-0]{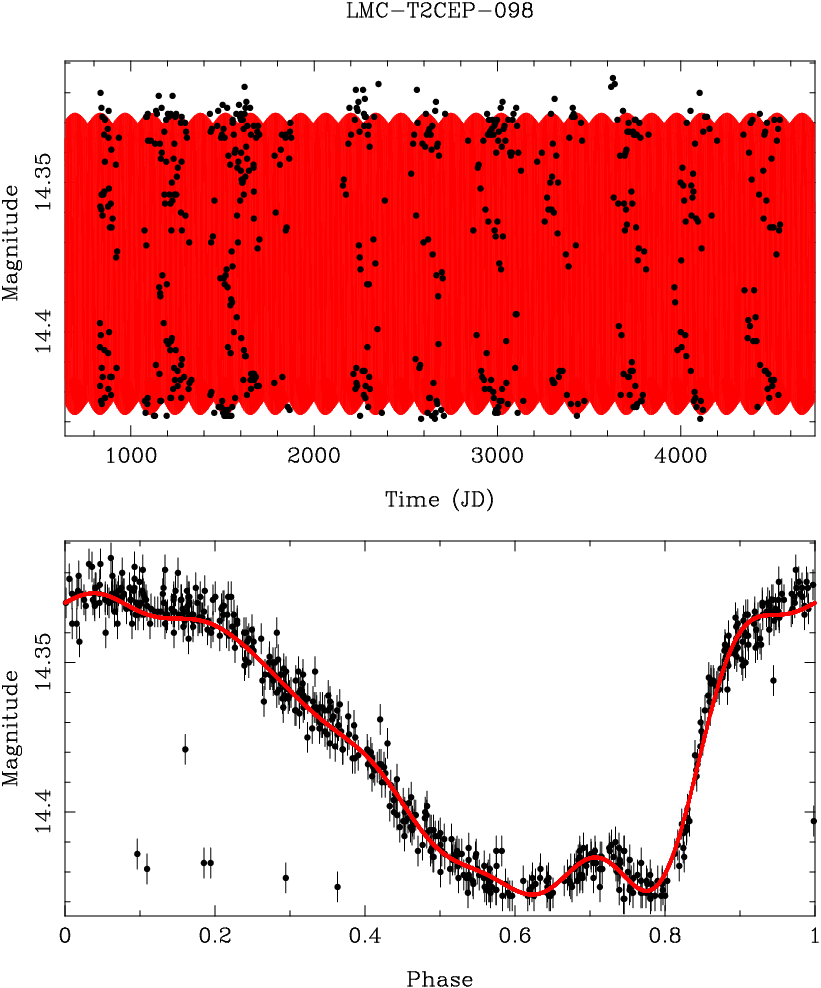}} 
\end{minipage}
\begin{minipage}{0.26\textwidth}
\resizebox{\hsize}{!}{\includegraphics[angle=-0]{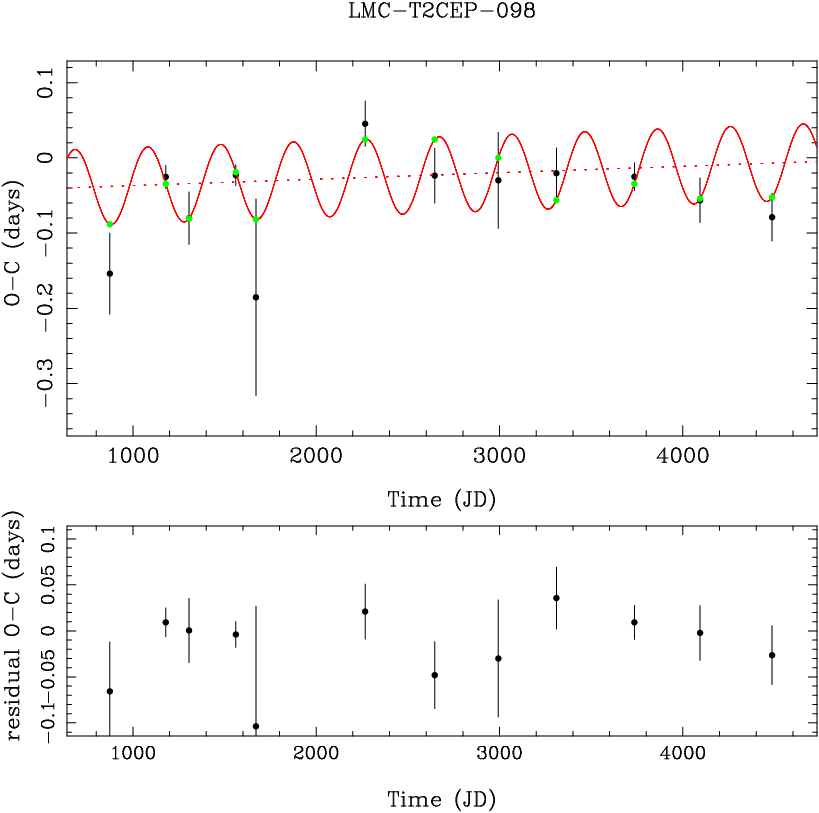}} 
\end{minipage}
\begin{minipage}{0.22\textwidth}
\resizebox{\hsize}{!}{\includegraphics[angle=-0]{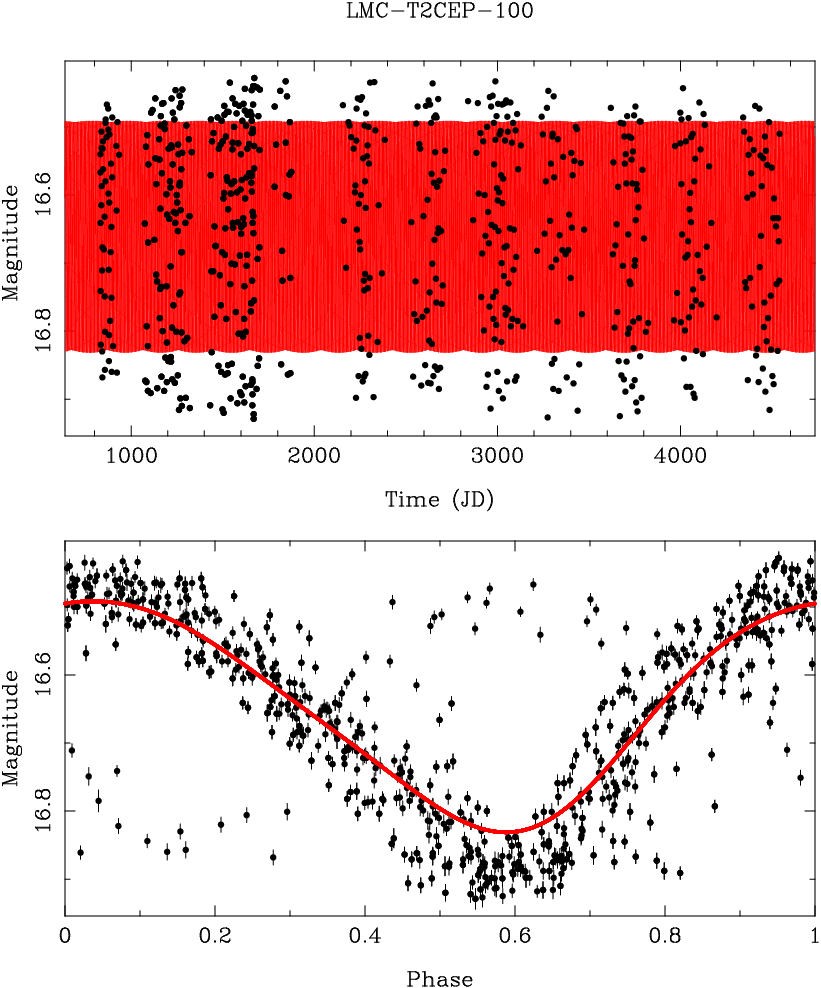}} 
\end{minipage}
\begin{minipage}{0.26\textwidth}
\resizebox{\hsize}{!}{\includegraphics[angle=-0]{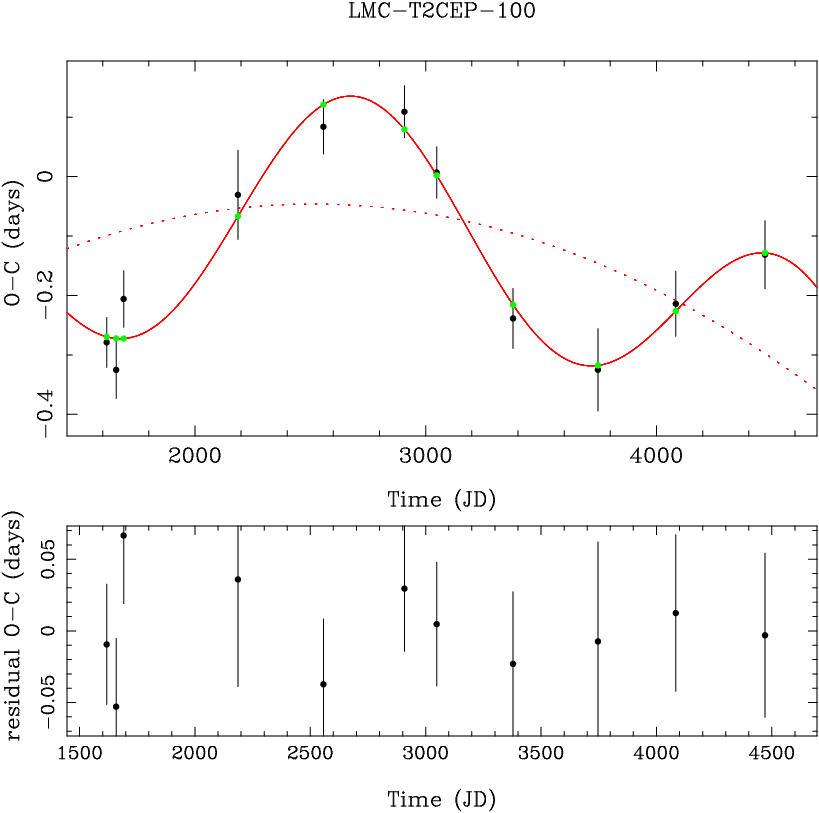}} 
\end{minipage}

\begin{minipage}{0.22\textwidth}
\resizebox{\hsize}{!}{\includegraphics[angle=-0]{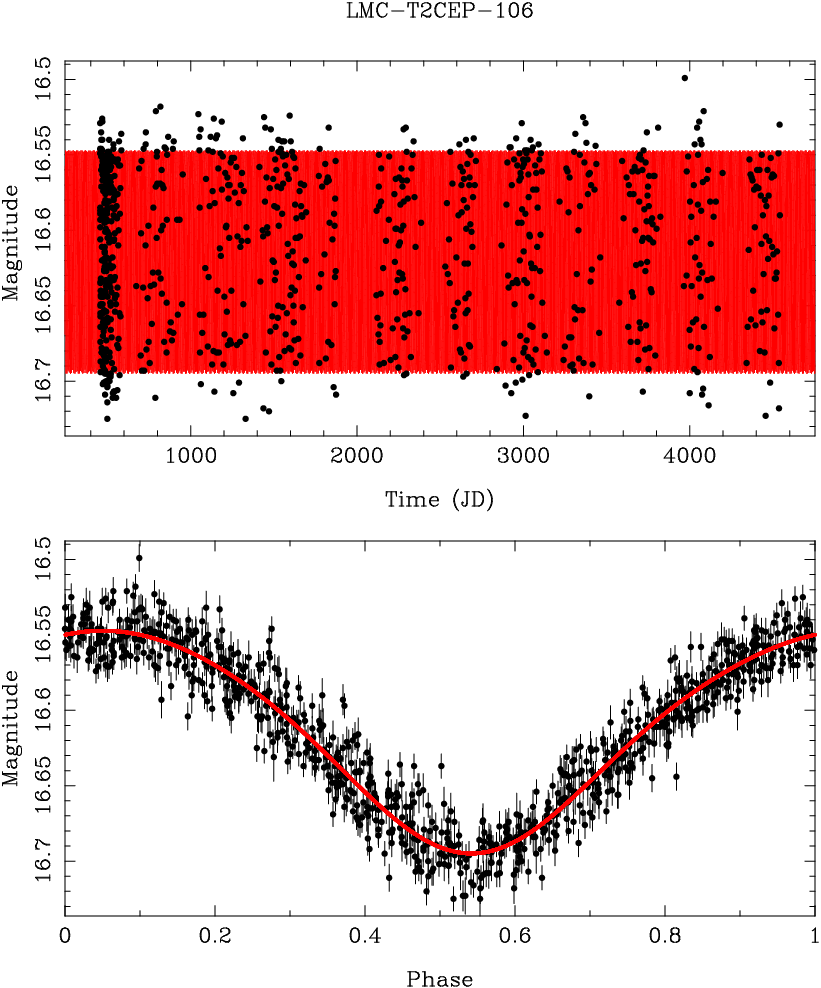}} 
\end{minipage}
\begin{minipage}{0.26\textwidth}
\resizebox{\hsize}{!}{\includegraphics[angle=-0]{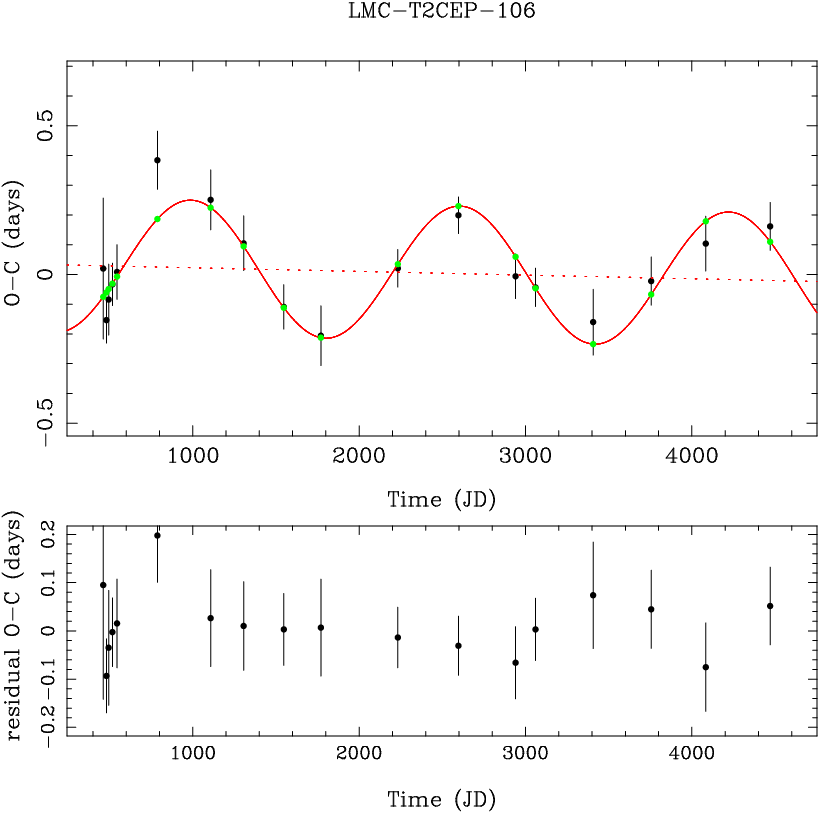}} 
\end{minipage}
\begin{minipage}{0.22\textwidth}
\resizebox{\hsize}{!}{\includegraphics[angle=-0]{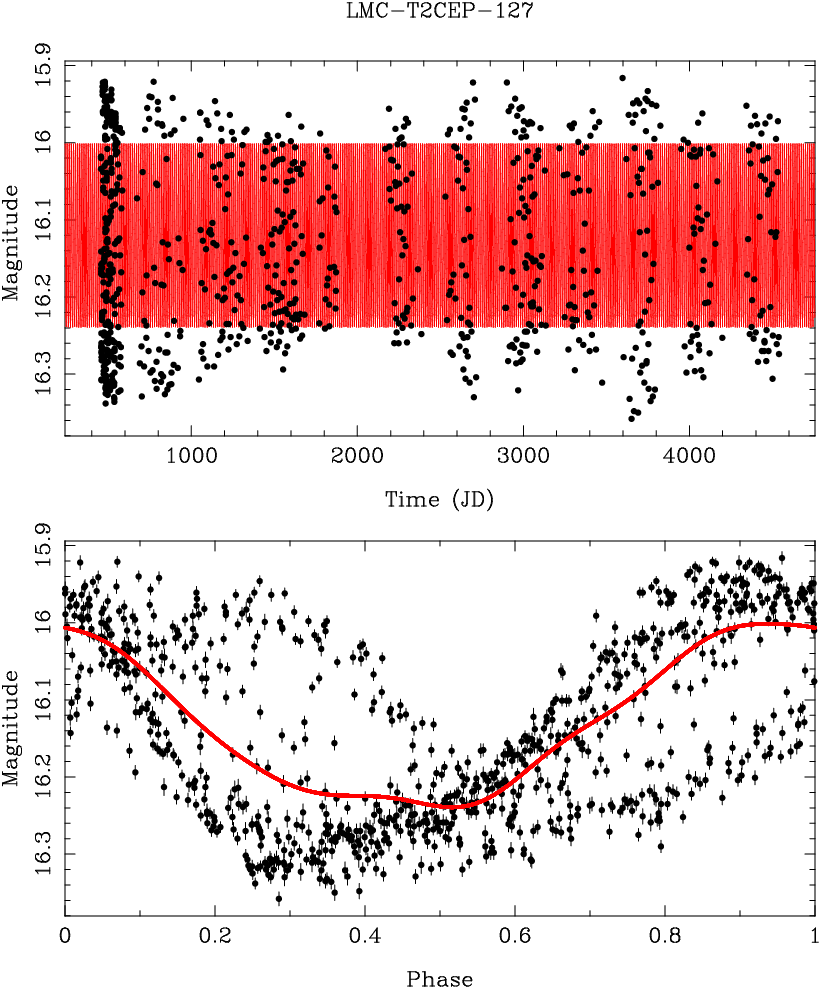}} 
\end{minipage}
\begin{minipage}{0.26\textwidth}
\resizebox{\hsize}{!}{\includegraphics[angle=-0]{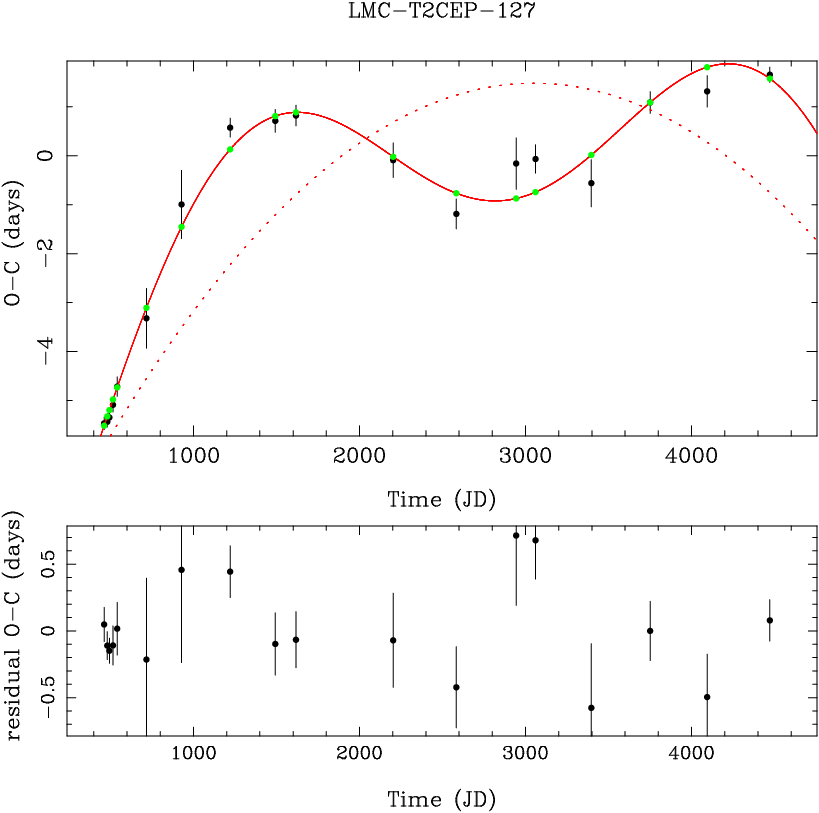}} 
\end{minipage}

\begin{minipage}{0.22\textwidth}
\resizebox{\hsize}{!}{\includegraphics[angle=-0]{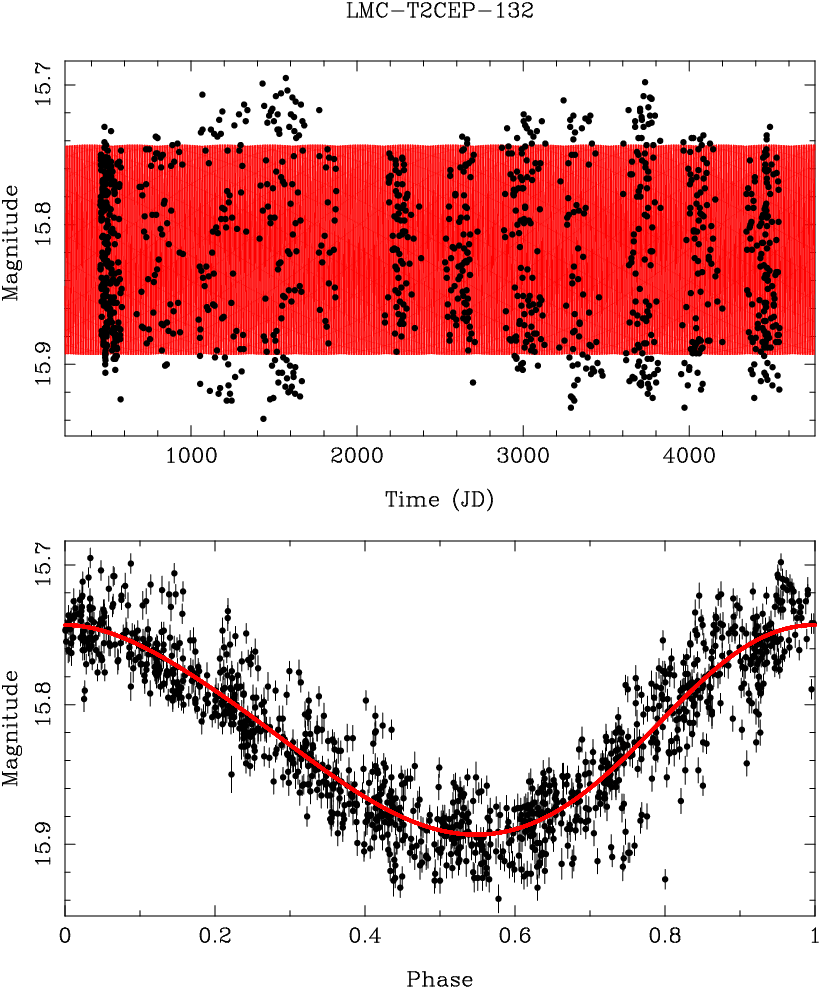}} 
\end{minipage}
\begin{minipage}{0.26\textwidth}
\resizebox{\hsize}{!}{\includegraphics[angle=-0]{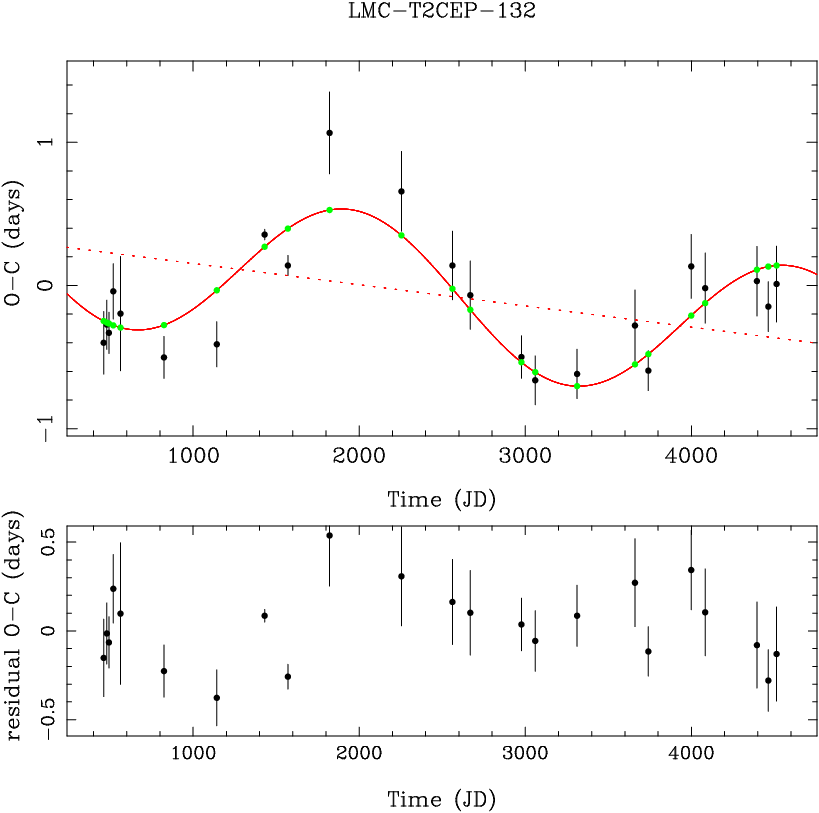}} 
\end{minipage}
\begin{minipage}{0.22\textwidth}
\resizebox{\hsize}{!}{\includegraphics[angle=-0]{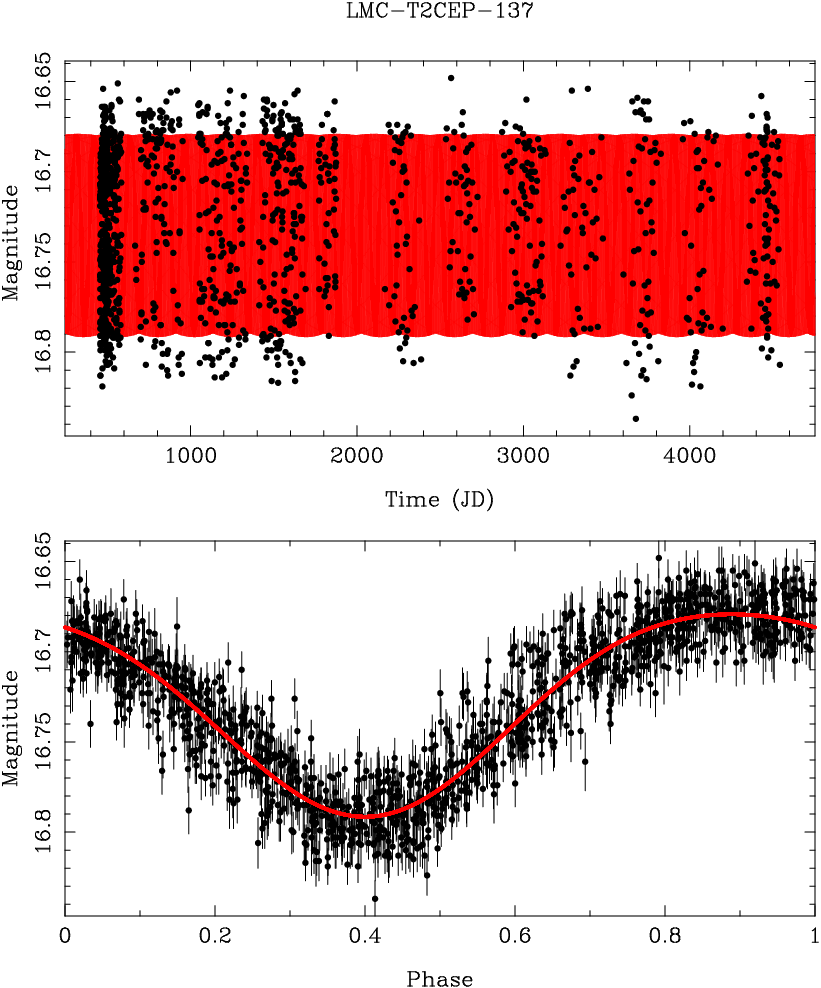}} 
\end{minipage}
\begin{minipage}{0.26\textwidth}
\resizebox{\hsize}{!}{\includegraphics[angle=-0]{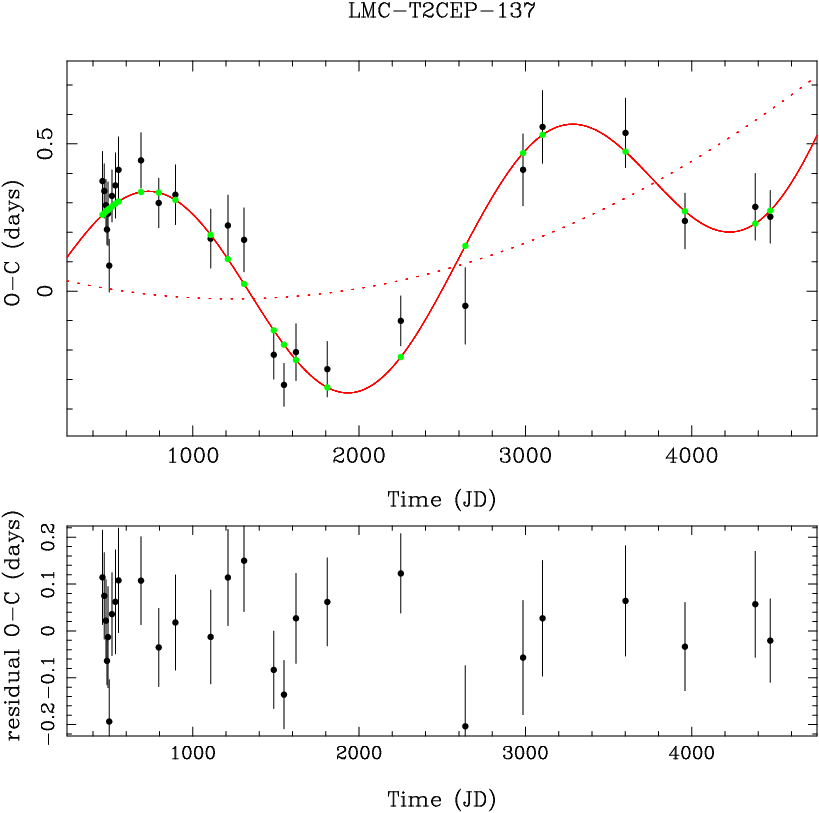}} 
\end{minipage}

\caption{Continued}
\end{figure*}

\setcounter{figure}{2}
\begin{figure*}
\centering

\begin{minipage}{0.22\textwidth}
\resizebox{\hsize}{!}{\includegraphics[angle=-0]{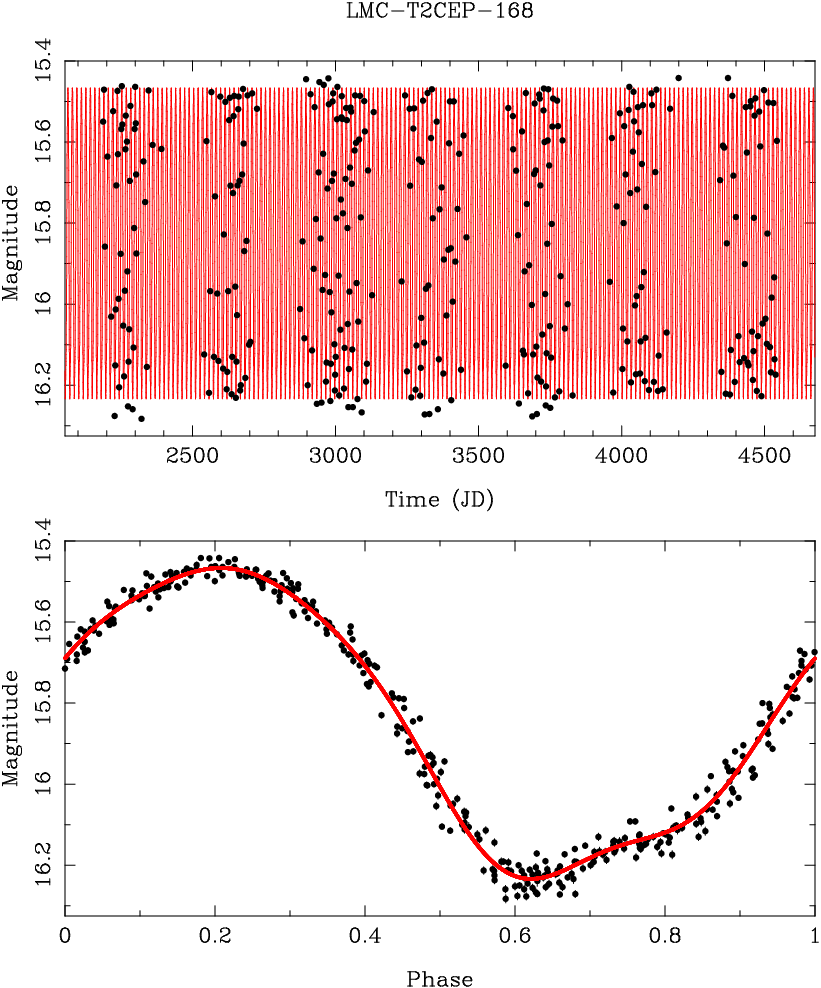}} 
\end{minipage}
\begin{minipage}{0.26\textwidth}
\resizebox{\hsize}{!}{\includegraphics[angle=-0]{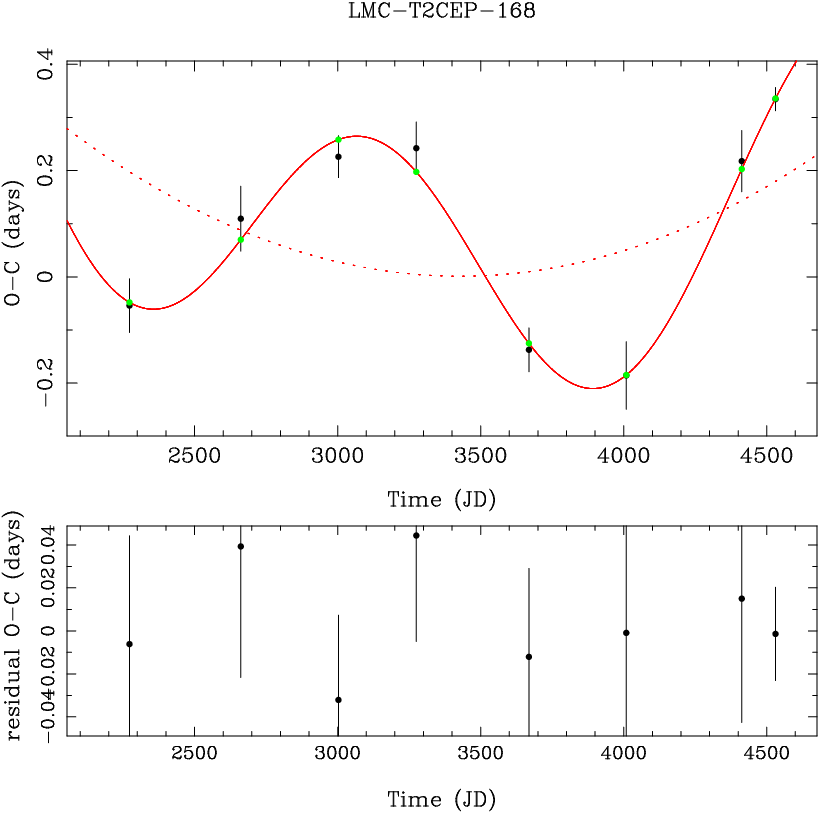}} 
\end{minipage}
\begin{minipage}{0.22\textwidth}
\resizebox{\hsize}{!}{\includegraphics[angle=-0]{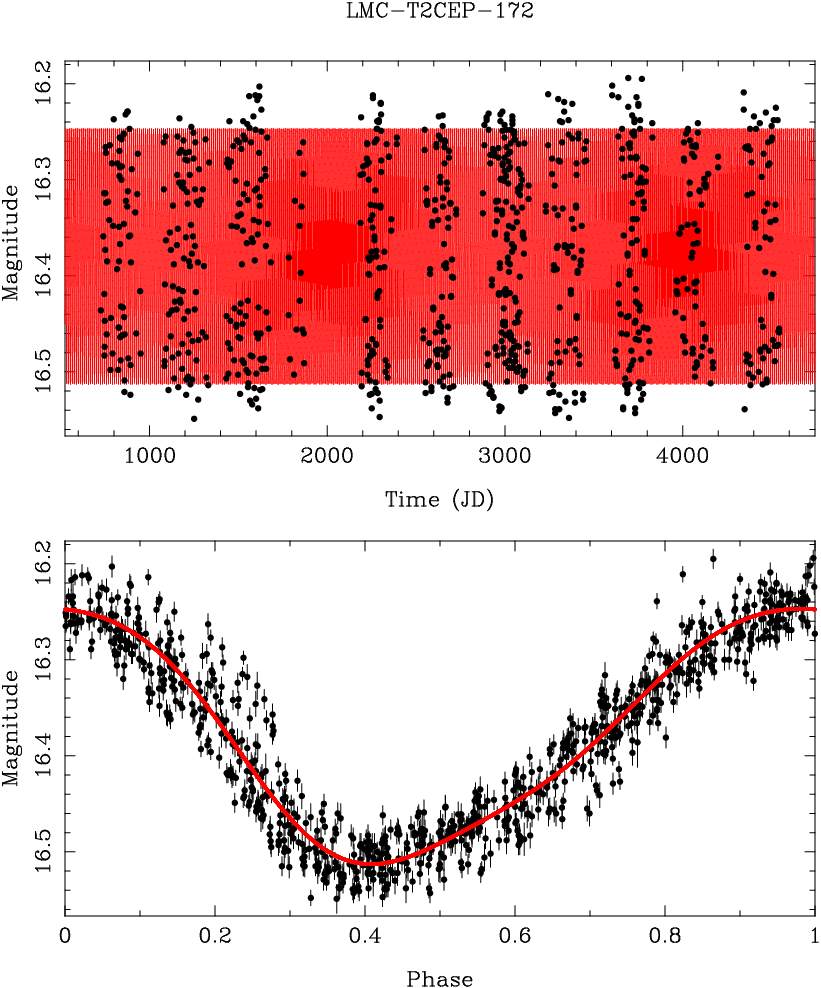}} 
\end{minipage}
\begin{minipage}{0.26\textwidth}
\resizebox{\hsize}{!}{\includegraphics[angle=-0]{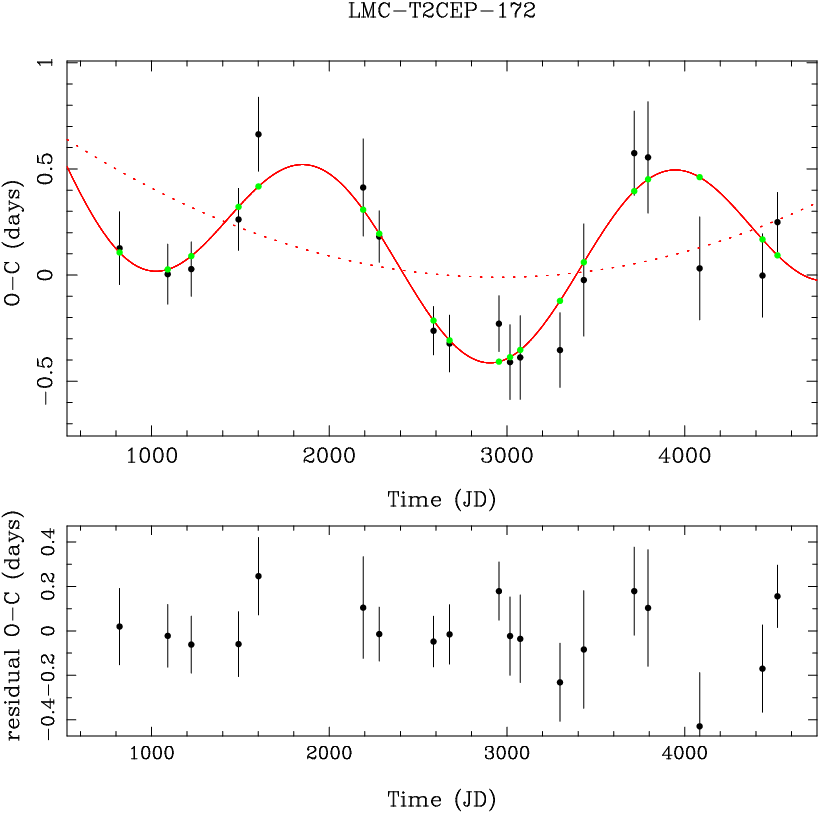}} 
\end{minipage}

\begin{minipage}{0.22\textwidth}
\resizebox{\hsize}{!}{\includegraphics[angle=-0]{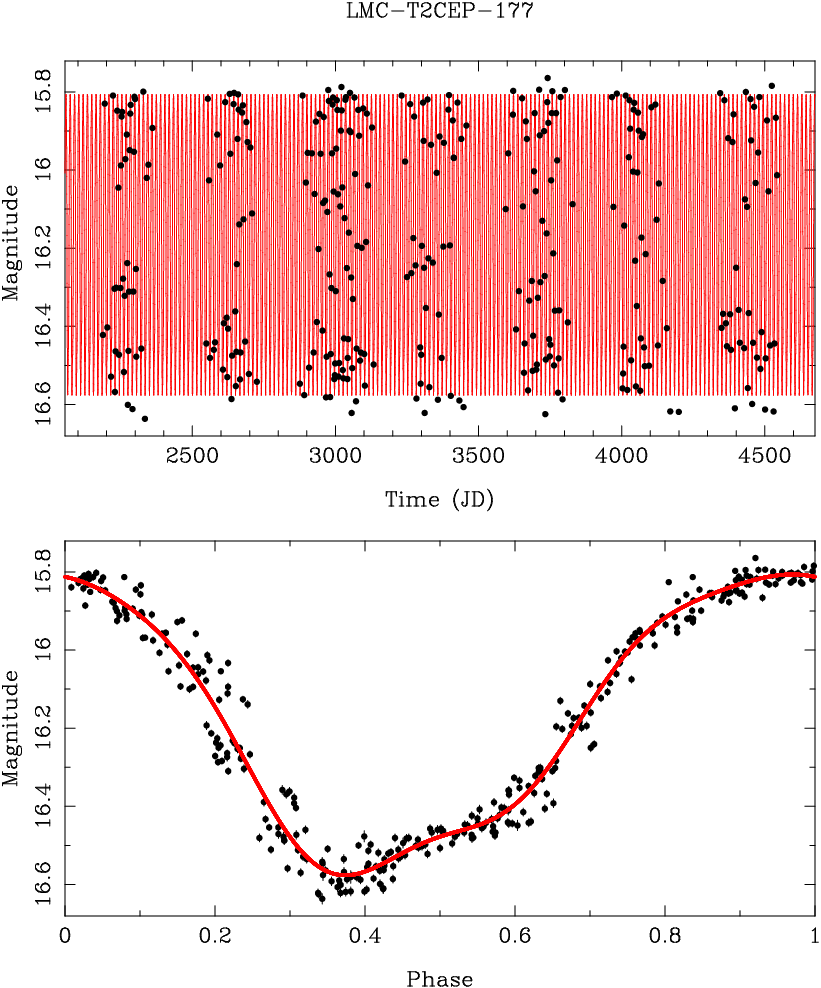}} 
\end{minipage}
\begin{minipage}{0.26\textwidth}
\resizebox{\hsize}{!}{\includegraphics[angle=-0]{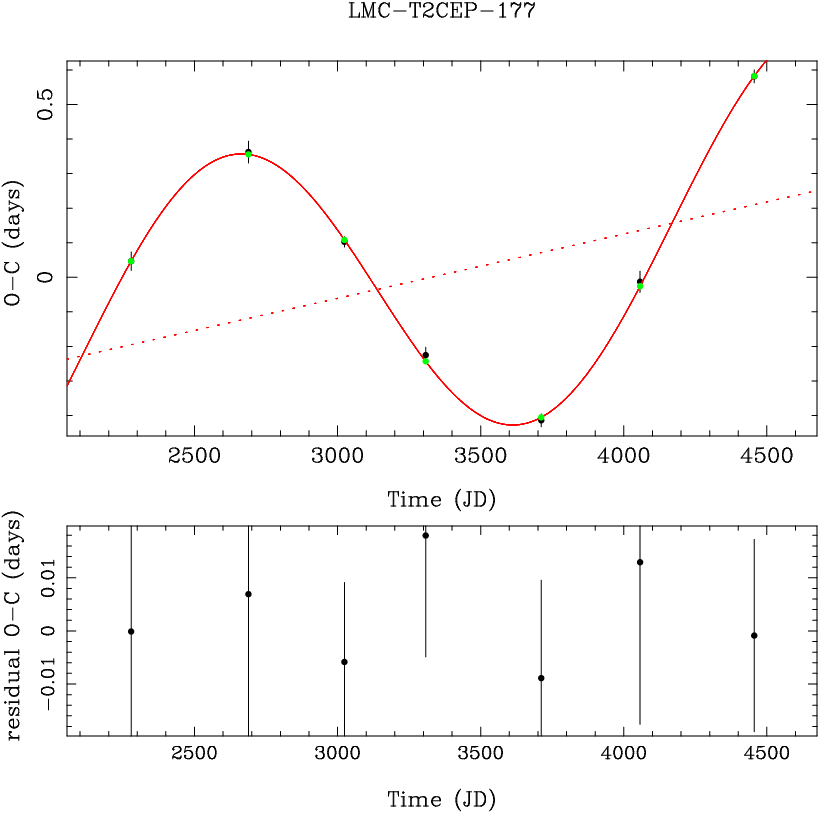}} 
\end{minipage}
\begin{minipage}{0.22\textwidth}
\resizebox{\hsize}{!}{\includegraphics[angle=-0]{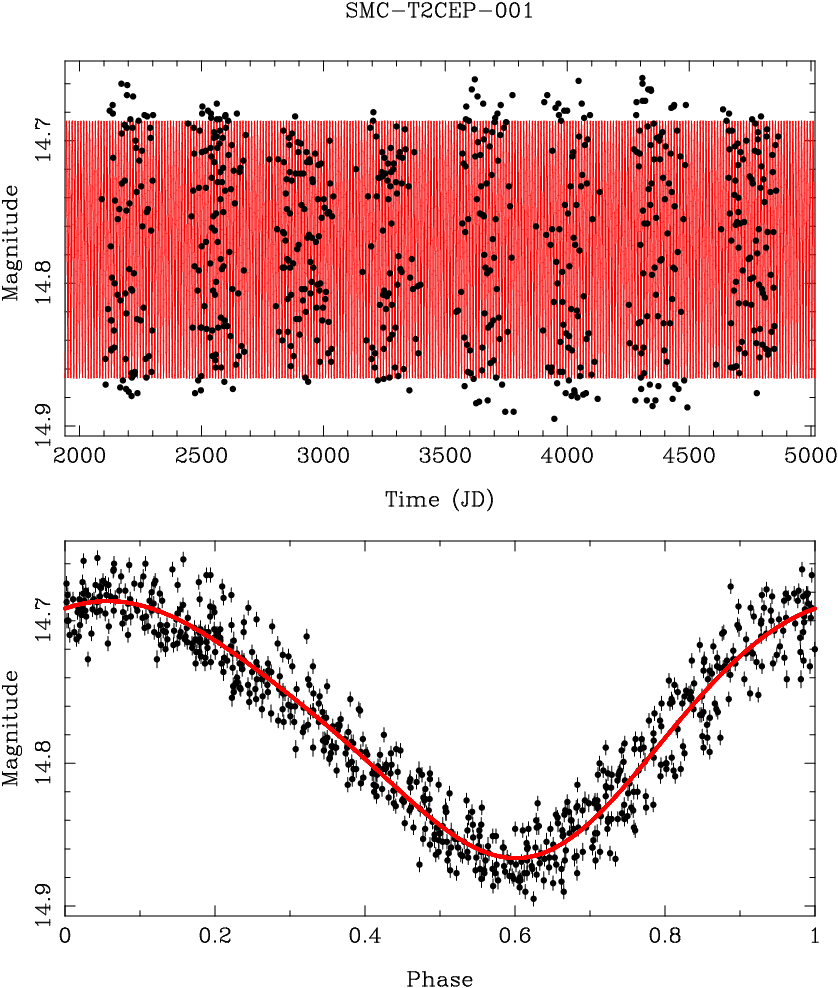}} 
\end{minipage}
\begin{minipage}{0.26\textwidth}
\resizebox{\hsize}{!}{\includegraphics[angle=-0]{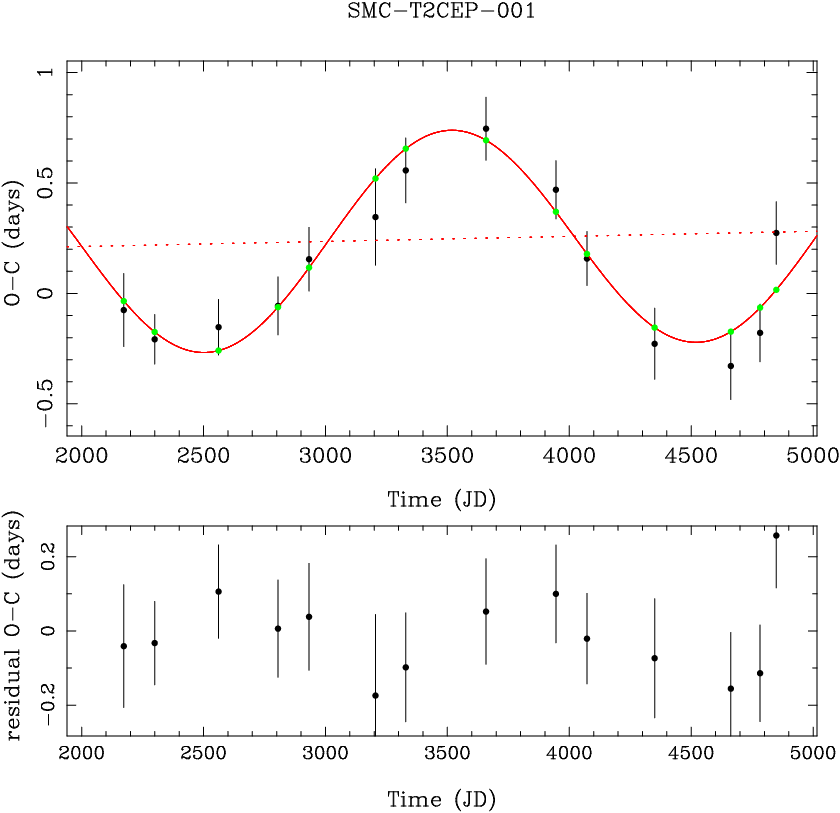}} 
\end{minipage}

\begin{minipage}{0.22\textwidth}
\resizebox{\hsize}{!}{\includegraphics[angle=-0]{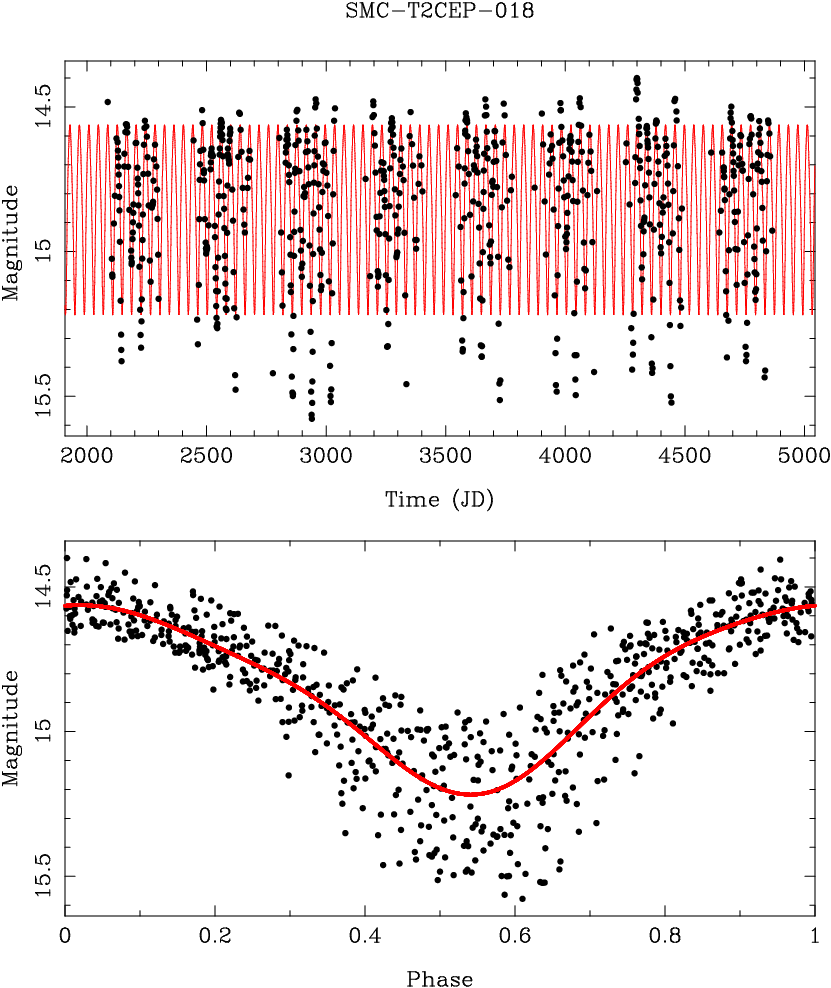}} 
\end{minipage}
\begin{minipage}{0.26\textwidth}
\resizebox{\hsize}{!}{\includegraphics[angle=-0]{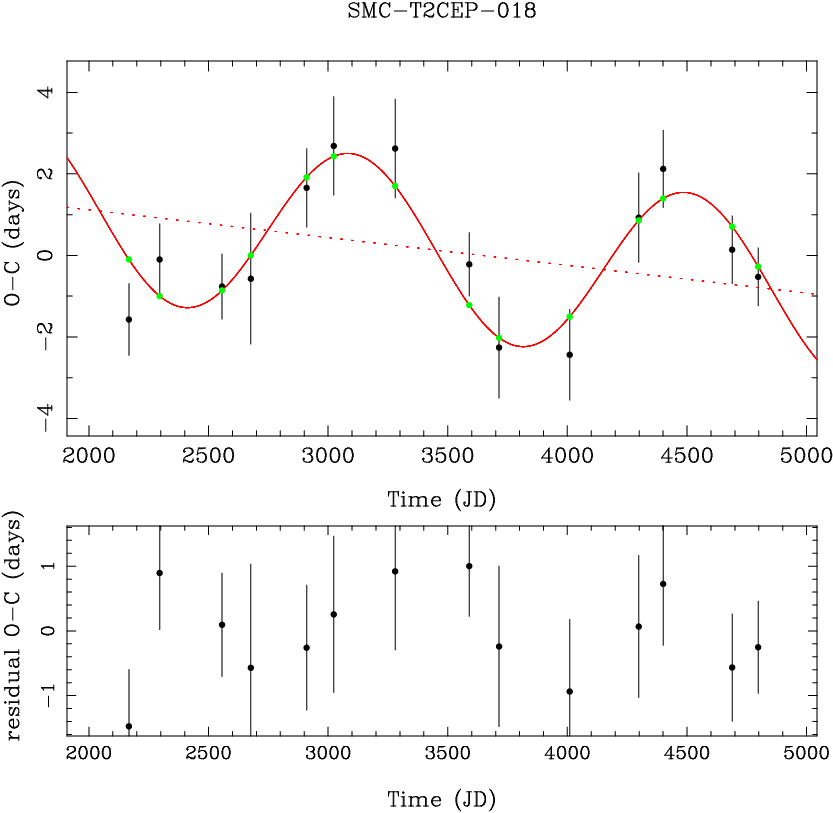}} 
\end{minipage}
\begin{minipage}{0.22\textwidth}
\resizebox{\hsize}{!}{\includegraphics[angle=-0]{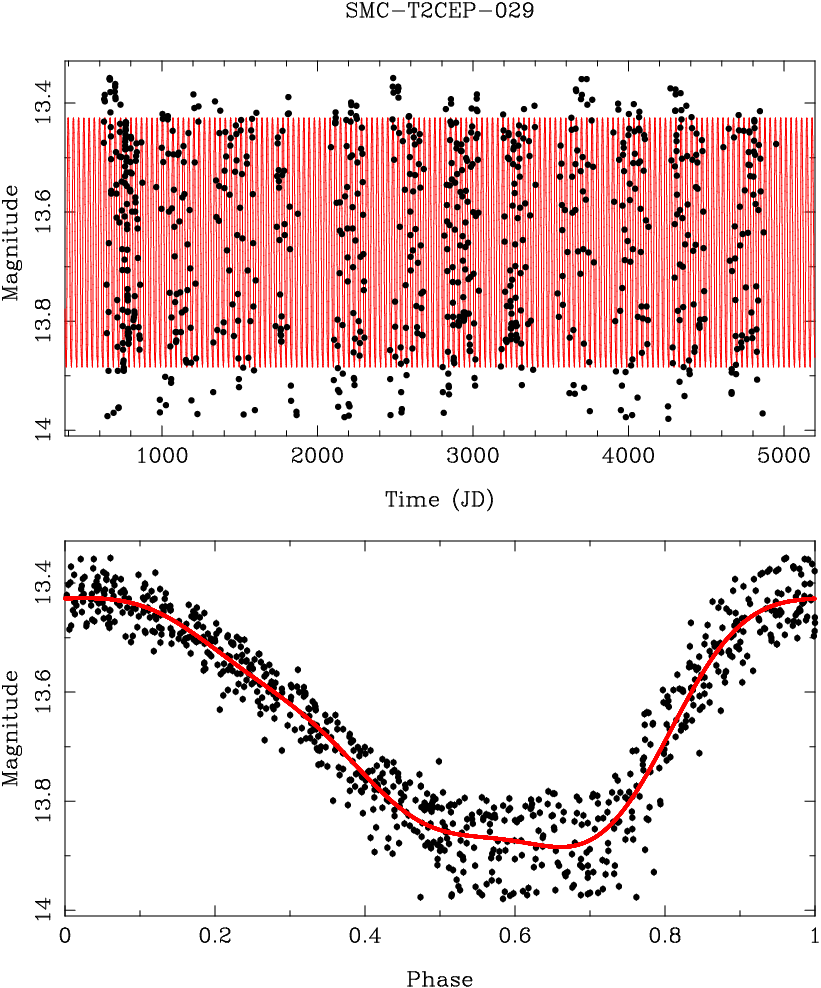}} 
\end{minipage}
\begin{minipage}{0.26\textwidth}
\resizebox{\hsize}{!}{\includegraphics[angle=-0]{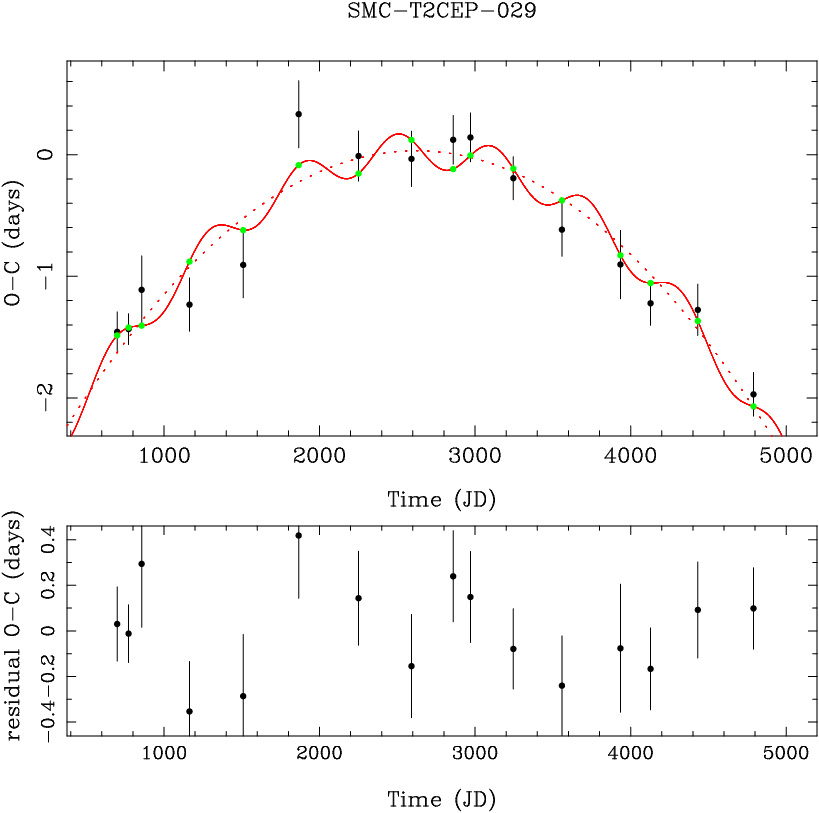}} 
\end{minipage}

\begin{minipage}{0.22\textwidth}
\resizebox{\hsize}{!}{\includegraphics[angle=-0]{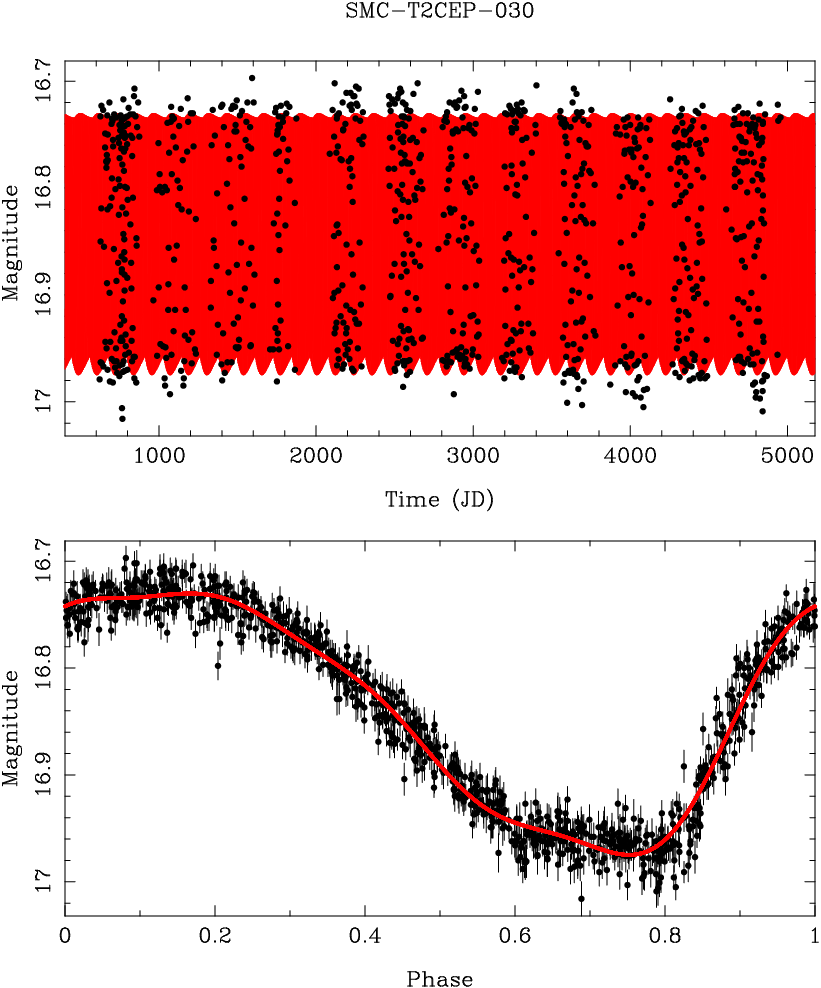}} 
\end{minipage}
\begin{minipage}{0.26\textwidth}
\resizebox{\hsize}{!}{\includegraphics[angle=-0]{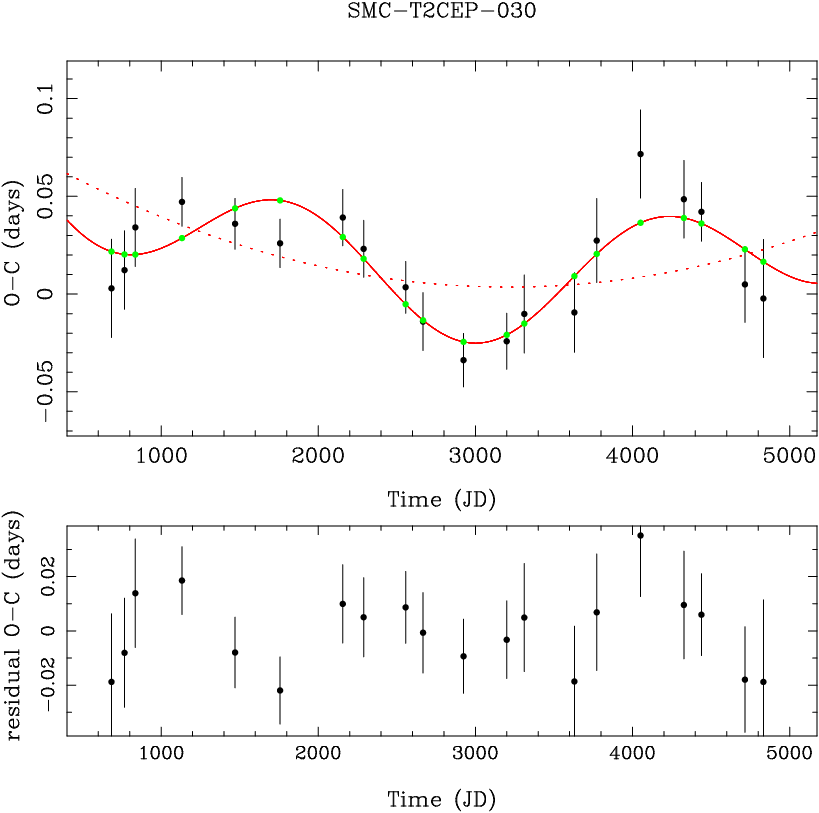}} 
\end{minipage}

\caption{Continued}
\end{figure*}

\begin{table*}
\caption{T2C with significant period change}

\begin{tabular}{lcrrlrr}
\hline
Name               & $\dot{P}$ &  Type & Period & Remarks \\ 
                   &   (s/yr)  &       &   (d)  &         \\
\hline

LMC-T2CEP-019 & $    84   \pm    6$   &  pWVir &  8.7 &     \\
LMC-T2CEP-026 &                       &   WVir & 13.6 & change in amplitude and period jump at JD $\sim$ 3800 \\
LMC-T2CEP-029 & $-10867   \pm  594$   &  RVTau & 21.2 & \\
LMC-T2CEP-034 & $   631   \pm   33$   &   WVir & 14.9 & \\
LMC-T2CEP-037 & $   236   \pm   23$   &   WVir &  6.9 & \\
LMC-T2CEP-049 & $    14.7 \pm    1.5$ &  BLHer &  3.2 & \\
LMC-T2CEP-072 & $  1532   \pm  160$   &   WVir & 14.5 & \\
LMC-T2CEP-074 & $  -875   \pm   91$   &   WVir &  9.0 & Binary (P= 1500d) ? \\
LMC-T2CEP-082 & $ -7715   \pm 1101$   &  RVTau & 35.1 & JD $<$ 4200 \\
LMC-T2CEP-099 & $   505   \pm   95$   &   WVir & 15.5 & Binary (P= 2300d) ? \\
LMC-T2CEP-103 & $  -303   \pm   32$   &   WVir & 12.9 &      \\ 
LMC-T2CEP-104 & $  -311   \pm  101$   &  RVTau & 24.9 &           \\
LMC-T2CEP-113 & $   -57.2 \pm    8.5$ &  BLHer &  3.1 &  \\
LMC-T2CEP-115 & $  1341   \pm  298  $ &  RVTau & 25.0 & \\
LMC-T2CEP-119 & $  9078   \pm 1751  $ &  RVTau & 33.8 & \\
LMC-T2CEP-126 & $   703   \pm   80  $ &   WVir & 16.3 & \\
LMC-T2CEP-135 & $  2163   \pm  450  $ &  RVTau & 26.5 & \\
LMC-T2CEP-139 & $   643   \pm   86  $ &   WVir & 14.8 & \\
LMC-T2CEP-143 & $  -900   \pm   70  $ &   WVir & 14.6 & \\
LMC-T2CEP-146 & $   704   \pm   54  $ &   WVir & 10.1 & \\
LMC-T2CEP-149 &                       &  RVTau & 42.5 &  decreasing amplitude and period jump at JD $\sim$ 3800 \\
LMC-T2CEP-152 & $  -111   \pm   15  $ &   WVir &  9.3 & \\
LMC-T2CEP-155 & $   158   \pm   41  $ &   WVir &  6.9 & \\
LMC-T2CEP-156 & $   708   \pm  140  $ &   WVir & 15.4 & alternatively, period jump near JD $\sim$ 3300  \\
LMC-T2CEP-162 & $ -1947   \pm  270  $ &  RVTau & 30.4 &  \\
LMC-T2CEP-170 & $  -133   \pm   32  $ &   WVir &  7.7 & \\
LMC-T2CEP-182 & $   244   \pm   38  $ &   WVir &  8.3 & \\
LMC-T2CEP-186 & $   714   \pm   95  $ &   WVir & 16.4 &  \\
LMC-T2CEP-190 & $ 29900   \pm 2400  $ &  RVTau & 38.6 & \\
LMC-T2CEP-191 & $-24500   \pm 2000  $ &  RVTau & 34.3 & \\
LMC-T2CEP-201 & $   153   \pm   25  $ &  pWVir & 11.0 & \\

SMC-ACEP-003 & $    -2.71 \pm   0.20 $ & ANCEP & 0.6 & \\

SMC-T2CEP-004 & $    89.8 \pm    7.9 $ &  WVir &  6.5 & \\
SMC-T2CEP-014 & $  -442   \pm   51   $ &  WVir & 13.9 & Fit for JD$>$ 1200. Different period before that date ? \\
SMC-T2CEP-015 & $    44.0 \pm    5.7 $ & BLHer &  2.6 & \\
SMC-T2CEP-019 &                        & RVTau & 40.9 & Amplitude variations. Period jump near JD $\sim$ 4000 ? \\
SMC-T2CEP-025 & $   196   \pm  26    $ & pWVir & 14.2 &  \\
SMC-T2CEP-032 & $  -943   \pm  69    $ &  WVir & 14.2 & Binary (P= 1500d) ? \\
SMC-T2CEP-034 &                        &  WVir & 20.1 & Period jump near JD $\sim$ 3050 ? \\
SMC-T2CEP-038 & $    11.3 \pm   1.9  $ & pWVir &  4.4 & \\
SMC-T2CEP-040 & $  -631   \pm  37    $ &  WVir & 16.1 &   \\
SMC-T2CEP-043 & $ -1121   \pm 249    $ & RVTau & 23.7 & \\

\hline
\end{tabular}

\label{Tab-LITE-Pdot}
\end{table*}

\begin{figure*}
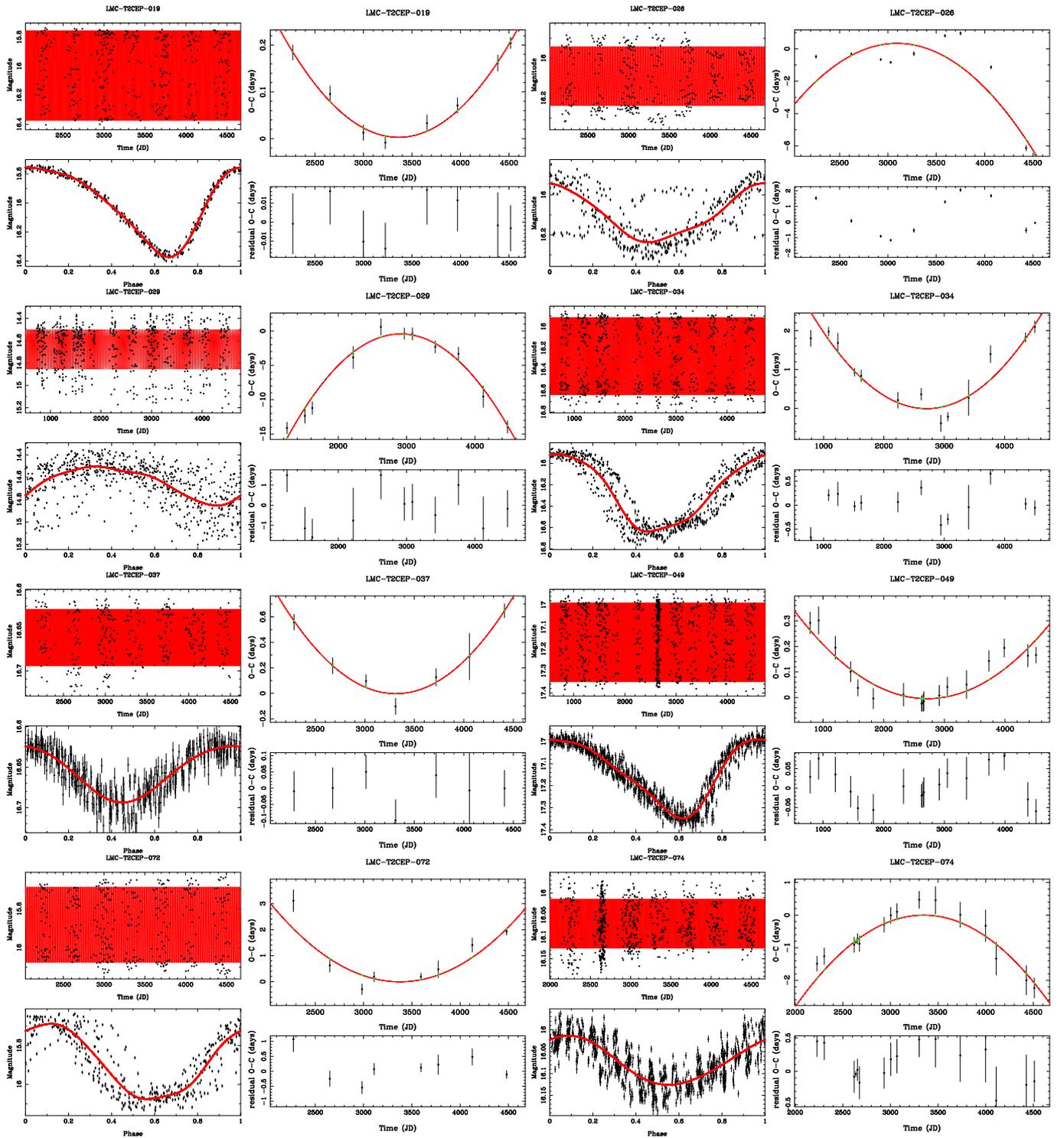


\centering

\begin{minipage}{0.22\textwidth}
\resizebox{\hsize}{!}{\includegraphics[angle=-0]{LITE/LMC-T2CEP-019_OLC.ps}} 
\end{minipage}
\begin{minipage}{0.26\textwidth}
\resizebox{\hsize}{!}{\includegraphics[angle=-0]{LITE/LMC-T2CEP-019_OmC.ps}} 
\end{minipage}
\begin{minipage}{0.22\textwidth}
\resizebox{\hsize}{!}{\includegraphics[angle=-0]{LITE/LMC-T2CEP-026_OLC.ps}} 
\end{minipage}
\begin{minipage}{0.26\textwidth}
\resizebox{\hsize}{!}{\includegraphics[angle=-0]{LITE/LMC-T2CEP-026_OmC.ps}} 
\end{minipage}

\begin{minipage}{0.22\textwidth}
\resizebox{\hsize}{!}{\includegraphics[angle=-0]{LITE/LMC-T2CEP-029_OLC.ps}} 
\end{minipage}
\begin{minipage}{0.26\textwidth}
\resizebox{\hsize}{!}{\includegraphics[angle=-0]{LITE/LMC-T2CEP-029_OmC.ps}} 
\end{minipage}
\begin{minipage}{0.22\textwidth}
\resizebox{\hsize}{!}{\includegraphics[angle=-0]{LITE/LMC-T2CEP-034_OLC.ps}} 
\end{minipage}
\begin{minipage}{0.26\textwidth}
\resizebox{\hsize}{!}{\includegraphics[angle=-0]{LITE/LMC-T2CEP-034_OmC.ps}} 
\end{minipage}

\begin{minipage}{0.22\textwidth}
\resizebox{\hsize}{!}{\includegraphics[angle=-0]{LITE/LMC-T2CEP-037_OLC.ps}} 
\end{minipage}
\begin{minipage}{0.26\textwidth}
\resizebox{\hsize}{!}{\includegraphics[angle=-0]{LITE/LMC-T2CEP-037_OmC.ps}} 
\end{minipage}
\begin{minipage}{0.22\textwidth}
\resizebox{\hsize}{!}{\includegraphics[angle=-0]{LITE/LMC-T2CEP-049_OLC.ps}} 
\end{minipage}
\begin{minipage}{0.26\textwidth}
\resizebox{\hsize}{!}{\includegraphics[angle=-0]{LITE/LMC-T2CEP-049_OmC.ps}} 
\end{minipage}

\begin{minipage}{0.22\textwidth}
\resizebox{\hsize}{!}{\includegraphics[angle=-0]{LITE/LMC-T2CEP-072_OLC.ps}} 
\end{minipage}
\begin{minipage}{0.26\textwidth}
\resizebox{\hsize}{!}{\includegraphics[angle=-0]{LITE/LMC-T2CEP-072_OmC.ps}} 
\end{minipage}
\begin{minipage}{0.22\textwidth}
\resizebox{\hsize}{!}{\includegraphics[angle=-0]{LITE/LMC-T2CEP-074_OLC.ps}} 
\end{minipage}
\begin{minipage}{0.26\textwidth}
\resizebox{\hsize}{!}{\includegraphics[angle=-0]{LITE/LMC-T2CEP-074_OmC.ps}} 
\end{minipage}

\caption{Light curves and O-C diagrams of the stars showing a significant change in period.
The left-hand panel shows the time series (top) and phased light curve (bottom).
The right-hand panel shows the O-C diagram (top) and the model and the residuals after subtracting the model (bottom).
}

\label{Fig-LITE-Pdot}
\end{figure*}

\setcounter{figure}{3}
\begin{figure*}
\centering

\begin{minipage}{0.22\textwidth}
\resizebox{\hsize}{!}{\includegraphics[angle=-0]{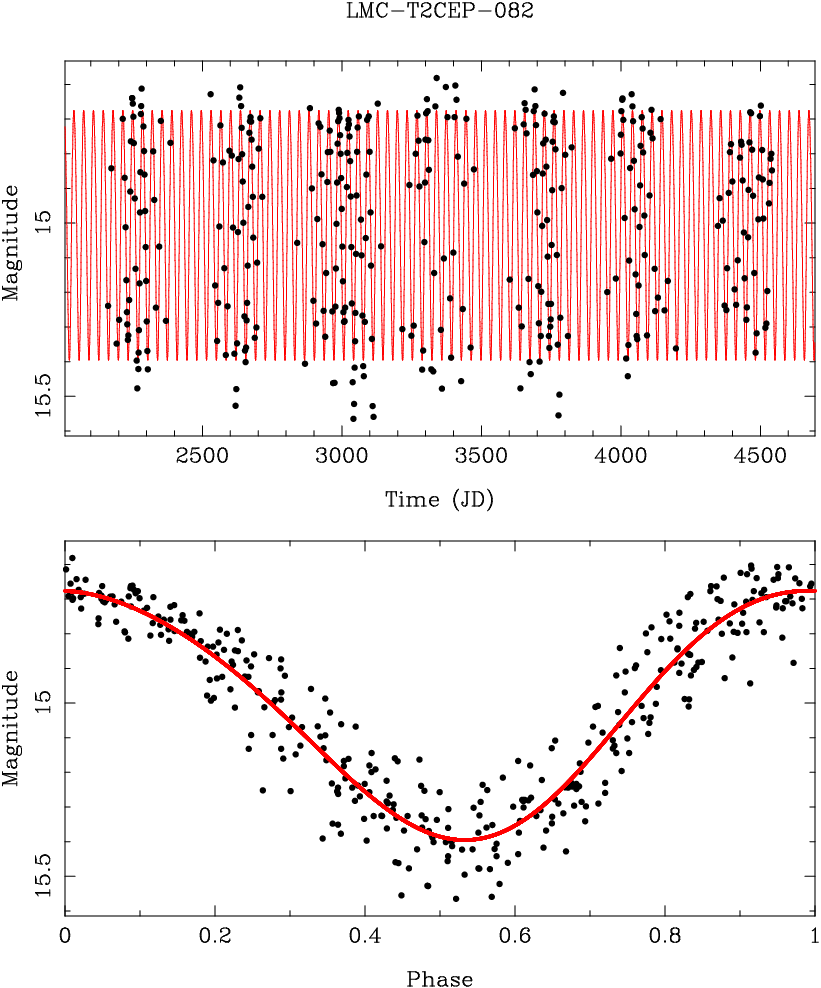}} 
\end{minipage}
\begin{minipage}{0.26\textwidth}
\resizebox{\hsize}{!}{\includegraphics[angle=-0]{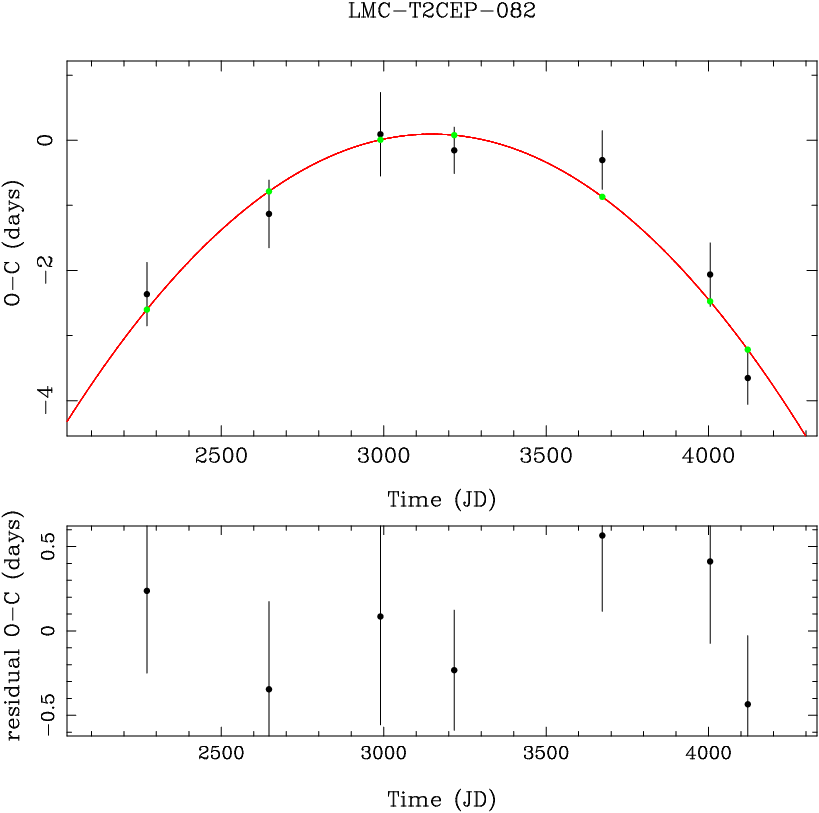}} 
\end{minipage}
\begin{minipage}{0.22\textwidth}
\resizebox{\hsize}{!}{\includegraphics[angle=-0]{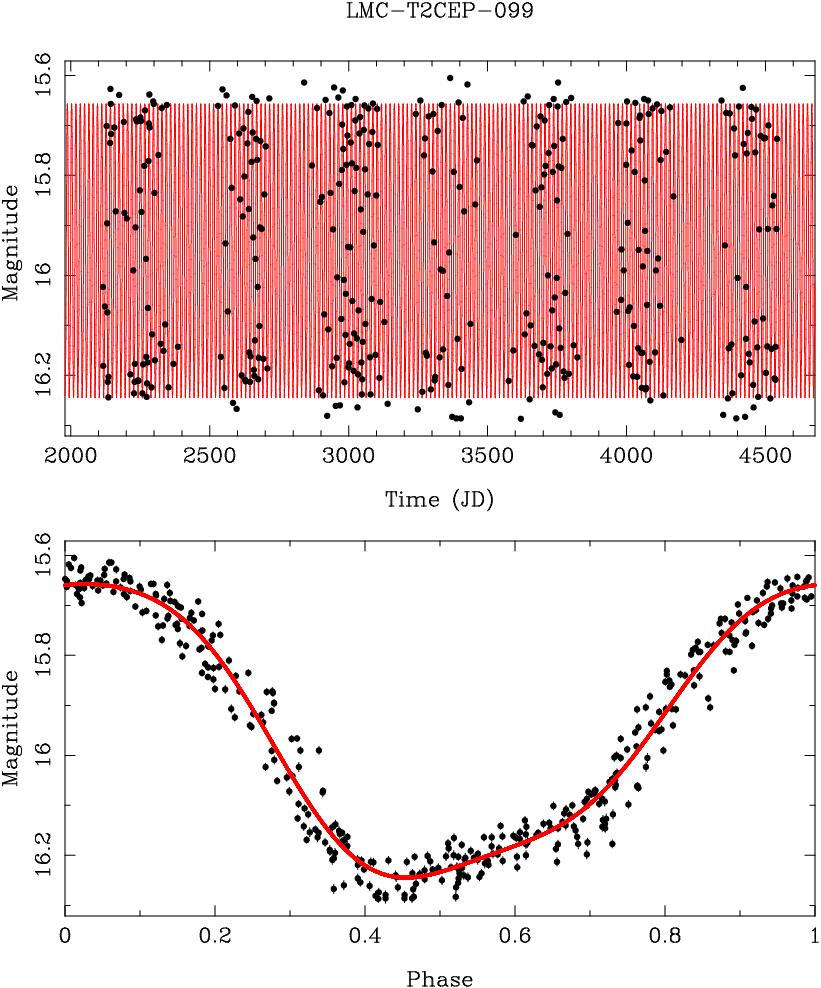}} 
\end{minipage}
\begin{minipage}{0.26\textwidth}
\resizebox{\hsize}{!}{\includegraphics[angle=-0]{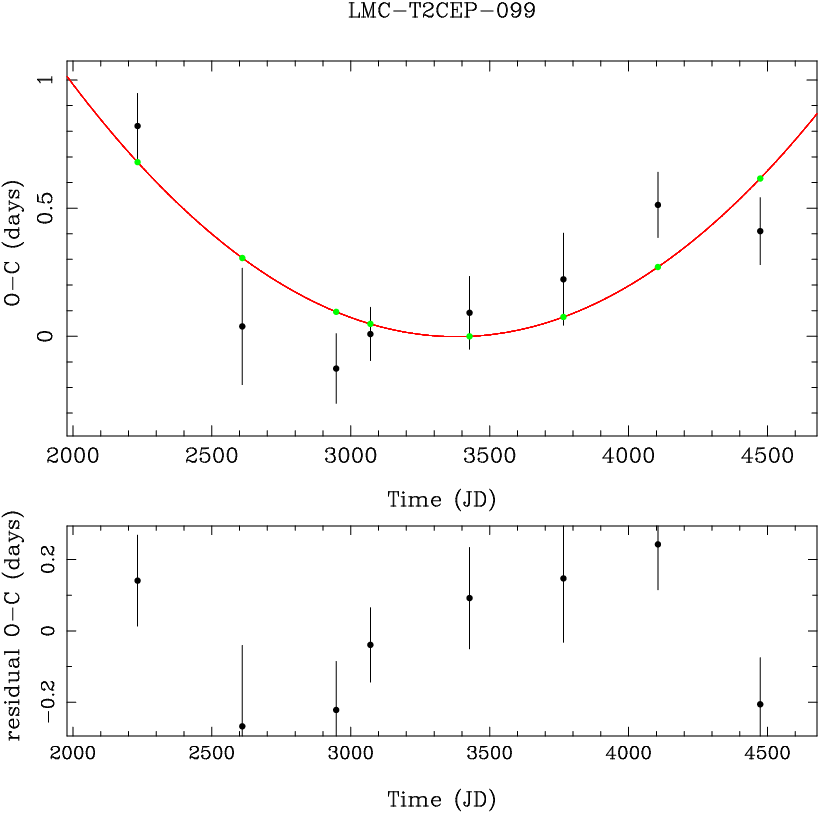}} 
\end{minipage}

\begin{minipage}{0.22\textwidth}
\resizebox{\hsize}{!}{\includegraphics[angle=-0]{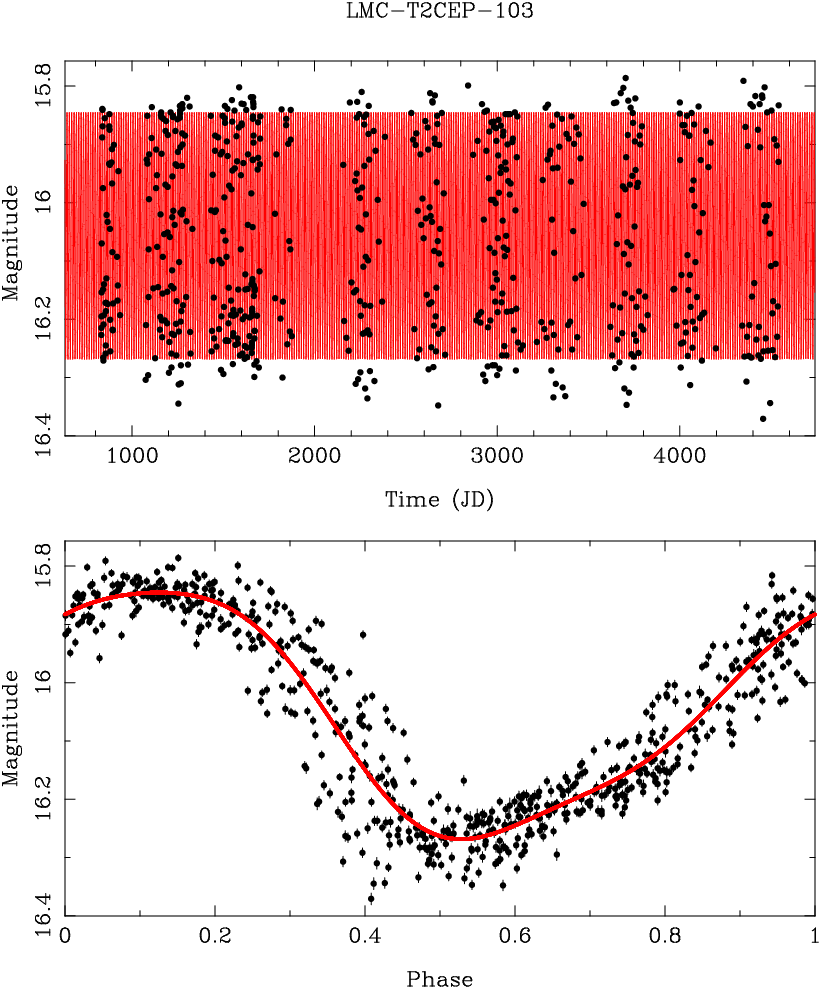}} 
\end{minipage}
\begin{minipage}{0.26\textwidth}
\resizebox{\hsize}{!}{\includegraphics[angle=-0]{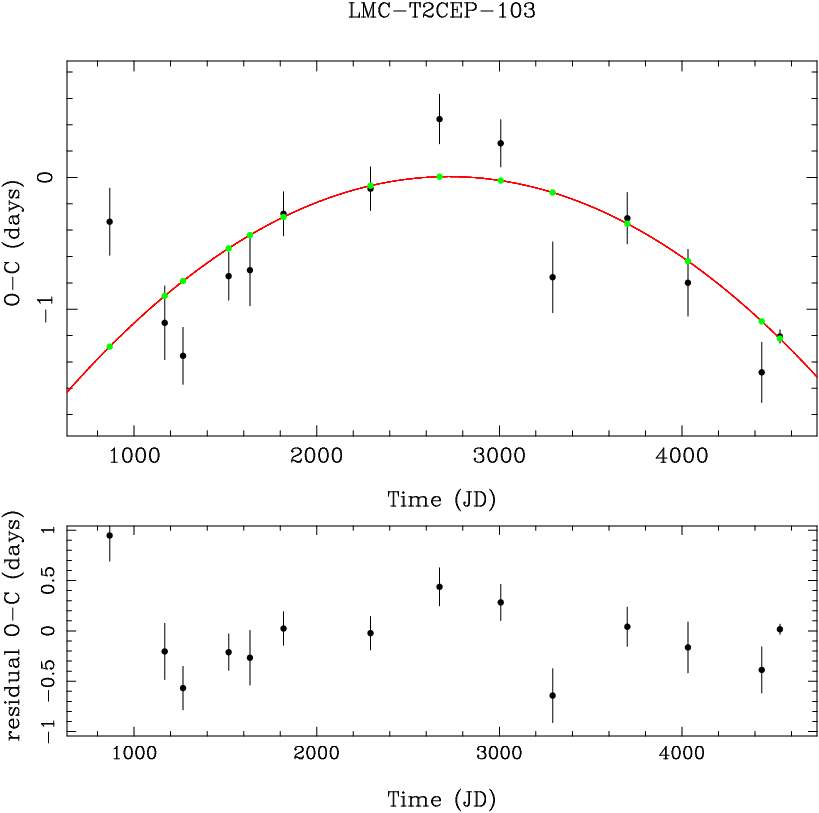}} 
\end{minipage}
\begin{minipage}{0.22\textwidth}
\resizebox{\hsize}{!}{\includegraphics[angle=-0]{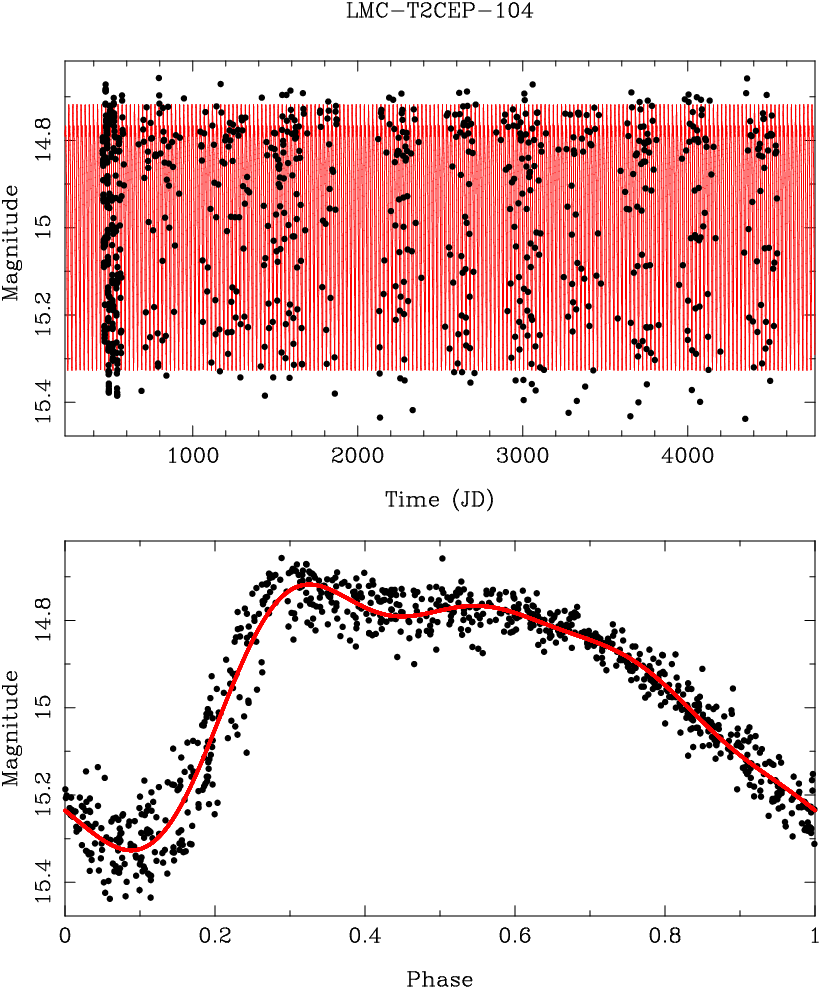}} 
\end{minipage}
\begin{minipage}{0.26\textwidth}
\resizebox{\hsize}{!}{\includegraphics[angle=-0]{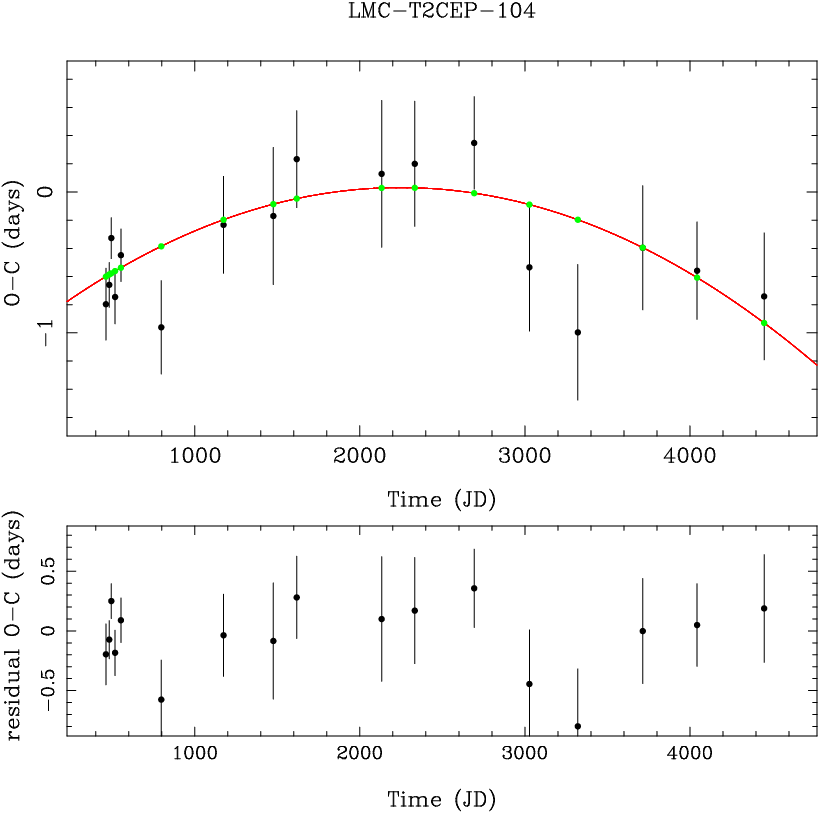}} 
\end{minipage}

\begin{minipage}{0.22\textwidth}
\resizebox{\hsize}{!}{\includegraphics[angle=-0]{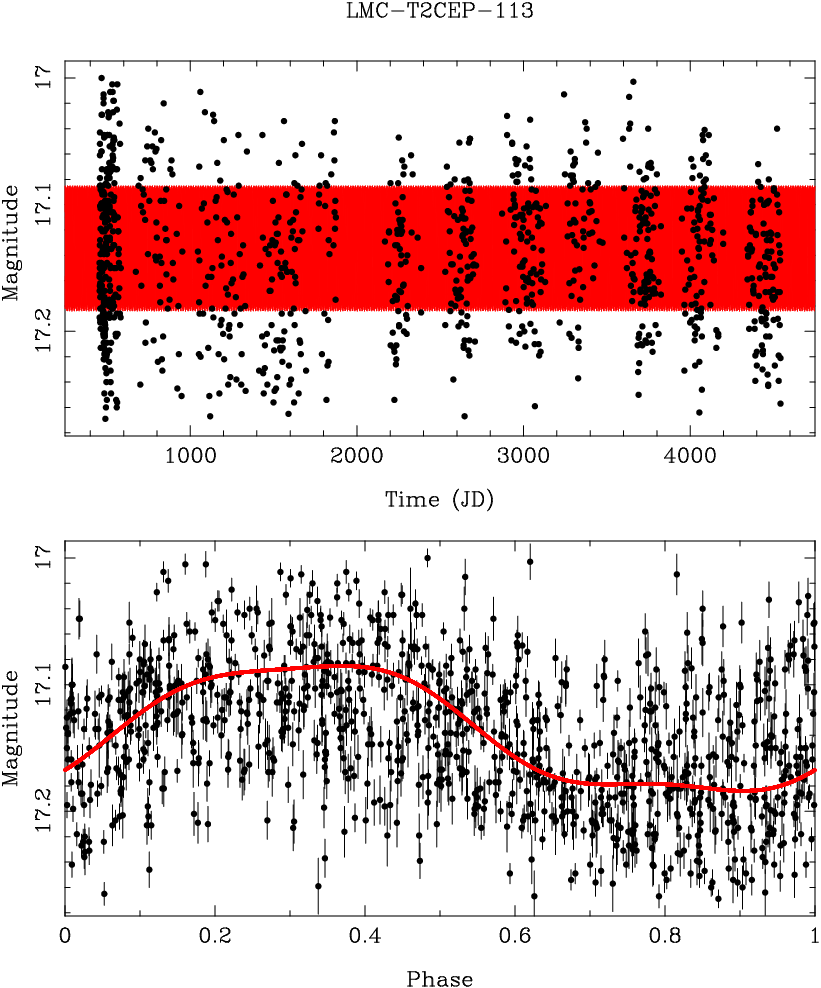}} 
\end{minipage}
\begin{minipage}{0.26\textwidth}
\resizebox{\hsize}{!}{\includegraphics[angle=-0]{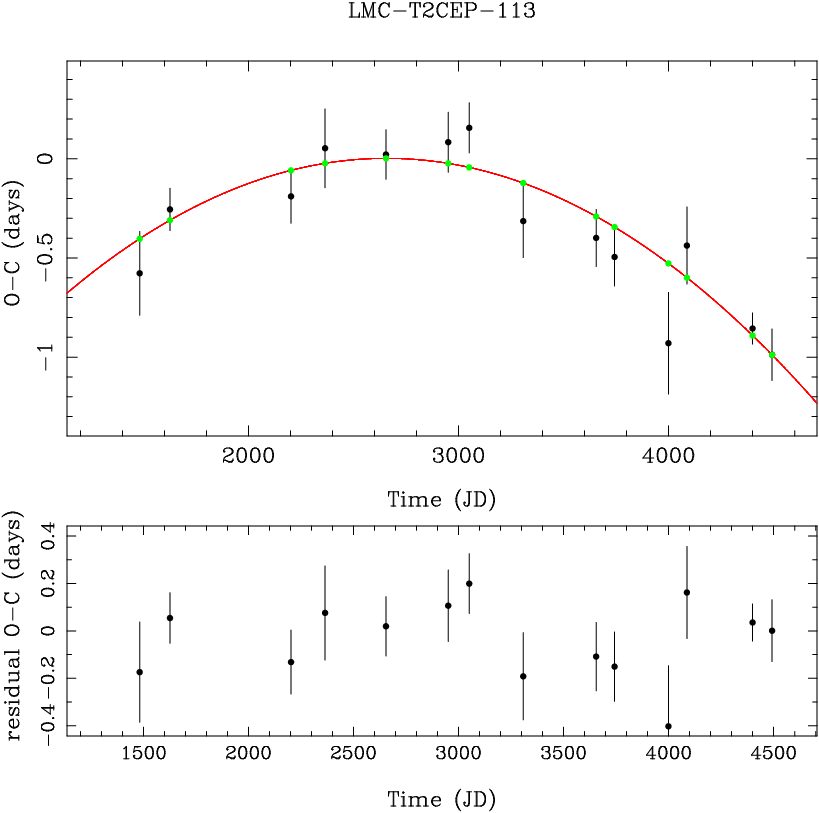}} 
\end{minipage}
\begin{minipage}{0.22\textwidth}
\resizebox{\hsize}{!}{\includegraphics[angle=-0]{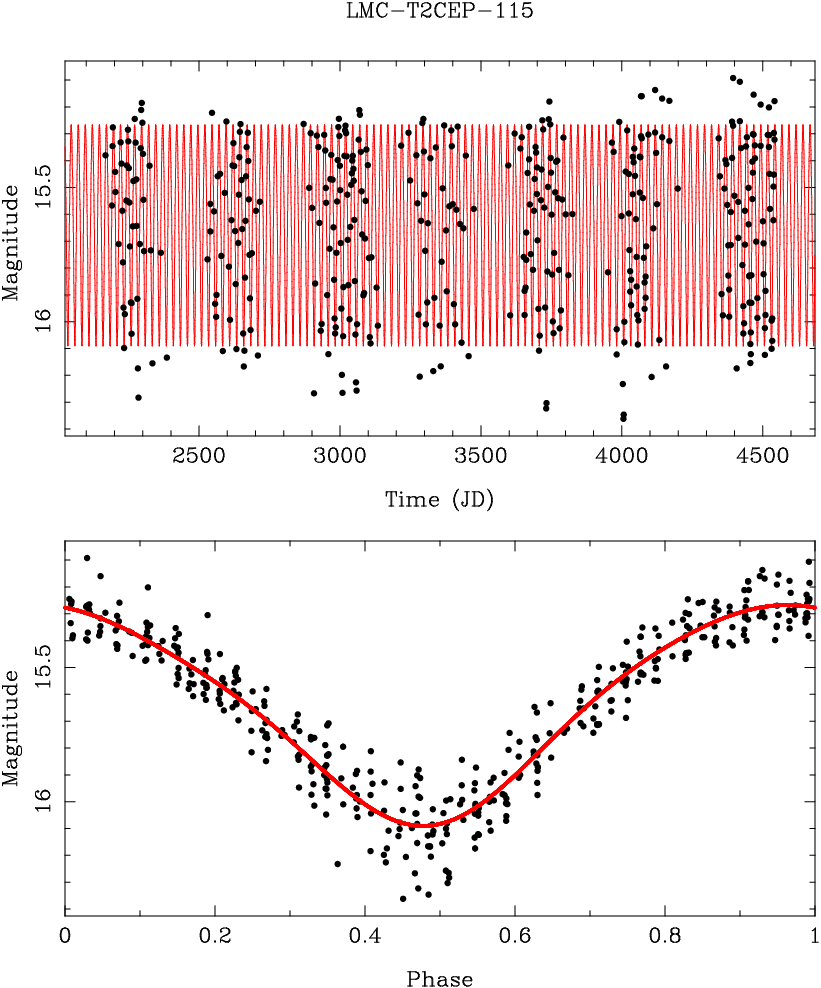}} 
\end{minipage}
\begin{minipage}{0.26\textwidth}
\resizebox{\hsize}{!}{\includegraphics[angle=-0]{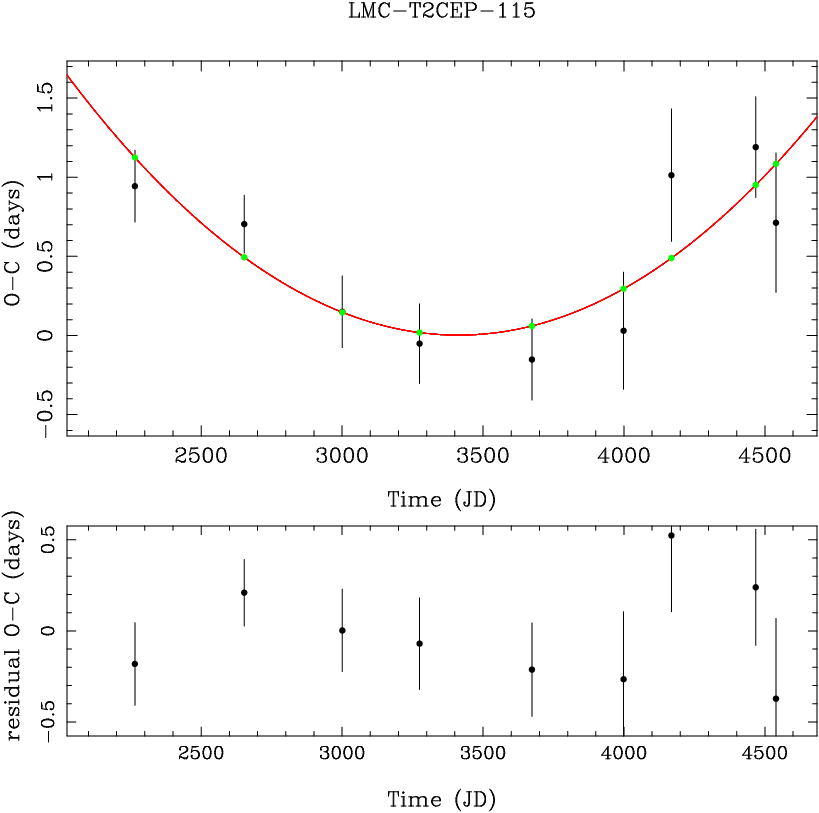}} 
\end{minipage}

\begin{minipage}{0.22\textwidth}
\resizebox{\hsize}{!}{\includegraphics[angle=-0]{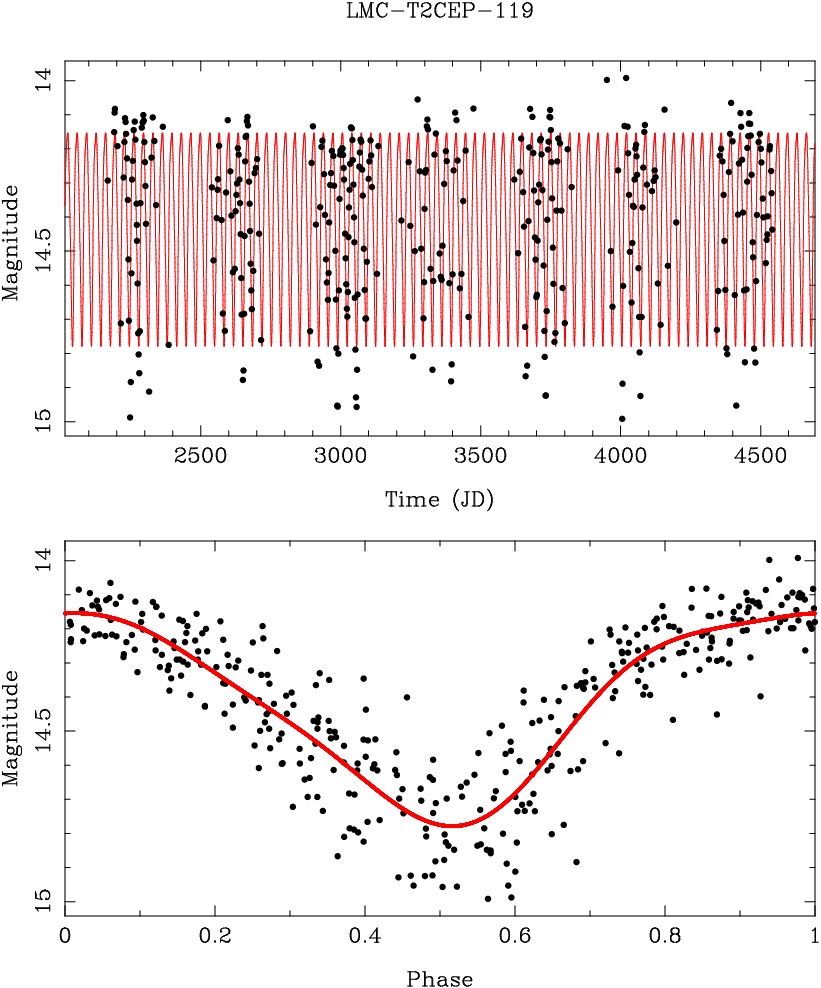}} 
\end{minipage}
\begin{minipage}{0.26\textwidth}
\resizebox{\hsize}{!}{\includegraphics[angle=-0]{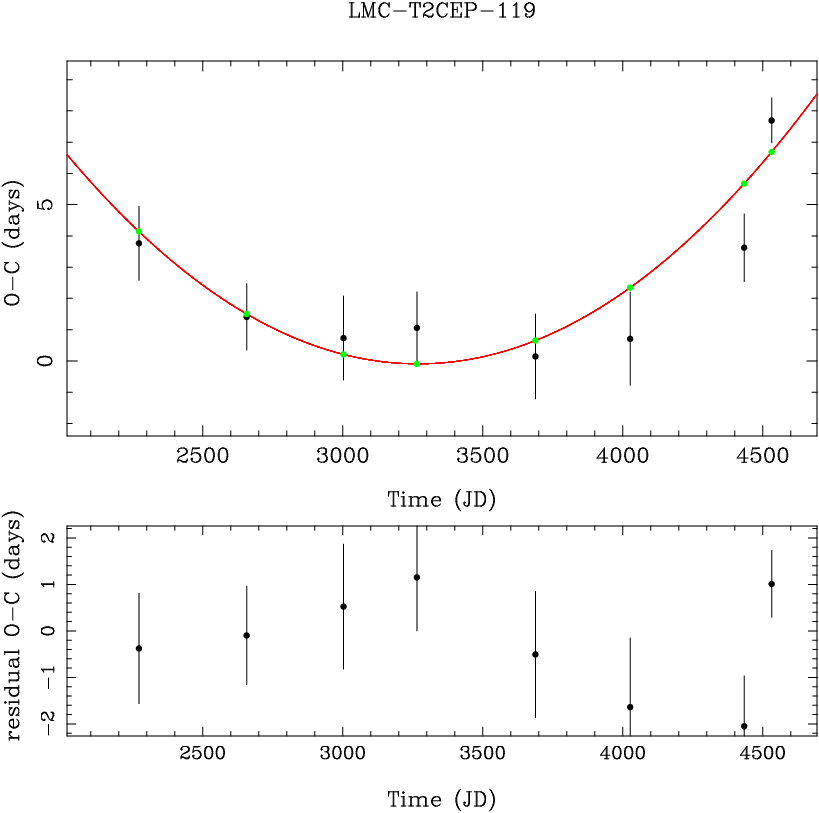}} 
\end{minipage}
\begin{minipage}{0.22\textwidth}
\resizebox{\hsize}{!}{\includegraphics[angle=-0]{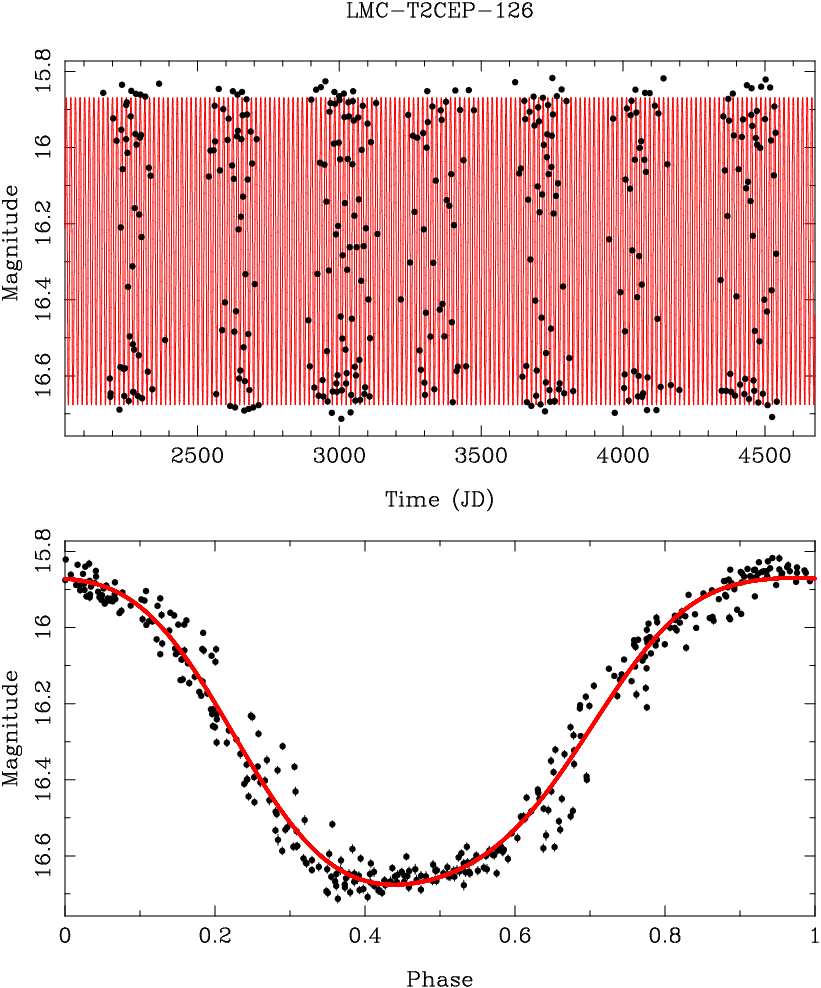}} 
\end{minipage}
\begin{minipage}{0.26\textwidth}
\resizebox{\hsize}{!}{\includegraphics[angle=-0]{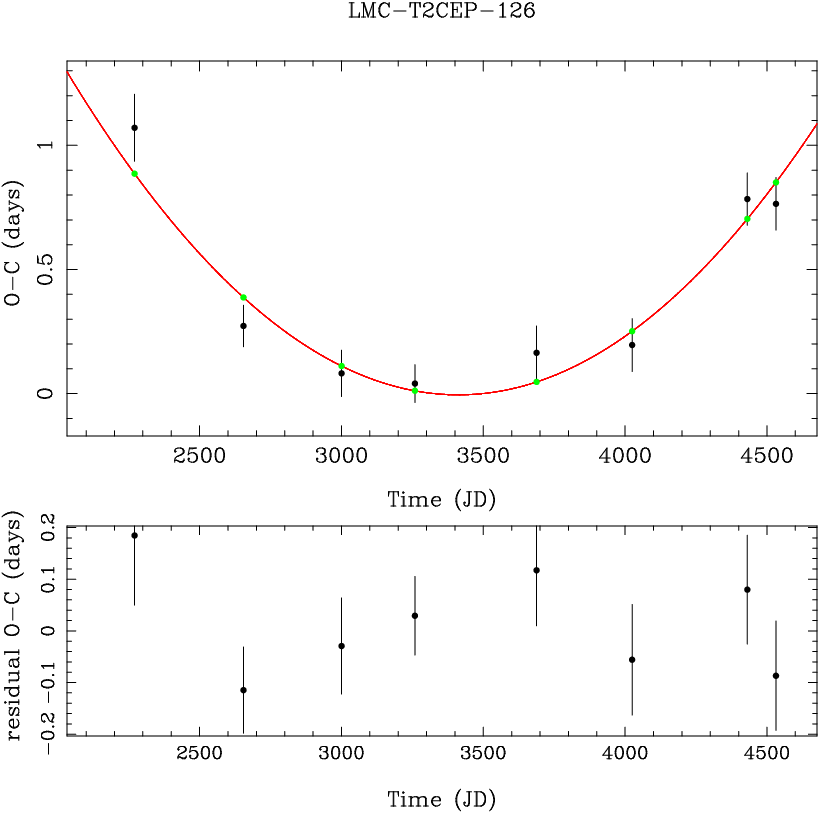}} 
\end{minipage}

\caption{Continued.}
\end{figure*}

\setcounter{figure}{3}
\begin{figure*}
\centering

\begin{minipage}{0.22\textwidth}
\resizebox{\hsize}{!}{\includegraphics[angle=-0]{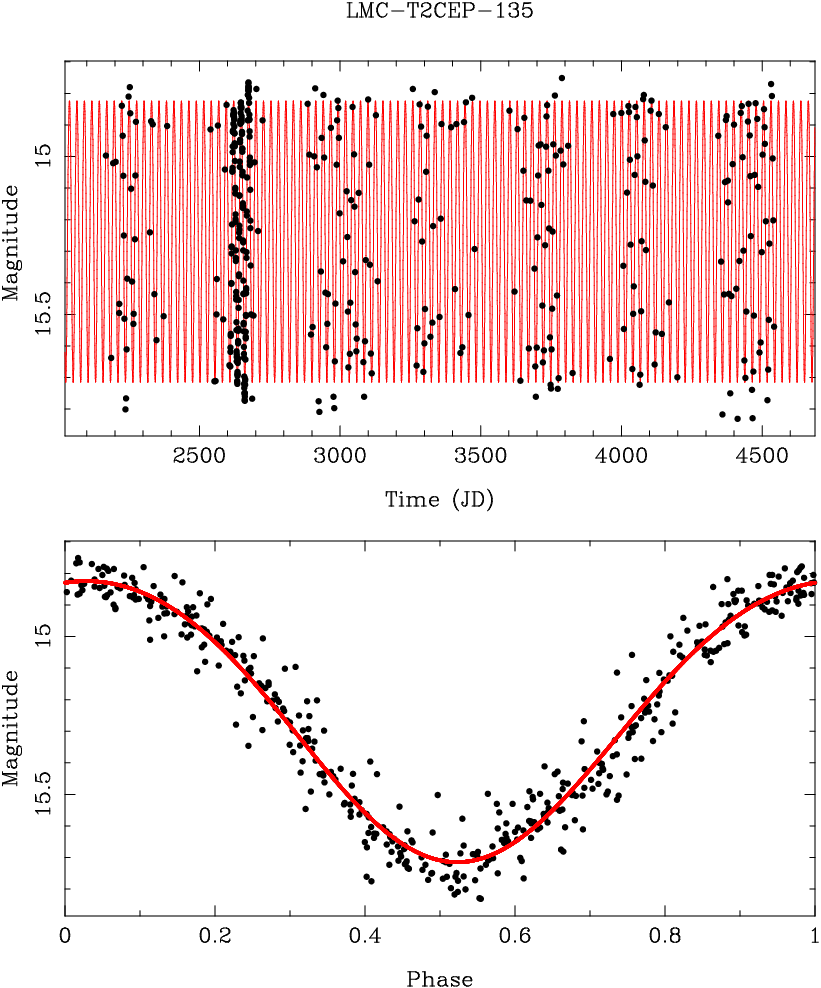}} 
\end{minipage}
\begin{minipage}{0.26\textwidth}
\resizebox{\hsize}{!}{\includegraphics[angle=-0]{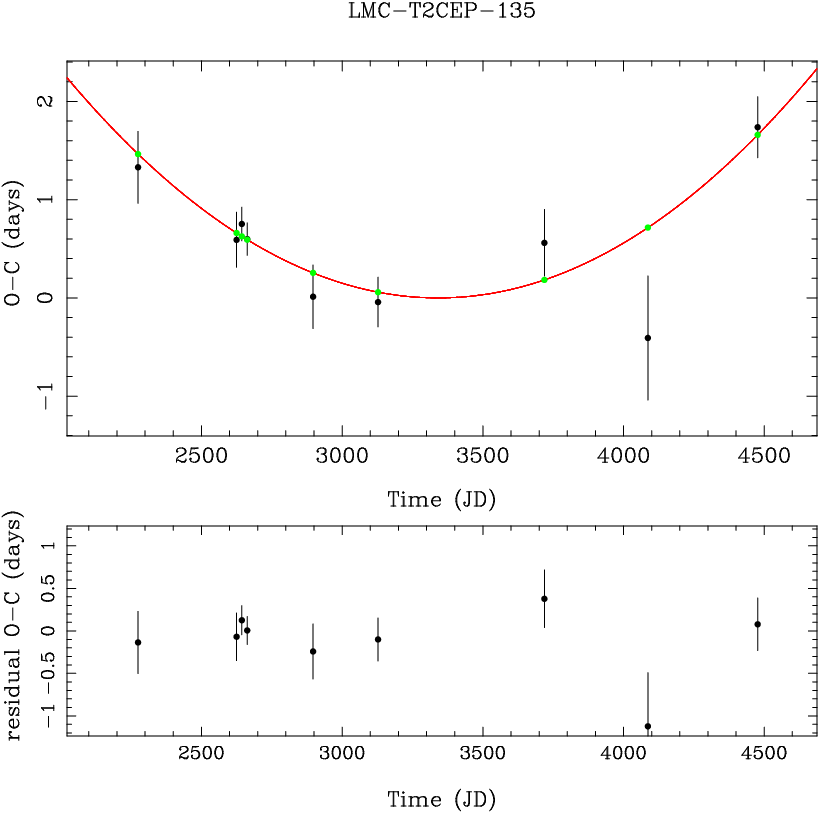}} 
\end{minipage}
\begin{minipage}{0.22\textwidth}
\resizebox{\hsize}{!}{\includegraphics[angle=-0]{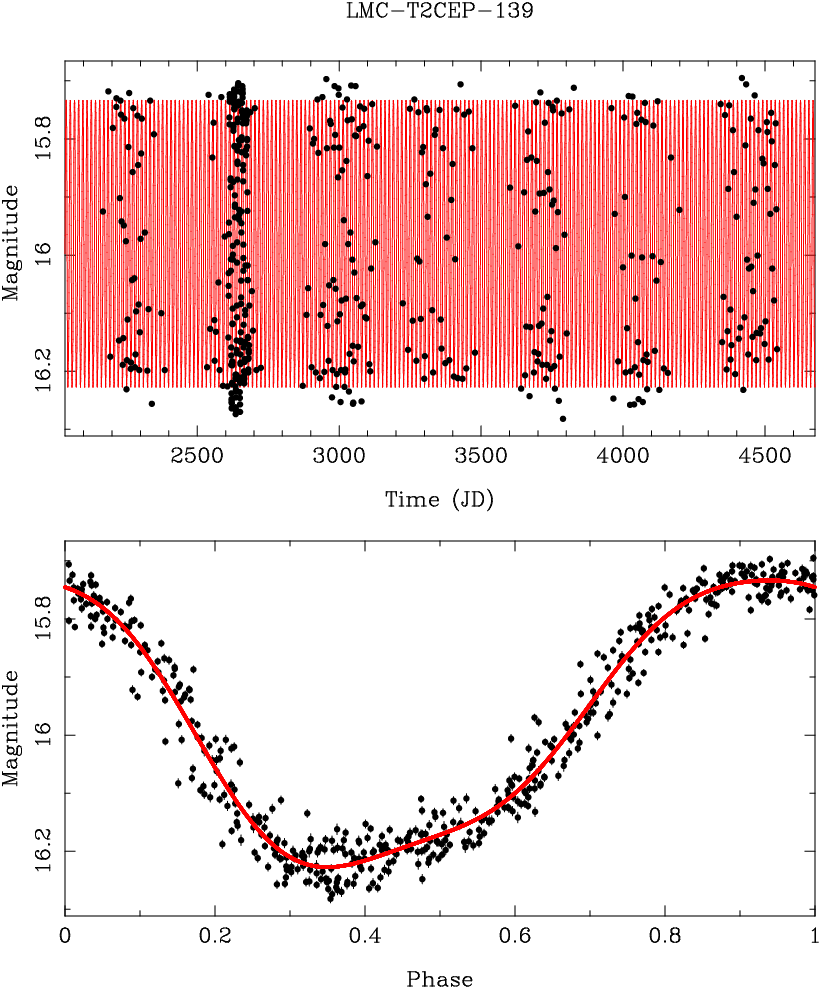}} 
\end{minipage}
\begin{minipage}{0.26\textwidth}
\resizebox{\hsize}{!}{\includegraphics[angle=-0]{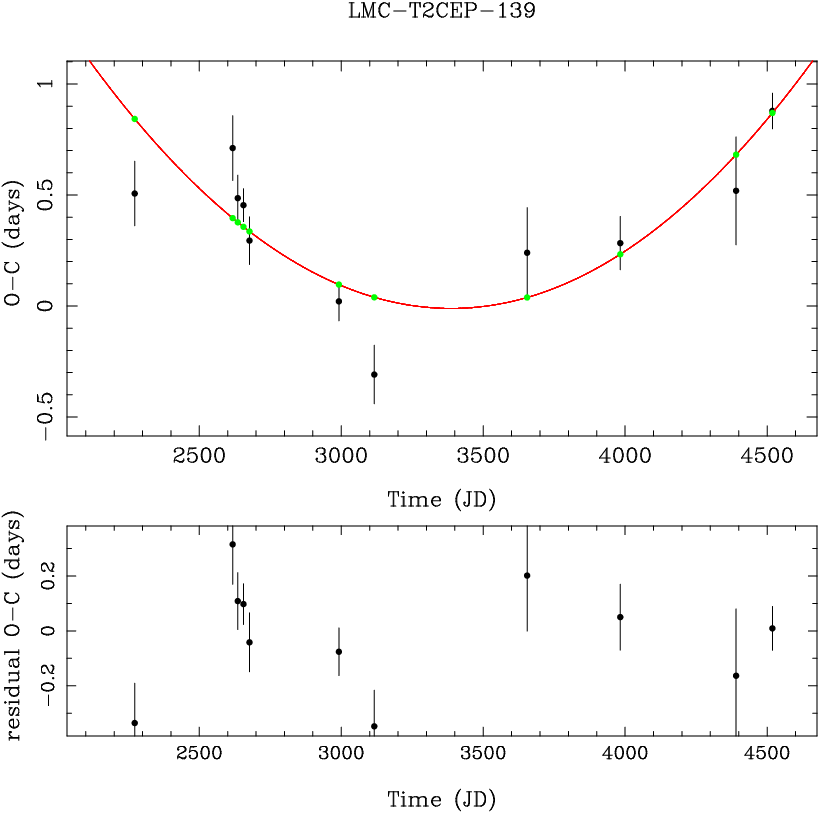}} 
\end{minipage}

\begin{minipage}{0.22\textwidth}
\resizebox{\hsize}{!}{\includegraphics[angle=-0]{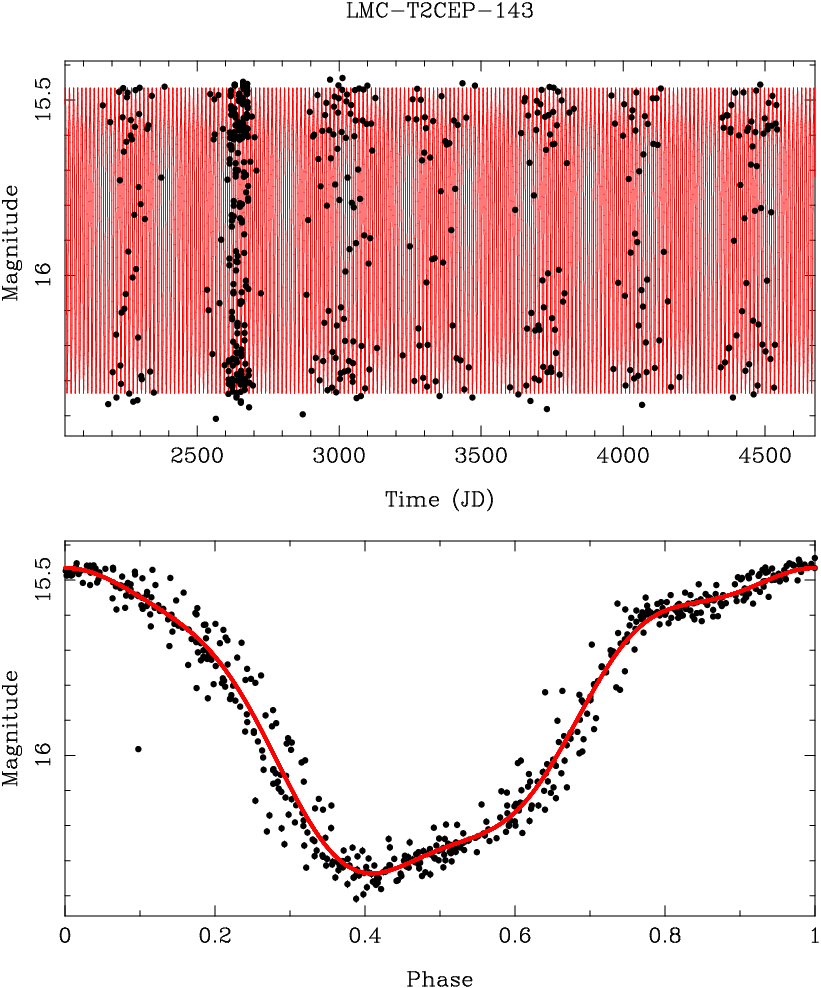}} 
\end{minipage}
\begin{minipage}{0.26\textwidth}
\resizebox{\hsize}{!}{\includegraphics[angle=-0]{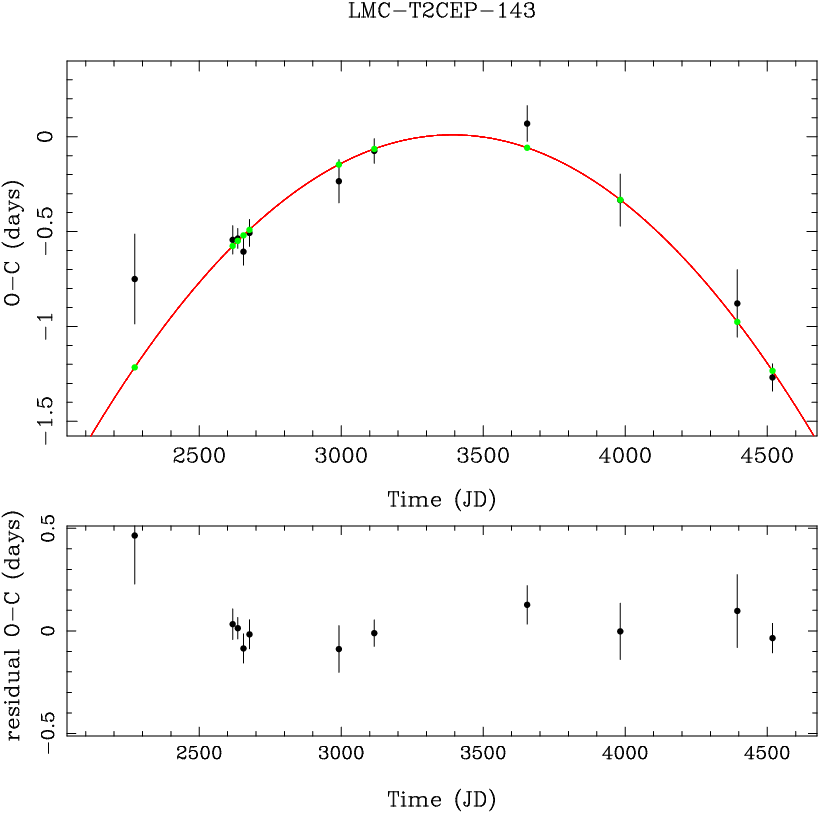}} 
\end{minipage}
\begin{minipage}{0.22\textwidth}
\resizebox{\hsize}{!}{\includegraphics[angle=-0]{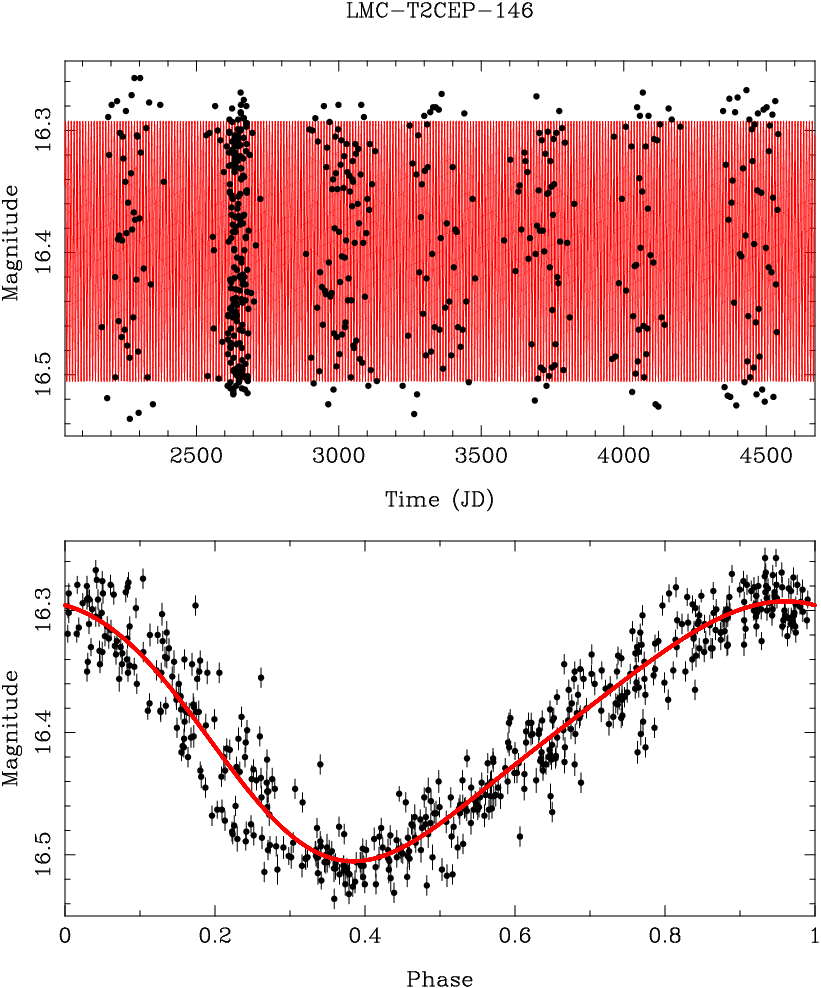}} 
\end{minipage}
\begin{minipage}{0.26\textwidth}
\resizebox{\hsize}{!}{\includegraphics[angle=-0]{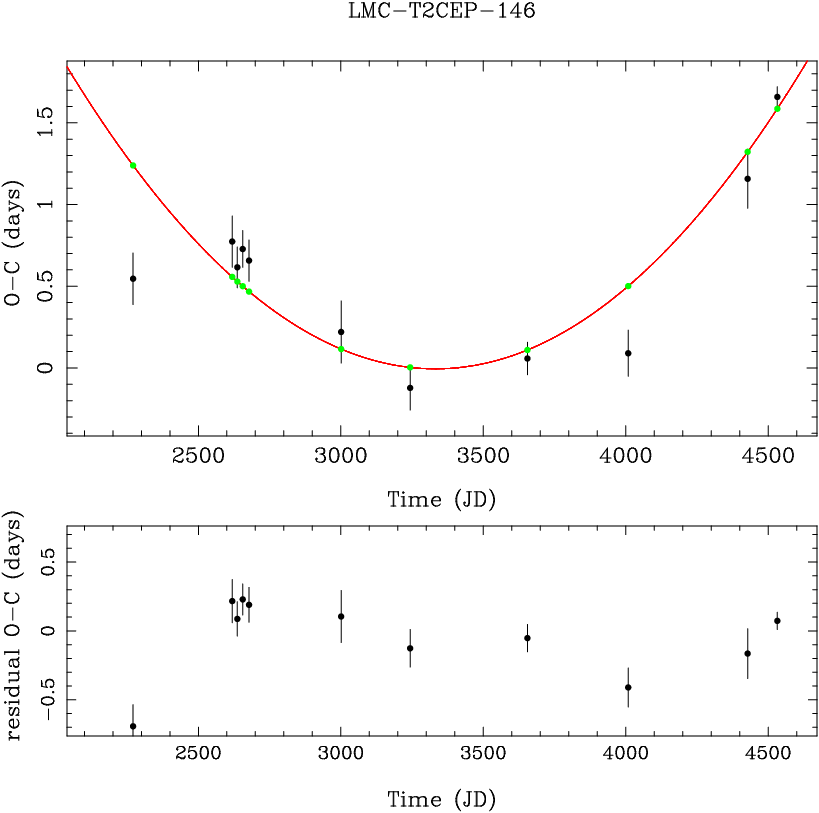}} 
\end{minipage}

\begin{minipage}{0.22\textwidth}
\resizebox{\hsize}{!}{\includegraphics[angle=-0]{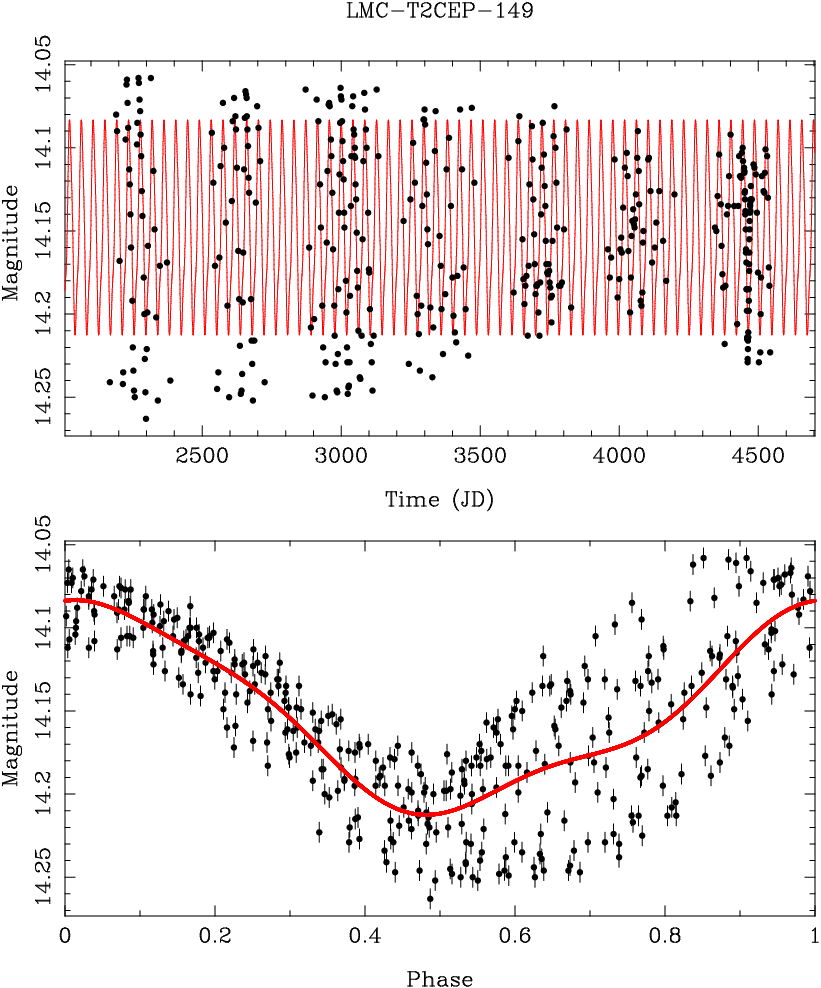}} 
\end{minipage}
\begin{minipage}{0.26\textwidth}
\resizebox{\hsize}{!}{\includegraphics[angle=-0]{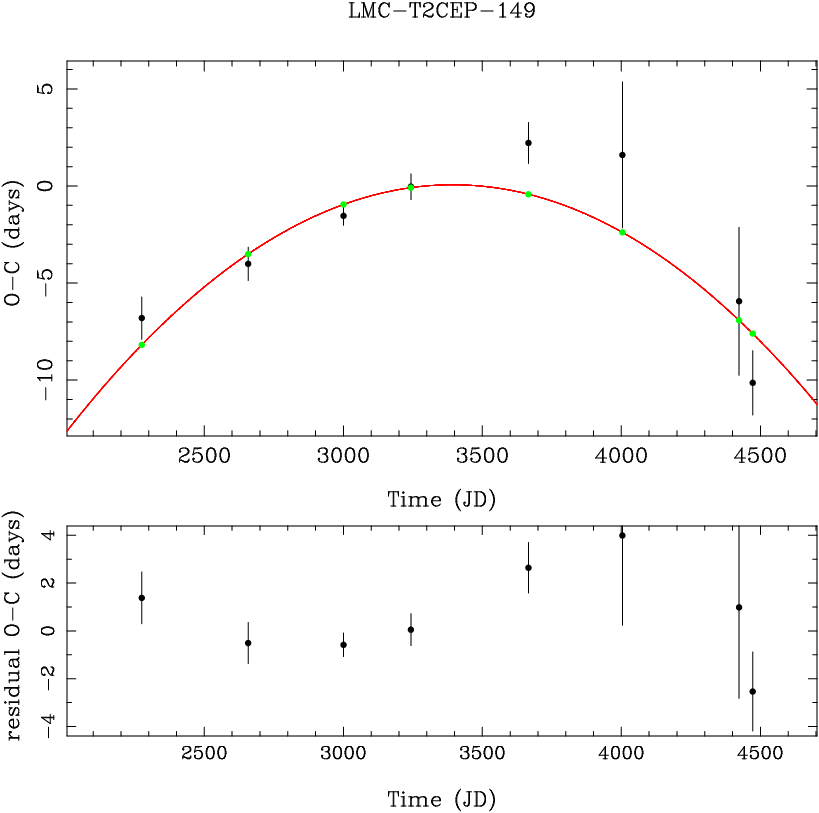}} 
\end{minipage}
\begin{minipage}{0.22\textwidth}
\resizebox{\hsize}{!}{\includegraphics[angle=-0]{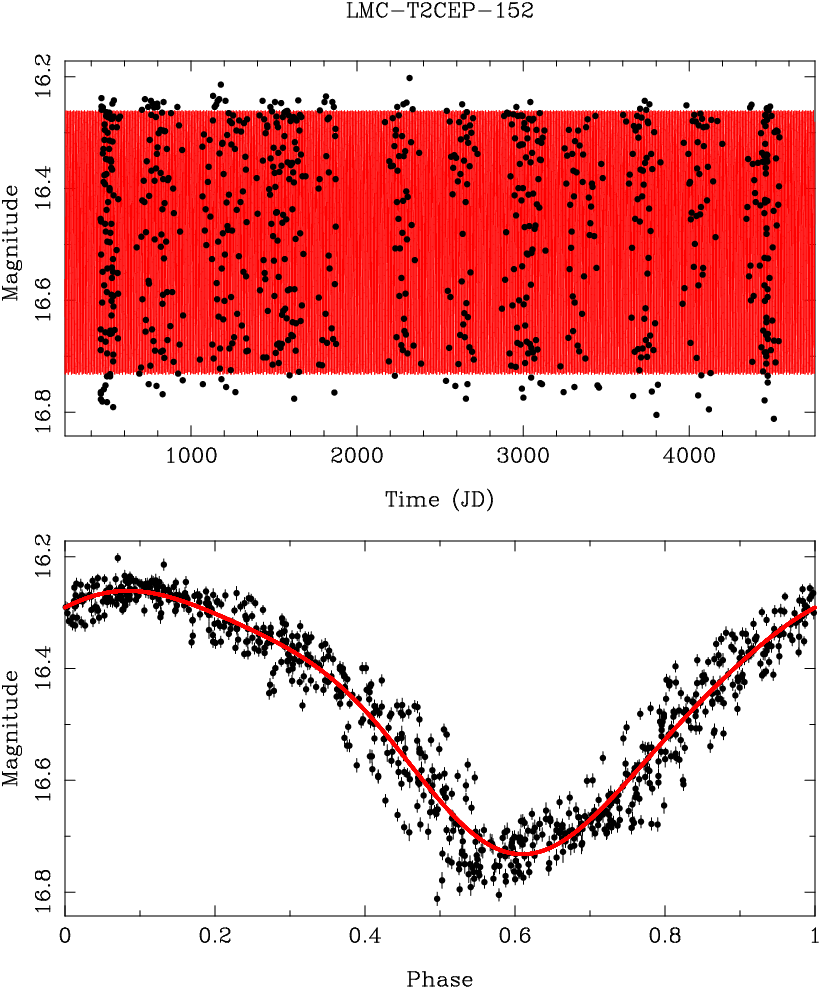}} 
\end{minipage}
\begin{minipage}{0.26\textwidth}
\resizebox{\hsize}{!}{\includegraphics[angle=-0]{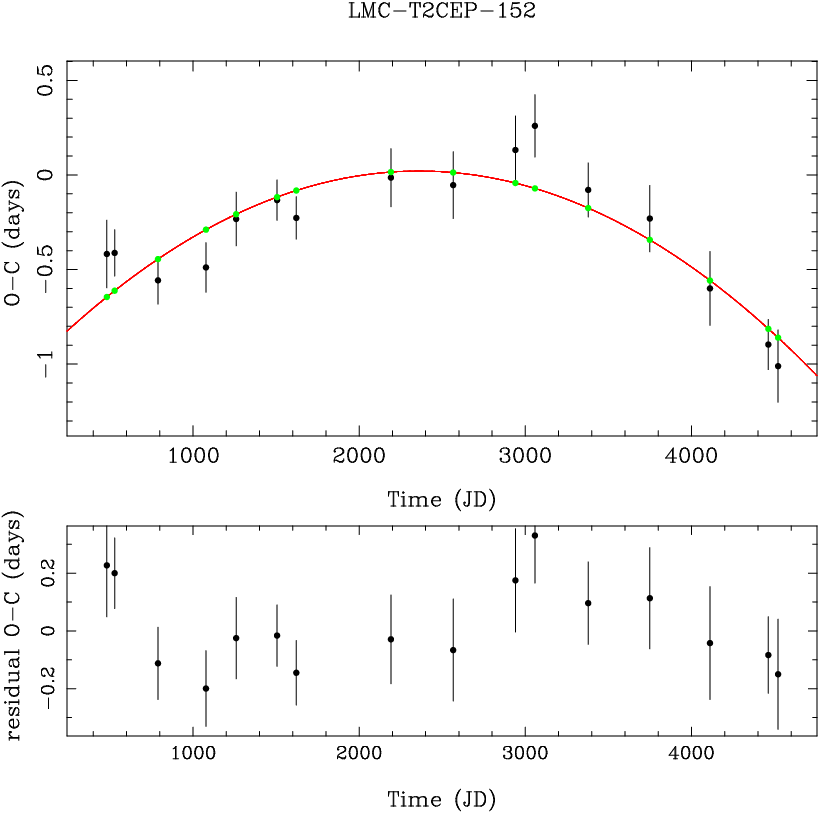}} 
\end{minipage}

\begin{minipage}{0.22\textwidth}
\resizebox{\hsize}{!}{\includegraphics[angle=-0]{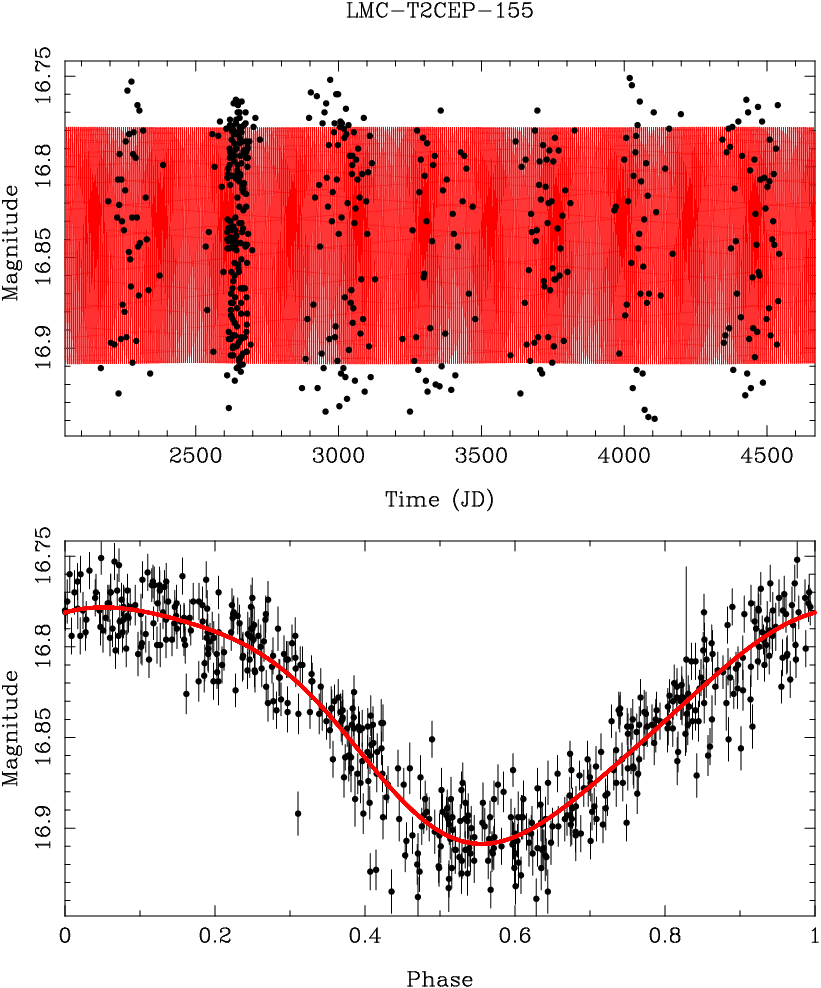}} 
\end{minipage}
\begin{minipage}{0.26\textwidth}
\resizebox{\hsize}{!}{\includegraphics[angle=-0]{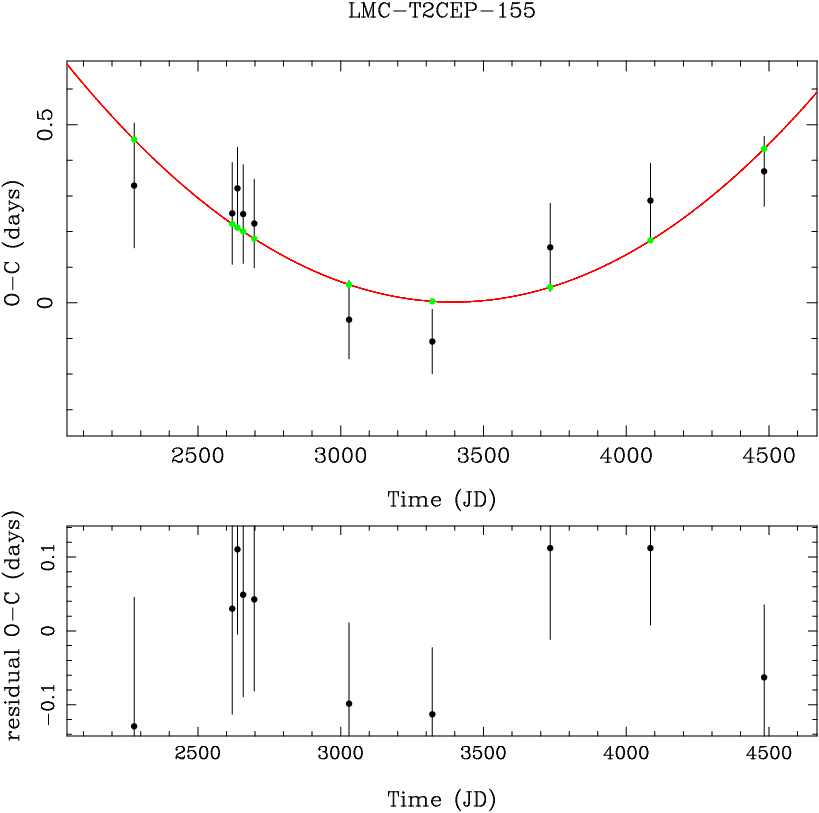}} 
\end{minipage}
\begin{minipage}{0.22\textwidth}
\resizebox{\hsize}{!}{\includegraphics[angle=-0]{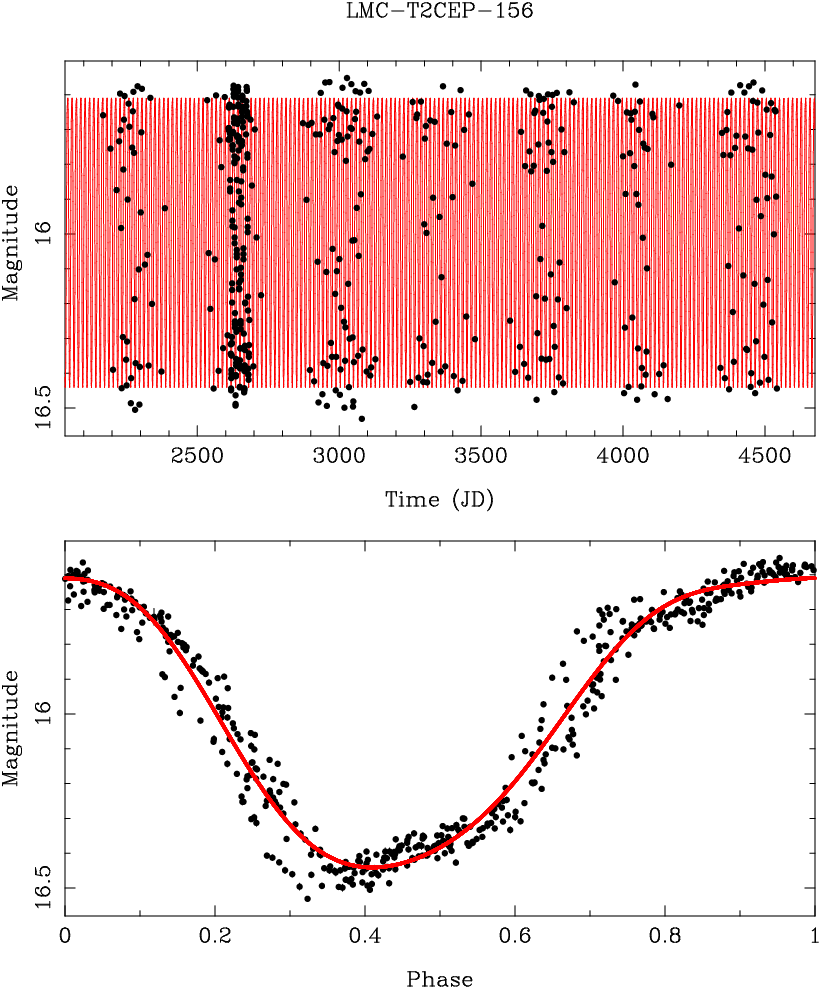}} 
\end{minipage}
\begin{minipage}{0.26\textwidth}
\resizebox{\hsize}{!}{\includegraphics[angle=-0]{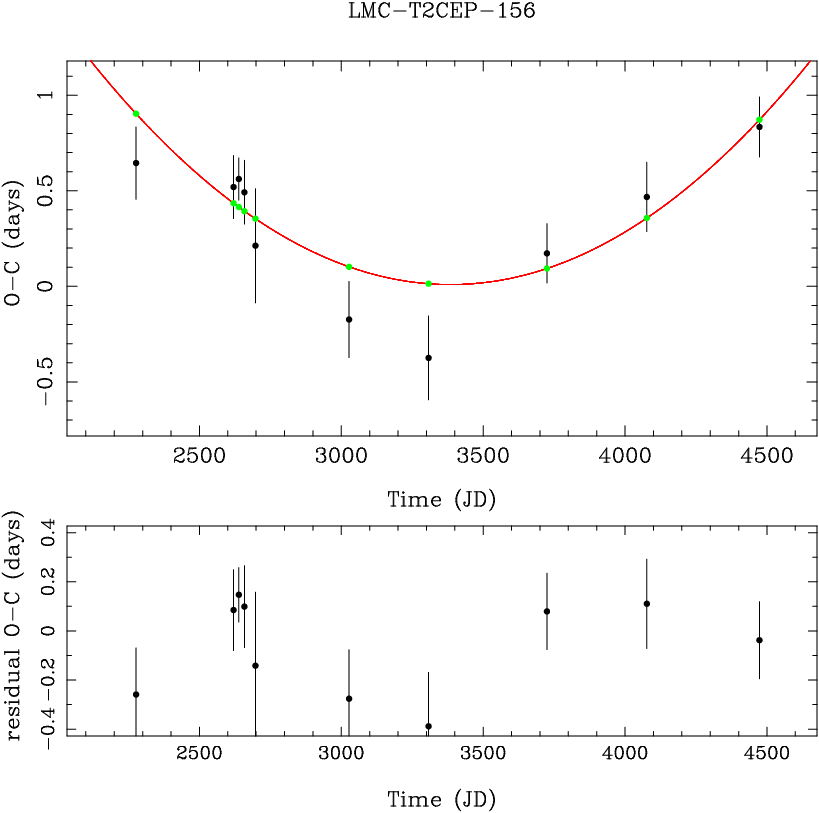}} 
\end{minipage}

\caption{Continued.}
\end{figure*}

\setcounter{figure}{3}
\begin{figure*}
\centering

\begin{minipage}{0.22\textwidth}
\resizebox{\hsize}{!}{\includegraphics[angle=-0]{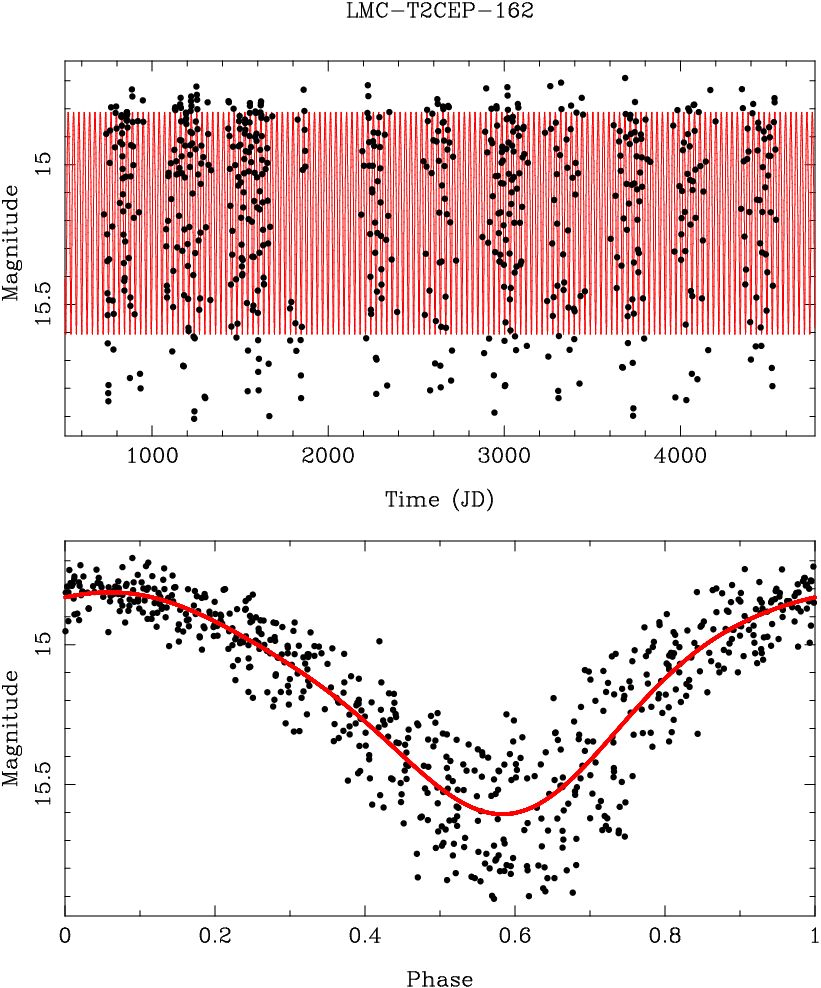}} 
\end{minipage}
\begin{minipage}{0.26\textwidth}
\resizebox{\hsize}{!}{\includegraphics[angle=-0]{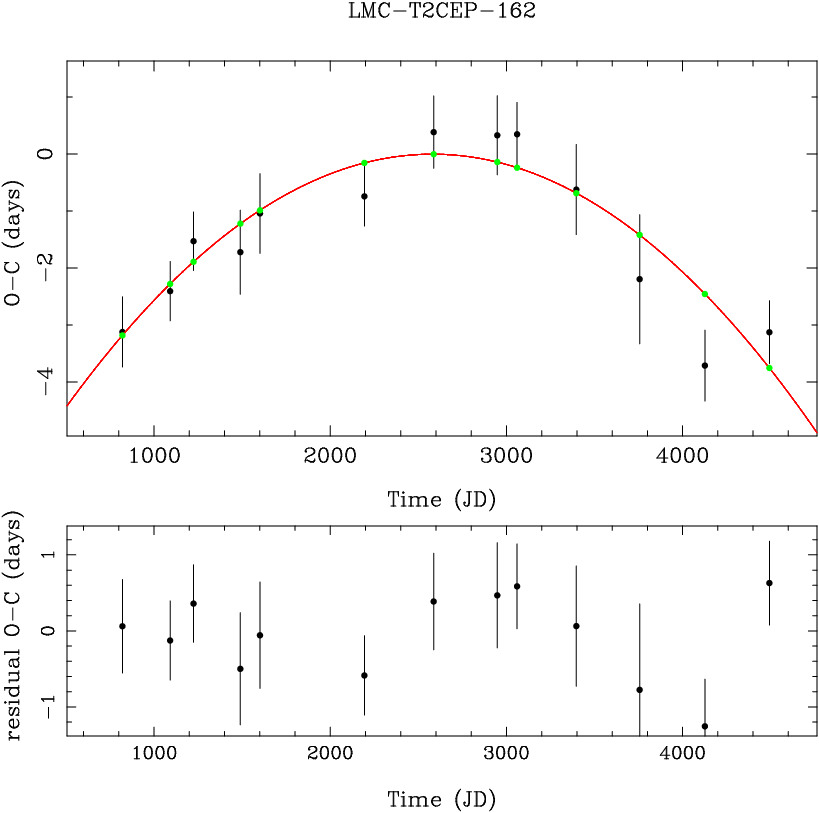}} 
\end{minipage}
\begin{minipage}{0.22\textwidth}
\resizebox{\hsize}{!}{\includegraphics[angle=-0]{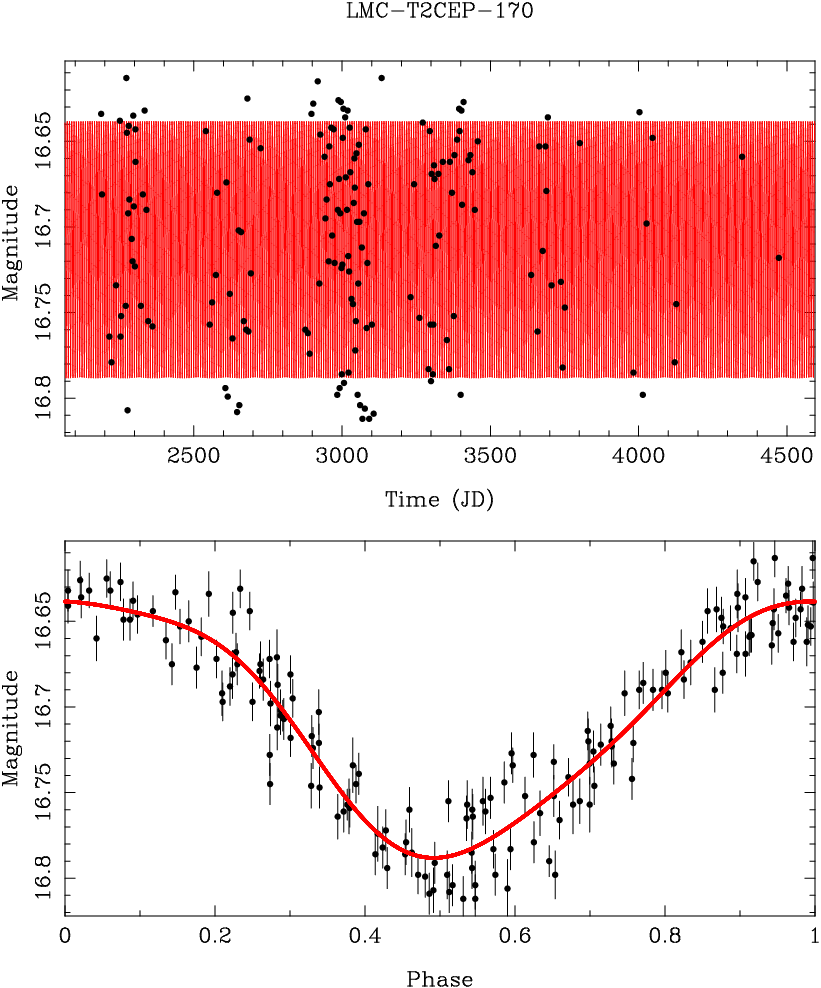}} 
\end{minipage}
\begin{minipage}{0.26\textwidth}
\resizebox{\hsize}{!}{\includegraphics[angle=-0]{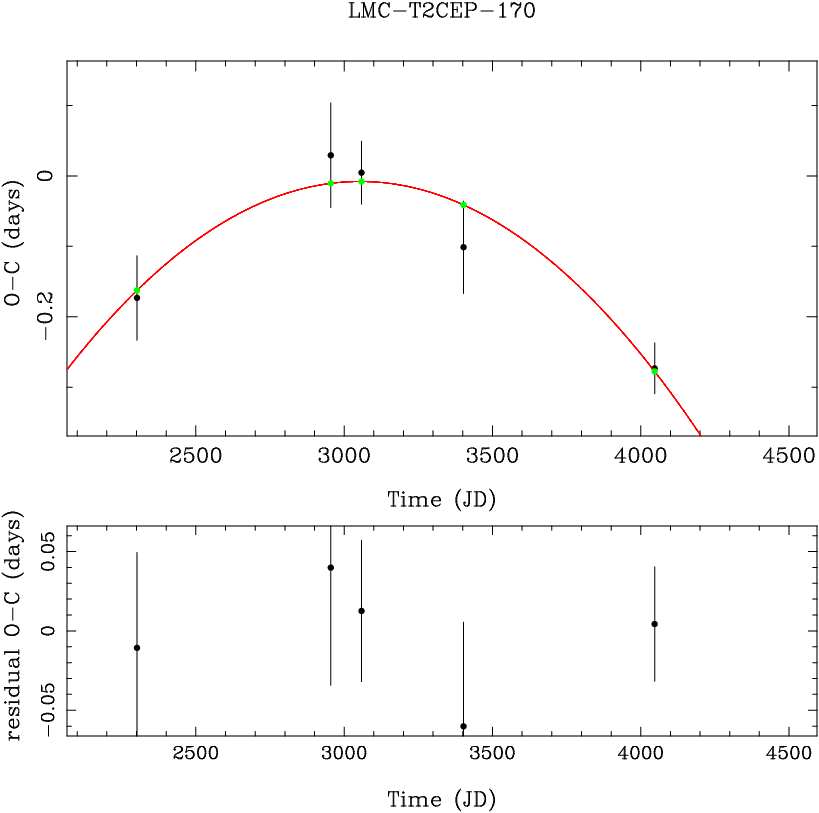}} 
\end{minipage}

\begin{minipage}{0.22\textwidth}
\resizebox{\hsize}{!}{\includegraphics[angle=-0]{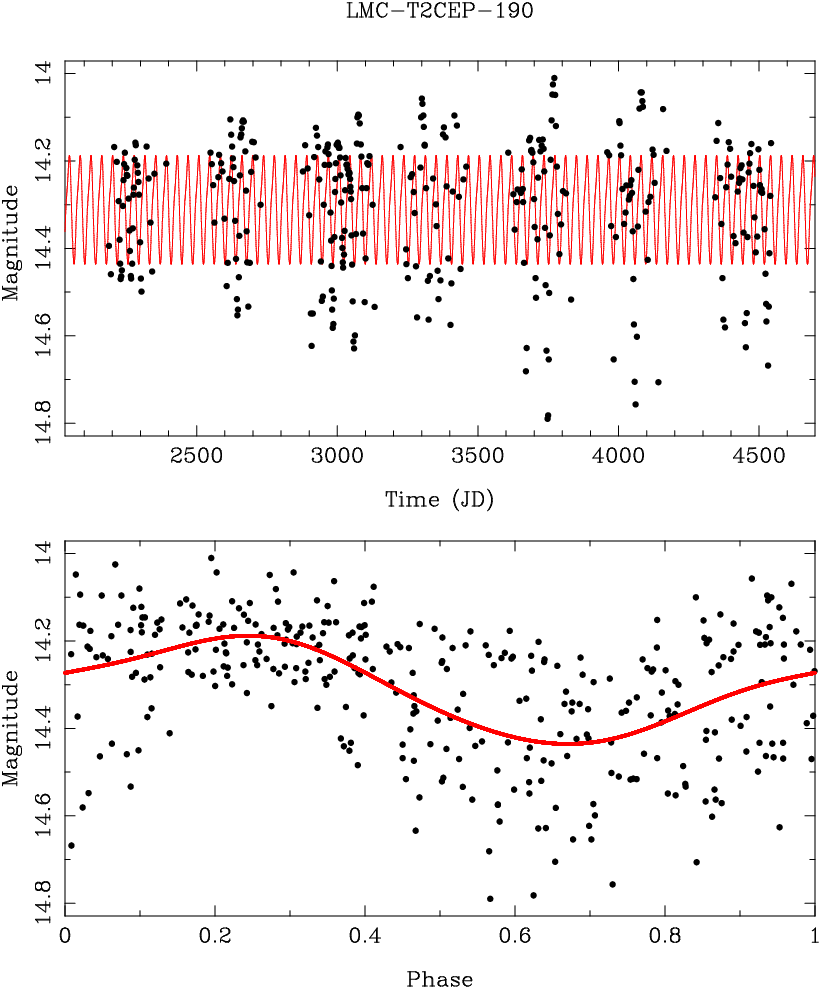}} 
\end{minipage}
\begin{minipage}{0.26\textwidth}
\resizebox{\hsize}{!}{\includegraphics[angle=-0]{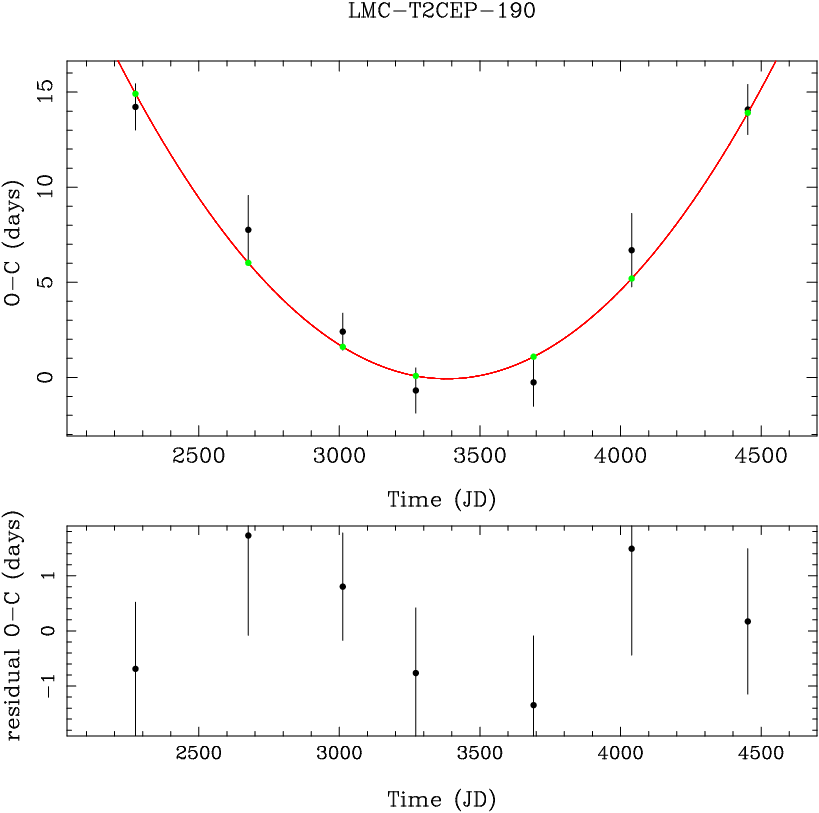}} 
\end{minipage}
\begin{minipage}{0.22\textwidth}
\resizebox{\hsize}{!}{\includegraphics[angle=-0]{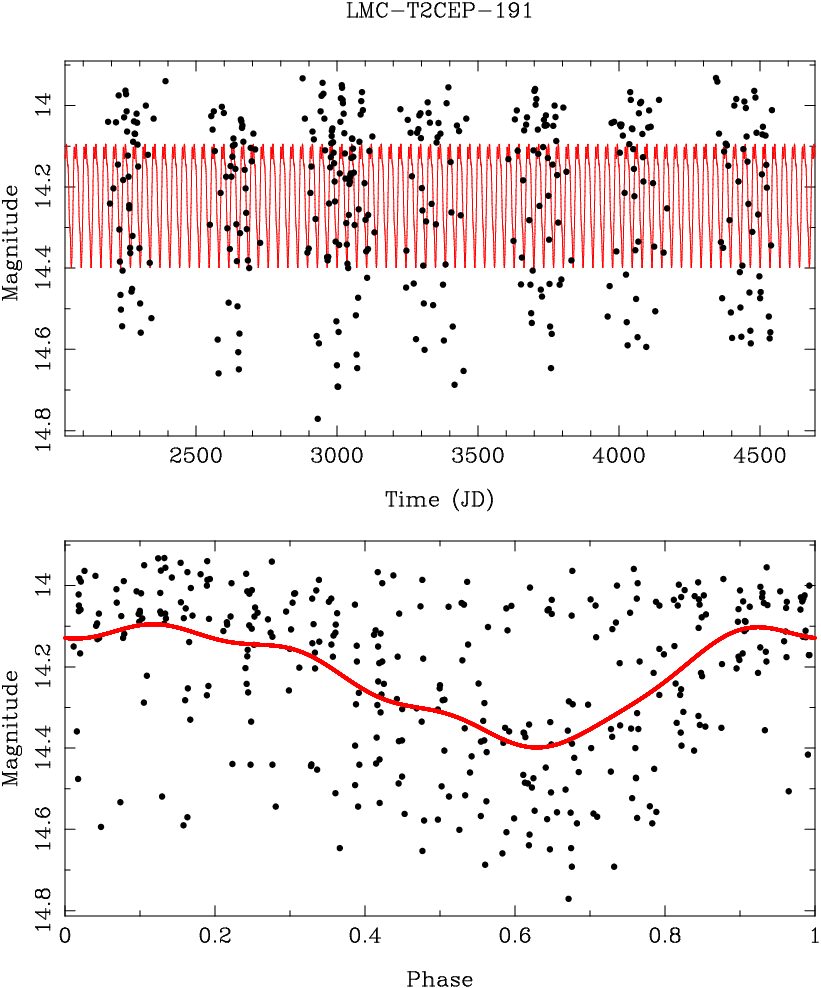}} 
\end{minipage}
\begin{minipage}{0.26\textwidth}
\resizebox{\hsize}{!}{\includegraphics[angle=-0]{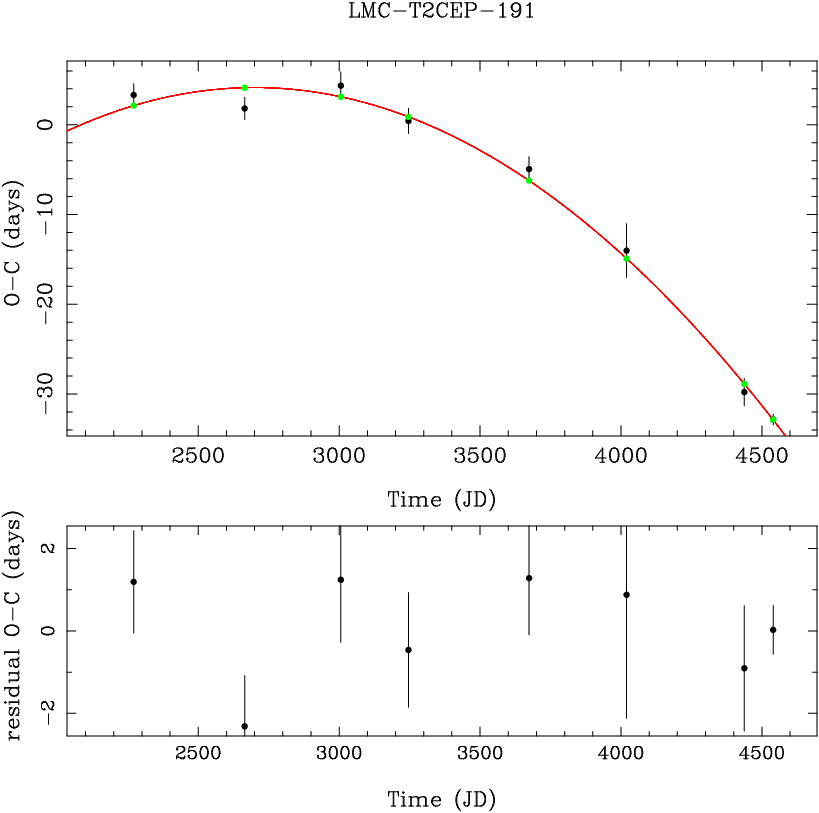}} 
\end{minipage}

\begin{minipage}{0.22\textwidth}
\resizebox{\hsize}{!}{\includegraphics[angle=-0]{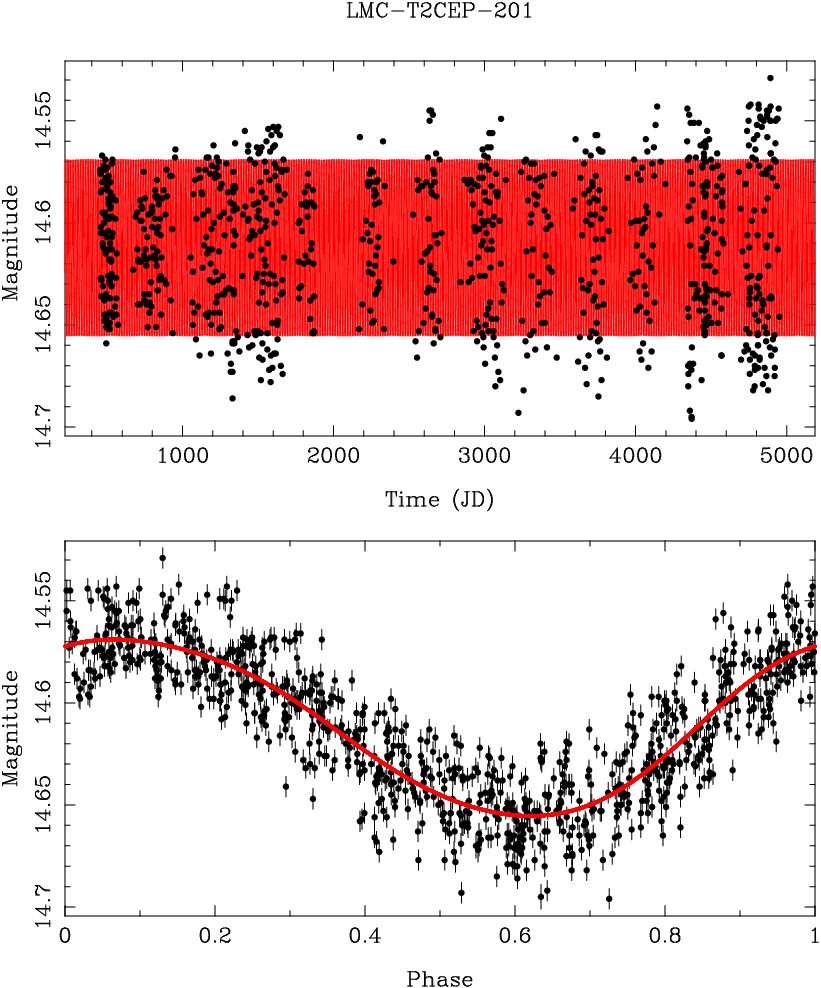}} 
\end{minipage}
\begin{minipage}{0.26\textwidth}
\resizebox{\hsize}{!}{\includegraphics[angle=-0]{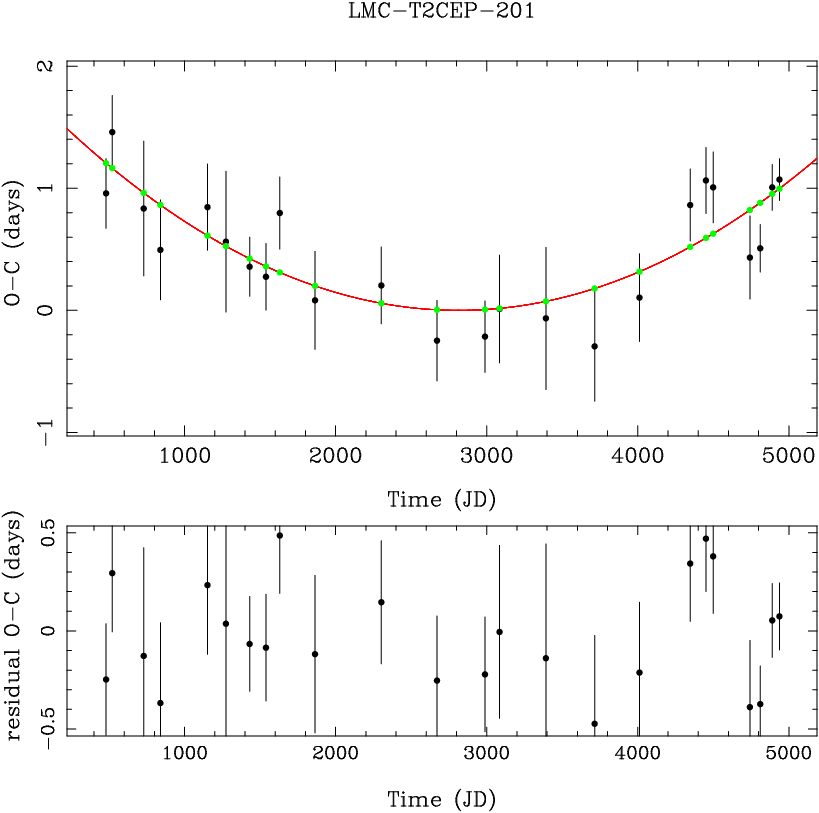}} 
\end{minipage}
\begin{minipage}{0.22\textwidth}
\resizebox{\hsize}{!}{\includegraphics[angle=-0]{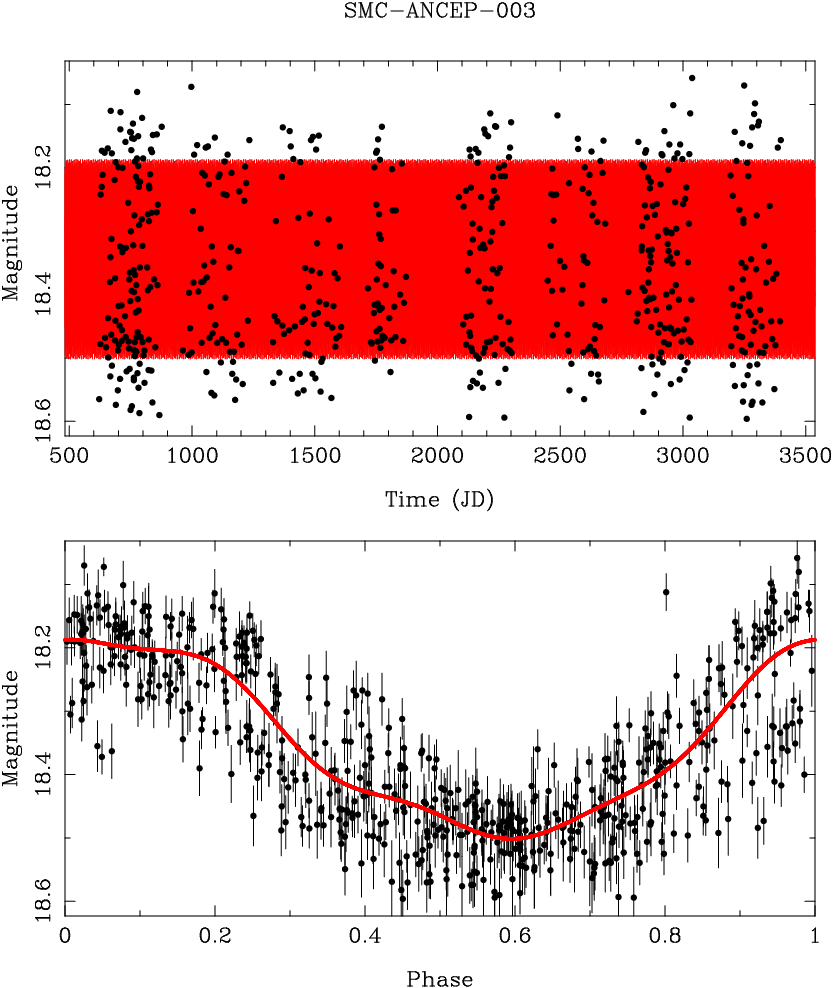}} 
\end{minipage}
\begin{minipage}{0.26\textwidth}
\resizebox{\hsize}{!}{\includegraphics[angle=-0]{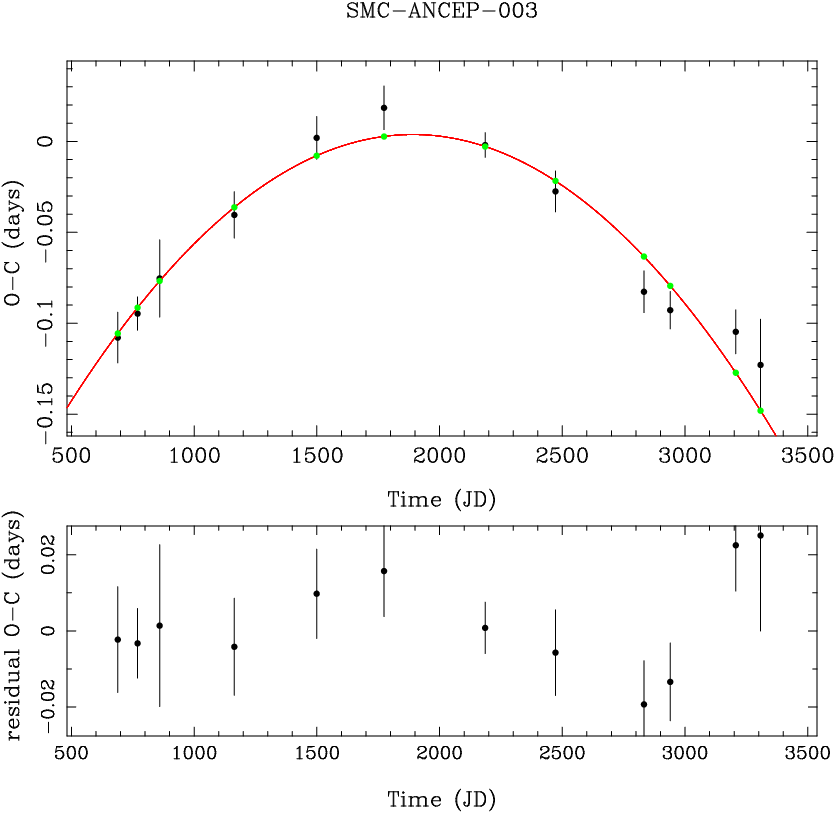}} 
\end{minipage}

\begin{minipage}{0.22\textwidth}
\resizebox{\hsize}{!}{\includegraphics[angle=-0]{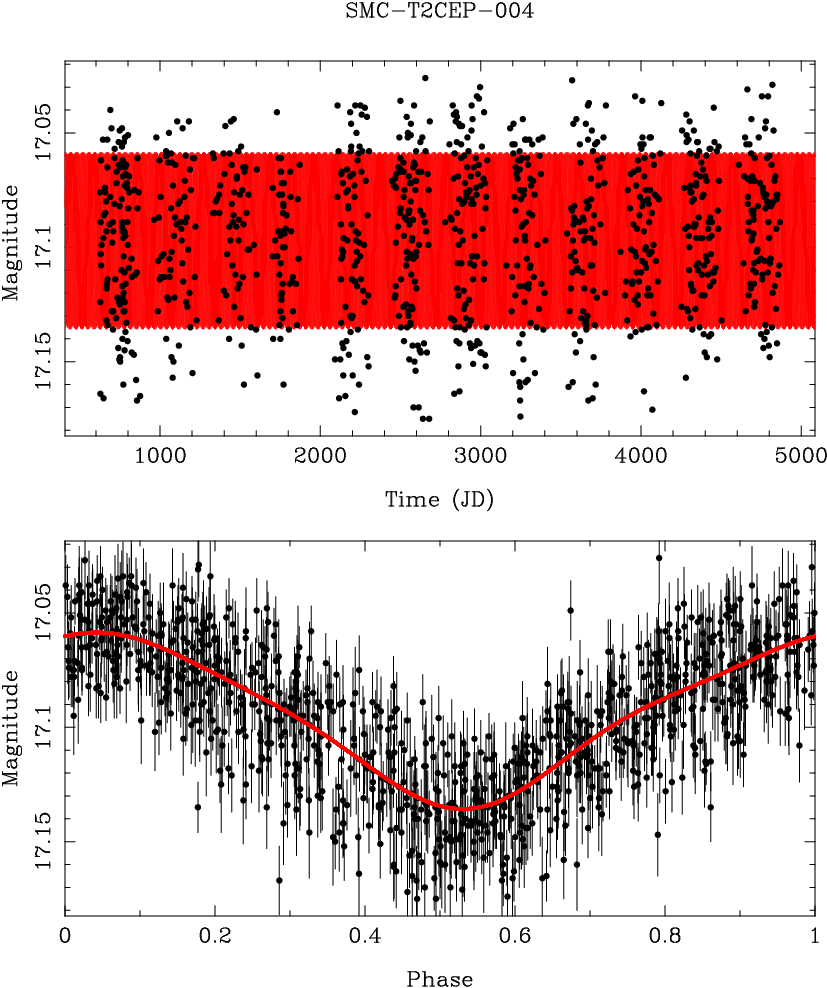}} 
\end{minipage}
\begin{minipage}{0.26\textwidth}
\resizebox{\hsize}{!}{\includegraphics[angle=-0]{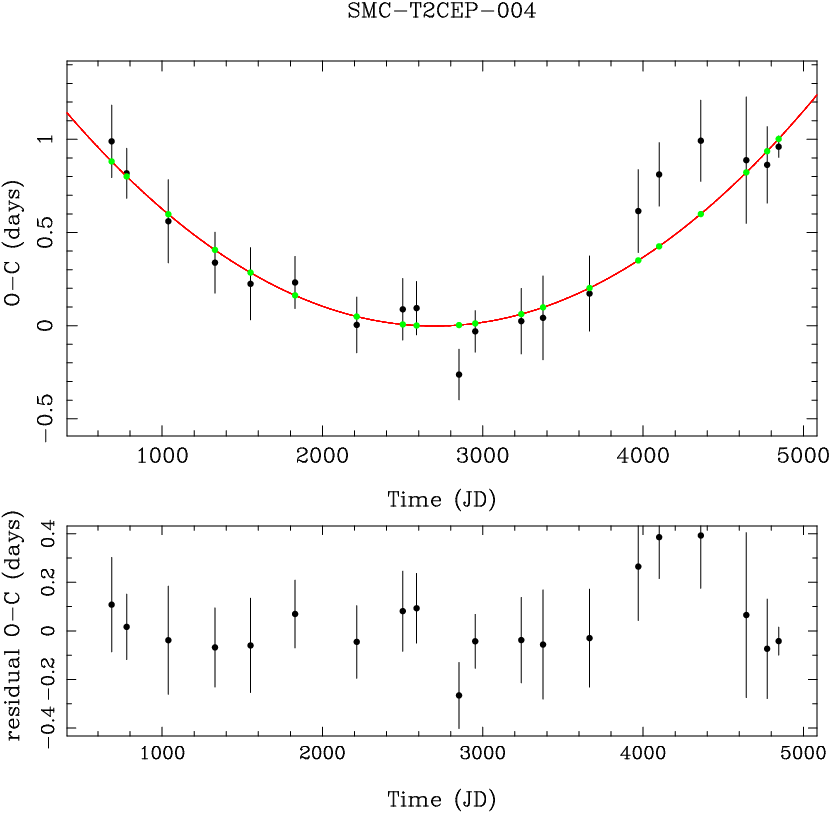}} 
\end{minipage}
\begin{minipage}{0.22\textwidth}
\resizebox{\hsize}{!}{\includegraphics[angle=-0]{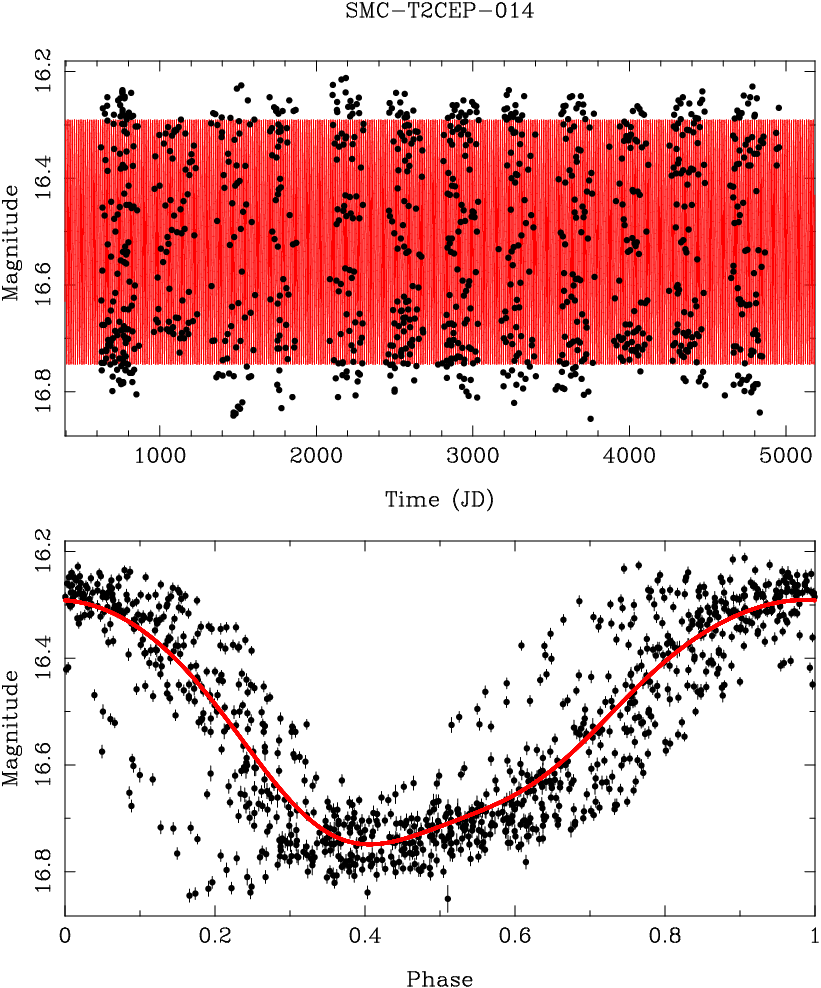}} 
\end{minipage}
\begin{minipage}{0.26\textwidth}
\resizebox{\hsize}{!}{\includegraphics[angle=-0]{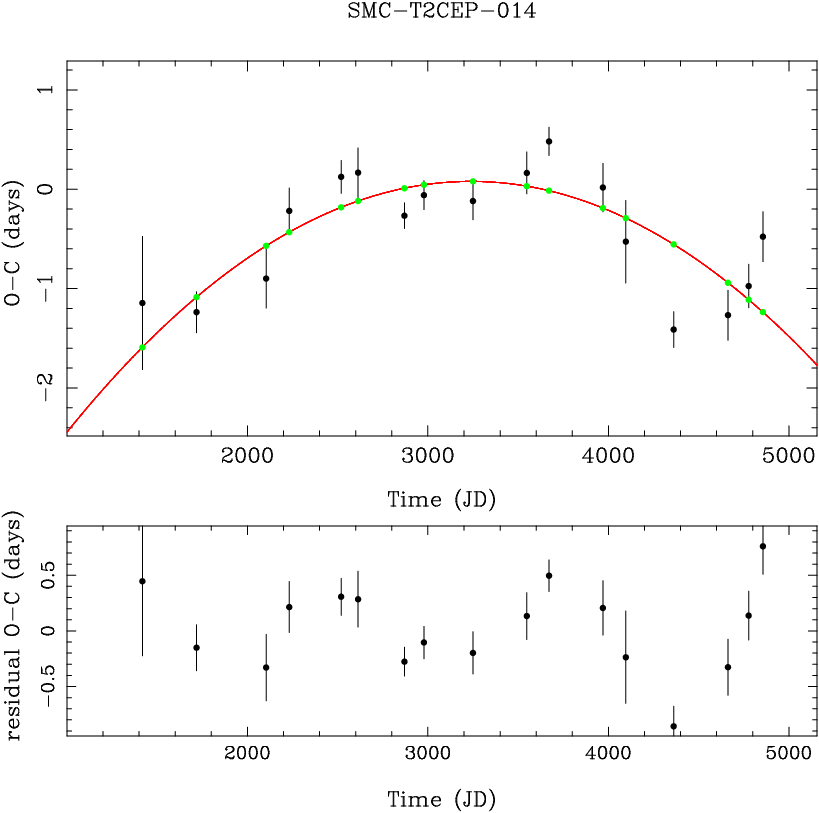}} 
\end{minipage}

\caption{Continued.}
\end{figure*}

\setcounter{figure}{3}
\begin{figure*}
\centering

\begin{minipage}{0.22\textwidth}
\resizebox{\hsize}{!}{\includegraphics[angle=-0]{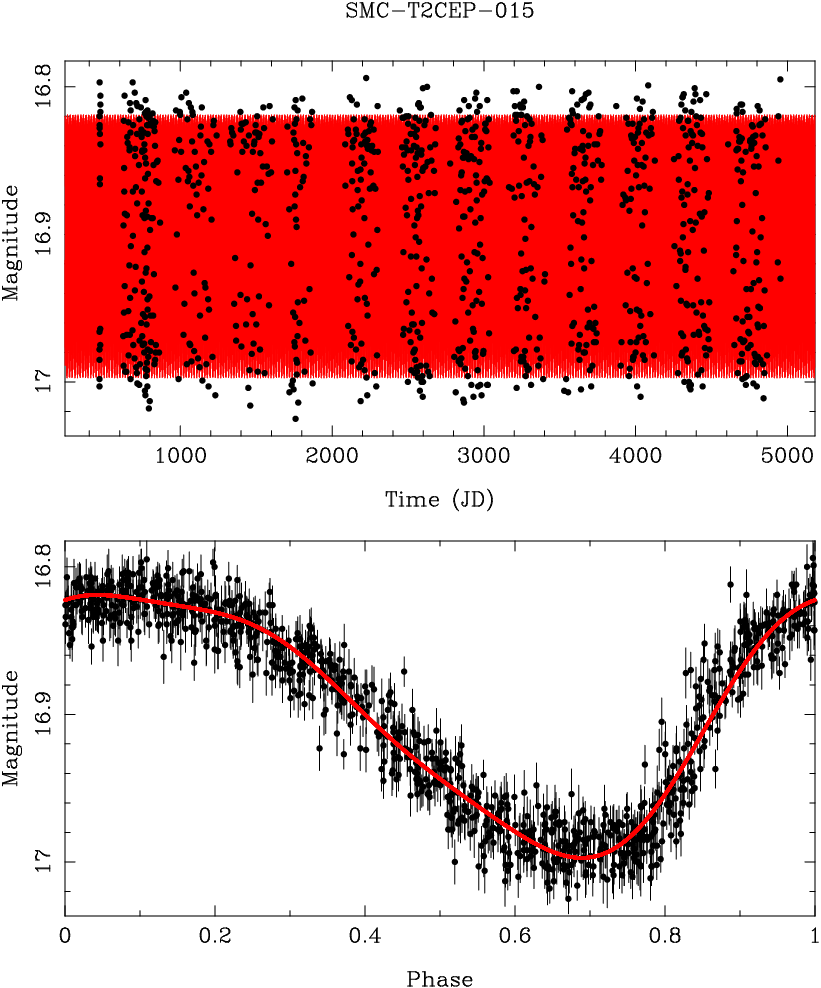}} 
\end{minipage}
\begin{minipage}{0.26\textwidth}
\resizebox{\hsize}{!}{\includegraphics[angle=-0]{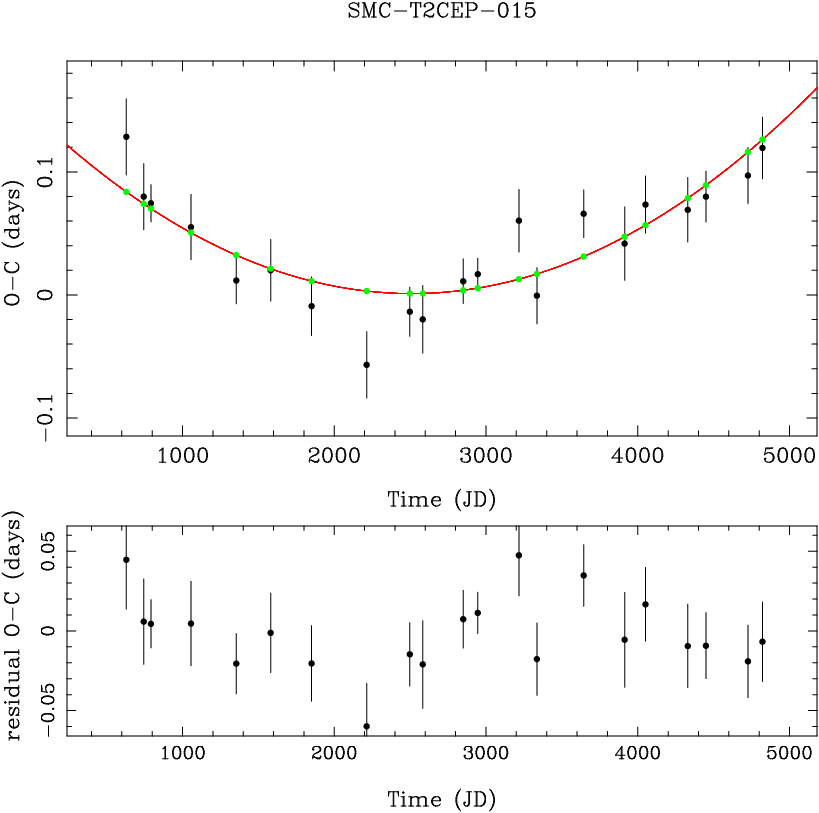}} 
\end{minipage}
\begin{minipage}{0.22\textwidth}
\resizebox{\hsize}{!}{\includegraphics[angle=-0]{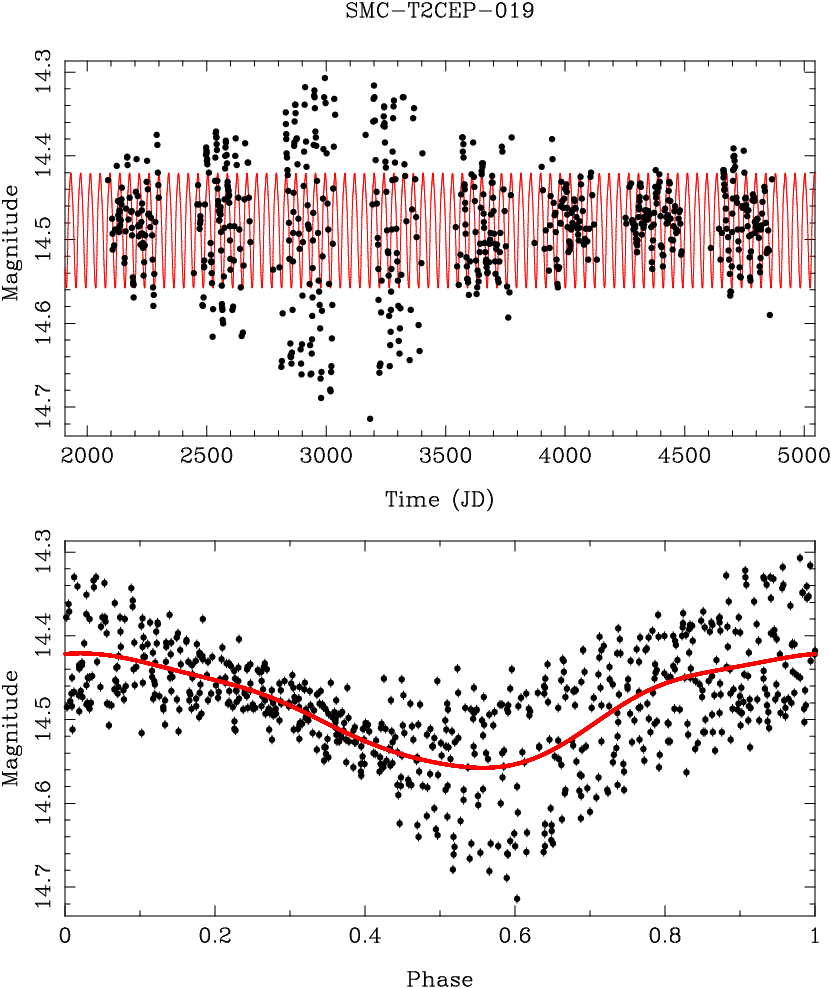}} 
\end{minipage}
\begin{minipage}{0.26\textwidth}
\resizebox{\hsize}{!}{\includegraphics[angle=-0]{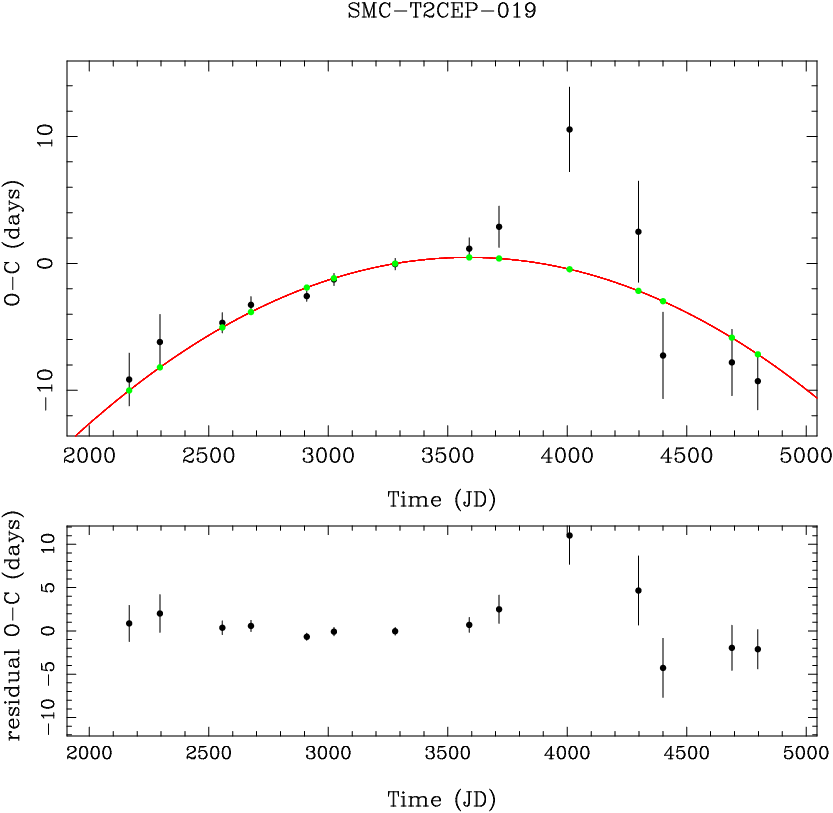}} 
\end{minipage}

\begin{minipage}{0.22\textwidth}
\resizebox{\hsize}{!}{\includegraphics[angle=-0]{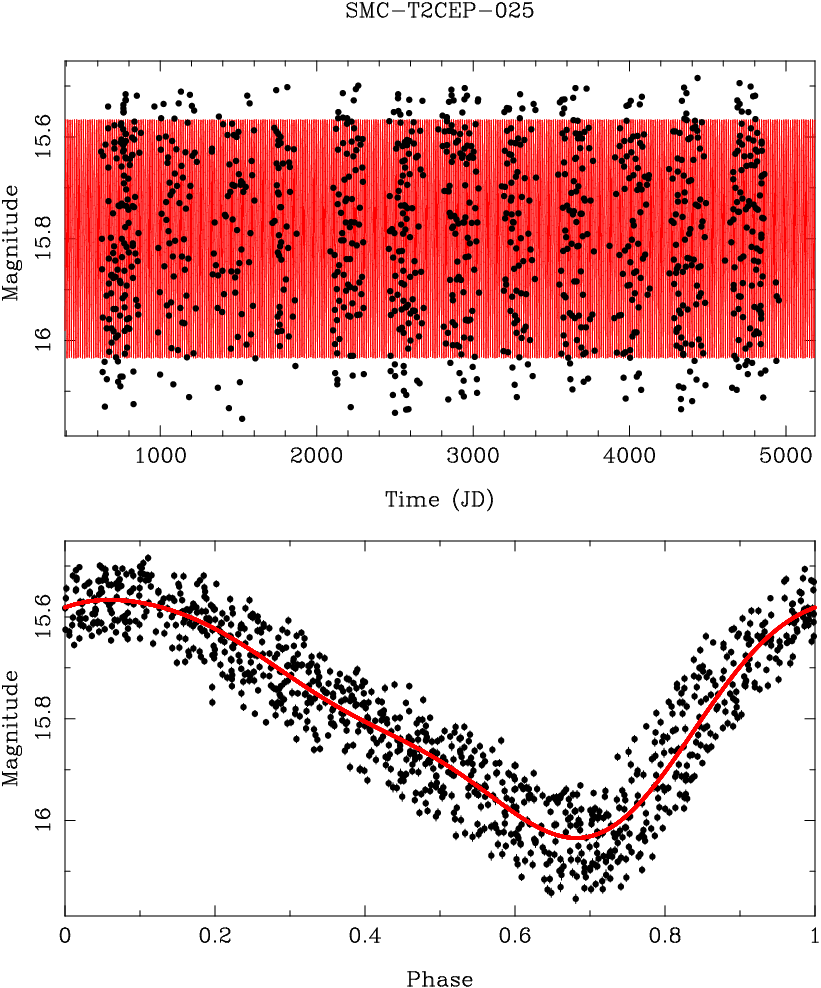}} 
\end{minipage}
\begin{minipage}{0.26\textwidth}
\resizebox{\hsize}{!}{\includegraphics[angle=-0]{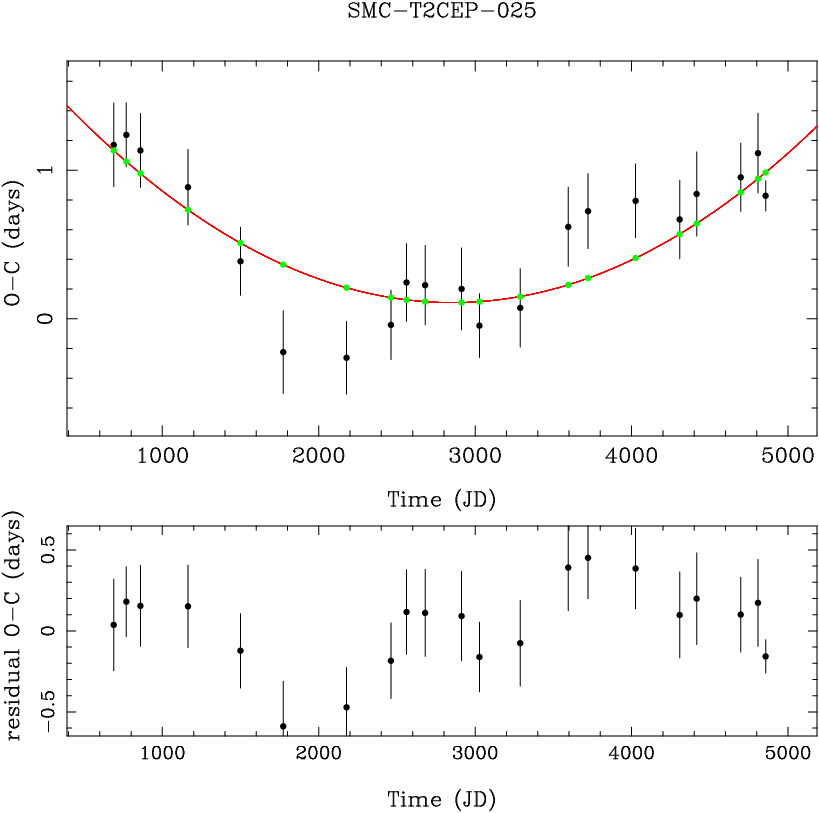}} 
\end{minipage}
\begin{minipage}{0.22\textwidth}
\resizebox{\hsize}{!}{\includegraphics[angle=-0]{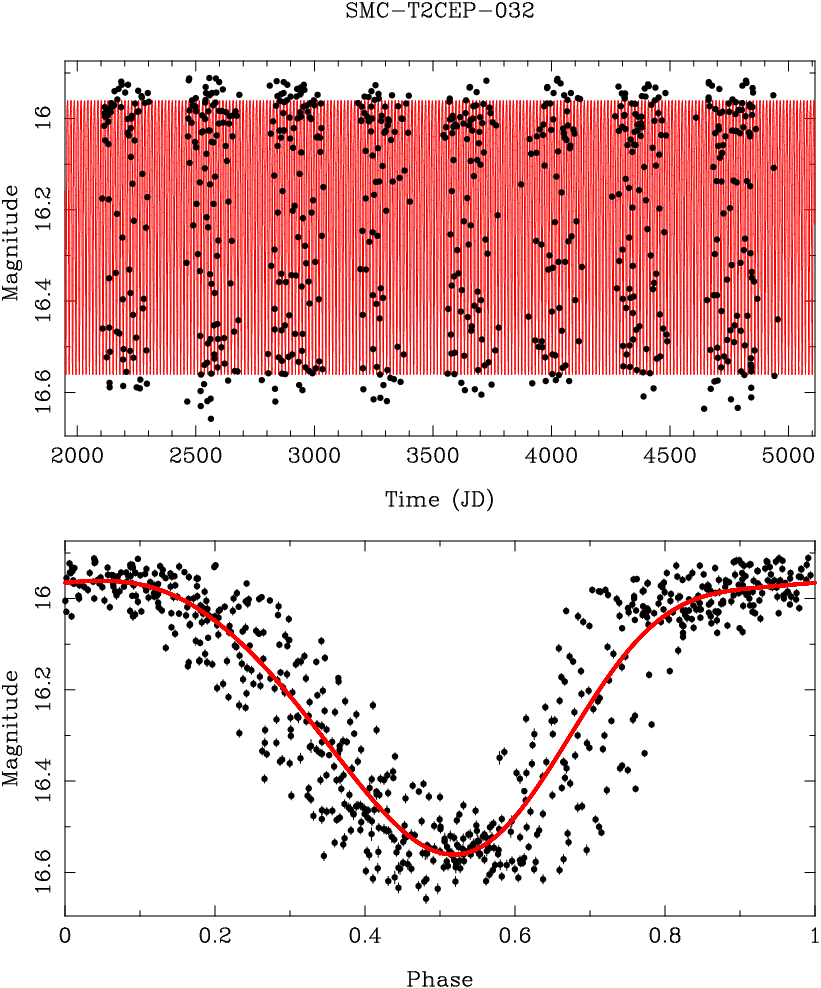}} 
\end{minipage}
\begin{minipage}{0.26\textwidth}
\resizebox{\hsize}{!}{\includegraphics[angle=-0]{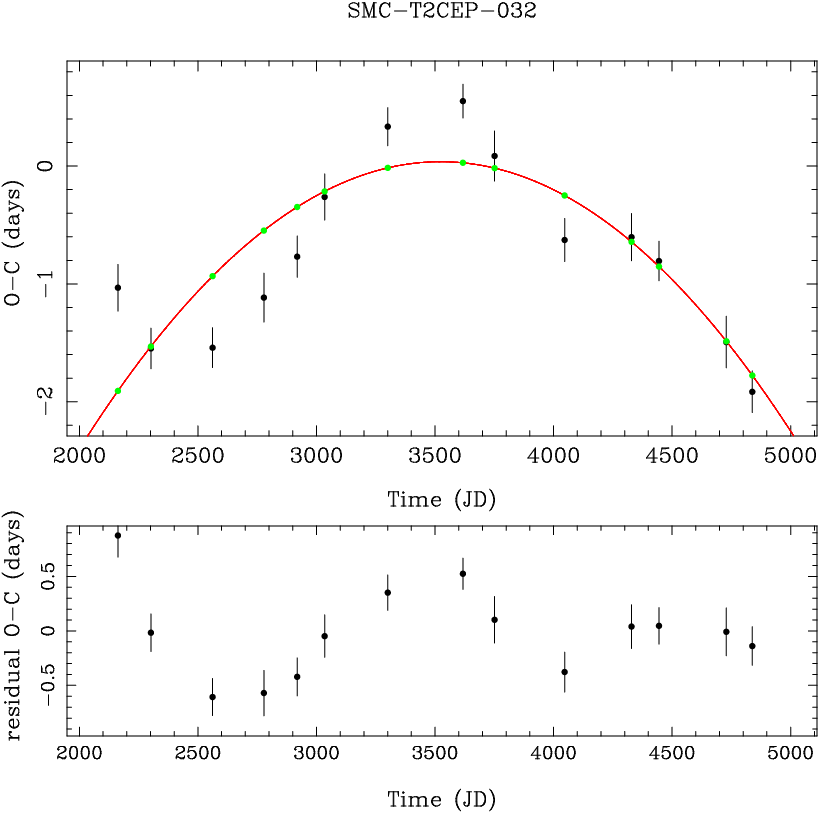}} 
\end{minipage}

\begin{minipage}{0.22\textwidth}
\resizebox{\hsize}{!}{\includegraphics[angle=-0]{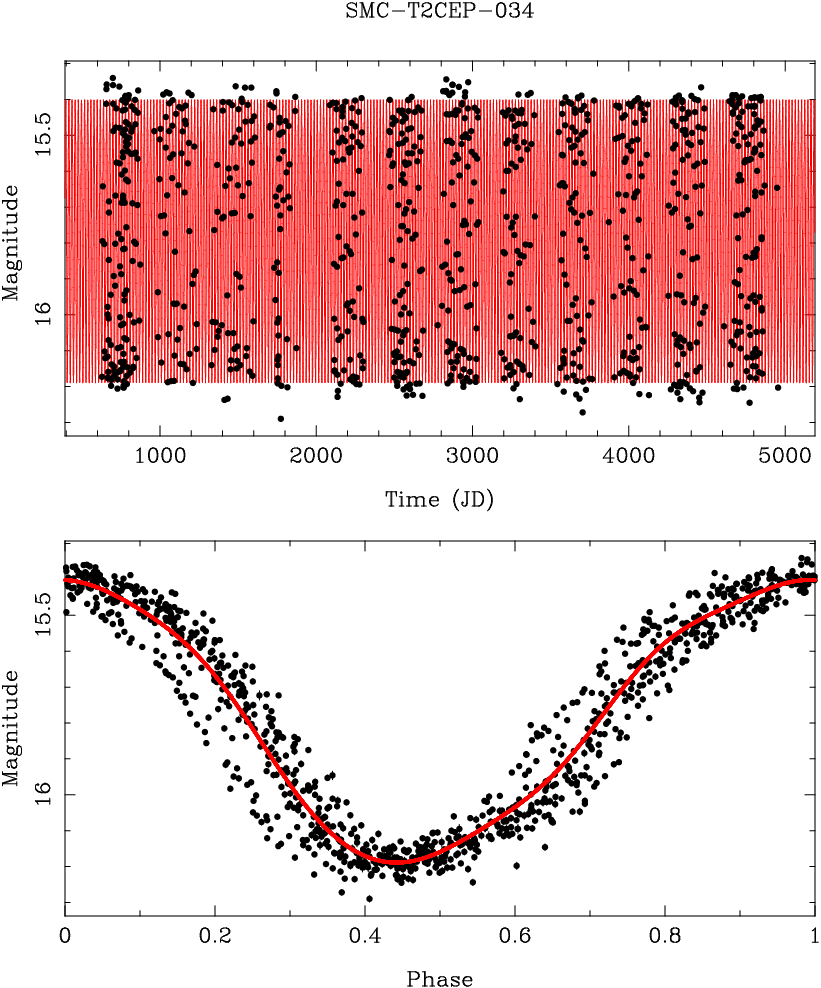}} 
\end{minipage}
\begin{minipage}{0.26\textwidth}
\resizebox{\hsize}{!}{\includegraphics[angle=-0]{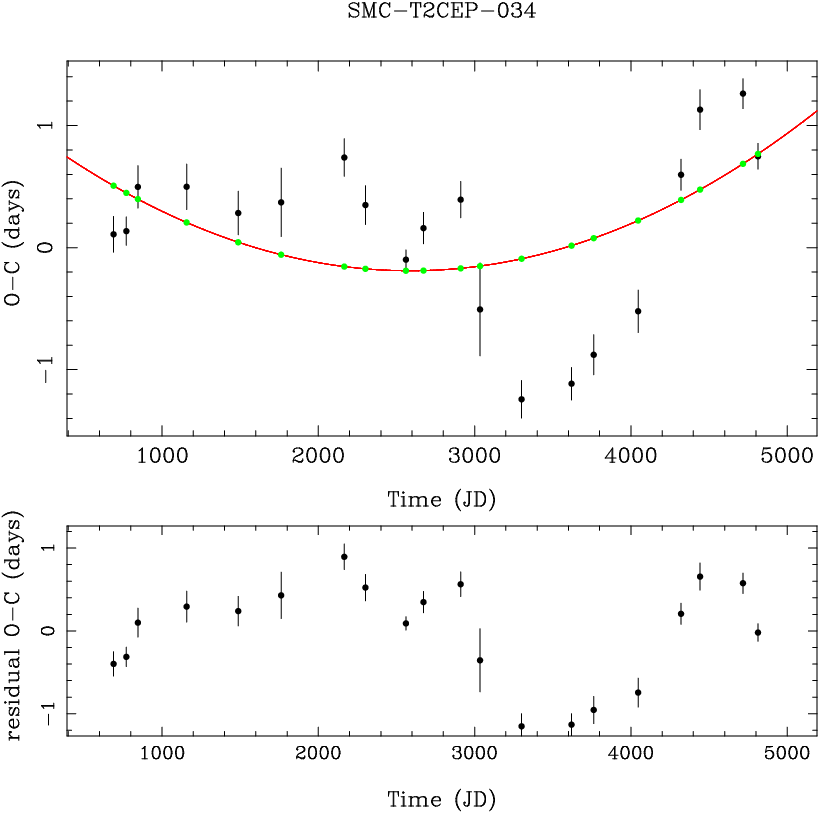}} 
\end{minipage}
\begin{minipage}{0.22\textwidth}
\resizebox{\hsize}{!}{\includegraphics[angle=-0]{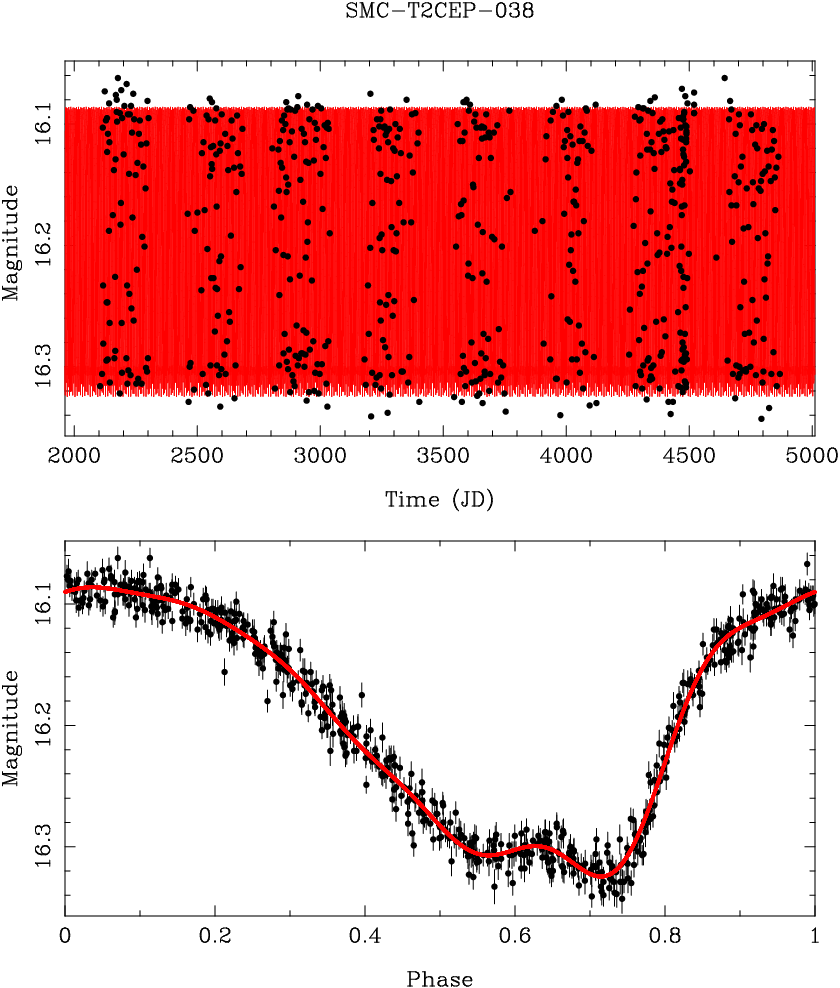}} 
\end{minipage}
\begin{minipage}{0.26\textwidth}
\resizebox{\hsize}{!}{\includegraphics[angle=-0]{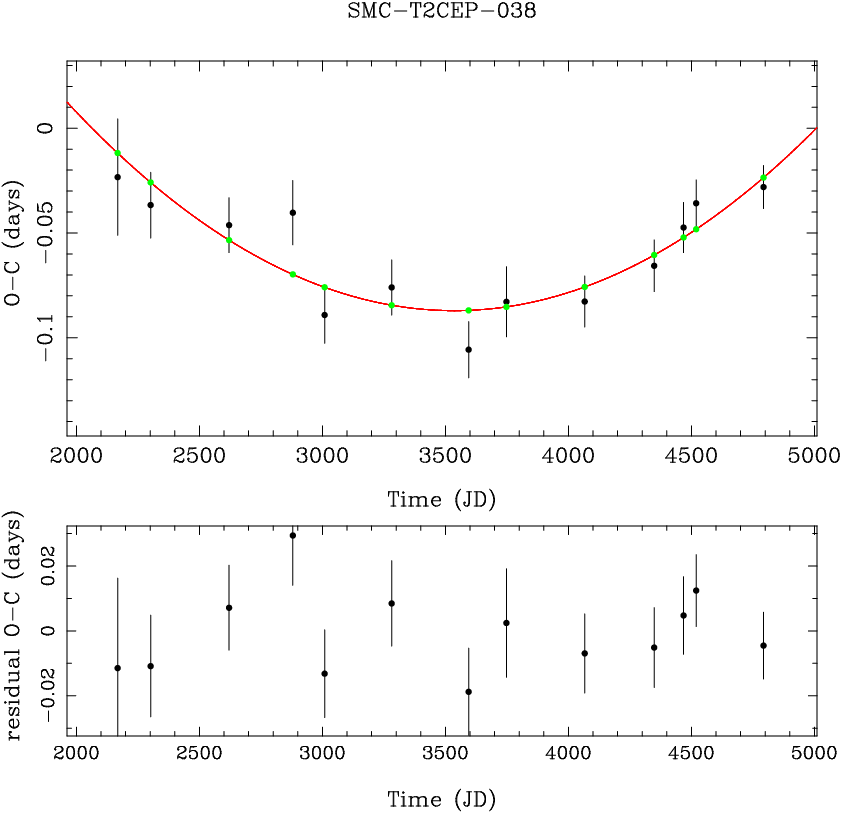}} 
\end{minipage}

\begin{minipage}{0.22\textwidth}
\resizebox{\hsize}{!}{\includegraphics[angle=-0]{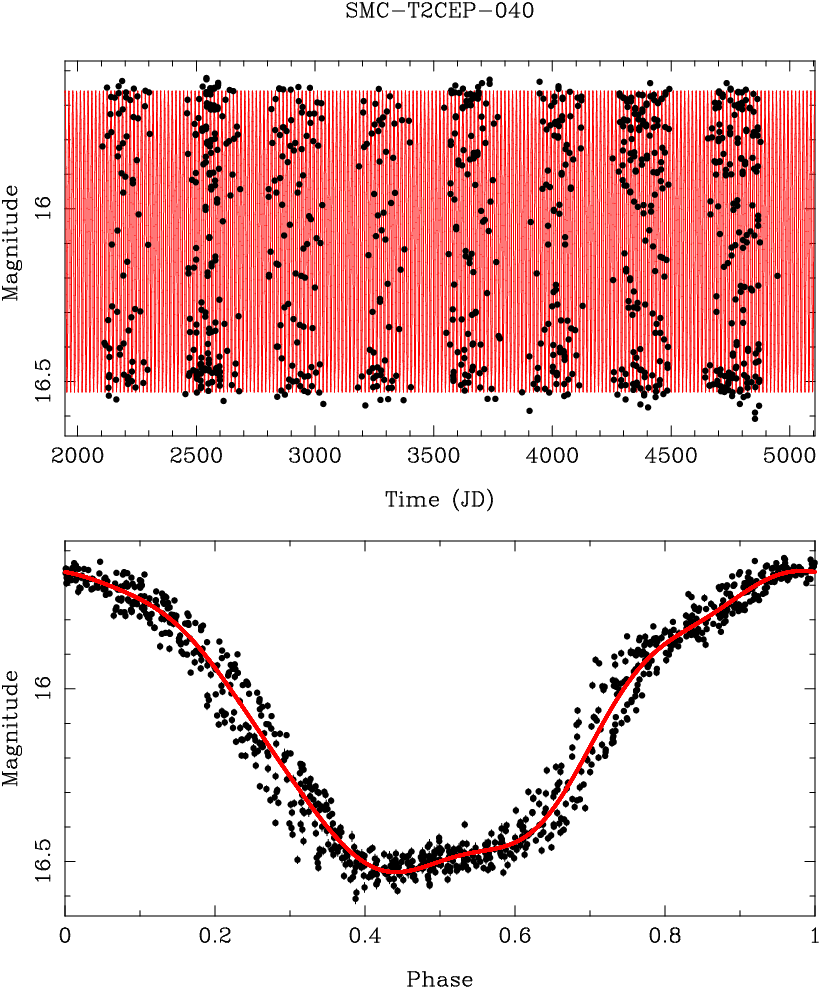}} 
\end{minipage}
\begin{minipage}{0.26\textwidth}
\resizebox{\hsize}{!}{\includegraphics[angle=-0]{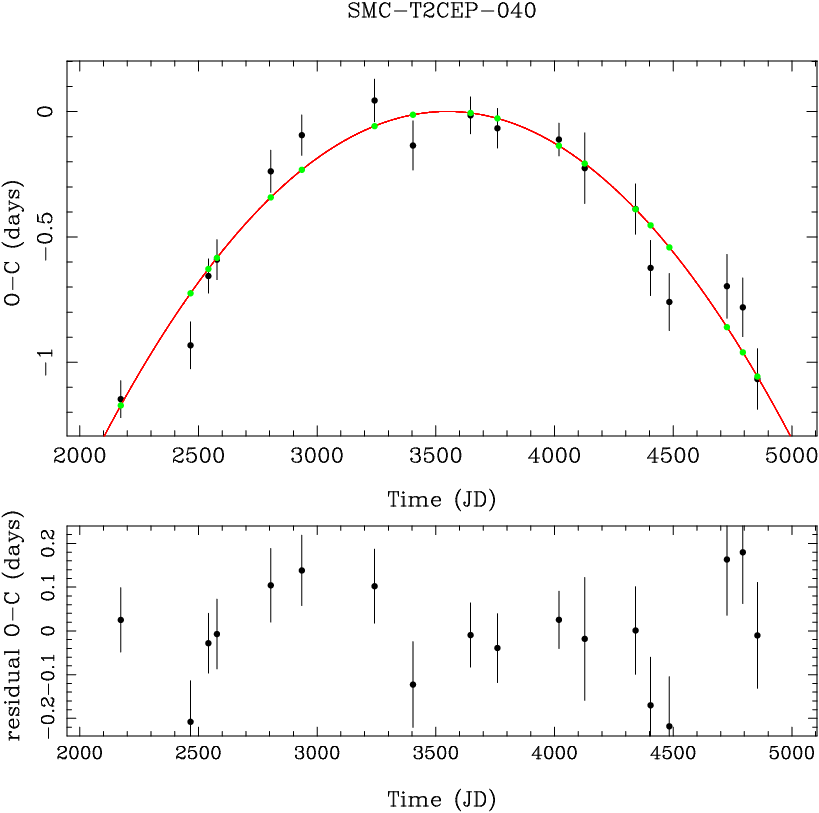}} 
\end{minipage}
\begin{minipage}{0.22\textwidth}
\resizebox{\hsize}{!}{\includegraphics[angle=-0]{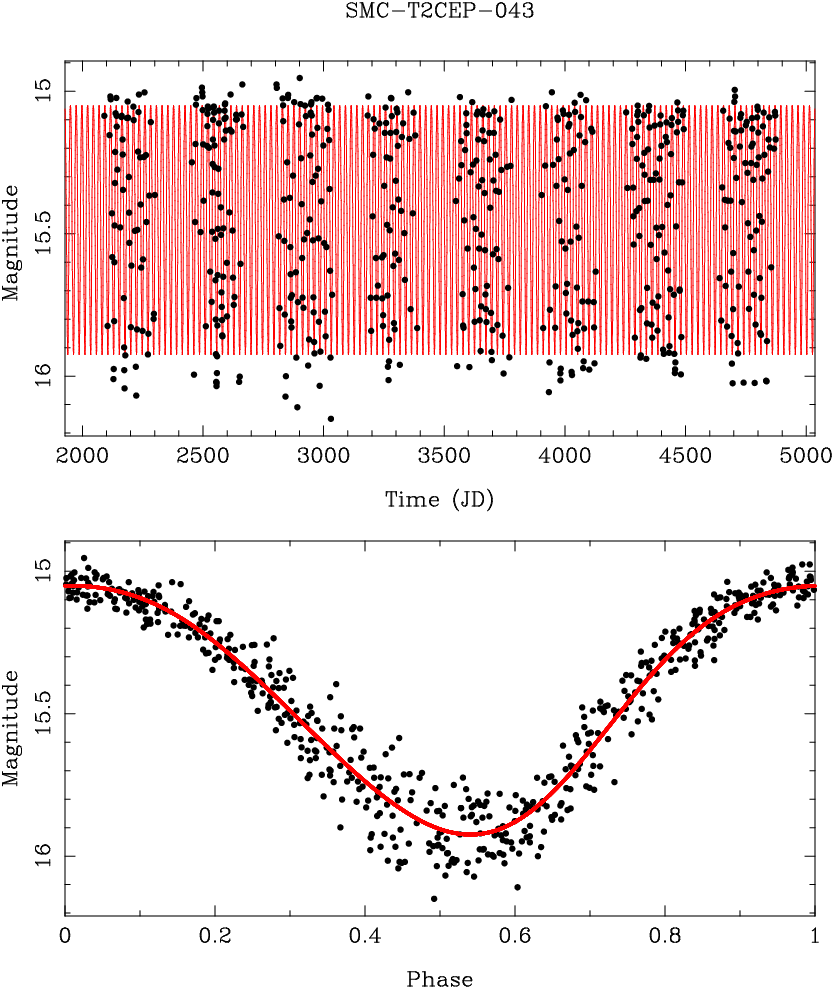}} 
\end{minipage}
\begin{minipage}{0.26\textwidth}
\resizebox{\hsize}{!}{\includegraphics[angle=-0]{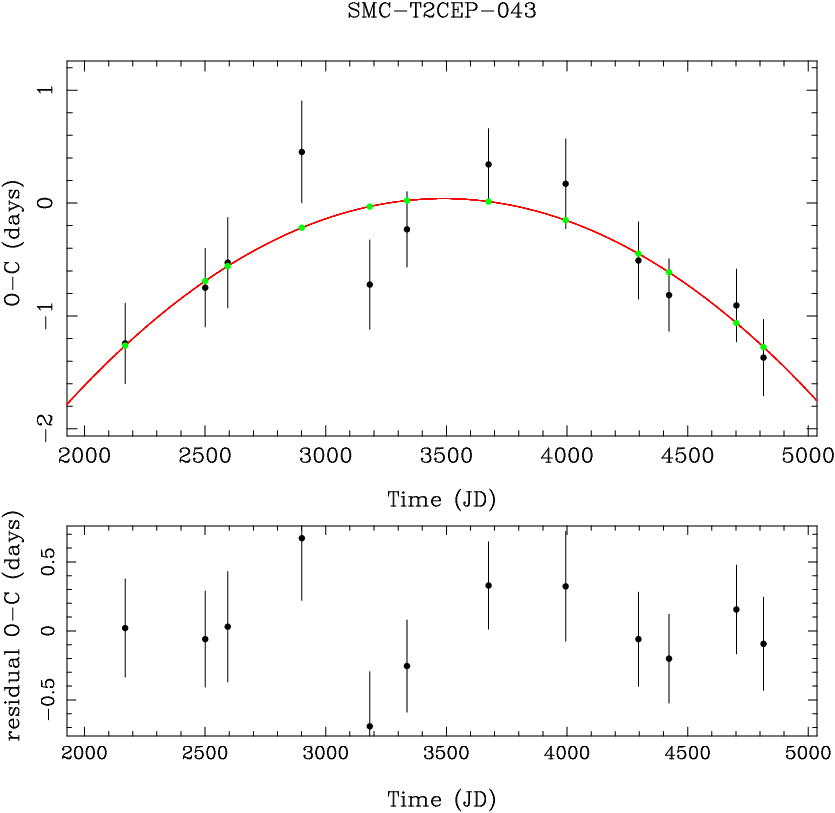}} 
\end{minipage}

\caption{Continued.}
\end{figure*}

\begin{table*}
\setlength{\tabcolsep}{1.6mm}
\caption{T2C investigated for the LITE with inconclusive results}

\begin{tabular}{llllll}
\hline
Name         &   Remarks & Name         &   Remarks & Name         &   Remarks \\ 

\hline
LMC-ACEP-024  &             & LMC-ACEP-058  &             & LMC-ACEP-070 & \\
LMC-ACEP-083  & $\dot{P}$ ? & LMC-T2CEP-002 & $\dot{P}$ ? & LMC-T2CEP-005 & $\dot{P}$ ? \\
LMC-T2CEP-012 &             & LMC-T2CEP-013 &             & LMC-T2CEP-015 & $\dot{P}$ ? \\
LMC-T2CEP-023 & known EB    & LMC-T2CEP-028 &             & LMC-T2CEP-032 & \\ 
LMC-T2CEP-035 &             & LMC-T2CEP-042 &             & LMC-T2CEP-051 & $\dot{P}$ ? Amplitude variations  \\
LMC-T2CEP-052 & known EB & LMC-T2CEP-056 &            & LMC-T2CEP-061 &  \\ 
LMC-T2CEP-067 &          & LMC-T2CEP-077 & claimed EB & LMC-T2CEP-078 & \\ 
LMC-T2CEP-083 &          & LMC-T2CEP-084 & known EB   & LMC-T2CEP-088 & \\ 
LMC-T2CEP-089 &          & LMC-T2CEP-093 & known EB   & LMC-T2CEP-101 & $\dot{P}$ ? \\
LMC-T2CEP-108 &          & LMC-T2CEP-117 &            & LMC-T2CEP-129 & irregular light curve \\
LMC-T2CEP-133 & $\dot{P}$ ? & LMC-T2CEP-134 & binary (P= 2200d) ? & LMC-T2CEP-145 &  \\
LMC-T2CEP-147 &             & LMC-T2CEP-150 &              & LMC-T2CEP-151 & \\
LMC-T2CEP-153 &             &  LMC-T2CEP-154 & $\dot{P}$ ? & LMC-T2CEP-164 & \\
LMC-T2CEP-169 &             &  LMC-T2CEP-174 &             & LMC-T2CEP-175 & $\dot{P}$ ? \\
LMC-T2CEP-176 & $\dot{P}$ ? & LMC-T2CEP-178 & $\dot{P}$ ?  & LMC-T2CEP-179 & $\dot{P}$ or binary (P= 1800d) ? \\
LMC-T2CEP-180 &             & LMC-T2CEP-181 &              & LMC-T2CEP-183 &       \\
LMC-T2CEP-184 &             & LMC-T2CEP-185 &              & LMC-T2CEP-192 &   \\
LMC-T2CEP-193 &             & LMC-T2CEP-199 &              & LMC-T2CEP-200 & known EB  \\ 
LMC-T2CEP-202 &             & LMC-T2CEP-203 &              &               & \\

SMC-T2CEP-005 & $\dot{P}$ ? & SMC-T2CEP-007 &                      & SMC-T2CEP-010 &     \\
SMC-T2CEP-012 &             & SMC-T2CEP-013 & binary (P= 1700d) ?  & SMC-T2CEP-020 &   \\
SMC-T2CEP-023 & known EB    & SMC-T2CEP-024 & Amplitude variations & SMC-T2CEP-028 & known EB \\
SMC-T2CEP-031 &             & SMC-T2CEP-036 &                      & SMC-T2CEP-041 &   \\

\hline
\end{tabular}

\label{Tab-LITE-Incon}
\end{table*}

\end{appendix}

\end{document}